\documentclass[smallextended,final]{svjour3}       

\smartqed  

\usepackage[notheorems]{rr-math}
\pdfoutput=1 

\usepackage[sort,round]{natbib}
\renewcommand{\cite}{\citep} 
\renewcommand{\citeauthor}{\citet}

\renewcommand{\tabcolsep}{2pt}

\usepackage{rotating}
\def\Rot#1#2{\rlap{\rotatebox{#1}{#2}~}}

\usepackage{textcase}
\usepackage{graphicx}
\usepackage{amsmath}
\usepackage{epstopdf}
\usepackage{array}
\usepackage{url}

\usepackage{stackengine}

\usepackage{algorithm}
\usepackage{algorithmicx}
\usepackage{algpseudocode}
\algnotext{EndFor}
\algnotext{EndIf}
\algnotext{EndProcedure}

\usepackage{multirow}
\usepackage{comment}
\usepackage{rotating}
\usepackage{graphicx}
\usepackage{amssymb,amsmath,multirow,rotate,color}
\usepackage{subfigure}
\usepackage{pifont}
\usepackage{setspace}
\usepackage{tabularx}
\usepackage{etex}
\reserveinserts{100}
\usepackage{epsfig}
\usepackage{graphics,dcolumn,bm,epic,eepic,float}
\usepackage{afterpage}
\usepackage{relsize}
\usepackage[table]{xcolor}
\usepackage{color}

\usepackage{paralist}
\usepackage{xspace}
\usepackage{booktabs}
\usepackage{mathdots}
\usepackage{lscape}

  \definecolor{snow}{rgb}{1.000000,0.980392,0.980392}
  \definecolor{ghost white}{rgb}{0.972549,0.972549,1.000000}
  \definecolor{GhostWhite}{rgb}{0.972549,0.972549,1.000000}
  \definecolor{white smoke}{rgb}{0.960784,0.960784,0.960784}
  \definecolor{WhiteSmoke}{rgb}{0.960784,0.960784,0.960784}
  \definecolor{gainsboro}{rgb}{0.862745,0.862745,0.862745}
  \definecolor{floral white}{rgb}{1.000000,0.980392,0.941176}
  \definecolor{FloralWhite}{rgb}{1.000000,0.980392,0.941176}
  \definecolor{old lace}{rgb}{0.992157,0.960784,0.901961}
  \definecolor{OldLace}{rgb}{0.992157,0.960784,0.901961}
  \definecolor{linen}{rgb}{0.980392,0.941176,0.901961}
  \definecolor{antique white}{rgb}{0.980392,0.921569,0.843137}
  \definecolor{AntiqueWhite}{rgb}{0.980392,0.921569,0.843137}
  \definecolor{papaya whip}{rgb}{1.000000,0.937255,0.835294}
  \definecolor{PapayaWhip}{rgb}{1.000000,0.937255,0.835294}
  \definecolor{blanched almond}{rgb}{1.000000,0.921569,0.803922}
  \definecolor{BlanchedAlmond}{rgb}{1.000000,0.921569,0.803922}
  \definecolor{bisque}{rgb}{1.000000,0.894118,0.768627}
  \definecolor{peach puff}{rgb}{1.000000,0.854902,0.725490}
  \definecolor{PeachPuff}{rgb}{1.000000,0.854902,0.725490}
  \definecolor{navajo white}{rgb}{1.000000,0.870588,0.678431}
  \definecolor{NavajoWhite}{rgb}{1.000000,0.870588,0.678431}
  \definecolor{moccasin}{rgb}{1.000000,0.894118,0.709804}
  \definecolor{cornsilk}{rgb}{1.000000,0.972549,0.862745}
  \definecolor{ivory}{rgb}{1.000000,1.000000,0.941176}
  \definecolor{lemon chiffon}{rgb}{1.000000,0.980392,0.803922}
  \definecolor{LemonChiffon}{rgb}{1.000000,0.980392,0.803922}
  \definecolor{seashell}{rgb}{1.000000,0.960784,0.933333}
  \definecolor{honeydew}{rgb}{0.941176,1.000000,0.941176}
  \definecolor{mint cream}{rgb}{0.960784,1.000000,0.980392}
  \definecolor{MintCream}{rgb}{0.960784,1.000000,0.980392}
  \definecolor{azure}{rgb}{0.941176,1.000000,1.000000}
  \definecolor{alice blue}{rgb}{0.941176,0.972549,1.000000}
  \definecolor{AliceBlue}{rgb}{0.941176,0.972549,1.000000}
  \definecolor{lavender}{rgb}{0.901961,0.901961,0.980392}
  \definecolor{lavender blush}{rgb}{1.000000,0.941176,0.960784}
  \definecolor{LavenderBlush}{rgb}{1.000000,0.941176,0.960784}
  \definecolor{misty rose}{rgb}{1.000000,0.894118,0.882353}
  \definecolor{MistyRose}{rgb}{1.000000,0.894118,0.882353}
  \definecolor{white}{rgb}{1.000000,1.000000,1.000000}
  \definecolor{black}{rgb}{0.000000,0.000000,0.000000}
  \definecolor{dark slate gray}{rgb}{0.184314,0.309804,0.309804}
  \definecolor{DarkSlateGray}{rgb}{0.184314,0.309804,0.309804}
  \definecolor{dark slate grey}{rgb}{0.184314,0.309804,0.309804}
  \definecolor{DarkSlateGrey}{rgb}{0.184314,0.309804,0.309804}
  \definecolor{dim gray}{rgb}{0.411765,0.411765,0.411765}
  \definecolor{DimGray}{rgb}{0.411765,0.411765,0.411765}
  \definecolor{dim grey}{rgb}{0.411765,0.411765,0.411765}
  \definecolor{DimGrey}{rgb}{0.411765,0.411765,0.411765}
  \definecolor{slate gray}{rgb}{0.439216,0.501961,0.564706}
  \definecolor{SlateGray}{rgb}{0.439216,0.501961,0.564706}
  \definecolor{slate grey}{rgb}{0.439216,0.501961,0.564706}
  \definecolor{SlateGrey}{rgb}{0.439216,0.501961,0.564706}
  \definecolor{light slate gray}{rgb}{0.466667,0.533333,0.600000}
  \definecolor{LightSlateGray}{rgb}{0.466667,0.533333,0.600000}
  \definecolor{light slate grey}{rgb}{0.466667,0.533333,0.600000}
  \definecolor{LightSlateGrey}{rgb}{0.466667,0.533333,0.600000}
  \definecolor{gray}{rgb}{0.745098,0.745098,0.745098}
  \definecolor{grey}{rgb}{0.745098,0.745098,0.745098}
  \definecolor{light grey}{rgb}{0.827451,0.827451,0.827451}
  \definecolor{LightGrey}{rgb}{0.827451,0.827451,0.827451}
  \definecolor{light gray}{rgb}{0.827451,0.827451,0.827451}
  \definecolor{LightGray}{rgb}{0.827451,0.827451,0.827451}
  \definecolor{midnight blue}{rgb}{0.098039,0.098039,0.439216}
  \definecolor{MidnightBlue}{rgb}{0.098039,0.098039,0.439216}
  \definecolor{navy}{rgb}{0.000000,0.000000,0.501961}
  \definecolor{navy blue}{rgb}{0.000000,0.000000,0.501961}
  \definecolor{NavyBlue}{rgb}{0.000000,0.000000,0.501961}
  \definecolor{cornflower blue}{rgb}{0.392157,0.584314,0.929412}
  \definecolor{CornflowerBlue}{rgb}{0.392157,0.584314,0.929412}
  \definecolor{dark slate blue}{rgb}{0.282353,0.239216,0.545098}
  \definecolor{DarkSlateBlue}{rgb}{0.282353,0.239216,0.545098}
  \definecolor{slate blue}{rgb}{0.415686,0.352941,0.803922}
  \definecolor{SlateBlue}{rgb}{0.415686,0.352941,0.803922}
  \definecolor{medium slate blue}{rgb}{0.482353,0.407843,0.933333}
  \definecolor{MediumSlateBlue}{rgb}{0.482353,0.407843,0.933333}
  \definecolor{light slate blue}{rgb}{0.517647,0.439216,1.000000}
  \definecolor{LightSlateBlue}{rgb}{0.517647,0.439216,1.000000}
  \definecolor{medium blue}{rgb}{0.000000,0.000000,0.803922}
  \definecolor{MediumBlue}{rgb}{0.000000,0.000000,0.803922}
  \definecolor{royal blue}{rgb}{0.254902,0.411765,0.882353}
  \definecolor{RoyalBlue}{rgb}{0.254902,0.411765,0.882353}
  \definecolor{blue}{rgb}{0.000000,0.000000,1.000000}
  \definecolor{dodger blue}{rgb}{0.117647,0.564706,1.000000}
  \definecolor{DodgerBlue}{rgb}{0.117647,0.564706,1.000000}
  \definecolor{deep sky blue}{rgb}{0.000000,0.749020,1.000000}
  \definecolor{DeepSkyBlue}{rgb}{0.000000,0.749020,1.000000}
  \definecolor{sky blue}{rgb}{0.529412,0.807843,0.921569}
  \definecolor{SkyBlue}{rgb}{0.529412,0.807843,0.921569}
  \definecolor{light sky blue}{rgb}{0.529412,0.807843,0.980392}
  \definecolor{LightSkyBlue}{rgb}{0.529412,0.807843,0.980392}
  \definecolor{steel blue}{rgb}{0.274510,0.509804,0.705882}
  \definecolor{SteelBlue}{rgb}{0.274510,0.509804,0.705882}
  \definecolor{light steel blue}{rgb}{0.690196,0.768627,0.870588}
  \definecolor{LightSteelBlue}{rgb}{0.690196,0.768627,0.870588}
  \definecolor{light blue}{rgb}{0.678431,0.847059,0.901961}
  \definecolor{LightBlue}{rgb}{0.678431,0.847059,0.901961}
  \definecolor{powder blue}{rgb}{0.690196,0.878431,0.901961}
  \definecolor{PowderBlue}{rgb}{0.690196,0.878431,0.901961}
  \definecolor{pale turquoise}{rgb}{0.686275,0.933333,0.933333}
  \definecolor{PaleTurquoise}{rgb}{0.686275,0.933333,0.933333}
  \definecolor{dark turquoise}{rgb}{0.000000,0.807843,0.819608}
  \definecolor{DarkTurquoise}{rgb}{0.000000,0.807843,0.819608}
  \definecolor{medium turquoise}{rgb}{0.282353,0.819608,0.800000}
  \definecolor{MediumTurquoise}{rgb}{0.282353,0.819608,0.800000}
  \definecolor{turquoise}{rgb}{0.250980,0.878431,0.815686}
  \definecolor{cyan}{rgb}{0.000000,1.000000,1.000000}
  \definecolor{light cyan}{rgb}{0.878431,1.000000,1.000000}
  \definecolor{LightCyan}{rgb}{0.878431,1.000000,1.000000}
  \definecolor{cadet blue}{rgb}{0.372549,0.619608,0.627451}
  \definecolor{CadetBlue}{rgb}{0.372549,0.619608,0.627451}
  \definecolor{medium aquamarine}{rgb}{0.400000,0.803922,0.666667}
  \definecolor{MediumAquamarine}{rgb}{0.400000,0.803922,0.666667}
  \definecolor{aquamarine}{rgb}{0.498039,1.000000,0.831373}
  \definecolor{dark green}{rgb}{0.000000,0.392157,0.000000}
  \definecolor{DarkGreen}{rgb}{0.000000,0.392157,0.000000}
  \definecolor{dark olive green}{rgb}{0.333333,0.419608,0.184314}
  \definecolor{DarkOliveGreen}{rgb}{0.333333,0.419608,0.184314}
  \definecolor{dark sea green}{rgb}{0.560784,0.737255,0.560784}
  \definecolor{DarkSeaGreen}{rgb}{0.560784,0.737255,0.560784}
  \definecolor{sea green}{rgb}{0.180392,0.545098,0.341176}
  \definecolor{SeaGreen}{rgb}{0.180392,0.545098,0.341176}
  \definecolor{medium sea green}{rgb}{0.235294,0.701961,0.443137}
  \definecolor{MediumSeaGreen}{rgb}{0.235294,0.701961,0.443137}
  \definecolor{light sea green}{rgb}{0.125490,0.698039,0.666667}
  \definecolor{LightSeaGreen}{rgb}{0.125490,0.698039,0.666667}
  \definecolor{pale green}{rgb}{0.596078,0.984314,0.596078}
  \definecolor{PaleGreen}{rgb}{0.596078,0.984314,0.596078}
  \definecolor{spring green}{rgb}{0.000000,1.000000,0.498039}
  \definecolor{SpringGreen}{rgb}{0.000000,1.000000,0.498039}
  \definecolor{lawn green}{rgb}{0.486275,0.988235,0.000000}
  \definecolor{LawnGreen}{rgb}{0.486275,0.988235,0.000000}
  \definecolor{green}{rgb}{0.000000,1.000000,0.000000}
  \definecolor{chartreuse}{rgb}{0.498039,1.000000,0.000000}
  \definecolor{medium spring green}{rgb}{0.000000,0.980392,0.603922}
  \definecolor{MediumSpringGreen}{rgb}{0.000000,0.980392,0.603922}
  \definecolor{green yellow}{rgb}{0.678431,1.000000,0.184314}
  \definecolor{GreenYellow}{rgb}{0.678431,1.000000,0.184314}
  \definecolor{lime green}{rgb}{0.196078,0.803922,0.196078}
  \definecolor{LimeGreen}{rgb}{0.196078,0.803922,0.196078}
  \definecolor{yellow green}{rgb}{0.603922,0.803922,0.196078}
  \definecolor{YellowGreen}{rgb}{0.603922,0.803922,0.196078}
  \definecolor{forest green}{rgb}{0.133333,0.545098,0.133333}
  \definecolor{ForestGreen}{rgb}{0.133333,0.545098,0.133333}
  \definecolor{olive drab}{rgb}{0.419608,0.556863,0.137255}
  \definecolor{OliveDrab}{rgb}{0.419608,0.556863,0.137255}
  \definecolor{dark khaki}{rgb}{0.741176,0.717647,0.419608}
  \definecolor{DarkKhaki}{rgb}{0.741176,0.717647,0.419608}
  \definecolor{khaki}{rgb}{0.941176,0.901961,0.549020}
  \definecolor{pale goldenrod}{rgb}{0.933333,0.909804,0.666667}
  \definecolor{PaleGoldenrod}{rgb}{0.933333,0.909804,0.666667}
  \definecolor{light goldenrod yellow}{rgb}{0.980392,0.980392,0.823529}
  \definecolor{LightGoldenrodYellow}{rgb}{0.980392,0.980392,0.823529}
  \definecolor{light yellow}{rgb}{1.000000,1.000000,0.878431}
  \definecolor{LightYellow}{rgb}{1.000000,1.000000,0.878431}
  \definecolor{yellow}{rgb}{1.000000,1.000000,0.000000}
  \definecolor{gold}{rgb}{1.000000,0.843137,0.000000}
  \definecolor{light goldenrod}{rgb}{0.933333,0.866667,0.509804}
  \definecolor{LightGoldenrod}{rgb}{0.933333,0.866667,0.509804}
  \definecolor{goldenrod}{rgb}{0.854902,0.647059,0.125490}
  \definecolor{dark goldenrod}{rgb}{0.721569,0.525490,0.043137}
  \definecolor{DarkGoldenrod}{rgb}{0.721569,0.525490,0.043137}
  \definecolor{rosy brown}{rgb}{0.737255,0.560784,0.560784}
  \definecolor{RosyBrown}{rgb}{0.737255,0.560784,0.560784}
  \definecolor{indian red}{rgb}{0.803922,0.360784,0.360784}
  \definecolor{IndianRed}{rgb}{0.803922,0.360784,0.360784}
  \definecolor{saddle brown}{rgb}{0.545098,0.270588,0.074510}
  \definecolor{SaddleBrown}{rgb}{0.545098,0.270588,0.074510}
  \definecolor{sienna}{rgb}{0.627451,0.321569,0.176471}
  \definecolor{peru}{rgb}{0.803922,0.521569,0.247059}
  \definecolor{burlywood}{rgb}{0.870588,0.721569,0.529412}
  \definecolor{beige}{rgb}{0.960784,0.960784,0.862745}
  \definecolor{wheat}{rgb}{0.960784,0.870588,0.701961}
  \definecolor{sandy brown}{rgb}{0.956863,0.643137,0.376471}
  \definecolor{SandyBrown}{rgb}{0.956863,0.643137,0.376471}
  \definecolor{tan}{rgb}{0.823529,0.705882,0.549020}
  \definecolor{chocolate}{rgb}{0.823529,0.411765,0.117647}
  \definecolor{firebrick}{rgb}{0.698039,0.133333,0.133333}
  \definecolor{brown}{rgb}{0.647059,0.164706,0.164706}
  \definecolor{dark salmon}{rgb}{0.913725,0.588235,0.478431}
  \definecolor{DarkSalmon}{rgb}{0.913725,0.588235,0.478431}
  \definecolor{salmon}{rgb}{0.980392,0.501961,0.447059}
  \definecolor{light salmon}{rgb}{1.000000,0.627451,0.478431}
  \definecolor{LightSalmon}{rgb}{1.000000,0.627451,0.478431}
  \definecolor{orange}{rgb}{1.000000,0.647059,0.000000}
  \definecolor{dark orange}{rgb}{1.000000,0.549020,0.000000}
  \definecolor{DarkOrange}{rgb}{1.000000,0.549020,0.000000}
  \definecolor{coral}{rgb}{1.000000,0.498039,0.313726}
  \definecolor{light coral}{rgb}{0.941176,0.501961,0.501961}
  \definecolor{LightCoral}{rgb}{0.941176,0.501961,0.501961}
  \definecolor{tomato}{rgb}{1.000000,0.388235,0.278431}
  \definecolor{orange red}{rgb}{1.000000,0.270588,0.000000}
  \definecolor{OrangeRed}{rgb}{1.000000,0.270588,0.000000}
  \definecolor{red}{rgb}{1.000000,0.000000,0.000000}
  \definecolor{hot pink}{rgb}{1.000000,0.411765,0.705882}
  \definecolor{HotPink}{rgb}{1.000000,0.411765,0.705882}
  \definecolor{deep pink}{rgb}{1.000000,0.078431,0.576471}
  \definecolor{DeepPink}{rgb}{1.000000,0.078431,0.576471}
  \definecolor{pink}{rgb}{1.000000,0.752941,0.796078}
  \definecolor{light pink}{rgb}{1.000000,0.713726,0.756863}
  \definecolor{LightPink}{rgb}{1.000000,0.713726,0.756863}
  \definecolor{pale violet red}{rgb}{0.858824,0.439216,0.576471}
  \definecolor{PaleVioletRed}{rgb}{0.858824,0.439216,0.576471}
  \definecolor{maroon}{rgb}{0.690196,0.188235,0.376471}
  \definecolor{medium violet red}{rgb}{0.780392,0.082353,0.521569}
  \definecolor{MediumVioletRed}{rgb}{0.780392,0.082353,0.521569}
  \definecolor{violet red}{rgb}{0.815686,0.125490,0.564706}
  \definecolor{VioletRed}{rgb}{0.815686,0.125490,0.564706}
  \definecolor{magenta}{rgb}{1.000000,0.000000,1.000000}
  \definecolor{violet}{rgb}{0.933333,0.509804,0.933333}
  \definecolor{plum}{rgb}{0.866667,0.627451,0.866667}
  \definecolor{orchid}{rgb}{0.854902,0.439216,0.839216}
  \definecolor{medium orchid}{rgb}{0.729412,0.333333,0.827451}
  \definecolor{MediumOrchid}{rgb}{0.729412,0.333333,0.827451}
  \definecolor{dark orchid}{rgb}{0.600000,0.196078,0.800000}
  \definecolor{DarkOrchid}{rgb}{0.600000,0.196078,0.800000}
  \definecolor{dark violet}{rgb}{0.580392,0.000000,0.827451}
  \definecolor{DarkViolet}{rgb}{0.580392,0.000000,0.827451}
  \definecolor{blue violet}{rgb}{0.541176,0.168627,0.886275}
  \definecolor{BlueViolet}{rgb}{0.541176,0.168627,0.886275}
  \definecolor{purple}{rgb}{0.627451,0.125490,0.941176}
  \definecolor{medium purple}{rgb}{0.576471,0.439216,0.858824}
  \definecolor{MediumPurple}{rgb}{0.576471,0.439216,0.858824}
  \definecolor{thistle}{rgb}{0.847059,0.749020,0.847059}
  \definecolor{snow1}{rgb}{1.000000,0.980392,0.980392}
  \definecolor{snow2}{rgb}{0.933333,0.913725,0.913725}
  \definecolor{snow3}{rgb}{0.803922,0.788235,0.788235}
  \definecolor{snow4}{rgb}{0.545098,0.537255,0.537255}
  \definecolor{seashell1}{rgb}{1.000000,0.960784,0.933333}
  \definecolor{seashell2}{rgb}{0.933333,0.898039,0.870588}
  \definecolor{seashell3}{rgb}{0.803922,0.772549,0.749020}
  \definecolor{seashell4}{rgb}{0.545098,0.525490,0.509804}
  \definecolor{AntiqueWhite1}{rgb}{1.000000,0.937255,0.858824}
  \definecolor{AntiqueWhite2}{rgb}{0.933333,0.874510,0.800000}
  \definecolor{AntiqueWhite3}{rgb}{0.803922,0.752941,0.690196}
  \definecolor{AntiqueWhite4}{rgb}{0.545098,0.513726,0.470588}
  \definecolor{bisque1}{rgb}{1.000000,0.894118,0.768627}
  \definecolor{bisque2}{rgb}{0.933333,0.835294,0.717647}
  \definecolor{bisque3}{rgb}{0.803922,0.717647,0.619608}
  \definecolor{bisque4}{rgb}{0.545098,0.490196,0.419608}
  \definecolor{PeachPuff1}{rgb}{1.000000,0.854902,0.725490}
  \definecolor{PeachPuff2}{rgb}{0.933333,0.796078,0.678431}
  \definecolor{PeachPuff3}{rgb}{0.803922,0.686275,0.584314}
  \definecolor{PeachPuff4}{rgb}{0.545098,0.466667,0.396078}
  \definecolor{NavajoWhite1}{rgb}{1.000000,0.870588,0.678431}
  \definecolor{NavajoWhite2}{rgb}{0.933333,0.811765,0.631373}
  \definecolor{NavajoWhite3}{rgb}{0.803922,0.701961,0.545098}
  \definecolor{NavajoWhite4}{rgb}{0.545098,0.474510,0.368627}
  \definecolor{LemonChiffon1}{rgb}{1.000000,0.980392,0.803922}
  \definecolor{LemonChiffon2}{rgb}{0.933333,0.913725,0.749020}
  \definecolor{LemonChiffon3}{rgb}{0.803922,0.788235,0.647059}
  \definecolor{LemonChiffon4}{rgb}{0.545098,0.537255,0.439216}
  \definecolor{cornsilk1}{rgb}{1.000000,0.972549,0.862745}
  \definecolor{cornsilk2}{rgb}{0.933333,0.909804,0.803922}
  \definecolor{cornsilk3}{rgb}{0.803922,0.784314,0.694118}
  \definecolor{cornsilk4}{rgb}{0.545098,0.533333,0.470588}
  \definecolor{ivory1}{rgb}{1.000000,1.000000,0.941176}
  \definecolor{ivory2}{rgb}{0.933333,0.933333,0.878431}
  \definecolor{ivory3}{rgb}{0.803922,0.803922,0.756863}
  \definecolor{ivory4}{rgb}{0.545098,0.545098,0.513726}
  \definecolor{honeydew1}{rgb}{0.941176,1.000000,0.941176}
  \definecolor{honeydew2}{rgb}{0.878431,0.933333,0.878431}
  \definecolor{honeydew3}{rgb}{0.756863,0.803922,0.756863}
  \definecolor{honeydew4}{rgb}{0.513726,0.545098,0.513726}
  \definecolor{LavenderBlush1}{rgb}{1.000000,0.941176,0.960784}
  \definecolor{LavenderBlush2}{rgb}{0.933333,0.878431,0.898039}
  \definecolor{LavenderBlush3}{rgb}{0.803922,0.756863,0.772549}
  \definecolor{LavenderBlush4}{rgb}{0.545098,0.513726,0.525490}
  \definecolor{MistyRose1}{rgb}{1.000000,0.894118,0.882353}
  \definecolor{MistyRose2}{rgb}{0.933333,0.835294,0.823529}
  \definecolor{MistyRose3}{rgb}{0.803922,0.717647,0.709804}
  \definecolor{MistyRose4}{rgb}{0.545098,0.490196,0.482353}
  \definecolor{azure1}{rgb}{0.941176,1.000000,1.000000}
  \definecolor{azure2}{rgb}{0.878431,0.933333,0.933333}
  \definecolor{azure3}{rgb}{0.756863,0.803922,0.803922}
  \definecolor{azure4}{rgb}{0.513726,0.545098,0.545098}
  \definecolor{SlateBlue1}{rgb}{0.513726,0.435294,1.000000}
  \definecolor{SlateBlue2}{rgb}{0.478431,0.403922,0.933333}
  \definecolor{SlateBlue3}{rgb}{0.411765,0.349020,0.803922}
  \definecolor{SlateBlue4}{rgb}{0.278431,0.235294,0.545098}
  \definecolor{RoyalBlue1}{rgb}{0.282353,0.462745,1.000000}
  \definecolor{RoyalBlue2}{rgb}{0.262745,0.431373,0.933333}
  \definecolor{RoyalBlue3}{rgb}{0.227451,0.372549,0.803922}
  \definecolor{RoyalBlue4}{rgb}{0.152941,0.250980,0.545098}
  \definecolor{blue1}{rgb}{0.000000,0.000000,1.000000}
  \definecolor{blue2}{rgb}{0.000000,0.000000,0.933333}
  \definecolor{blue3}{rgb}{0.000000,0.000000,0.803922}
  \definecolor{blue4}{rgb}{0.000000,0.000000,0.545098}
  \definecolor{DodgerBlue1}{rgb}{0.117647,0.564706,1.000000}
  \definecolor{DodgerBlue2}{rgb}{0.109804,0.525490,0.933333}
  \definecolor{DodgerBlue3}{rgb}{0.094118,0.454902,0.803922}
  \definecolor{DodgerBlue4}{rgb}{0.062745,0.305882,0.545098}
  \definecolor{SteelBlue1}{rgb}{0.388235,0.721569,1.000000}
  \definecolor{SteelBlue2}{rgb}{0.360784,0.674510,0.933333}
  \definecolor{SteelBlue3}{rgb}{0.309804,0.580392,0.803922}
  \definecolor{SteelBlue4}{rgb}{0.211765,0.392157,0.545098}
  \definecolor{DeepSkyBlue1}{rgb}{0.000000,0.749020,1.000000}
  \definecolor{DeepSkyBlue2}{rgb}{0.000000,0.698039,0.933333}
  \definecolor{DeepSkyBlue3}{rgb}{0.000000,0.603922,0.803922}
  \definecolor{DeepSkyBlue4}{rgb}{0.000000,0.407843,0.545098}
  \definecolor{SkyBlue1}{rgb}{0.529412,0.807843,1.000000}
  \definecolor{SkyBlue2}{rgb}{0.494118,0.752941,0.933333}
  \definecolor{SkyBlue3}{rgb}{0.423529,0.650980,0.803922}
  \definecolor{SkyBlue4}{rgb}{0.290196,0.439216,0.545098}
  \definecolor{LightSkyBlue1}{rgb}{0.690196,0.886275,1.000000}
  \definecolor{LightSkyBlue2}{rgb}{0.643137,0.827451,0.933333}
  \definecolor{LightSkyBlue3}{rgb}{0.552941,0.713726,0.803922}
  \definecolor{LightSkyBlue4}{rgb}{0.376471,0.482353,0.545098}
  \definecolor{SlateGray1}{rgb}{0.776471,0.886275,1.000000}
  \definecolor{SlateGray2}{rgb}{0.725490,0.827451,0.933333}
  \definecolor{SlateGray3}{rgb}{0.623529,0.713726,0.803922}
  \definecolor{SlateGray4}{rgb}{0.423529,0.482353,0.545098}
  \definecolor{LightSteelBlue1}{rgb}{0.792157,0.882353,1.000000}
  \definecolor{LightSteelBlue2}{rgb}{0.737255,0.823529,0.933333}
  \definecolor{LightSteelBlue3}{rgb}{0.635294,0.709804,0.803922}
  \definecolor{LightSteelBlue4}{rgb}{0.431373,0.482353,0.545098}
  \definecolor{LightBlue1}{rgb}{0.749020,0.937255,1.000000}
  \definecolor{LightBlue2}{rgb}{0.698039,0.874510,0.933333}
  \definecolor{LightBlue3}{rgb}{0.603922,0.752941,0.803922}
  \definecolor{LightBlue4}{rgb}{0.407843,0.513726,0.545098}
  \definecolor{LightCyan1}{rgb}{0.878431,1.000000,1.000000}
  \definecolor{LightCyan2}{rgb}{0.819608,0.933333,0.933333}
  \definecolor{LightCyan3}{rgb}{0.705882,0.803922,0.803922}
  \definecolor{LightCyan4}{rgb}{0.478431,0.545098,0.545098}
  \definecolor{PaleTurquoise1}{rgb}{0.733333,1.000000,1.000000}
  \definecolor{PaleTurquoise2}{rgb}{0.682353,0.933333,0.933333}
  \definecolor{PaleTurquoise3}{rgb}{0.588235,0.803922,0.803922}
  \definecolor{PaleTurquoise4}{rgb}{0.400000,0.545098,0.545098}
  \definecolor{CadetBlue1}{rgb}{0.596078,0.960784,1.000000}
  \definecolor{CadetBlue2}{rgb}{0.556863,0.898039,0.933333}
  \definecolor{CadetBlue3}{rgb}{0.478431,0.772549,0.803922}
  \definecolor{CadetBlue4}{rgb}{0.325490,0.525490,0.545098}
  \definecolor{turquoise1}{rgb}{0.000000,0.960784,1.000000}
  \definecolor{turquoise2}{rgb}{0.000000,0.898039,0.933333}
  \definecolor{turquoise3}{rgb}{0.000000,0.772549,0.803922}
  \definecolor{turquoise4}{rgb}{0.000000,0.525490,0.545098}
  \definecolor{cyan1}{rgb}{0.000000,1.000000,1.000000}
  \definecolor{cyan2}{rgb}{0.000000,0.933333,0.933333}
  \definecolor{cyan3}{rgb}{0.000000,0.803922,0.803922}
  \definecolor{cyan4}{rgb}{0.000000,0.545098,0.545098}
  \definecolor{DarkSlateGray1}{rgb}{0.592157,1.000000,1.000000}
  \definecolor{DarkSlateGray2}{rgb}{0.552941,0.933333,0.933333}
  \definecolor{DarkSlateGray3}{rgb}{0.474510,0.803922,0.803922}
  \definecolor{DarkSlateGray4}{rgb}{0.321569,0.545098,0.545098}
  \definecolor{aquamarine1}{rgb}{0.498039,1.000000,0.831373}
  \definecolor{aquamarine2}{rgb}{0.462745,0.933333,0.776471}
  \definecolor{aquamarine3}{rgb}{0.400000,0.803922,0.666667}
  \definecolor{aquamarine4}{rgb}{0.270588,0.545098,0.454902}
  \definecolor{DarkSeaGreen1}{rgb}{0.756863,1.000000,0.756863}
  \definecolor{DarkSeaGreen2}{rgb}{0.705882,0.933333,0.705882}
  \definecolor{DarkSeaGreen3}{rgb}{0.607843,0.803922,0.607843}
  \definecolor{DarkSeaGreen4}{rgb}{0.411765,0.545098,0.411765}
  \definecolor{SeaGreen1}{rgb}{0.329412,1.000000,0.623529}
  \definecolor{SeaGreen2}{rgb}{0.305882,0.933333,0.580392}
  \definecolor{SeaGreen3}{rgb}{0.262745,0.803922,0.501961}
  \definecolor{SeaGreen4}{rgb}{0.180392,0.545098,0.341176}
  \definecolor{PaleGreen1}{rgb}{0.603922,1.000000,0.603922}
  \definecolor{PaleGreen2}{rgb}{0.564706,0.933333,0.564706}
  \definecolor{PaleGreen3}{rgb}{0.486275,0.803922,0.486275}
  \definecolor{PaleGreen4}{rgb}{0.329412,0.545098,0.329412}
  \definecolor{SpringGreen1}{rgb}{0.000000,1.000000,0.498039}
  \definecolor{SpringGreen2}{rgb}{0.000000,0.933333,0.462745}
  \definecolor{SpringGreen3}{rgb}{0.000000,0.803922,0.400000}
  \definecolor{SpringGreen4}{rgb}{0.000000,0.545098,0.270588}
  \definecolor{green1}{rgb}{0.000000,1.000000,0.000000}
  \definecolor{green2}{rgb}{0.000000,0.933333,0.000000}
  \definecolor{green3}{rgb}{0.000000,0.803922,0.000000}
  \definecolor{green4}{rgb}{0.000000,0.545098,0.000000}
  \definecolor{chartreuse1}{rgb}{0.498039,1.000000,0.000000}
  \definecolor{chartreuse2}{rgb}{0.462745,0.933333,0.000000}
  \definecolor{chartreuse3}{rgb}{0.400000,0.803922,0.000000}
  \definecolor{chartreuse4}{rgb}{0.270588,0.545098,0.000000}
  \definecolor{OliveDrab1}{rgb}{0.752941,1.000000,0.243137}
  \definecolor{OliveDrab2}{rgb}{0.701961,0.933333,0.227451}
  \definecolor{OliveDrab3}{rgb}{0.603922,0.803922,0.196078}
  \definecolor{OliveDrab4}{rgb}{0.411765,0.545098,0.133333}
  \definecolor{DarkOliveGreen1}{rgb}{0.792157,1.000000,0.439216}
  \definecolor{DarkOliveGreen2}{rgb}{0.737255,0.933333,0.407843}
  \definecolor{DarkOliveGreen3}{rgb}{0.635294,0.803922,0.352941}
  \definecolor{DarkOliveGreen4}{rgb}{0.431373,0.545098,0.239216}
  \definecolor{khaki1}{rgb}{1.000000,0.964706,0.560784}
  \definecolor{khaki2}{rgb}{0.933333,0.901961,0.521569}
  \definecolor{khaki3}{rgb}{0.803922,0.776471,0.450980}
  \definecolor{khaki4}{rgb}{0.545098,0.525490,0.305882}
  \definecolor{LightGoldenrod1}{rgb}{1.000000,0.925490,0.545098}
  \definecolor{LightGoldenrod2}{rgb}{0.933333,0.862745,0.509804}
  \definecolor{LightGoldenrod3}{rgb}{0.803922,0.745098,0.439216}
  \definecolor{LightGoldenrod4}{rgb}{0.545098,0.505882,0.298039}
  \definecolor{LightYellow1}{rgb}{1.000000,1.000000,0.878431}
  \definecolor{LightYellow2}{rgb}{0.933333,0.933333,0.819608}
  \definecolor{LightYellow3}{rgb}{0.803922,0.803922,0.705882}
  \definecolor{LightYellow4}{rgb}{0.545098,0.545098,0.478431}
  \definecolor{yellow1}{rgb}{1.000000,1.000000,0.000000}
  \definecolor{yellow2}{rgb}{0.933333,0.933333,0.000000}
  \definecolor{yellow3}{rgb}{0.803922,0.803922,0.000000}
  \definecolor{yellow4}{rgb}{0.545098,0.545098,0.000000}
  \definecolor{gold1}{rgb}{1.000000,0.843137,0.000000}
  \definecolor{gold2}{rgb}{0.933333,0.788235,0.000000}
  \definecolor{gold3}{rgb}{0.803922,0.678431,0.000000}
  \definecolor{gold4}{rgb}{0.545098,0.458824,0.000000}
  \definecolor{goldenrod1}{rgb}{1.000000,0.756863,0.145098}
  \definecolor{goldenrod2}{rgb}{0.933333,0.705882,0.133333}
  \definecolor{goldenrod3}{rgb}{0.803922,0.607843,0.113725}
  \definecolor{goldenrod4}{rgb}{0.545098,0.411765,0.078431}
  \definecolor{DarkGoldenrod1}{rgb}{1.000000,0.725490,0.058824}
  \definecolor{DarkGoldenrod2}{rgb}{0.933333,0.678431,0.054902}
  \definecolor{DarkGoldenrod3}{rgb}{0.803922,0.584314,0.047059}
  \definecolor{DarkGoldenrod4}{rgb}{0.545098,0.396078,0.031373}
  \definecolor{RosyBrown1}{rgb}{1.000000,0.756863,0.756863}
  \definecolor{RosyBrown2}{rgb}{0.933333,0.705882,0.705882}
  \definecolor{RosyBrown3}{rgb}{0.803922,0.607843,0.607843}
  \definecolor{RosyBrown4}{rgb}{0.545098,0.411765,0.411765}
  \definecolor{IndianRed1}{rgb}{1.000000,0.415686,0.415686}
  \definecolor{IndianRed2}{rgb}{0.933333,0.388235,0.388235}
  \definecolor{IndianRed3}{rgb}{0.803922,0.333333,0.333333}
  \definecolor{IndianRed4}{rgb}{0.545098,0.227451,0.227451}
  \definecolor{sienna1}{rgb}{1.000000,0.509804,0.278431}
  \definecolor{sienna2}{rgb}{0.933333,0.474510,0.258824}
  \definecolor{sienna3}{rgb}{0.803922,0.407843,0.223529}
  \definecolor{sienna4}{rgb}{0.545098,0.278431,0.149020}
  \definecolor{burlywood1}{rgb}{1.000000,0.827451,0.607843}
  \definecolor{burlywood2}{rgb}{0.933333,0.772549,0.568627}
  \definecolor{burlywood3}{rgb}{0.803922,0.666667,0.490196}
  \definecolor{burlywood4}{rgb}{0.545098,0.450980,0.333333}
  \definecolor{wheat1}{rgb}{1.000000,0.905882,0.729412}
  \definecolor{wheat2}{rgb}{0.933333,0.847059,0.682353}
  \definecolor{wheat3}{rgb}{0.803922,0.729412,0.588235}
  \definecolor{wheat4}{rgb}{0.545098,0.494118,0.400000}
  \definecolor{tan1}{rgb}{1.000000,0.647059,0.309804}
  \definecolor{tan2}{rgb}{0.933333,0.603922,0.286275}
  \definecolor{tan3}{rgb}{0.803922,0.521569,0.247059}
  \definecolor{tan4}{rgb}{0.545098,0.352941,0.168627}
  \definecolor{chocolate1}{rgb}{1.000000,0.498039,0.141176}
  \definecolor{chocolate2}{rgb}{0.933333,0.462745,0.129412}
  \definecolor{chocolate3}{rgb}{0.803922,0.400000,0.113725}
  \definecolor{chocolate4}{rgb}{0.545098,0.270588,0.074510}
  \definecolor{firebrick1}{rgb}{1.000000,0.188235,0.188235}
  \definecolor{firebrick2}{rgb}{0.933333,0.172549,0.172549}
  \definecolor{firebrick3}{rgb}{0.803922,0.149020,0.149020}
  \definecolor{firebrick4}{rgb}{0.545098,0.101961,0.101961}
  \definecolor{brown1}{rgb}{1.000000,0.250980,0.250980}
  \definecolor{brown2}{rgb}{0.933333,0.231373,0.231373}
  \definecolor{brown3}{rgb}{0.803922,0.200000,0.200000}
  \definecolor{brown4}{rgb}{0.545098,0.137255,0.137255}
  \definecolor{salmon1}{rgb}{1.000000,0.549020,0.411765}
  \definecolor{salmon2}{rgb}{0.933333,0.509804,0.384314}
  \definecolor{salmon3}{rgb}{0.803922,0.439216,0.329412}
  \definecolor{salmon4}{rgb}{0.545098,0.298039,0.223529}
  \definecolor{LightSalmon1}{rgb}{1.000000,0.627451,0.478431}
  \definecolor{LightSalmon2}{rgb}{0.933333,0.584314,0.447059}
  \definecolor{LightSalmon3}{rgb}{0.803922,0.505882,0.384314}
  \definecolor{LightSalmon4}{rgb}{0.545098,0.341176,0.258824}
  \definecolor{orange1}{rgb}{1.000000,0.647059,0.000000}
  \definecolor{orange2}{rgb}{0.933333,0.603922,0.000000}
  \definecolor{orange3}{rgb}{0.803922,0.521569,0.000000}
  \definecolor{orange4}{rgb}{0.545098,0.352941,0.000000}
  \definecolor{DarkOrange1}{rgb}{1.000000,0.498039,0.000000}
  \definecolor{DarkOrange2}{rgb}{0.933333,0.462745,0.000000}
  \definecolor{DarkOrange3}{rgb}{0.803922,0.400000,0.000000}
  \definecolor{DarkOrange4}{rgb}{0.545098,0.270588,0.000000}
  \definecolor{coral1}{rgb}{1.000000,0.447059,0.337255}
  \definecolor{coral2}{rgb}{0.933333,0.415686,0.313726}
  \definecolor{coral3}{rgb}{0.803922,0.356863,0.270588}
  \definecolor{coral4}{rgb}{0.545098,0.243137,0.184314}
  \definecolor{tomato1}{rgb}{1.000000,0.388235,0.278431}
  \definecolor{tomato2}{rgb}{0.933333,0.360784,0.258824}
  \definecolor{tomato3}{rgb}{0.803922,0.309804,0.223529}
  \definecolor{tomato4}{rgb}{0.545098,0.211765,0.149020}
  \definecolor{OrangeRed1}{rgb}{1.000000,0.270588,0.000000}
  \definecolor{OrangeRed2}{rgb}{0.933333,0.250980,0.000000}
  \definecolor{OrangeRed3}{rgb}{0.803922,0.215686,0.000000}
  \definecolor{OrangeRed4}{rgb}{0.545098,0.145098,0.000000}
  \definecolor{red1}{rgb}{1.000000,0.000000,0.000000}
  \definecolor{red2}{rgb}{0.933333,0.000000,0.000000}
  \definecolor{red3}{rgb}{0.803922,0.000000,0.000000}
  \definecolor{red4}{rgb}{0.545098,0.000000,0.000000}
  \definecolor{DeepPink1}{rgb}{1.000000,0.078431,0.576471}
  \definecolor{DeepPink2}{rgb}{0.933333,0.070588,0.537255}
  \definecolor{DeepPink3}{rgb}{0.803922,0.062745,0.462745}
  \definecolor{DeepPink4}{rgb}{0.545098,0.039216,0.313726}
  \definecolor{HotPink1}{rgb}{1.000000,0.431373,0.705882}
  \definecolor{HotPink2}{rgb}{0.933333,0.415686,0.654902}
  \definecolor{HotPink3}{rgb}{0.803922,0.376471,0.564706}
  \definecolor{HotPink4}{rgb}{0.545098,0.227451,0.384314}
  \definecolor{pink1}{rgb}{1.000000,0.709804,0.772549}
  \definecolor{pink2}{rgb}{0.933333,0.662745,0.721569}
  \definecolor{pink3}{rgb}{0.803922,0.568627,0.619608}
  \definecolor{pink4}{rgb}{0.545098,0.388235,0.423529}
  \definecolor{LightPink1}{rgb}{1.000000,0.682353,0.725490}
  \definecolor{LightPink2}{rgb}{0.933333,0.635294,0.678431}
  \definecolor{LightPink3}{rgb}{0.803922,0.549020,0.584314}
  \definecolor{LightPink4}{rgb}{0.545098,0.372549,0.396078}
  \definecolor{PaleVioletRed1}{rgb}{1.000000,0.509804,0.670588}
  \definecolor{PaleVioletRed2}{rgb}{0.933333,0.474510,0.623529}
  \definecolor{PaleVioletRed3}{rgb}{0.803922,0.407843,0.537255}
  \definecolor{PaleVioletRed4}{rgb}{0.545098,0.278431,0.364706}
  \definecolor{maroon1}{rgb}{1.000000,0.203922,0.701961}
  \definecolor{maroon2}{rgb}{0.933333,0.188235,0.654902}
  \definecolor{maroon3}{rgb}{0.803922,0.160784,0.564706}
  \definecolor{maroon4}{rgb}{0.545098,0.109804,0.384314}
  \definecolor{VioletRed1}{rgb}{1.000000,0.243137,0.588235}
  \definecolor{VioletRed2}{rgb}{0.933333,0.227451,0.549020}
  \definecolor{VioletRed3}{rgb}{0.803922,0.196078,0.470588}
  \definecolor{VioletRed4}{rgb}{0.545098,0.133333,0.321569}
  \definecolor{magenta1}{rgb}{1.000000,0.000000,1.000000}
  \definecolor{magenta2}{rgb}{0.933333,0.000000,0.933333}
  \definecolor{magenta3}{rgb}{0.803922,0.000000,0.803922}
  \definecolor{magenta4}{rgb}{0.545098,0.000000,0.545098}
  \definecolor{orchid1}{rgb}{1.000000,0.513726,0.980392}
  \definecolor{orchid2}{rgb}{0.933333,0.478431,0.913725}
  \definecolor{orchid3}{rgb}{0.803922,0.411765,0.788235}
  \definecolor{orchid4}{rgb}{0.545098,0.278431,0.537255}
  \definecolor{plum1}{rgb}{1.000000,0.733333,1.000000}
  \definecolor{plum2}{rgb}{0.933333,0.682353,0.933333}
  \definecolor{plum3}{rgb}{0.803922,0.588235,0.803922}
  \definecolor{plum4}{rgb}{0.545098,0.400000,0.545098}
  \definecolor{MediumOrchid1}{rgb}{0.878431,0.400000,1.000000}
  \definecolor{MediumOrchid2}{rgb}{0.819608,0.372549,0.933333}
  \definecolor{MediumOrchid3}{rgb}{0.705882,0.321569,0.803922}
  \definecolor{MediumOrchid4}{rgb}{0.478431,0.215686,0.545098}
  \definecolor{DarkOrchid1}{rgb}{0.749020,0.243137,1.000000}
  \definecolor{DarkOrchid2}{rgb}{0.698039,0.227451,0.933333}
  \definecolor{DarkOrchid3}{rgb}{0.603922,0.196078,0.803922}
  \definecolor{DarkOrchid4}{rgb}{0.407843,0.133333,0.545098}
  \definecolor{purple1}{rgb}{0.607843,0.188235,1.000000}
  \definecolor{purple2}{rgb}{0.568627,0.172549,0.933333}
  \definecolor{purple3}{rgb}{0.490196,0.149020,0.803922}
  \definecolor{purple4}{rgb}{0.333333,0.101961,0.545098}
  \definecolor{MediumPurple1}{rgb}{0.670588,0.509804,1.000000}
  \definecolor{MediumPurple2}{rgb}{0.623529,0.474510,0.933333}
  \definecolor{MediumPurple3}{rgb}{0.537255,0.407843,0.803922}
  \definecolor{MediumPurple4}{rgb}{0.364706,0.278431,0.545098}
  \definecolor{thistle1}{rgb}{1.000000,0.882353,1.000000}
  \definecolor{thistle2}{rgb}{0.933333,0.823529,0.933333}
  \definecolor{thistle3}{rgb}{0.803922,0.709804,0.803922}
  \definecolor{thistle4}{rgb}{0.545098,0.482353,0.545098}
  \definecolor{gray0}{rgb}{0.000000,0.000000,0.000000}
  \definecolor{grey0}{rgb}{0.000000,0.000000,0.000000}
  \definecolor{gray1}{rgb}{0.011765,0.011765,0.011765}
  \definecolor{grey1}{rgb}{0.011765,0.011765,0.011765}
  \definecolor{gray2}{rgb}{0.019608,0.019608,0.019608}
  \definecolor{grey2}{rgb}{0.019608,0.019608,0.019608}
  \definecolor{gray3}{rgb}{0.031373,0.031373,0.031373}
  \definecolor{grey3}{rgb}{0.031373,0.031373,0.031373}
  \definecolor{gray4}{rgb}{0.039216,0.039216,0.039216}
  \definecolor{grey4}{rgb}{0.039216,0.039216,0.039216}
  \definecolor{gray5}{rgb}{0.050980,0.050980,0.050980}
  \definecolor{grey5}{rgb}{0.050980,0.050980,0.050980}
  \definecolor{gray6}{rgb}{0.058824,0.058824,0.058824}
  \definecolor{grey6}{rgb}{0.058824,0.058824,0.058824}
  \definecolor{gray7}{rgb}{0.070588,0.070588,0.070588}
  \definecolor{grey7}{rgb}{0.070588,0.070588,0.070588}
  \definecolor{gray8}{rgb}{0.078431,0.078431,0.078431}
  \definecolor{grey8}{rgb}{0.078431,0.078431,0.078431}
  \definecolor{gray9}{rgb}{0.090196,0.090196,0.090196}
  \definecolor{grey9}{rgb}{0.090196,0.090196,0.090196}
  \definecolor{gray10}{rgb}{0.101961,0.101961,0.101961}
  \definecolor{grey10}{rgb}{0.101961,0.101961,0.101961}
  \definecolor{gray11}{rgb}{0.109804,0.109804,0.109804}
  \definecolor{grey11}{rgb}{0.109804,0.109804,0.109804}
  \definecolor{gray12}{rgb}{0.121569,0.121569,0.121569}
  \definecolor{grey12}{rgb}{0.121569,0.121569,0.121569}
  \definecolor{gray13}{rgb}{0.129412,0.129412,0.129412}
  \definecolor{grey13}{rgb}{0.129412,0.129412,0.129412}
  \definecolor{gray14}{rgb}{0.141176,0.141176,0.141176}
  \definecolor{grey14}{rgb}{0.141176,0.141176,0.141176}
  \definecolor{gray15}{rgb}{0.149020,0.149020,0.149020}
  \definecolor{grey15}{rgb}{0.149020,0.149020,0.149020}
  \definecolor{gray16}{rgb}{0.160784,0.160784,0.160784}
  \definecolor{grey16}{rgb}{0.160784,0.160784,0.160784}
  \definecolor{gray17}{rgb}{0.168627,0.168627,0.168627}
  \definecolor{grey17}{rgb}{0.168627,0.168627,0.168627}
  \definecolor{gray18}{rgb}{0.180392,0.180392,0.180392}
  \definecolor{grey18}{rgb}{0.180392,0.180392,0.180392}
  \definecolor{gray19}{rgb}{0.188235,0.188235,0.188235}
  \definecolor{grey19}{rgb}{0.188235,0.188235,0.188235}
  \definecolor{gray20}{rgb}{0.200000,0.200000,0.200000}
  \definecolor{grey20}{rgb}{0.200000,0.200000,0.200000}
  \definecolor{gray21}{rgb}{0.211765,0.211765,0.211765}
  \definecolor{grey21}{rgb}{0.211765,0.211765,0.211765}
  \definecolor{gray22}{rgb}{0.219608,0.219608,0.219608}
  \definecolor{grey22}{rgb}{0.219608,0.219608,0.219608}
  \definecolor{gray23}{rgb}{0.231373,0.231373,0.231373}
  \definecolor{grey23}{rgb}{0.231373,0.231373,0.231373}
  \definecolor{gray24}{rgb}{0.239216,0.239216,0.239216}
  \definecolor{grey24}{rgb}{0.239216,0.239216,0.239216}
  \definecolor{gray25}{rgb}{0.250980,0.250980,0.250980}
  \definecolor{grey25}{rgb}{0.250980,0.250980,0.250980}
  \definecolor{gray26}{rgb}{0.258824,0.258824,0.258824}
  \definecolor{grey26}{rgb}{0.258824,0.258824,0.258824}
  \definecolor{gray27}{rgb}{0.270588,0.270588,0.270588}
  \definecolor{grey27}{rgb}{0.270588,0.270588,0.270588}
  \definecolor{gray28}{rgb}{0.278431,0.278431,0.278431}
  \definecolor{grey28}{rgb}{0.278431,0.278431,0.278431}
  \definecolor{gray29}{rgb}{0.290196,0.290196,0.290196}
  \definecolor{grey29}{rgb}{0.290196,0.290196,0.290196}
  \definecolor{gray30}{rgb}{0.301961,0.301961,0.301961}
  \definecolor{grey30}{rgb}{0.301961,0.301961,0.301961}
  \definecolor{gray31}{rgb}{0.309804,0.309804,0.309804}
  \definecolor{grey31}{rgb}{0.309804,0.309804,0.309804}
  \definecolor{gray32}{rgb}{0.321569,0.321569,0.321569}
  \definecolor{grey32}{rgb}{0.321569,0.321569,0.321569}
  \definecolor{gray33}{rgb}{0.329412,0.329412,0.329412}
  \definecolor{grey33}{rgb}{0.329412,0.329412,0.329412}
  \definecolor{gray34}{rgb}{0.341176,0.341176,0.341176}
  \definecolor{grey34}{rgb}{0.341176,0.341176,0.341176}
  \definecolor{gray35}{rgb}{0.349020,0.349020,0.349020}
  \definecolor{grey35}{rgb}{0.349020,0.349020,0.349020}
  \definecolor{gray36}{rgb}{0.360784,0.360784,0.360784}
  \definecolor{grey36}{rgb}{0.360784,0.360784,0.360784}
  \definecolor{gray37}{rgb}{0.368627,0.368627,0.368627}
  \definecolor{grey37}{rgb}{0.368627,0.368627,0.368627}
  \definecolor{gray38}{rgb}{0.380392,0.380392,0.380392}
  \definecolor{grey38}{rgb}{0.380392,0.380392,0.380392}
  \definecolor{gray39}{rgb}{0.388235,0.388235,0.388235}
  \definecolor{grey39}{rgb}{0.388235,0.388235,0.388235}
  \definecolor{gray40}{rgb}{0.400000,0.400000,0.400000}
  \definecolor{grey40}{rgb}{0.400000,0.400000,0.400000}
  \definecolor{gray41}{rgb}{0.411765,0.411765,0.411765}
  \definecolor{grey41}{rgb}{0.411765,0.411765,0.411765}
  \definecolor{gray42}{rgb}{0.419608,0.419608,0.419608}
  \definecolor{grey42}{rgb}{0.419608,0.419608,0.419608}
  \definecolor{gray43}{rgb}{0.431373,0.431373,0.431373}
  \definecolor{grey43}{rgb}{0.431373,0.431373,0.431373}
  \definecolor{gray44}{rgb}{0.439216,0.439216,0.439216}
  \definecolor{grey44}{rgb}{0.439216,0.439216,0.439216}
  \definecolor{gray45}{rgb}{0.450980,0.450980,0.450980}
  \definecolor{grey45}{rgb}{0.450980,0.450980,0.450980}
  \definecolor{gray46}{rgb}{0.458824,0.458824,0.458824}
  \definecolor{grey46}{rgb}{0.458824,0.458824,0.458824}
  \definecolor{gray47}{rgb}{0.470588,0.470588,0.470588}
  \definecolor{grey47}{rgb}{0.470588,0.470588,0.470588}
  \definecolor{gray48}{rgb}{0.478431,0.478431,0.478431}
  \definecolor{grey48}{rgb}{0.478431,0.478431,0.478431}
  \definecolor{gray49}{rgb}{0.490196,0.490196,0.490196}
  \definecolor{grey49}{rgb}{0.490196,0.490196,0.490196}
  \definecolor{gray50}{rgb}{0.498039,0.498039,0.498039}
  \definecolor{grey50}{rgb}{0.498039,0.498039,0.498039}
  \definecolor{gray51}{rgb}{0.509804,0.509804,0.509804}
  \definecolor{grey51}{rgb}{0.509804,0.509804,0.509804}
  \definecolor{gray52}{rgb}{0.521569,0.521569,0.521569}
  \definecolor{grey52}{rgb}{0.521569,0.521569,0.521569}
  \definecolor{gray53}{rgb}{0.529412,0.529412,0.529412}
  \definecolor{grey53}{rgb}{0.529412,0.529412,0.529412}
  \definecolor{gray54}{rgb}{0.541176,0.541176,0.541176}
  \definecolor{grey54}{rgb}{0.541176,0.541176,0.541176}
  \definecolor{gray55}{rgb}{0.549020,0.549020,0.549020}
  \definecolor{grey55}{rgb}{0.549020,0.549020,0.549020}
  \definecolor{gray56}{rgb}{0.560784,0.560784,0.560784}
  \definecolor{grey56}{rgb}{0.560784,0.560784,0.560784}
  \definecolor{gray57}{rgb}{0.568627,0.568627,0.568627}
  \definecolor{grey57}{rgb}{0.568627,0.568627,0.568627}
  \definecolor{gray58}{rgb}{0.580392,0.580392,0.580392}
  \definecolor{grey58}{rgb}{0.580392,0.580392,0.580392}
  \definecolor{gray59}{rgb}{0.588235,0.588235,0.588235}
  \definecolor{grey59}{rgb}{0.588235,0.588235,0.588235}
  \definecolor{gray60}{rgb}{0.600000,0.600000,0.600000}
  \definecolor{grey60}{rgb}{0.600000,0.600000,0.600000}
  \definecolor{gray61}{rgb}{0.611765,0.611765,0.611765}
  \definecolor{grey61}{rgb}{0.611765,0.611765,0.611765}
  \definecolor{gray62}{rgb}{0.619608,0.619608,0.619608}
  \definecolor{grey62}{rgb}{0.619608,0.619608,0.619608}
  \definecolor{gray63}{rgb}{0.631373,0.631373,0.631373}
  \definecolor{grey63}{rgb}{0.631373,0.631373,0.631373}
  \definecolor{gray64}{rgb}{0.639216,0.639216,0.639216}
  \definecolor{grey64}{rgb}{0.639216,0.639216,0.639216}
  \definecolor{gray65}{rgb}{0.650980,0.650980,0.650980}
  \definecolor{grey65}{rgb}{0.650980,0.650980,0.650980}
  \definecolor{gray66}{rgb}{0.658824,0.658824,0.658824}
  \definecolor{grey66}{rgb}{0.658824,0.658824,0.658824}
  \definecolor{gray67}{rgb}{0.670588,0.670588,0.670588}
  \definecolor{grey67}{rgb}{0.670588,0.670588,0.670588}
  \definecolor{gray68}{rgb}{0.678431,0.678431,0.678431}
  \definecolor{grey68}{rgb}{0.678431,0.678431,0.678431}
  \definecolor{gray69}{rgb}{0.690196,0.690196,0.690196}
  \definecolor{grey69}{rgb}{0.690196,0.690196,0.690196}
  \definecolor{gray70}{rgb}{0.701961,0.701961,0.701961}
  \definecolor{grey70}{rgb}{0.701961,0.701961,0.701961}
  \definecolor{gray71}{rgb}{0.709804,0.709804,0.709804}
  \definecolor{grey71}{rgb}{0.709804,0.709804,0.709804}
  \definecolor{gray72}{rgb}{0.721569,0.721569,0.721569}
  \definecolor{grey72}{rgb}{0.721569,0.721569,0.721569}
  \definecolor{gray73}{rgb}{0.729412,0.729412,0.729412}
  \definecolor{grey73}{rgb}{0.729412,0.729412,0.729412}
  \definecolor{gray74}{rgb}{0.741176,0.741176,0.741176}
  \definecolor{grey74}{rgb}{0.741176,0.741176,0.741176}
  \definecolor{gray75}{rgb}{0.749020,0.749020,0.749020}
  \definecolor{grey75}{rgb}{0.749020,0.749020,0.749020}
  \definecolor{gray76}{rgb}{0.760784,0.760784,0.760784}
  \definecolor{grey76}{rgb}{0.760784,0.760784,0.760784}
  \definecolor{gray77}{rgb}{0.768627,0.768627,0.768627}
  \definecolor{grey77}{rgb}{0.768627,0.768627,0.768627}
  \definecolor{gray78}{rgb}{0.780392,0.780392,0.780392}
  \definecolor{grey78}{rgb}{0.780392,0.780392,0.780392}
  \definecolor{gray79}{rgb}{0.788235,0.788235,0.788235}
  \definecolor{grey79}{rgb}{0.788235,0.788235,0.788235}
  \definecolor{gray80}{rgb}{0.800000,0.800000,0.800000}
  \definecolor{grey80}{rgb}{0.800000,0.800000,0.800000}
  \definecolor{gray81}{rgb}{0.811765,0.811765,0.811765}
  \definecolor{grey81}{rgb}{0.811765,0.811765,0.811765}
  \definecolor{gray82}{rgb}{0.819608,0.819608,0.819608}
  \definecolor{grey82}{rgb}{0.819608,0.819608,0.819608}
  \definecolor{gray83}{rgb}{0.831373,0.831373,0.831373}
  \definecolor{grey83}{rgb}{0.831373,0.831373,0.831373}
  \definecolor{gray84}{rgb}{0.839216,0.839216,0.839216}
  \definecolor{grey84}{rgb}{0.839216,0.839216,0.839216}
  \definecolor{gray85}{rgb}{0.850980,0.850980,0.850980}
  \definecolor{grey85}{rgb}{0.850980,0.850980,0.850980}
  \definecolor{gray86}{rgb}{0.858824,0.858824,0.858824}
  \definecolor{grey86}{rgb}{0.858824,0.858824,0.858824}
  \definecolor{gray87}{rgb}{0.870588,0.870588,0.870588}
  \definecolor{grey87}{rgb}{0.870588,0.870588,0.870588}
  \definecolor{gray88}{rgb}{0.878431,0.878431,0.878431}
  \definecolor{grey88}{rgb}{0.878431,0.878431,0.878431}
  \definecolor{gray89}{rgb}{0.890196,0.890196,0.890196}
  \definecolor{grey89}{rgb}{0.890196,0.890196,0.890196}
  \definecolor{gray90}{rgb}{0.898039,0.898039,0.898039}
  \definecolor{grey90}{rgb}{0.898039,0.898039,0.898039}
  \definecolor{gray91}{rgb}{0.909804,0.909804,0.909804}
  \definecolor{grey91}{rgb}{0.909804,0.909804,0.909804}
  \definecolor{gray92}{rgb}{0.921569,0.921569,0.921569}
  \definecolor{grey92}{rgb}{0.921569,0.921569,0.921569}
  \definecolor{gray93}{rgb}{0.929412,0.929412,0.929412}
  \definecolor{grey93}{rgb}{0.929412,0.929412,0.929412}
  \definecolor{gray94}{rgb}{0.941176,0.941176,0.941176}
  \definecolor{grey94}{rgb}{0.941176,0.941176,0.941176}
  \definecolor{gray95}{rgb}{0.949020,0.949020,0.949020}
  \definecolor{grey95}{rgb}{0.949020,0.949020,0.949020}
  \definecolor{gray96}{rgb}{0.960784,0.960784,0.960784}
  \definecolor{grey96}{rgb}{0.960784,0.960784,0.960784}
  \definecolor{gray97}{rgb}{0.968627,0.968627,0.968627}
  \definecolor{grey97}{rgb}{0.968627,0.968627,0.968627}
  \definecolor{gray98}{rgb}{0.980392,0.980392,0.980392}
  \definecolor{grey98}{rgb}{0.980392,0.980392,0.980392}
  \definecolor{gray99}{rgb}{0.988235,0.988235,0.988235}
  \definecolor{grey99}{rgb}{0.988235,0.988235,0.988235}
  \definecolor{gray100}{rgb}{1.000000,1.000000,1.000000}
  \definecolor{grey100}{rgb}{1.000000,1.000000,1.000000}
  \definecolor{dark grey}{rgb}{0.662745,0.662745,0.662745}
  \definecolor{DarkGrey}{rgb}{0.662745,0.662745,0.662745}
  \definecolor{dark gray}{rgb}{0.662745,0.662745,0.662745}
  \definecolor{DarkGray}{rgb}{0.662745,0.662745,0.662745}
  \definecolor{dark blue}{rgb}{0.000000,0.000000,0.545098}
  \definecolor{DarkBlue}{rgb}{0.000000,0.000000,0.545098}
  \definecolor{dark cyan}{rgb}{0.000000,0.545098,0.545098}
  \definecolor{DarkCyan}{rgb}{0.000000,0.545098,0.545098}
  \definecolor{dark magenta}{rgb}{0.545098,0.000000,0.545098}
  \definecolor{DarkMagenta}{rgb}{0.545098,0.000000,0.545098}
  \definecolor{dark red}{rgb}{0.545098,0.000000,0.000000}
  \definecolor{DarkRed}{rgb}{0.545098,0.000000,0.000000}
  \definecolor{light green}{rgb}{0.564706,0.933333,0.564706}
  \definecolor{LightGreen}{rgb}{0.564706,0.933333,0.564706}

\graphicspath{{.}{./images/}}

\definecolor{dkgreen}{rgb}{0,0.6,0}
\definecolor{gray}{rgb}{0.5,0.5,0.5}

\definecolor{lightgraydark}{rgb}{0.6,0.6,0.6}
\definecolor{lightgraydarker}{rgb}{0.7,0.7,0.7}

\definecolor{lightblue}{rgb}{0.5,0.90,1.0}
\definecolor{lightgreen}{rgb}{0.5,0.92,0.5}
\definecolor{verylightred}{rgb}{0.98,0.5,0.5}
\definecolor{lightyellow}{rgb}{1,0.90,0.40}

\definecolor{lightgray}{rgb}{0.83,0.83,0.83}
\definecolor{verylightgray}{rgb}{0.65,0.65,0.65}
\definecolor{verylightblue}{rgb}{0.5,0.80,1.0}
\definecolor{verylightgreen}{RGB}	{204,255,204}

\definecolor{verylightyellow}{RGB}		{255,255,204}

\definecolor{color1and29}{rgb}{1,0.1725,0}
\definecolor{color2and29}{rgb}{1,0.3016,0}
\definecolor{color3and29}{rgb}{1,0.4306,0}
\definecolor{color4and29}{rgb}{1,0.5633,0}
\definecolor{color5and29}{rgb}{1,0.7221,0}
\definecolor{color6and29}{rgb}{1,0.8809,0}
\definecolor{color7and29}{rgb}{0.8824,0.9705,0}
\definecolor{color8and29}{rgb}{0.4118,0.8524,0}
\definecolor{color9and29}{rgb}{0,0.749,0}
\definecolor{color1and30}{rgb}{1,0.1725,0}
\definecolor{color2and30}{rgb}{1,0.2854,0}
\definecolor{color3and30}{rgb}{1,0.3822,0}
\definecolor{color4and30}{rgb}{1,0.479,0}
\definecolor{color5and30}{rgb}{1,0.603,0}
\definecolor{color6and30}{rgb}{1,0.7221,0}
\definecolor{color7and30}{rgb}{1,0.8412,0}
\definecolor{color8and30}{rgb}{1,0.9802,0}
\definecolor{color9and30}{rgb}{0.7059,0.9262,0}
\definecolor{color10and30}{rgb}{0.3529,0.8376,0}
\definecolor{color11and30}{rgb}{0,0.749,0}

\definecolor{bio-color1and2}{rgb}{1,0.1725,0}
\definecolor{bio-color2and2}{rgb}{1,0.7221,0}
\definecolor{bio-color3and2}{rgb}{0,0.749,0}
\definecolor{ca-color1and4}{rgb}{1,0.1725,0}
\definecolor{ca-color2and4}{rgb}{1,0.7221,0}
\definecolor{ca-color3and4}{rgb}{0,0.749,0}
\definecolor{ia-color1and7}{rgb}{1,0.1725,0}
\definecolor{ia-color2and7}{rgb}{1,0.4306,0}
\definecolor{ia-color3and7}{rgb}{1,0.7221,0}
\definecolor{ia-color4and7}{rgb}{0.8824,0.9705,0}
\definecolor{ia-color5and7}{rgb}{0,0.749,0}
\definecolor{soc-color1and11}{rgb}{1,0.1725,0}
\definecolor{soc-color2and11}{rgb}{1,0.3177,0}
\definecolor{soc-color3and11}{rgb}{1,0.4628,0}
\definecolor{soc-color4and11}{rgb}{1,0.6229,0}
\definecolor{soc-color5and11}{rgb}{1,0.8015,0}
\definecolor{soc-color6and11}{rgb}{1,0.9802,0}
\definecolor{soc-color7and11}{rgb}{0.5294,0.8819,0}
\definecolor{soc-color8and11}{rgb}{0,0.749,0}
\definecolor{socfbcolor1and8}{rgb}{1,0.1725,0}
\definecolor{socfbcolor2and8}{rgb}{1,0.5112,0}
\definecolor{socfbcolor3and8}{rgb}{1,0.9206,0}
\definecolor{socfbcolor4and8}{rgb}{0,0.749,0}
\definecolor{tech-color1and2}{rgb}{1,0.1725,0}
\definecolor{tech-color2and2}{rgb}{1,0.7221,0}
\definecolor{tech-color3and2}{rgb}{0,0.749,0}
\definecolor{web-color1and4}{rgb}{1,0.1725,0}
\definecolor{web-color2and4}{rgb}{1,0.5112,0}
\definecolor{web-color3and4}{rgb}{1,0.9206,0}
\definecolor{web-color4and4}{rgb}{0,0.749,0}

\definecolor{plotblue}{RGB}	{30,144,255}
\definecolor{plotgreen}{RGB}	{50,205,50}
\definecolor{plotred}{RGB}	{220,20,60}

\definecolor{myyellow}{RGB}{255,255,204}
\definecolor{myred}{RGB}{255,204,204}
\definecolor{myblue}{RGB}{0,200,255}
\definecolor{mygreen}{RGB}{80,220,80}

\definecolor{gray}{RGB}{20,20,20}
\definecolor{greencm}{RGB}{0,153,0}

\definecolor{theblue}{RGB}{0,0,180}

\definecolor{matlabgreen}{rgb}{0,0.6,0}
\definecolor{matlabgray}{rgb}{0.5,0.5,0.5}
\definecolor{matlabmauve}{rgb}{0.58,0,0.82}

\newcommand{\cmark}{\ding{51}}%
\newcommand{\xmark}{\ding{55}}%
\newcommand{\cm}{ {\color{greencm}\cmark}}

\newcommand{\xm}{ {\color{red}\xmark}}

\newcommand{\bspacing}{\begin{spacing}{1.1}}
\newcommand{\espacing}{\end{spacing}}

\newcommand{\bcenum}{\begin{compactenum}[$\star$ \leftmargin=0em]}
\newcommand{\ecenum}{\end{compactenum}}

\newcommand{\benum}{\begin{compactenum}[$\star$ \leftmargin=0em]}
\newcommand{\eenum}{\end{compactenum}}

\newcommand{\benumbullet}{\begin{compactenum}[$\bullet$ \leftmargin=0em]}
\newcommand{\eenumbullet}{\end{compactenum}}

\algrenewcommand{\alglinenumber}[1]{\scriptsize #1} 
\newcommand{\algrule}[1][.2pt]{\par\vskip.5\baselineskip\hrule height #1\par\vskip.5\baselineskip} 

\newlength{\arrayrulewidthOriginal}

\usepackage{adjustbox}
\usepackage{array}

\newcolumntype{H}{>{\setbox0=\hbox\bgroup}c<{\egroup}@{}}

\newcommand{\hboldline}{\noalign{\hrule height 0.5mm}}

\newcommand{\hboldlinesmall}{\noalign{\hrule height 0.4mm}}

\newcommand*\hrulefillvar[1][0.4pt]{\leavevmode\leaders\hrule height#1\hfill\kern0pt}

\newcommand{\algfontsize}{\footnotesize}

\newcommand{\figszhist}{\includegraphics[width=0.33\linewidth]}
\newcommand{\hspacehist}{{\hspace{-4mm}}}

\newcommand\TTT{\rule{0pt}{3.2ex}}
\newcommand\BBB{\rule[-1.4ex]{0pt}{0pt}}

\newcommand\TTTT{\rule{0pt}{3.2ex}}
\newcommand\BBBB{\rule[-3.4ex]{0pt}{0pt}}

\newcommand\TT{\rule{0pt}{2.3ex}}
\newcommand\BB{\rule[-1.0ex]{0pt}{0pt}}

\newcommand\TTZZ{\rule{0pt}{2.3ex}}
\newcommand\BBZZ{\rule[-1.0ex]{0pt}{0pt}}

\colorlet{TufteRed}{red!80!black}
\definecolor{halfgray}{gray}{0.55}

\definecolor{subtleblue}     {rgb}{0.02,0.04,0.48}
\definecolor{subtlered}      {rgb}{0.65,0.04,0.07} 
\definecolor{subtlegreen}    {rgb}{0.06,0.44,0.08}
\definecolor{subtledarkblue} {rgb}{0,.1,.6}
\definecolor{lightsubtleblue}{rgb}{0,.4,.6}
\definecolor{ecru}           {rgb}{1.0,.98823,.95686}

\definecolor{mygreen}{rgb}{0,0.6,0}
\definecolor{mygray}{rgb}{0.5,0.5,0.5}
\definecolor{mymauve}{rgb}{0.58,0,0.82}

\def\helvetica{cmss10 }
\def\helveticabold{cmssbx10 }
\def\helveticaboldoblique{cmssbx10 }
\def\helveticaoblique{cmssi10 }

\def\frutiger{cmss10 }

\def\frutigerbold{cmssbx10 }

\let\frutigerboldcondensed\helveticabold

\helvetica at 6pt

=\frutigerbold at 26pt
= \helvetica at 6pt

  =\frutiger at8pt
  =\helveticaoblique at8pt
  =\frutigerbold at 8pt

=\frutigerboldcondensed at 22pt 
  =\frutiger at 7pt 

  =\frutiger at 6pt
  =\frutiger at 6pt

=\frutigerbold at 8pt
=\helveticaboldoblique at 8pt

=\frutigerbold at 8pt
=\frutigerbold at9pt

  =\frutigerbold at 8pt
  =\frutiger at 7pt
  =\frutigerbold at 7pt  
  =\frutigerbold at 9pt
=\frutiger at 8pt

=\frutiger at 9pt
\frutigerbold at 9pt
\frutigerbold at 9pt

=\frutiger at 7pt
=\helveticabold at 7pt
=\helveticaboldoblique at 7pt

=\frutigerbold at 6pt

\providecommand{\bound}{\ensuremath{B}}

\providecommand{\X}{\ensuremath{{{\rm X}}}}

\providecommand{\bwd}{\ensuremath{{{\sf \it b}}}}

\providecommand{\Vb}{\ensuremath{{{ V_{\bwd}}}}}

\providecommand{\db}{\ensuremath{{{ d_{\bwd}}}}}

\providecommand{\Hg}{\ensuremath{{H}}}
\providecommand{\maxcolor}{\ensuremath{{\mathtt{max}}}}

\providecommand{\dbmax}{ \ensuremath{{{ \theta_{\pi} }} } } 

\providecommand{\col}{ \ensuremath{{{ \sf col}} } } 

\providecommand{\slo}{\ensuremath{{{\textsc{slo}}}}}
\providecommand{\id}{\ensuremath{{{\textsc{id}}}}}

\providecommand{\N}{\ensuremath{N}}
\providecommand{\Nr}{\ensuremath{N_{R}}}

\providecommand{\lb}{\ensuremath{\tilde{\omega}}}
\providecommand{\mc}{\ensuremath{\omega}}

\providecommand{\coloring}{\ensuremath{{\rm \chi}} } 
\providecommand{\localcoloring}{\ensuremath{{\rm \chi_{\ell}}} } 

\providecommand{\coloringnum}{\ensuremath{\rm \max}} 
\providecommand{\colorindex}{\ensuremath{{\sf color}}}
\providecommand{\used}{\ensuremath{{\sf used}}}

\providecommand{\T}{\ensuremath{{\rm T}}}
\providecommand{\K}{\ensuremath{{\rm K}}}
\providecommand{\tr}{\ensuremath{{\rm tr}}}
\renewcommand{\d}{\ensuremath{{\rm d}}}
\providecommand{\Chromatic}{\ensuremath{\chi}}

\providecommand{\f}{\ensuremath{f}}
\providecommand{\fs}{\ensuremath{f^{\star}}}

\providecommand{\global}{\ensuremath{\rm G}}

\providecommand{\p}{\ensuremath{p}}

\providecommand{\dmax}{\ensuremath{\Delta}}

\usepackage{amsthm}

\newtheoremstyle{probstyle}
  {1.4mm}
  {1.4mm}
  {\rm}
  {}
 {\small \sc}
  {:}
  { }
  {}
\theoremstyle{probstyle}

\newtheoremstyle{probst}
  {8pt}
  {8pt}
  {\rm }
  {}
 {\small \rm \bf}
  {}
  {\newline}
  {\thmname{#1}\thmnumber{ #2}\; \sc \thmnote{ #3}}
\theoremstyle{probst}

\usepackage{blindtext}

\newtheoremstyle{mystyle}
  {1.5mm}
  {1.5mm}
  {\itshape}
  {}
  {\bfseries}
  {:}
  { }
  {}
\theoremstyle{mystyle}

\newtheorem{definition}{Definition}[section]

\newtheoremstyle{newstyle}
  {1.0mm}
  {1.0mm}
  {\itshape}
  {}
  {\sf \it \bfseries}
  {:}
  { }
  {}
\theoremstyle{newstyle}

\newtheoremstyle{nonumberplain}
  {-0.2}
  {}
  {\normalfont\upshape}
  {1mm}
  {\normalfont\small\sffamily\sc}
  {:}
  { }
  {}
\theoremstyle{nonumberplain}

\usepackage[breaklinks]{hyperref}

\usepackage{url}

\begin{document}
\vspace{-25mm}

\title{Coloring Large Complex Networks}
\titlerunning{Coloring Large Complex Networks}        

\author{Ryan A. Rossi \and Nesreen K. Ahmed}

\authorrunning{R.A. Rossi and N.K. Ahmed } 

\institute{Ryan A. Rossi and Nesreen K. Ahmed \at
Department of Computer Science, Purdue University \\
West Lafayette, IN 47907 USA \\
\email{\{rrossi, nkahmed\}@purdue.edu} 
} 

\date{}

\maketitle

\vspace{-5mm}
\begin{abstract}
Given a large social or information network, how can we \textit{partition} the vertices into sets (i.e., colors) such that no two vertices linked by an edge are in the same set while minimizing the number of sets used.
Despite the obvious practical importance of graph coloring, existing works have not systematically investigated or designed methods for large complex networks.
In this work, we develop a unified framework for coloring large complex networks that consists of two main coloring variants that effectively balances the tradeoff between accuracy and efficiency.
Using this framework as a fundamental basis, we propose coloring methods designed for the scale and structure of complex networks.
In particular, the methods leverage triangles, triangle-cores, and other egonet properties and their combinations.
We systematically compare the proposed methods across a wide range of networks (e.g., social, web, biological networks) and find a significant improvement over previous approaches in nearly all cases.
Additionally, the solutions obtained are nearly optimal and sometimes provably optimal for certain classes of graphs (e.g., collaboration networks).
We also propose a parallel algorithm for the problem of coloring neighborhood subgraphs and make several key observations.
Overall, the coloring methods are shown to be (i) accurate with solutions close to optimal, (ii) fast and scalable for large networks, and (iii) flexible for use in a variety of applications.
\keywords{network coloring \and unified framework \and greedy methods \and neighborhood coloring \and triangle-core ordering \and social networks} 
\end{abstract}

\section{Introduction}
\label{sec:intro}
We study the problem of graph coloring for complex networks such as social and information networks.
Our focus is on designing (i) accurate coloring methods that are (ii) fast for large-scale networks of massive size.
These requirements lead us to introduce a unified coloring framework that can serve as a basis for investigating and comparing the proposed methods.

Graph coloring is an important fundamental problem in combinatorial optimization with numerous applications including 
timetabling and scheduling~\cite{budiono2012pure}, 
frequency assignment~\cite{sivarajan1989channel,banerjee1996practical}, 
register allocation~\cite{chaitin1982register}, and more recently to study networks of human subjects~\cite{kearns2006experimental,chaudhuri2008network}, among many others~\cite{colbourn2010handbook,moscibroda2008coloring,ni2011coloring,capar2012secret,schneider2011distributed,grohe2013dimension}.
The graph coloring problem consists of assigning colors to vertices such that no two adjacent vertices are assigned identical colors, while minimizing the number of colors.
However, in general, the coloring problem is known to be computationally intractable (NP-hard), even to \textit{approximate} it within $n^{1-\epsilon}$~\cite{garey1979computers}.
Nevertheless, coloring lies at the heart of many applications where the goal is to partition a set of entities into classes where two related entities are not in the same class while also minimizing the number of classes used.

Despite its practical importance in a variety of domains (e.g., engineering, scientific computing), coloring algorithms for complex networks such as social, biological and information networks have received considerably less attention.
Majority of work focuses on graphs that are relatively small, synthetic, or from other domains.
However, these real-world networks (e.g., social networks) are usually sparse with complex structural patterns~\cite{newman2003social,boccaletti2006complex,barabasi2004network,davidson2013network,kleinberg2000navigation,adamic2001search}, while also massive in size and growing at a tremendous rate over time.
For instance, the web graph has well over 1 trillion pages, whereas social networks such as Facebook have hundreds of millions of users.
Unfortunately, coloring algorithms suitable for these large sparse real-world networks have been largely ignored, even despite the significance of coloring and its potential for use in a wide variety of applications.
Furthermore, due to the aforementioned reasons, 
there has yet to be a systematic investigation of coloring and its potential applications.

In terms of social networks, coloring has been used for finding roles (see \cite{everett1991role}), but that work is limited to extremely small instances and does not scale to the requirements of modern social and information networks present in the age of big data.
Others have used coloring to study small controlled groups of human subjects and their behavior~\cite{kearns2006experimental,chaudhuri2008network}.
Nevertheless, coloring methods for large sparse networks have not been proposed, nor has coloring been used for applications in these large networks.

The age of big network data has given rise to numerous opportunities and potential applications for graph coloring including descriptive and predictive modeling tasks.
A few of the possibilities are discussed below.
For instance,
the number of colors, distribution of the size of independent sets, and other properties derived from coloring are useful in tasks such as relational classification (as features)~\cite{sen2008collective,de2008probabilistic}, graph similarity~\cite{berlingerio2013network}, anomaly detection~\cite{akoglu2010oddball,aggarwal2011outlier}, network analysis~\cite{chaoji2008integrated,sun2008two,kang2011spectral,wang2010active}, or for evaluating graph generators, among many other tasks~\cite{sharara2012stability}.
Additionally, vertex or edge induced neighborhoods may also be colored to study various questions; similar to the work of~\citet{ugander2013subgraph} which used neighborhood motifs instead.
Independent sets are also seemingly useful in many applications.
One such application is network sampling, where
vertices/edges may be selected from a large independent set to ensure good network expansion (and of course independence), 
and may be useful for estimating properties efficiently in the age of big data~\cite{al2009musk,ahmed2014graph}.
Indeed, such a sampling strategy would also be particularly useful for machine learning problems such as relational active learning~\cite{sharma2013most}, see the work of~\citet{bilgic2010active}.
It is also easy to find applications in other problem domains, e.g., network A/B testing~\cite{ugander2013graph} which requires running randomized experiments on two independently sampled universes, A and B, to test the effectiveness of new products and marketing campaigns.

Although some recent work has used coloring in small social networks~\cite{enemark2011does,mossel2010reaching},
there has not been any systematic evaluation or comparison of coloring methods for large complex networks of various types.
Further, this recent work also used only small networks.
Moreover, the majority of previous work used a single coloring method and therefore lacked any evaluation or comparison to other coloring methods.
Due to this, the properties and behavior of coloring algorithms for social and information networks are not well understood and are left largely unexplored.
This work attempts to fill this gap by developing a variety of techniques that exploit the structure of these large networks while also being fast and scalable for partitioning the vertices into independent sets.

More specifically,
we address the theoretically and practically important problem of graph coloring with a focus on coloring large complex networks such as social, biological and technological networks.
For this purpose, we develop a flexible framework that serves as a foundation for coloring real-world graphs.
The framework is designed to be fast, scalable, and accurate across a wide variety of networks (i.e., social, biological).
To satisfy these requirements, we relax the constraint of using the minimum number of colors, and instead focus on balancing the competing tradeoffs of \textit{accuracy} and \textit{performance}.
This relaxation provides us a framework that scales linearly with the graph size, while also accurate as demonstrated in Section~\ref{sec:results}.
Using this framework, we propose three classes of coloring methods designed specifically for the scale and the underlying structure of these complex networks.
These include social-based methods, multi-property methods, and egonet-based coloring methods (See Table~\ref{tab:selection-criterion}).
We also adapt previous coloring methods/heuristics that have been widely used on small and/or dense graphs from other domains~\cite{gebremedhin2011colpack,leighton1979graph,matula1983smallest,coleman1983estimation,welsh1967upper,mccormick1983optimal} and unify them under the greedy coloring framework.
This provides us with a basis for comparing our proposed techniques with those traditionally used.
We also develop static and dynamic ordering techniques for coloring based on triangle counts, triangle-cores~\cite{zhang2012extracting,rossi2014pakdd}, and a variety of egonet properties, and demonstrate the effectiveness of these methods using a large collection of networks from a variety of domains including social, biological, and technological networks.

The dynamic triangle ordering techniques proposed here are likely to be of use in other applications and/or problems such as for improving community detection~\cite{blondel2008fast,fortunato2010community}, distance queries~\cite{jiang2014hop}, the maximum clique problem~\cite{prosser2012,carraghan1990exact}, and numerous other problems that rely on an appropriate vertex/edge ordering. 

We also formulated the problem of coloring neighborhood subgraphs and proposed a parallel algorithm that leverages our previous methods.
One key finding is that neighborhoods that are colored using a relatively few number of colors are not well connected, with low clustering and a small number of triangles.
While neighborhood colorings that use a relatively large number of colors have large clustering coefficients and usually contain large cliques.
Nevertheless, we also find linear speedups and many other interesting results (See Section~\ref{sec:neigh-coloring} for further details).

In addition to the technical contributions,
the other aim of this work is a large-scale investigation of coloring methods for these types of networks.
In particular, we compare the three classes of our proposed coloring methods to a wide variety of previous methods that are considered state-of-the-art for 
relatively small and/or dense graphs from other domains.
Using our unified framework as a basis, we systematically evaluate our proposed coloring methods (with past methods) on over 100 networks from a variety of types~\cite{networkrepository2013} including social, biological, and information networks\footnote{In the spirit of reproducible research, the large 100+ collection of benchmark graphs used in this article are available for download at \url{http://www.networkrepository.com}~\cite{networkrepository2013}}.

The types of graphs differ in their size, semantics, structure, and the underlying process governing their formation. 
Overall, we find a significant improve over the previously proposed methods in nearly all cases.
Moreover, the solutions obtained are nearly optimal and sometimes provably optimal for certain classes of graphs (e.g., collaboration networks).
Additionally, the large-scale investigation on 100+ networks revealed a number of useful and insightful observations.
One main finding of this work is that despite the pessimistic theoretical results previously mentioned, large sparse networks found in the real-world can be colored fast and accurately using the proposed methods.

The remainder of this article is organized as follows.
Preliminaries are given in Section~\ref{sec:background}.
Section~\ref{sec:framework} introduces the framework along with the proposed methods while Section~\ref{sec:recolor} proposes the more accurate recolor variant.
In Section~\ref{sec:bounds}, we derive the lower and upper bounds used throughout the remainder of the article.
Section~\ref{sec:results} demonstrates the effectiveness of the proposed methods on over a hundred networks.
Next, Section~\ref{sec:neigh-coloring} formulates the neighborhood coloring problem and proposes a parallel algorithm for coloring neighborhood subgraphs.
We also provide numerous results indicating the scalability and utility of our approach.
Finally, Section~\ref{sec:conclusion} concludes.

\section{Background}
\label{sec:background}

Networks are ubiquitous and can be used to represent data in various domains, from social, biological, and information domains. Facebook is a good live example of a real-world network, where vertices represent people, and edges represent relationships/communications among them. In this section, we start by defining the fundamental graph properties used in the problem of coloring networks. 

Assume $G=(V,E)$ is an undirected graph used to represent some network, such that $V$ is the set of vertices, and $E$ is the set of edges. We use the term $index(v)$ to refer to the index of a vertex $v$. This index represents the unique identifier of a vertex $v$ as it appears in the graph $G$. One simple example of an index could be the unique \emph{userid} assigned to each user by online social network providers (e.g, Facebook). Similarly, we use $d(v)$ to represent the vertex degree, such that $d(v)$ is the number of adjacent vertices (i.e, neighbors) to $v$ in the graph. The concept of a vertex degree could simply describe the number of friends of a Facebook user. 

Another property that proved to be useful particularly in social networks, is transitivity. A transitive edge would mean that if $u$ is connected to $v$ and $v$ is connected to $w$, then $u$ is connected to $w$. In this case $uvw$ represents a triangle in $G$. We use the term $tr(v)$ to refer to the number of triangles incident to a vertex $v$. In common parlance, for a user $x$ in a social network, the number of pairs of friends of $x$ that are also friends themselves would represent the number of triangles. The concept of transitivity can be also generalized to subgraphs with more than three vertices. In this case, every vertex in the subgraph is connected by an edge to every other. These types of subgraphs is typically called cliques. Note that cliques are maximal subgraphs, means that no other vertex in the network can be a member of the clique while preserving the same property that every vertex in the clique is connected to every other. In social networks, the occurrence of cliques indicates highly connected subgroups of users, such as co-workers. 

Cliques are one example of the more generic concept of network groups. In networks, vertices can be divided into   various types of groups or communities that help to explain the underlying network structure. In this section, we introduce two fundamental concepts of network groups related to the problem of coloring networks ($k$-core, and $k$ \emph{triangle-core}). 

A $k$-core is a maximal subgraph of $G$, such that every vertex in the subgraph is connected to at least $k$ others in the subgraph~\cite{matula1983smallest}. The concept of $k$-core was first introduced in ~\cite{Szekeres-1968-coloring}. $k$-cores are useful for various applications in network analysis, such as finding communities and cliques~\cite{rossi2014pmc-www}. A simple algorithm to find the $k$-core of the graph $G$ is to start with the whole graph, and remove any vertices that have degree less than $k$. Clearly, the removed vertices cannot be members of a $k$-core (i.e, a core with order $k$) under any conditions. Note that by removing these vertices, naturally, the connected vertices to the removed ones will reduce their degrees as well. Therefore, the procedure continues until there are no vertices in the graph with degree less than $k$. The output of this procedure is the $k$-core (or $k$-cores) of $G$. 

This procedure can also be repeatedly used to compute the core decomposition of the graph -- this means computing the core number of each vertex $v$. The core number of a vertex (denoted by $K(v)$) is defined as the highest order $k$ of a maximum $k$-core that $v$ can possibly belong to. While simple to implement, this procedure has a worst case runtime of $O(|E|.|V|.log|V|)$. However, the runtime can be efficiently reduced to $O(|V|+|E|)$ by another implementation--which we use in this paper (see more details in~\cite{batagelj2003m}). 

\begin{figure}[b!]
\centering
\includegraphics[width=2.0in,bb=0 0 367 259]{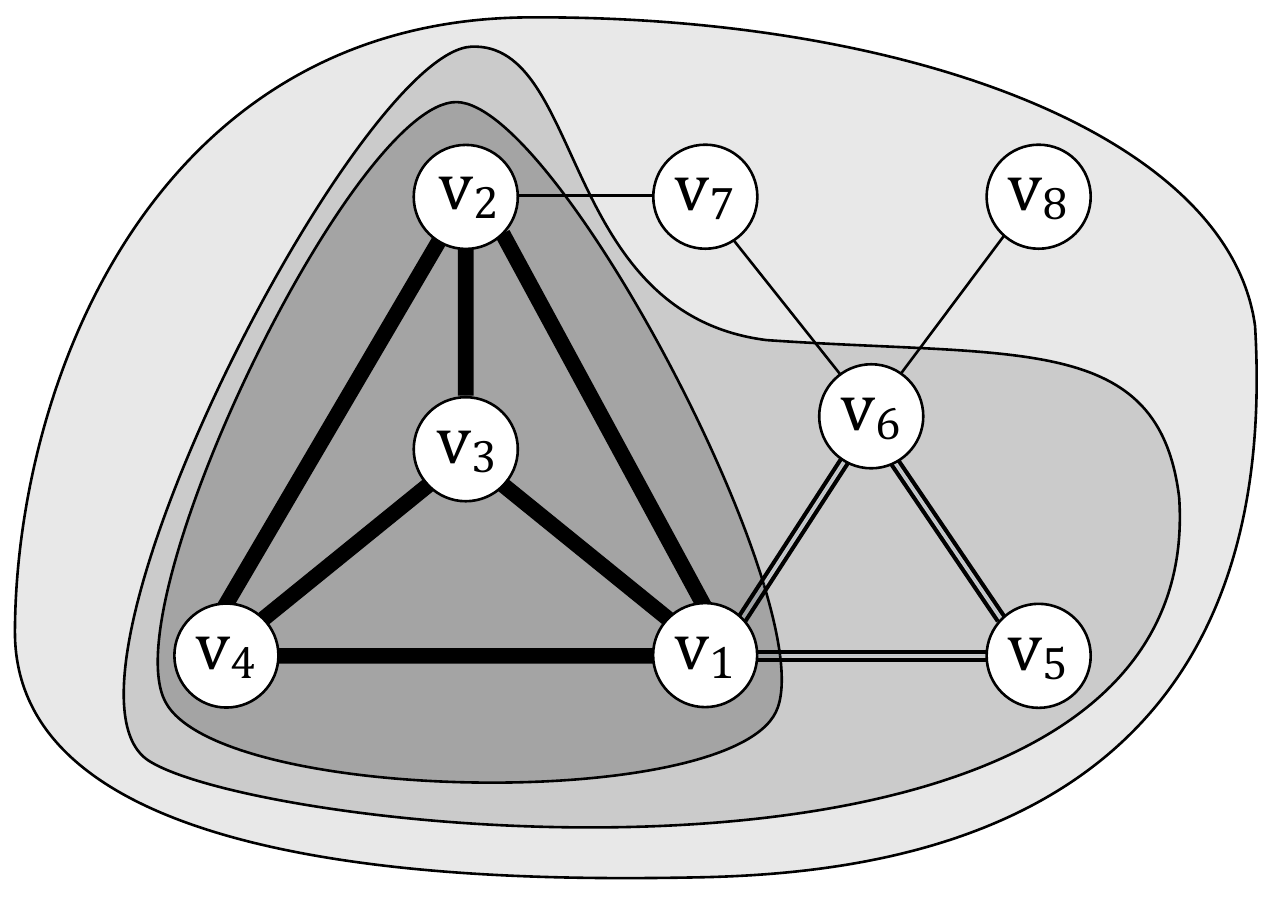}
\caption{
Triangle cores 4, 3, and 2.
A $k$ triangle-core is an edge-induced subgraph of $G$ such that each edge participates in $k - 2$ triangles.
Hence, each clique of size $k$ is contained within a $k$ triangle-core of $G$.
Similarly, the $k$ triangle-core is contained within the $(k - 1)$-core (i.e., the k-1 core from the k-core decomposition).
}\label{fig:triangle-core-example}
\end{figure}

The concept of $k$ \emph{triangle-core} has recently emerged in network analysis research, it was first proposed in~\cite{cohen2009graph}, and improved in~\cite{zhang2012extracting,rossi2014pakdd}.
A $k$ \emph{triangle-core} is an edge-induced subgraph of $G$ such that each edge participates in at least $k-2$ triangles and $k \geq 2$.
A subgraph $H_k = (V|E(F))$ induced by the edge-set $F$ is a \emph{maximal triangle core} of order $k$ if $\forall (u,v) \in F : tr_H(u,v) \geq k-2$, and $H_k$ is the maximum subgraph with this property.
Most importantly, we define the \emph{triangle core number} denoted $T(u,v)$ of an edge $e=(u,v) \in E$ to be the highest order $k$ of a maximum triangle $k$-core that $e$ can possibly belong to. 
See Figure~\ref{fig:triangle-core-example} for further intuition. Computing the triangle core numbers of each edge $e$ in the graph $G$ is called the triangle core decomposition of $G$. In Section~\ref{sec:ordering}, we provide an efficient algorithm for computing the triangle core decomposition with runtime $O(|E|^{3/2})$. 

\section{Greedy Coloring Framework}
\label{sec:framework}
In this section, we present a scalable fast framework for coloring large complex networks and introduce the variations designed for the structure of these large complex networks found in the real-world.

\subsection{Problem Definition}
Let $G=(V,E)$ be an undirected graph. 
A clique is a set of vertices any two of which are adjacent.
The \textit{maximum size of a clique} in $G$ is denoted $\mc(G)$.
An \textit{independent set} $C$ is a set of vertices any two of which are \textit{non-adjacent}, thus, $\forall (v,u) \in C$ iff $(v,u) \not \in E$.
The graph coloring problem consists of assigning a color to each vertex in a graph $G$ such that no adjacent vertices share the same color, minimizing the number of colors used.
More formally, 

\begin{definition}[{\bf Graph Coloring Problem}]
Given a graph $G$, find 
a mapping $\phi : V \rightarrow \{1,...,k\}$ where $\phi(v_i) \not = \phi(v_j)$ for each edge $(v_i,v_j) \in E$.
such that $k$ (the number of colors) is minimum.
\end{definition}

This problem may also be viewed as a partitioning of vertices $V$ into independent sets $C_1, C_2,...,C_k$ where $\{1,2,...,k\}$ are called colors and the sets $C_1,...,C_k$ are referred to as color classes.
Thus, the graph coloring problem is to find the minimum number $k$ of independent sets (or color classes/partitions) required to color the graph $G$.
Nevertheless, graph coloring is NP-hard to solve optimally (on general graphs), and for all $\epsilon > 0$, it is even NP-hard to \textit{approximate} to within $n^{1-\epsilon}$ where $n$ is the number of vertices~\cite{garey1979computers}.

In this work, we relax the strict requirement of partitioning the vertices into the minimum number of independent sets to allow for colorings that are close to the optimal.
This relaxation gives rise to fast linear-time coloring algorithms that perform well in practice (See Section~\ref{sec:results}).
Motivated by this, we describe general conditions for greedy coloring that can serve as a unifying framework in the study of these algorithms.
More formally, we define the greedy coloring framework as follows:

\begin{figure}[t!]
\centering
\includegraphics[width=3.3in,bb=0 0 734 540] {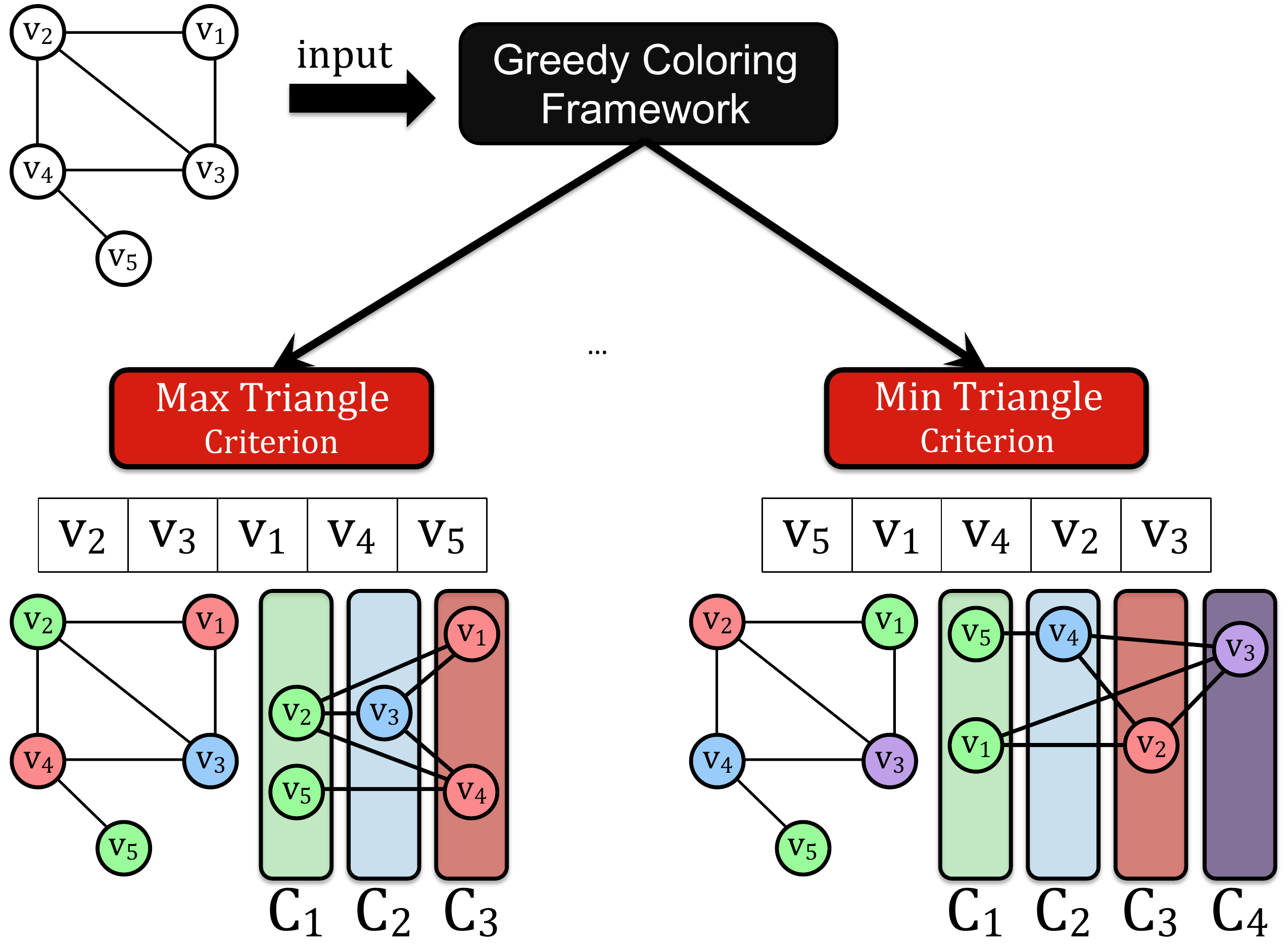}
\caption{
{Greedy Coloring Framework.}
In this graph, we use the number of triangles incident to a vertex $v \in V$ as the selection criterion. On the left, vertices are ordered from largest to smallest using the number of triangles, which results in a greedy coloring that utilizes only three colors.
For this graph, this coloring is also optimal and thus $\Chromatic(G,\pi) = \coloring(G)$.
However, when vertices are ordered from smallest to largest (on the right) results in a coloring that uses four colors.
As an aside, $\dmax(G)+1$ is the maximum number of colors that can be used from any greedy coloring method from the framework and thus $\coloring(G,\pi) \leq \dmax(G)+1$.
In this graph, $\dmax(G)+1 = 4$ and thus, the ordering used on the right is also the worst possible coloring that can be obtained. Notice that in this example, we used the vertex index as the tie breaking strategy, i.e., $v_i$ is ordered before $v_j$ if $i < j$.
We also note that if the proposed repair coloring scheme (Section~\ref{sec:recolor}) were used in the minimum triangle selection criterion, then only three colors would be needed.
Other selection criterion (e.g., degree) may lead to a different vertex ranking and as a consequence the greedy coloring framework may result in an entirely different coloring.
For instance, ranking the vertices by max degree gives $\{v_2, v_3, v_4, v_1, v_5\}$ which differs from the ranking given by max triangle counts.
Further, if two nodes have equal degrees, then we break ties using triangle counts (known as a tie-breaking strategy).
}
\label{fig:greedy-coloring-framework}
\end{figure}

\begin{definition}[{\bf Framework}]
\label{def1}
Given a graph $G=(V,E)$ and a vertex property $f(\cdot)$, the greedy coloring framework
selects the next (uncolored) vertex $v$ to be colored such that 
\[ v = \argmax_{v_i} \f(v_i) \]
The selected vertex $v$ is then assigned to the smallest permissible color.
This process is repeated until all vertices are colored.
\end{definition}

The main intuition of the greedy coloring framework is to color the vertices that are more constrained in their choice of color as early as possible, giving more freedom to the coloring algorithm to use fewer colors, and thus result in a tighter upper bound on the exact number of colors.
As an aside, selecting the vertex that minimizes $f(v)$ usually results in a coloring that uses significantly more colors than the latter.
Notice that a fundamental property of the above greedy coloring framework is that it is both fast and efficient, thus, providing us with a natural basis for investigating the coloring of large real-world networks, which is precisely the scope of this work.

The above definition of the framework uses a selection criterion as the basis for coloring.
Instead, we replace the selection criterion with the more general notion of a vertex ordering.
More specifically, given a graph $G=(V,E)$ and a vertex ordering 
\[ \pi = \{v_1,v_2,...,v_i,...,v_n\} \] 
\noindent of $V$, let $\coloring(G,\pi)$ denote the number of colors used by a \textit{greedy coloring method} that uses the vertex ordering $\pi$ of $G$.
Hence, the greedy coloring framework selects the next vertex to color based on the vertex ordering.
This formalization allows for a more precise characterization of the framework that depends on three components:
\smallskip
\begin{compactenum}[\leftmargin=4em 1.]
\item A graph property $\f(G)$ for selecting the vertices to color
\item The direction in which vertices are selected (e.g., smallest to largest). For instance, $\pi = \{v_1,...,v_n\}$ is from \textit{max to min} if $\f(v_1) \geq \cdots \geq \f(v_n)$, or \textit{min to max} if $\f(v_1) \leq \cdots \leq \f(v_n)$.
\item A tie-breaking strategy for the case when the graph property assigns the same value to two vertices. Suppose $\f(v) = \f(u)$, then $v$ is before $u$ in the ordering $\pi$ if $\fs(v) > \fs(u)$ where $\fs(\cdot)$ is \emph{another} graph property used to break-ties.
\end{compactenum}
\smallskip

Notice that two vertex orderings $\pi_1$ and $\pi_2$ from the graph property $\f(G)$ may significantly differ in the number of colors used in a greedy coloring (i.e., $\coloring(G,\pi_1) \not = \coloring(G,\pi_2) + \epsilon$).
This is due to the direction of the ordering (smallest to largest) and tie-breaking strategy selected.
Consequently, a specific graph property $\f(\cdot)$ defines a class of orderings where the order direction (from max to min) and tie-breaking strategy ($\fs(\cdot)$) represent a specific member of that class of orderings.
Note that in general $\f(G)$ can be thought simply as a function for obtaining an ordering $\pi$.

In addition, we also define a few relationships between the graph parameters introduced thus far.
Clearly, $\coloring(G,\pi)$ from a greedy coloring method is an upper bound on the \emph{exact number} of colors required, denoted by $\Chromatic(G)$, i.e., the minimum number of colors required for coloring $G$.
Further, let $\omega(G)$ be the size of the maximum clique in $G$, which is also a lower bound on the minimum number of colors required to color $G$.
This gives the following relationship:
\[ \mc(G) \leq \Chromatic(G) \leq \coloring(G,\pi) \leq \dmax(G)+1 \]
\noindent where $\dmax(G)$ is the maximum degree of $G$.

An example of the framework is shown in Figure~\ref{fig:greedy-coloring-framework}.
This illustration uses a proposed triangle selection criterion, which is shown later in Section~\ref{sec:results} to be extremely effective for large social and information networks.

\begin{table}[t!]
\small
\caption{
\textbf{Methods used as selection criterion.}
The previously proposed methods are unified under the framework and categorized into three general classes (i.e., index-based, degree-based methods, and degree distance-2-based methods). Many of these greedy coloring methods are considered the state-of-the-art for small and/or relatively dense graphs from other domains~\cite{gebremedhin2011colpack}, and thus used as a baseline for evaluating our methods. However, this work proposes three main classes of methods for large complex networks including social-based methods, multi-property methods, and egonet-based methods.}
\label{tab:selection-criterion}
\begin{tabular*}{1.0\linewidth}
{@{\extracolsep{\fill}} H r p{80mm} H}
\hline
\toprule
\TTT \BBB  
&
\textbf{Name}
& 
{
\textbf{Property}
\; $\f(\cdot)$} &
\textbf{Definition} 
\\
\bottomrule \hline
\TT \BB
\TTT \BBB & \textsc{natural}  & $f(v) = index(v)$, select next uncolored vertex in the order in which vertices appear in $G$ \\
\TTT \BBB & \textsc{rand}  & $f(v) \sim Uni(1,|V|)$, select the next uncolored vertex uniformly at random from the uncolored vertices\\
\midrule
\TTT \BBB & \multicolumn{2}{l}{\textbf{\footnotesize 
 Degree distance-1 methods}} \\
\midrule
\TTT \BBB  & \textsc{deg} & $\f(v) = d(v)$, no. adjacent vertices of $v$ in $G$ (i.e., degree ) \\
\TTT \BBB  & \textsc{dlf} & $\f(v) = $ no. {\emph uncolored} adjacent vertices of $v$ \\
\TTT \BBB  & \textsc{ido} & $\f(v) = $ no. {\emph colored} adjacent vertices of $v$ (i.e., $|\N_{c}(v)|$) \\
\TTT \BBB & \textsc{kcore (\textsc{slo})}  & $\f(v) = K(v)$, k-core number of $v$ \\ 
\midrule
\TTT \BBB & \multicolumn{2}{l}{\textbf{\footnotesize 
 Degree distance-2 methods}} \\
\midrule
\TTT \BBB &\textsc{dist-two-deg} & $\f(v) = |\N_{hops=2}(v)|$, no. unique vertices {\emph 2-hops} away of $v$ in $G$ \\ 
\TTT \BBB &\textsc{dist-two-dlf} & $\f(v) = $ no. unique uncolored vertices {\emph 2-hops} away of $v$ \\ 
\TTT \BBB &\textsc{dist-two-ido} & $\f(v) = $ no. unique colored vertices {\emph 2-hops} away of $v$ \\
\midrule
\TTT \BBB & \multicolumn{2}{l}{\textbf{\footnotesize 
 Social-based methods}} \\
\midrule
\TTT \BBB & \textsc{tri}  & $\f(v) = tr(v)$, no. triangles of $v$ in $G$ \\ 
\TTT \BBB & \textsc{tcore-max}  & $\f(v) = \max_{w \in \N(v)} \T(v,w)$, triangle core number of $v$  \\ 
\midrule
\TTT \BBB & \multicolumn{2}{l}{\textbf{\footnotesize 
 Multi-property methods}} \\
\midrule
\TTT \BBB & \textsc{kcore-deg}  & $f(v) = K(v) \cdot d(v)$ \\ 
\TTT \BBB & \textsc{tri-deg}  & $f(v) = tr(v) \cdot d(v)$ \\ 
\TTT \BBB & \textsc{tri-kcore}  & $f(v) = tr(v) \cdot K(v)$ \\ 
\TTT \BBB & \textsc{tri-kcore-deg}  & $f(v) = tr(v) \cdot K(v) \cdot d(v)$ \\
\midrule
\TTT \BBB & \multicolumn{2}{l}{\textbf{\footnotesize 
 Egonet-based methods}} \\
\midrule
\TTT \BBB &\textsc{deg-vol}   & $\f(v) = \sum_{w \in \N(v)} \d(w)$ \\ 
\TTT \BBB & \textsc{kcore-vol} & $\f(v) = \sum_{w \in \N(v)} \K(w)$ \\ 
\TTT \BBB & \textsc{tri-vol}  &  $\f(v) = \sum_{w \in \N(v)} tr(w)$ \\ 
\TTT \BBB & \textsc{tcore-vol}  & $\f(v) = \sum_{w \in \N(v)} \T(v,w)$ \\ 
\midrule
\TTT \BBB & \textsc{kcore-deg-vol}  & $f(v) = \sum_{w \in \N(v)} tr(w) \cdot d(w)$ \\ 
\TTT \BBB & \textsc{tri-kcore-vol}  & $f(v) = \sum_{w \in \N(v)} tr(w) \cdot K(w)$ \\ 
\TTT \BBB & \textsc{tri-kc-deg-vol}  & $f(v) = \sum_{w \in \N(v)} tr(w) \cdot K(w) \cdot d(w)$ \\ 
\bottomrule \hline
\end{tabular*}
\end{table}

\subsection{Ordering Techniques}
\label{sec:ordering}

In this section, we first review the previous methods used for coloring relatively small and/or dense graphs from other domains (see~\citet{gebremedhin2011colpack}), which are unified under our coloring framework.
Many are considered state-of-the-art greedy coloring techniques and shown to perform reasonably well for those types of graphs. 
Despite the past success of these methods, they are not as well suited for large sparse complex networks (e.g., social, information, and technological networks) as demonstrated in this work.
As a result, we propose three classes of methods for greedy coloring based on well-known fundamental properties of these large complex networks.
In particular, we propose social-based methods, multi-property, and methods based on egonet properties, which are shown later in Section~\ref{sec:results} to be more effective than the state-of-the-art techniques used in coloring graphs from other domains.
A summary and categorization of these methods are provided in Table~\ref{tab:selection-criterion}. 

\begin{figure}[b!]
\vspace{-4mm}
\begin{center}
\begin{minipage}{1.0\linewidth}
\begin{algorithm}[H]
\small
\caption{\small Basic Greedy Coloring 
}
\label{alg:greedy-coloring}
{\algfontsize
\bspacing
\begin{algorithmic}[1]
\Procedure {GreedyColoring}{$G$, $\pi$}
\State Initialize data structures 						\label{algline:initialize}			
\For{$v \in \pi$ in order} \label{algline:ordering}											
		\For{$w \in \N(v)$}			 $\used(\colorindex(w)) \leftarrow v$		
									\label{algline:mark}
		\EndFor
		\State $k \leftarrow \min\{ i > 0 : \used(i) \not= v \}$  								
									\label{algline:smallest-color}
		\If{$k > \coloringnum$} 
				$\coloringnum \leftarrow k$																		
		\EndIf
		\State $\colorindex(v) \leftarrow k$ 			\label{algline:assign-color}								
\EndFor
\EndProcedure
\end{algorithmic}
\espacing
}
\end{algorithm}
\end{minipage}
\end{center}
\vspace{-4mm}
\end{figure}

\smallskip \noindent {\normalsize \theoremfont Index-based Methods:} 
The simplest arbitrary ordering techniques under the \textit{sequential greedy coloring framework} are natural ordering (\textsc{natural}) and random ordering (\textsc{rand}). The natural ordering (\textsc{natural}) method selects the vertices to be colored in their natural order as they appear in the input graph $G$, i.e., $v_1,v_2,...,v_n$. We also define the random ordering (\textsc{rand}) as the method that selects the vertices to be colored randomly. Therefore, the (\textsc{rand}) method selects a vertex by drawing an \emph{uncolored} vertex uniformly at random without replacement from $V$.

\smallskip \noindent {\normalsize \theoremfont  Degree Methods:} 
The four simplest, yet most popular ordering methods under the \textit{sequential greedy coloring framework} (Section~\ref{sec:framework}) are all based on \textit{vertex degree}.
Specifically, we use the degree ordering \textsc{deg}, the incidence degree ordering (\textsc{ido}), the dynamic-largest-first (\textsc{dlf}), and the k-core ordering (\textsc{kcore}) (a.k.a smallest-last ordering (\textsc{slo})). 
First, the degree ordering (\textsc{deg})~\cite{welsh1967upper} orders vertices from largest to smallest by their \emph{static degree} as it appears in $G$. Second, the incidence-degree ordering (\textsc{ido})~\cite{coleman1983estimation} dynamically orders vertices from largest to smallest by their \emph{back degree}, such that the back degree of $v$ is the number of its \emph{colored} neighbors. In this case, the incidence-degree method initially starts with all vertices with back degree equal to zero, and initially selects an arbitrary vertex $v$ to color. Then, all the neighbors of $v$ will increase their back degree by one, and the next vertex with largest back degree will be selected for coloring. This process continues until all vertices are colored. Third, in contrast to the incidence-degree method (\textsc{ido}), the dynamic-largest-first (\textsc{dlf})~\cite{gebremedhin2011colpack} dynamically orders the vertices by their \emph{forward degree} from largest to smallest, where the forward degree of $v$ is the number of its \emph{uncolored} neighbors. Thus, the dynamic-largest-first method initially starts with all vertices with forward degree equal to their original degree in $G$, and selects the first vertex $v$ to color, such that $v$ has the maximum degree in $G$ (i.e, $d(v)=\dmax(G)$). Consequently, all the neighbors of $v$ will decrease their forward degree by one, and the vertex with the largest forward degree will be selected next to be colored.

Finally, the $k$-core ordering (\textsc{kcore}) (also known as the smallest-last ordering (\textsc{slo}) ~\cite{matula1983smallest}) orders the vertices from lowest to highest by their $k$-core number (refer to Section~\ref{sec:background} for definition). 
The $k$-core ordering method (a.k.a smallest-last ordering) was proposed in~\cite{matula1983smallest}, based on the concept of $k$-core decomposition, to find a vertex ordering of a finite graph $G$ that optimizes the coloring number of the ordering in linear time, by repeatedly removing the vertex of smallest degree. The $k$-core ordering dynamically orders the vertices by their \emph{forward degree} from smallest to largest, where the forward degree of $v$ is the number of its \emph{uncolored} neighbors. The method initially starts with all vertices with forward degree equal to their original degree in $G$, and selects the first vertex $v$ to color, such that $v$ has the smallest degree in $G$ (i.e, $\d(v)=\delta(G)$). Thus, all the neighbors of $v$ will decrease their forward degree by one, and the vertex with the next smallest forward degree will be selected for coloring. The output of this method is the vertex ordering for the coloring number, which is equivalent to ordering vertices by their $k$-core number as defined in~\cite{Szekeres-1968-coloring}. 

\begin{table}[b!]
\vspace{-0mm}
\centering
\footnotesize
\caption{
\textbf{Dynamic Ordering Methods.}
Summary of the main dynamic degree-based and dynamic triangle-based ordering methods.
Note that \textsc{slt} and \textsc{tcore} are used interchangeably.
For convenience, let $e$ denote an edge $(v,u)$.
}
\label{table:dynamic-ordering-techniques}
\scriptsize
\begin{tabularx}{.70\linewidth}
{l c !{\vrule width 0.40mm}  c | c | c}
\TTT \BBB  & & \multicolumn{3}{c}{\textbf{\normalsize \sf Operations}} \\

\multicolumn{2}{c !{\vrule width 0.40mm}}{\textbf{Methods}}  \TTT \BBB   
& \textsf{Initialization}
& \textsf{Find} 
& \textsf{Update}
\\ \hboldlinesmall

\multirow{4}{*}{\Rot{90}{\textbf{\footnotesize Degree}}}
\hspace{1mm}
& \id
& \TTTT \BBBB  $\db(v) = 0$ 
& $v=\stackMath\stackunder[3pt]{\max}{\scriptstyle w \in \Vb} \; \db(w)$
& $\db(w) \leftarrow \db(w) + 1$
\\ \cline{2-5}

& \slo
& \TTTT \BBBB  $\db(v) = d(v)$ 
& $v = \stackMath\stackunder[3pt]{\min}{\scriptstyle w \in \Vb} \; \db(w)$
& $\db(w) \leftarrow \db(w) - 1$
\\ \hboldlinesmall

\multirow{7}{*}{\Rot{90}{\textbf{\footnotesize Triangles}}}
& \textsc{it}
& \TTTT \BBBB  $T(e_i) = 0$ 
& $e_i = \stackMath\stackunder[3pt]{\max}{\scriptstyle e_j \in E_b}\; T(e_j)$
& $\T(e_j) \leftarrow \T(e_j) + 1$
\\ \cline{2-5}

& \textsc{slt}
& \TTTT \BBBB  $T(e_i) = \tr(e_i)$ 
& $e_i = \stackMath\stackunder[3pt]{\min}{\scriptstyle e_j \in \E_b}\; T(e_j)$
& $\T(e_j) \leftarrow \T(e_j) - 1$
\\ \cline{2-5}

& \textsc{lft}
& \TTTT \BBBB  $T(e_i) = \tr(e_i)$ 
& $e_i = \stackMath\stackunder[3pt]{\max}{\scriptstyle e_j \in \E_b}\; T(e_j)$
& $\T(e_j) \leftarrow \T(e_j) - 1$
\\ \hboldlinesmall
\end{tabularx}
\end{table}

These methods (including \textsc{kcore}) were found to be superior to others, especially for forests and a few types of planar graphs~\cite{gebremedhin2011colpack}.
We also use these as baselines for evaluating our proposed methods (see Section~\ref{sec:results}).

\smallskip \noindent {\normalsize \theoremfont Distance-2 Degree Methods:}
We note that the degree-based methods were defined on the \emph{1-hop} away neighbors of each vertex $v \in V$. 
These methods can also be extended for the unique \emph{2-hop} away neighbors of each vertex $v \in V$~\cite{mccormick1983optimal}, we call these methods distance-2 degree ordering (\textsc{dist-two-deg}), distance-2 incidence degree ordering (\textsc{dist-two-id}), distance-2 dynamic largest first ordering (\textsc{dist-two-dlf}), and distance-2 $k$-core ordering (\textsc{dist-two-kcore}) respectively.

\smallskip \noindent {\normalsize \theoremfont Social-based Methods:}
While the degree-based methods were shown to perform well in the past, in this paper, we compare them to other social-based orderings such as triangle ordering (\textsc{tri}), and triangle-core ordering (\textsc{tcore}). 

First, the triangle ordering (\textsc{tri}) method orders vertices from largest to smallest by the number of triangles they participate in, i.e. $f(v)=tr(v)$ where $tr(v)$ can be computed fast and in parallel using Alg~\ref{alg:parallel-vertex-triangles}.
Other triangle-based quantities such as clustering coefficient may also be used and computed fast and efficiently using Alg~\ref{alg:parallel-vertex-triangles}. 
Thus, the triangle ordering initially selects the vertex $v$ with the largest number of triangles centered around it. This process continues until all vertices are colored. The intuition behind triangles in social networks is that vertices tend to cluster, and therefore, triangles were extensively used to measure the number of vertices adjacent to $v$ that are also linked together (as explained in Section~\ref{sec:background}). We conjecture that ordering vertices from largest to smallest by their triangle number would give a chance to those vertices that are more constrained in their choices of color to be colored first than those that have more freedom (as we explained earlier).

{
\algblockdefx[parallel]{ParFor}{EndPar}
[1][]{$\mathtt{\bf parallel} \mathbf{for each}$ #1}
{$\textbf{parallel end}$}
\begin{figure}[h!]
\vspace{-4mm}
\begin{center}
\begin{minipage}{1.0\linewidth}
\vspace{-2mm}
\begin{algorithm}[H]
\small
\caption{Parallel Vertex Triangle Counting}
\label{alg:parallel-vertex-triangles}
{\algfontsize
\begin{algorithmic}[1]
\bspacing
\Procedure {ParallelVertexTriangles}{$G=\left (V,E\right )$}
\State Initialize arrays
\For{{\bf each} $v \in V$ {\bf in parallel}}
		\For{{\bf each} $u \in N(v)$} 	$X(u) \leftarrow v$ \Comment{s.t. $v > 0$} 
		\EndFor
		\For{{\bf each} $u \in N(v)$}	
  				\For{{\bf each} $w \in N(u)$}
  						\If{$v = w$} \textbf{continue} \EndIf
  						\If{$X(w) = v$}  $\tr(v) \leftarrow \tr(v) + 1$
  						\EndIf
				\EndFor
		\EndFor
\EndFor
\EndProcedure
\espacing
\end{algorithmic}}
\end{algorithm}
\end{minipage}
\end{center}
\vspace{-4mm}
\end{figure}

Second, the triangle-core ordering (\textsc{tcore}) method orders vertices from largest to smallest by their triangle core number (as explained in Section~\ref{sec:background}). 
Using the triangle core numbers, we obtain an ordering and use it to determine the next vertex $v$ (or edge) to color, using the criteria: $\f(v) = \max_{w \in \N(v)} \T(v,w)$, where $\N(v)$ is the set of neighbors of vertex $v$, and $\T(v,w)$ is the triangle core number of the edge $(v,w) \in E$.
Notice that triangle core ordering is comparable to $k$-core ordering, however, instead of removing a vertex and its edges at each iteration, we remove an edge and its triangles.
This gives rise to a variety of ordering methods based on the fundamental notion of removing edges and their triangles.
We call these dynamic triangle ordering methods and provide a summary of the main ones in Table~\ref{table:dynamic-ordering-techniques} as well as a comparison with a few of the dynamic degree-based methods.
Let us note that any edge-based quantity may be used for ordering vertices (and vice-versa).
For instance, \textsc{tcore-max} defined in Table~\ref{tab:selection-criterion} computes for every vertex $v$ in the graph, the maximum triangle core number among the (1-hop)-away-neighbors of $v$.

The proposed triangle ordering template is shown in Alg~\ref{alg:triangle-ordering-template} and the key operations are also summarized in Table~\ref{table:dynamic-ordering-techniques}.
The backward (or forward) triangle counts are initialized in Line~\ref{alg-line:initialize}.
For \textsc{slt}, \textsc{ParallelEdgeTriangles} shown in Alg~\ref{alg:parallel-triangles} is used to initialize the triangle counts. 
Next, line~\ref{alg-line:bucket-sort} adds $(v,u)$ to the bucket consisting of the edges with $T(v,u)$ triangles which is denoted $\mathtt{bin}[T(v,u)]$.
Hence, the edges are ordered in $O(|E|)$ time using a bucket sort.
Note that if \textsc{it} is used then this step can be skipped since each edge $(v,u)$ is initialized as $T(v,u) = 0$.

The triangle ordering begins in line~\ref{alg-line:while} by ensuring $|E| > 0$ where $E$ initially consists of all edges in $G$.
At each iteration, a single edge $(v,u)$ is removed from $E$.
Line~\ref{alg-line:find} finds the edge $(v,u)$ with the smallest $T(v,u)$ or largest $T(v,u)$, see Table~\ref{table:dynamic-ordering-techniques} for the variants.
The neighbors of $u$ that remain in $E$ are marked in line~\ref{alg-line:mark-neighbors} with the unique edge identifier $e_i$ of $(v,u)$ (to avoid resetting the array).
In line~\ref{alg-line:for-edge-neighbors}, we iterate over the triangles that $(v,u)$ participates, i.e., the pairs of edges $(v,w)$ and $(u,w)$ that form a triangle with $(v,u)$.
Since the neighbors of $u$ are marked in $X$, then a triangle is verified by checking if each neighbor $w$ of $v$ has been marked in $X$, if so then $u,v,w$ must form a triangle.
Line~\ref{alg-line:remove} sets $\mathtt{bin}[T(v,w)] \leftarrow \mathtt{bin}[T(v,w)] \setminus (v,w)$ and $\mathtt{bin}[T(u,w)] \leftarrow \mathtt{bin}[T(u,w)] \setminus (u,w)$, removing $(v,w)$ and $(u,w)$ from their previous bins.
Next, the triangle counts of $(v,w)$ and $(u,w)$ are updated in line \ref{alg-line:update-triangles} using an update rule from Table~\ref{table:dynamic-ordering-techniques}.
Afterwards, line~\ref{alg-line:update-bin} adds the edges to the appropriate bin, i.e., $\mathtt{bin}[T(v,w)] \leftarrow \mathtt{bin}[T(v,w)] \cup (v,w)$ and $\mathtt{bin}[T(u,w)] \leftarrow \mathtt{bin}[T(u,w)] \cup (u,w)$.
This is repeated for each pair of edges $(v,w)$ and $(u,w)$ that form a triangle with $(v,u)$.
Finally, line~\ref{alg-line:remove-edge} implicitly removes the edge $(v,u)$ from $E$.

\begin{figure}[h!]
\vspace{-6mm}
\begin{center}
\begin{minipage}{1.0\linewidth}
\begin{algorithm}[H]
\small
\caption{Dynamic Triangle Ordering Template}
\label{alg:triangle-ordering-template}
{\algfontsize
\bspacing
\begin{algorithmic}[1]
\For{{\bf each} $(v,u) \in E$ {\bf in parallel}}
		\State $T(v,u) \leftarrow \textsf{Initialize}(v,u)$
		\label{alg-line:initialize} 
		
		\State $\mathtt{bin}[T(v,u)] \leftarrow \mathtt{bin}[T(v,u)] \cup (v,u)$ 
		\label{alg-line:bucket-sort}
\EndFor
\While{$|E| > 0$}													\label{alg-line:while} 	
		\State \textsf{Find} the edge $(v,u)$ with  $\min \{ T(v,u)\}$ (or $\max \{T(v,u)\}$) \label{alg-line:find}
		\State Add the edge $(v,u)$ to the back of $\pi$ \label{alg-line:ordering}
		\For{{\bf each} $w \in \N(v)$ that remain} 	$X(w) \leftarrow e_i$ 				\label{alg-line:mark-neighbors}
		\EndFor
  		\For{{\bf each} $w \in \N(u)$ such that $X(w) = e_i$} 																		\label{alg-line:for-edge-neighbors} 		
  								\State Remove $(v,w)$ and $(u,w)$ from $\mathtt{bin}$ \label{alg-line:remove}
    							\State \textsf{Update} $T(v,w)$ and $T(u,w)$		\label{alg-line:update-triangles}
    							\State Add $(v,w)$ and $(u,w)$ to the appropriate bin \label{alg-line:update-bin}
  		\EndFor
  		\State $E \leftarrow E \setminus \{(v,u)\}$	\label{alg-line:remove-edge}
\EndWhile
\State {\bf return} $\pi$
\end{algorithmic}
\espacing
}
\end{algorithm}
\vspace{-8mm}
\begin{algorithm}[H]
\small
\caption{Parallel Edge Triangle Counting}
\label{alg:parallel-triangles}
{\algfontsize
\begin{algorithmic}[1]
\bspacing
\Procedure {ParallelEdgeTriangles}{$G=\left (V,E\right )$}
\State Initialize arrays
\For{{\bf each} $(v,u) \in E$ {\bf in parallel}}
		\For{{\bf each} $w \in N(v)$} 	$X(w) \leftarrow \text{edge pos of (v,u)}$ \EndFor
  		\For{{\bf each} $w \in N(u)$}
  				\If{$v = w$} \textbf{continue} \EndIf
  				\If{$X(w) = \text{edge pos of (v,u)}$} 
  						\State $\tr(v,u) \leftarrow \tr(v,u) + 1$
  				\EndIf
		\EndFor
\EndFor
\EndProcedure
\espacing
\end{algorithmic}}
\end{algorithm}
\end{minipage}
\end{center}
\vspace{-6mm}
\end{figure}

\smallskip \noindent {\normalsize \foliofont\theoremfont Egonet-based Methods:}
An egonet is the induced subgraph centered around a vertex $v$ and consists of $v$ and all its neighbors $\N(v)$. Assume we are given an arbitrary \textit{graph property} $\f(\cdot)$ (e.g., triangle-cores, number of triangles) computed over the set of neighbors of $v$, i.e., $\N(v)$, we define an egonet ordering criterion for a vertex $v$ as $\sum_{w \in \N(v)} \f(w)$. In addition, besides using the ${\rm sum}$ operator over the egonet, one may use other relational aggregators such as $\min$, ${\rm max}$, ${\rm var}$, ${\rm avg}$, among many others.

\smallskip \noindent {\normalsize \foliofont\theoremfont Multi-property Methods:}
We also propose ordering techniques that utilize multiple graph properties.
For instance, the vertex to be colored next may be selected based on the product of the vertex degree and $k$-core number, i.e., $f(v) = K(v) \cdot d(v)$.

\subsection{Algorithm and Implementation}
\label{sec:algorithm}

This section describes the algorithms and implementation.
The graph is stored using $O(|E| + |V|)$ space in a structure similar to Compressed Sparse Column (CSC) format used for sparse matrices~\cite{tewarson1973sparse}.
If the graph is small and/or dense enough, then it is also stored as an adjacency matrix for constant time edge lookups.
Besides the graph, the algorithm uses two additional data structures.
In particular, let $\colorindex$ be an array of length $n$ that stores the color assigned to each vertex, i.e., $\colorindex(v)$ returns the color class assigned to $v$.
Additionally, we also have another array to mark the colors that have been assigned to the neighbors of a given vertex and thus we denote it as $\used$ to refer to the colors ``used'' by the neighbors.

The algorithmic framework for greedy coloring is shown in Alg~\ref{alg:greedy-coloring}.
For the purpose of generalization, we assume the vertex ordering $\pi$ is given as input and computed using a technique from Section~\ref{sec:ordering}.

The algorithm starts by initializing each entry of $\colorindex$ with $0$.
We also initialize each of the entries in $\used$ to be an integer $x \not\in V$ (i.e., an integer that does not match a vertex id).
The greedy algorithm starts by selecting the next vertex $v_i$ in the ordering $\pi$ to color.
For each vertex $v_i$ in order, we first iterate over the neighbors of $v_i$ denoted $w \in N(v)$, and set $\used(\colorindex(w)) = v_i$ as shown in Line~\ref{algline:mark}.
This essentially marks the colors that have been used by the neighbors.
Afterwards, we sequentially search for the minimal $k$ such that $\used(k) \not= v_i$ (in Line~\ref{algline:assign-color}).
Line~\ref{algline:smallest-color} assigns this color to $v_i$, hence $\colorindex(v_i) = k$.
Upon termination, $\colorindex$ is a valid coloring and the number of colors is $\coloring(G,\pi) = \argmax_{v \in P} \colorindex(v)$.
We denote $\coloring(G,\pi)$ as the number of colors from a greedy coloring algorithm that uses the ordering $\pi$ of $V$, which is easily computed in $O(1)$ time by maintaining the max color assigned to any vertex.

Note that in Line~\ref{algline:mark}, the color of $w$ (a positive integer) is given as an index into the $\used$ array and marked with the label of vertex $v_i$.
This trick allows us to avoid re-initializing the $\used$ array after each iteration over a vertex $v_i \in \pi$ -- the outer for loop.
Hence, if $w$ has not yet been assigned a color, i.e., $\colorindex(w) = 0$, then $used(0)$ is assigned the label of $v_i$, and since $0$ is an invalid color, it is effectively ignored.
In addition, each entry in $\used(k)$, $1 \leq k \leq \dmax+1$ must initially be assigned an integer $x \not \in \pi$.

\subsection{Complexity}
\label{sec:complexity}

The storage cost is only linear in the size of the graph, since CSC takes $O(|E| + |V|)$ space, the vertex-indexed array $\colorindex$ costs $O(|V|)$, and $\used$ costs $O(\dmax + 1)$ space.
For the ordering methods, degree and random take $O(|V|)$ time, whereas the other ``dynamic degree-based'' techniques such as \textsc{kcore} have a runtime of $O(|E|)$ time.
The other ordering techniques that utilize triangles and triangle-cores take $O(|E|^{3/2})$ time in the worst-case, but are shown to be much faster in practice.
Importantly, we also parallelize the triangle-based ordering methods by computing triangles independently for each vertex or edge.
We also note the distance two ordering methods are just as hard as the triangle ordering methods, yet perform much worse as shown in Section~\ref{sec:results}.
Finally, the greedy coloring framework has a runtime of  $O(|V| + |E|)$ and $O(|E|)$ for connected graphs.

\begin{figure}[h!]
\centering
\includegraphics[width=2.4in,bb=0 0 266 266]{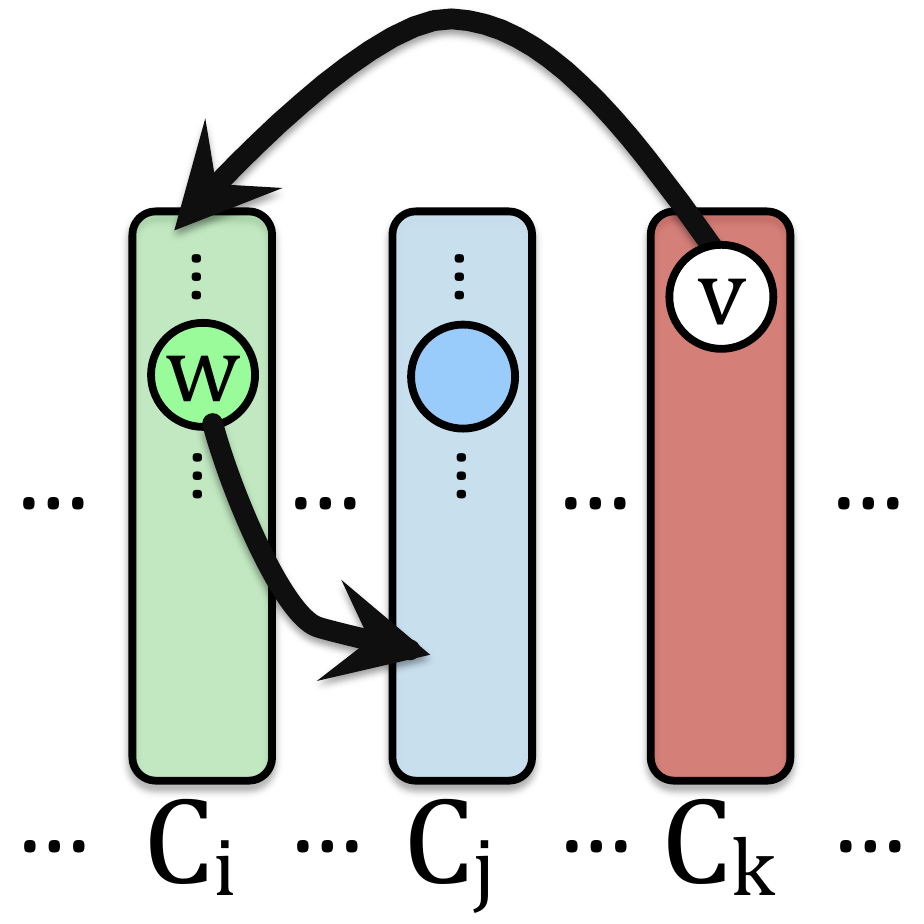}
\caption{
{Repair Coloring.}
Suppose $v$ is the vertex to be recolored since it is assigned to a new color class $C_k$, then we
find a color class $C_i$ where $v$ is adjacent to a single vertex $w$ (i.e., $\N(v) \cap C_i = \{w\}$).
Now, we find a color class $C_j$ s.t.  $j > i$ and $w$ is not adjacent to any vertex in $C_j$, i.e. $|\N(w) \cap C_j| = \emptyset$.
If such a color class exists, then $w$ is removed from $C_i$ and assigned to $C_j$.
As a result of this reassignment, $v$ can now be assigned to the $C_i$ color class, therefore reducing the number of colors by 1.
}\label{fig:recolor}
\end{figure}

\section{Recolor Variant}
\label{sec:recolor}

This section proposes another coloring variant that attempts to recolor vertices to reduce the number of colors.
The variant is effective while also fast for large real-world networks.

\subsection{Algorithm}
The recoloring variant is shown in Alg.~\ref{alg:greedy-coloring-repair}.
This variant proceeds in a similar manner as the basic coloring algorithm from Section~\ref{sec:algorithm}.
The difference is that if a vertex is assigned an entirely new color $k$ (i.e., number of colors used in the coloring increases), then an attempt is made to reduce the number of colors.
Using this as a basis for \textsc{recolor} ensures that the algorithm is fast, taking advantage of only the most promising situations that arise.

Suppose the next vertex $v$ in the ordering is assigned a new color $k$ and thus $C_k = \{v\}$, then we attempt to reduce the number of colors by reassigning an adjacent vertex $u$ that was assigned a previous color $i$ such that $i < k$.
Hence, if $|C_i \cap N(v)| = 1$, then $C_i$ contains a single adjacent vertex of $v$ (i.e., a single conflict), and thus, we attempt to recolor $u$ by assigning it to the minimum color $j$ such that $i < j < k$ and $C_j \cap N(u) = \emptyset$.
This arises due to the nature of the sequential greedy coloring and is formalized as follows:
Given vertices $v$ and $u$ assigned to the $i^{th}$ and the $j^{th}$ colors, respectively, where $v$ is colored first and $i < j$, 
then since $v$ is assigned the minimum possible color, then we know the colors less than $i$ are invalid, however, $v$ could potentially be assigned the colors $i+1,...,k$, since these colors arose after $v$ was assigned a color.

The key intuition of the \textsc{recolor} variant is illustrated in Figure~\ref{fig:recolor}.
In the start of the example, notice that $v$ is assigned to a new color class $C_k$ (i.e., contains only $v$).
Therefore, the \textsc{recolor} method is called, which attempts to find $v$ another color class denoted $C_i$ where $C_i < C_k$.
For this, we search for a color class $C_i$ that contains a single adjacent vertex denoted $w$ (known as a conflict).
Intuitively, we may assign $v$ to $C_i$ if we can find $w$ another ``valid'' color class denoted $C_j$.
Notice that $i < j < k$ such that the color class $C_i$ appeared before $C_j$ and so forth.
In other words, $v$ can be assigned to $C_i$ if there exists a valid color class $C_j$ for which $w$ can be assigned.
If such a $C_j$ exists, then the number of colors is decreased by one.

\begin{figure}[h!]
\vspace{-6mm}
\begin{center}
\begin{minipage}{1.00\linewidth}
\centering
\begin{algorithm}[H]
\small
\caption{\small Fast Greedy Recoloring
}
\label{alg:greedy-coloring-repair}
{\algfontsize
\begin{algorithmic}[1]
\bspacing
\fontsize{7.9}{8.5}\selectfont
\Procedure {GreedyRecolor}{$G$, $\pi$}
\State Initialize data structures 						
\For{$v \in \pi$ in order} 									
		\For{$w \in \N(v)$}			 $\used(\colorindex(w)) \leftarrow v$		 
		\EndFor
		\State $k \leftarrow \min\{ i > 0 : \used(i) \not= v \}$  										
		\State $\colorindex(v) \leftarrow k$ 																	
		\If{$k > \coloringnum$} 		
		 			\If{\textsc{recolor}($\colorindex$, $v$, $k$)}  		$k \leftarrow k - 1$		 		
		 			\EndIf		 			
		 			\State $\coloringnum \leftarrow k$ 		 		 			
		 \EndIf
\EndFor
\EndProcedure
\algrule[0.5pt]
\Procedure {recolor}{$\colorindex$, $v$, $k$}\label{alg:recolor}
\State Initialize \textsf{conflicts} to be $0$ 				\label{algline:init-conflicts}	   
\For{$w \in \N(v)$}			 
	\State $\mathsf{conflicts}(\colorindex(w)) \leftarrow \mathsf{conflicts}(\colorindex(w)) + 1$	\label{algline:find-conflicts}	  
	\State $\used(\colorindex(w)) \leftarrow 	w$ 				\label{algline:mark-conflicting-vertex}			
\EndFor
\For{$i = 1$ {\bf to} $(k-1)$} 																				 
		\If{$\mathsf{conflicts}(i) = 1$} 					\label{algline:find-single-conflict}				 
					\State 	$w \leftarrow \used(i)$ 			\label{algline:candidate-vertex}						
							\For{$u \in \N(w)$}			 $\used(\colorindex(u)) \leftarrow w$		 
							\EndFor
							\State $c \leftarrow \min\{ j > i: \used(j) \not= w \}$ 						
							\If{$c < \colorindex(v)$} 														 
										\State $\colorindex(v) \leftarrow \colorindex(w)$  {\bf and} $\colorindex(w) \leftarrow c$							
										\State {\bf return true} 			
							\EndIf															
		\EndIf
\EndFor
\State {\bf return false} 							
\EndProcedure
\espacing
\end{algorithmic}}
\end{algorithm}
\end{minipage}
\end{center}
\vspace{-3mm}
\end{figure}

\section{Bounds} 
\label{sec:bounds}
Lower and upper bounds on the minimum number of colors are useful for a number of reasons (see Section~\ref{sec:results-bounds}).
In this section, we first provide a fast parallel method for computing a lower bound that is especially tight for large sparse networks.
Next, we summarize the upper bounds used in this work, which are also shown to be strong, and in many cases matching that of the lower bound, and thus allowing us to verify the coloring from one of our methods.

\begin{figure}[h!]
\begin{center}
\includegraphics[width=0.6\linewidth,bb=0 0 468 381]{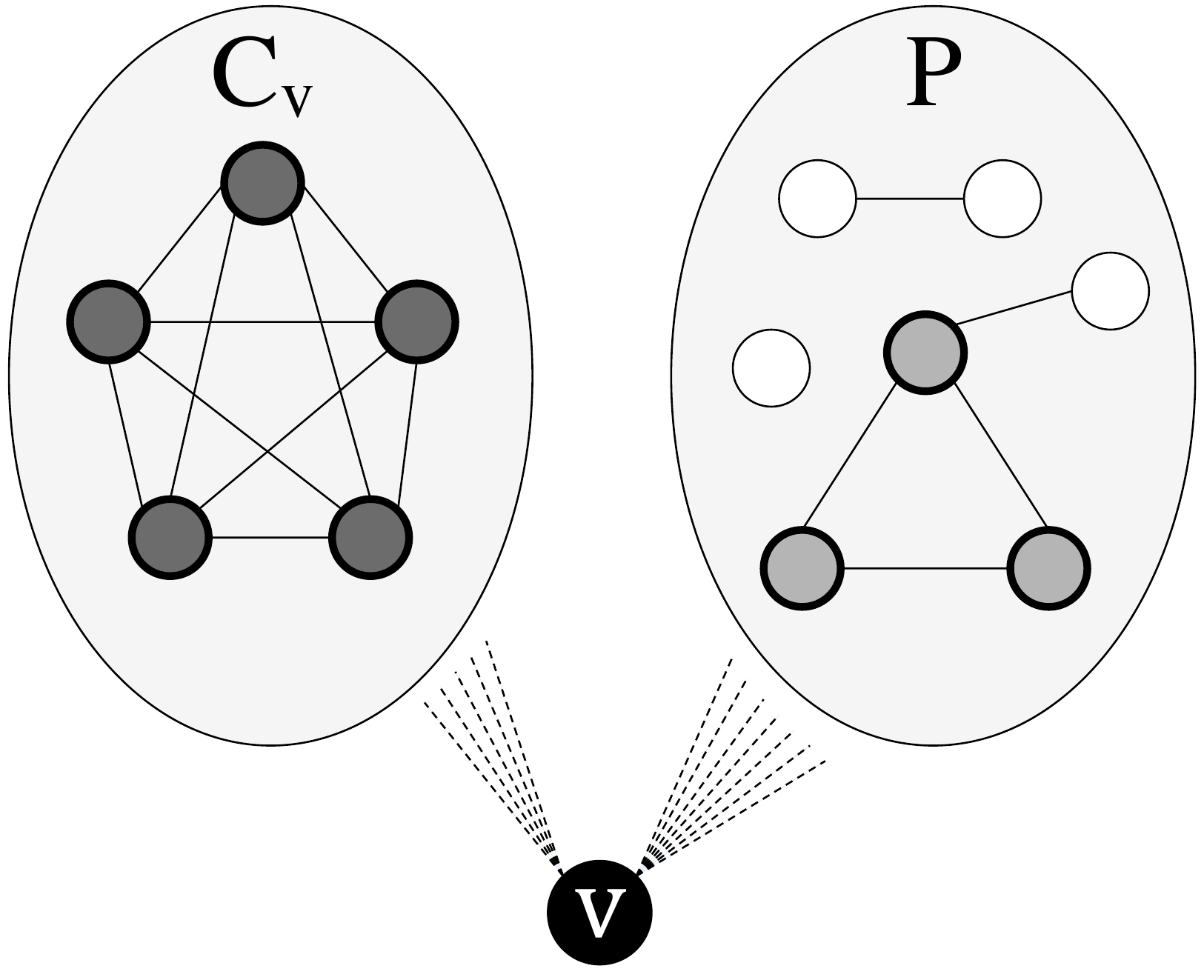}
\footnotesize
\caption{
\footnotesize
\textbf{Clique Invariant and Fast Heuristic Clique Finder}.
Recall $C_v$ is the clique being constructed, whereas $P$ is the set of potential vertices that could be added to $C_v$ to form a clique of $|C_v|+1$.
Further, after a vertex $u$ from $P$ is added to $C_v$, we must then remove $u$ from $P$ and compute the intersection $P \cap \Nr(u)$.
The result of this intersection depends intrinsically on how well $u$ is connected to the vertices in $P$.
In the ideal case, the heuristic is guaranteed to find the largest possible clique as long as the vertices in $P$ that form the largest clique among each other are added to $C_v$.
For instance, the largest clique in the above example is $|C_v|+3=8$ formed by adding the three vertices forming a 3-clique (triangle) in $P$ to $C_v$, 
whereas if $u \in P$ with 0 degree is added to $C_v$, then $|C_v|+1=6$, since $P_{t} \cap \Nr(u) = \emptyset$.
}
\label{fig:heuristic-clique-invariant}
\end{center}
\end{figure}

\subsection{Lower Bounds}
\label{sec:lowerbounds}

Let $\tilde{\omega}(G)$ be the size of a large clique from a heuristic clique finder and thus a \textit{lower-bound} on the size of the maximum clique $\mc(G)$.
As previously mentioned,  $\tilde{\omega}(G) \leq \omega(G) \leq \Chromatic(G)$.
Since the maximum clique problem is known to be NP-hard, we use a fast parallel heuristic clique finder tuned specifically for large sparse complex networks.
Our approach is shown in Alg.~\ref{alg:heuristic} and found to be efficient while also useful for obtaining a large clique that is often of maximum or near-optimal size (i.e., $\tilde{\omega}(G)$ is close to $\omega(G)$)  for many types of large real-world networks.

Given a graph $G=(V,E)$, the heuristic obtains a vertex ordering $\pi = \{v_1,...,v_n\}$ and searches each vertex $v_i$ in the ordering $\pi$ for a large clique in $\N(v_i)$.
For convenience, let $\Nr(v)$ be the reduced neighborhood of $v$ defined formally as,
\[ N_R(v) = G( \{ v \} \cup \{ u : (u,v) \in E, \bound(u) \geq |C_{\max}|, u \not \in X) \} \]
\noindent where $|C_{\max}|$ is the largest clique found thus far, $\bound(u)$ is a vertex upper bound\footnote{The local vertex upper bound for $u$ denoted by $\bound(u)$ is typically the maximum k-core number of the vertex $u$ denoted by $\K(u)$}, and $\X$ is a vertex-index array of pruned vertices (i.e., $O(1)$ time check).
Thus, let $P \leftarrow \Nr(v)$ be the set of potential vertices and initially we set $C_v \leftarrow \emptyset$.
At each step in the heuristic, a vertex $u \in P$ is selected according to a \textit{greedy selection criterion} $\f(\cdot)$ such that $u \leftarrow \max_{w \in P} \f(w)$ where $\f(\cdot)$ is a graph property.
The selected vertex $u$ is added to $C_v \leftarrow C_v \cup \{u\}$ and $P_{t+1} \leftarrow P_{t} \cap \Nr(u)$ where $t$ denotes the iteration (or depth of the search tree).
The local clique search terminates if $|P_{t}| + |C_v| \leq C_{max}$, since this indicates that a clique of a larger size cannot be found from searching further.
See Figure~\ref{fig:heuristic-clique-invariant} for a simple example.
Notice that $C_v$ is the clique being built and grows by a single vertex each iteration, whereas $P_{t+1}$ are the potential vertices remaining after adding $u$ to $C_v$.
Hence, $|P_{t+1}| < |P_{t}| < |P_{t-1}|$ is monotonically decreasing with respect to $t$.
It is clear from Figure~\ref{fig:heuristic-clique-invariant} that $|P_{t+1}|$ and thus the size of the clique $|C_v|$ strongly depends on $u$ selected by the greedy selection criterion.
In Figure~\ref{fig:heuristic-clique-invariant}, suppose the vertex without edges to other vertices in $P$ is selected and added to $C_v$, then $P_{t+1} \leftarrow \emptyset$ and the search terminates.
The proposed heuristic clique finder is equivalent to searching down a single branch in a greedy fashion.

Let us also point out that Alg.~\ref{alg:heuristic} is extremely flexible.
For instance, the vertices in $G$ (globally) and $P$ (locally) are ordered by their k-core numbers (see Line~\ref{algline:global-order} and \ref{algline:local-order}), but any ordering from Table~\ref{tab:selection-criterion} may be used.
In addition, while Alg.~\ref{alg:heuristic} is presented using vertex k-core numbers for pruning (Line~\ref{algline:global-pruning}), one may also leverage stronger bounds such as the triangle-core numbers (See~\cite{rossi2014pakdd}).
We used k-core numbers for ordering and pruning since these are relatively tight bounds while also efficient to compute for large networks.
Later in Section~\ref{sec:results}, we demonstrate the tightness of these bounds on large sparse real-world networks (See Table~\ref{table:stats-sparse-ca} and~\ref{table:stats-sparse-social}).

\algrenewcommand{\alglinenumber}[1]{\scriptsize #1}
\begin{algorithm}[b!]
\centering
\caption{Fast Heuristic Clique Finder}
\label{alg:heuristic}{\small
\begin{algorithmic}[1]
\Procedure {HeuristicClique}{$G=\left (V,E\right ), K$}
\State Set $C_{\max} = \{\}$ 
\For{{\bf each} $v \in V$ in decr. k-core order in $\mathbf{parallel}$} \label{algline:global-order}
			\If{$\K(v) \geq |C_{\max}|$}		
					\State Let $P$ be the neighs. of $v$ with core numbers $\geq$ $|C_{\max}|$ \label{algline:global-pruning}
					\State Set $C_v = \{\}$
					\For{each vertex $u \in P$ by decreasing core number} \label{algline:local-order}	
							\If{$C_v \cup \{u\}$ is a clique} 
							    \State Add $u$ to $C_v$ 
							\EndIf
					\EndFor
					\If{$|C_v| > |C_{\max}|$} 
					      \State Set $C_{\max} = C_{v}$
					\EndIf
			\EndIf
\EndFor
\State \textbf{return} $C_{\max}$, a large clique in $G$
\EndProcedure
\end{algorithmic}}
\end{algorithm}

\smallskip \noindent {\theoremfont Complexity:}
The runtime of the heuristic is $O(|E| \cdot \K(G))$ since it takes $\sum_{v \in V} \d(v) = 2|E| = O(|E|)$ to form the initial set of neighbors for each vertex.
The \textsc{HeuristicClique} is essentially a greedy depth-first search where the depth is at most $\K(G)$.
As an aside, if $\T(G)$ is used instead, then the heuristic is computed in $O(|E| \cdot \T(G))$.
Observe that at each step, the greedy selection criterion $u \leftarrow \max_{v \in P} \f(v)$ is evaluated in $O(1)$ time by pre-ordering the vertices prior to searching.
The runtime of the ordering is $O(|P|)$ using bucket sort.
A global bound on the depth of the search tree for any vertex neighborhood is clearly $\K(G)$ and for a specific vertex $v$ is no larger than $\K(v)$.
In practice, the heuristic is fast and usually terminates after only a few iterations due to the removal of vertices from $P$ via the strong upper bounds.

\smallskip \noindent {\theoremfont Parallel Algorithm}
The vertex neighborhoods are searched in $\mathsf{parallel}$ for a large clique.
Each worker (i.e., processing unit, core) is assigned dynamically a block $\beta$ of vertices to search. 
The workers maintain a vertex neighborhood subgraph for the vertex currently being searched.
In addition, the workers share a vertex-indexed array $X$ of pruned vertices and the largest clique $C_{\max}$ found among all the workers.
If a worker finds a clique $C_v$ larger than $C_{\max}$, i.e., $|C_v| > |C_{\max}|$ (max so far among all workers), 
then a lock is obtained, and $C_{\max} \leftarrow C_v$ and the updated $C_{\max}$ is immediately sent to all workers.
As an aside, this immediate sharing of $C_{\max}$ typically leads to a significant speedup, 
since the updated $C_{\max}$ allows for the workers to further prune their search space including entire vertices.

\subsection{Upper Bounds}
\label{sec:upperbounds}

A simple, but not very useful upper bound on the Chromatic number $\Chromatic(G)$ is given by the maximum degree: $\Chromatic(G) \leq \dmax(G)+1$.
A stronger upper bound is given by the \textit{maximum k-core number of $G$} denoted by $\K(G)$.
This gives the following relationship:
\[ \Chromatic(G) \leq \K(G)+1 \leq \dmax(G)+1 \]
In this work, we observe that this upper bound is significantly stronger than the maximum degree on nearly all large sparse networks.

Since  $\coloring(G,\pi)$ depends on an ordering $\pi$ then no relationship exists between $\coloring(G,\slo)$ from \textsc{slo} and $\coloring(G,\pi)$ where $\pi$ gave rise to $\dbmax(G)$.
Nevertheless, suppose the vertices are colored using \textsc{slo} resulting in $\coloring(G,\pi)$ colors, then using $\K(G)$ gives the following relationship:
\[ \tilde{\omega}(G) \leq \omega(G) \leq \Chromatic(G) \leq \coloring(G,\pi) \leq \K(G)+1 \leq \dmax(G)+1 \]
\noindent where $\omega(G)$ is the maximum clique in $G$ and $\tilde{\omega}(G)$ is a large clique in $G$ from the fast heuristic clique finder in Section~\ref{sec:lowerbounds}.
In other words, if a greedy coloring method uses $\pi$ from \textsc{slo} then the resulting coloring of $G$ must use at most $\K(G)+1$ colors.
Furthermore, $\K(G)+1$ is also known as the coloring number denoted $\col(G)$~\cite{erdHos1966chromatic}\footnote{Also referred to as degeneracy~\cite{erdHos1966chromatic}, maximum k-core number~\cite{batagelj2003m}, linkage~\cite{matula1983smallest}, among others.}.

The above relationship can be further strengthened using the notion of the \textit{maximum triangle core number of $G$} denoted $\T(G)$.
This gives rise to the following relationship:
\[ \mc(G) \leq \Chromatic(G) \leq \T(G) \leq \K(G)+1 \leq \dmax(G)+1 \]

\begin{table}[htb]
\caption{
\textbf{Network Statistics and Coloring Bounds.}
From the large collection of 100+ graphs used in our experiments, we selected a small representative set from the various types (e.g., web, social networks) to study relationships between key network statistics and lower and upper bounds on the Chromatic number.
Here, $\rho$ is the density, $\bar{d}$ is the average degree, and ${\rm r}$ is the assortativity coefficient.
We also study the following triangle related statistics: $\kappa$ is the global clustering coefficient, $|T|$ is the total number of triangles, and $tr_{\rm avg}$ and $tr_{\rm \max}$ are the maximum and average number of triangles incident on a vertex, respectively.
Using these fundamental network statistics as a basis, we analyze the relationships between these characteristic network properties and our derived bounds on the Chromatic number.
The lower bound from the heuristic clique finder is denoted $\tilde{\omega}$.
For the upper bounds, we denote $K$ as the maximum k-core (i.e., the largest degree for a $k$-core to exist), and similarly, we also upper bound the Chromatic number using the notion of the maximum triangle-core, which we denote by $T$.
Finally, we also include the maximum and minimum number of colors from a coloring method in our framework, which we denote $\coloring_{\max}$ and $\coloring_{\min}$, respectively.
}
\vspace{1mm}
\label{table:stats-sparse-ca}
\centering\small\scriptsize
\fontsize{6.0}{6.5}\selectfont
\noindent
\begin{tabularx}{\linewidth}{ rr XXXHc rHcHX cccc cc}
\toprule
& & 
\multicolumn{10}{c}{\textbf{Graph measures}}& 
\multicolumn{4}{c}{\textbf{Bounds}}& 
\multicolumn{2}{c}{\textbf{Colors}}\\ 
 &
\textbf{graph}& 
$|V|$ & 
$|E|$ & 
$|T|$ & 
$\rho$ & 
$\bar{d}$ & 
${\rm r}$ & 
$\bar{\kappa}$ & 
$\kappa$ & 
$\tr_{\rm avg}$ & 
$\tr_{\rm max}$ & 
$\dmax$ & 
$K$+1 & 
$T$ & 
$\tilde{\omega}$ & 
$\coloring_{\min}$ &
$\coloring_{\max}$
\\ \midrule
\multirow{4}{*}{\rotatebox{90}{\textbf { \sc bio}}}&    \textsf{bio-celegans}&  453 &  2K &  9.8K &  0.02 &  8 &  -0.23 &  0.24 &  0.12 &  22 &  870 &  237 &  11 &  9 &  9 &  10 &  16  \\ 
	 &    \textsf{bio-diseasome}&  516 &  1.1K &  4K &  $10^{-3}$ &  4 &  0.07 &  0.52 &  0.43 &  8 &  152 &  50 &  11 &  11 &  10 &  11 &  12  \\ 
	 &    \textsf{bio-dmela}&  7.3K &  25.5K &  8.6K &  $10^{-4}$ &  6 &  -0.05 &  0.00 &  0.01 &  1 &  225 &  190 &  12 &  7 &  7 &  8 &  15  \\ 
	 &    \textsf{bio-yeast}&  1.4K &  1.9K &  618 &  $10^{-3}$ &  2 &  -0.21 &  0.04 &  0.05 &  0 &  18 &  56 &  6 &  6 &  5 &  6 &  8  \\ 
\midrule 
\multirow{13}{*}{\rotatebox{90}{\textbf { \sc collaboration}}}&    \textsf{ca-AstroPh}&  17.9K &  196.9K &  4M &  $10^{-3}$ &  22 &  0.20 &  0.37 &  0.32 &  226 &  11.2K &  504 &  57 &  57 &  56 &  57 &  64  \\ 
	 &   \cellcolor{verylightblue}\textsf{ca-CSphd}& \cellcolor{verylightblue}1.8K & \cellcolor{verylightblue}1.7K & \cellcolor{verylightblue}24 & \cellcolor{verylightblue}$10^{-4}$ & \cellcolor{verylightblue}1 & \cellcolor{verylightblue}-0.20 & \cellcolor{verylightblue}0.00 & \cellcolor{verylightblue}0.00 & \cellcolor{verylightblue}0 & \cellcolor{verylightblue}4 & \cellcolor{verylightblue}46 & \cellcolor{verylightblue}3 & \cellcolor{verylightblue}3 & \cellcolor{verylightblue}3 & \cellcolor{verylightblue}3 & \cellcolor{verylightblue}5  \\ 
	 &   \cellcolor{verylightblue}\textsf{ca-CondMat}& \cellcolor{verylightblue}21.3K & \cellcolor{verylightblue}91.2K & \cellcolor{verylightblue}513.1K & \cellcolor{verylightblue}$10^{-4}$ & \cellcolor{verylightblue}8 & \cellcolor{verylightblue}0.13 & \cellcolor{verylightblue}0.44 & \cellcolor{verylightblue}0.26 & \cellcolor{verylightblue}24 & \cellcolor{verylightblue}1.6K & \cellcolor{verylightblue}279 & \cellcolor{verylightblue}26 & \cellcolor{verylightblue}26 & \cellcolor{verylightblue}26 & \cellcolor{verylightblue}26 & \cellcolor{verylightblue}29  \\ 
	 &   \cellcolor{verylightblue}\textsf{ca-Erdos992}& \cellcolor{verylightblue}6.1K & \cellcolor{verylightblue}7.5K & \cellcolor{verylightblue}4.8K & \cellcolor{verylightblue}$10^{-4}$ & \cellcolor{verylightblue}2 & \cellcolor{verylightblue}-0.44 & \cellcolor{verylightblue}0.05 & \cellcolor{verylightblue}0.04 & \cellcolor{verylightblue}1 & \cellcolor{verylightblue}99 & \cellcolor{verylightblue}61 & \cellcolor{verylightblue}8 & \cellcolor{verylightblue}8 & \cellcolor{verylightblue}8 & \cellcolor{verylightblue}8 & \cellcolor{verylightblue}12  \\ 
	 &   \cellcolor{verylightblue}\textsf{ca-GrQc}& \cellcolor{verylightblue}4.1K & \cellcolor{verylightblue}13.4K & \cellcolor{verylightblue}143.3K & \cellcolor{verylightblue}$10^{-3}$ & \cellcolor{verylightblue}6 & \cellcolor{verylightblue}0.64 & \cellcolor{verylightblue}0.39 & \cellcolor{verylightblue}0.63 & \cellcolor{verylightblue}34 & \cellcolor{verylightblue}1.1K & \cellcolor{verylightblue}81 & \cellcolor{verylightblue}44 & \cellcolor{verylightblue}44 & \cellcolor{verylightblue}44 & \cellcolor{verylightblue}44 & \cellcolor{verylightblue}45  \\ 
	 &   \cellcolor{verylightgreen}\textsf{ca-HepPh}& \cellcolor{verylightgreen}11.2K & \cellcolor{verylightgreen}117.6K & \cellcolor{verylightgreen}10M & \cellcolor{verylightgreen}$10^{-3}$ & \cellcolor{verylightgreen}20 & \cellcolor{verylightgreen}0.63 & \cellcolor{verylightgreen}0.40 & \cellcolor{verylightgreen}0.66 & \cellcolor{verylightgreen}899 & \cellcolor{verylightgreen}39.6K & \cellcolor{verylightgreen}491 & \cellcolor{verylightgreen}239 & \cellcolor{verylightgreen}239 & \cellcolor{verylightgreen}239 & \cellcolor{verylightgreen}239 & \cellcolor{verylightgreen}239  \\ 
	 &   \cellcolor{verylightblue}\textsf{ca-MathSciNet}& \cellcolor{verylightblue}332.6K & \cellcolor{verylightblue}820.6K & \cellcolor{verylightblue}1.7M & \cellcolor{verylightblue}$10^{-5}$ & \cellcolor{verylightblue}4 & \cellcolor{verylightblue}0.10 & \cellcolor{verylightblue}0.30 & \cellcolor{verylightblue}0.14 & \cellcolor{verylightblue}5 & \cellcolor{verylightblue}1.5K & \cellcolor{verylightblue}496 & \cellcolor{verylightblue}25 & \cellcolor{verylightblue}25 & \cellcolor{verylightblue}25 & \cellcolor{verylightblue}25 & \cellcolor{verylightblue}28  \\ 
	 &   \cellcolor{verylightgreen}\textsf{ca-citeseer}& \cellcolor{verylightgreen}227.3K & \cellcolor{verylightgreen}814.1K & \cellcolor{verylightgreen}8.1M & \cellcolor{verylightgreen}$10^{-5}$ & \cellcolor{verylightgreen}7 & \cellcolor{verylightgreen}0.07 & \cellcolor{verylightgreen}0.52 & \cellcolor{verylightgreen}0.46 & \cellcolor{verylightgreen}36 & \cellcolor{verylightgreen}5.3K & \cellcolor{verylightgreen}1.3K & \cellcolor{verylightgreen}87 & \cellcolor{verylightgreen}87 & \cellcolor{verylightgreen}87 & \cellcolor{verylightgreen}87 & \cellcolor{verylightgreen}87  \\ 
	 &   \cellcolor{verylightgreen}\textsf{ca-dblp10}& \cellcolor{verylightgreen}226.4K & \cellcolor{verylightgreen}716.4K & \cellcolor{verylightgreen}4.7M & \cellcolor{verylightgreen}$10^{-5}$ & \cellcolor{verylightgreen}6 & \cellcolor{verylightgreen}0.30 & \cellcolor{verylightgreen}0.49 & \cellcolor{verylightgreen}0.38 & \cellcolor{verylightgreen}21 & \cellcolor{verylightgreen}5.9K & \cellcolor{verylightgreen}238 & \cellcolor{verylightgreen}75 & \cellcolor{verylightgreen}75 & \cellcolor{verylightgreen}75 & \cellcolor{verylightgreen}75 & \cellcolor{verylightgreen}75  \\ 
	 &   \cellcolor{verylightgreen}\textsf{ca-dblp12}& \cellcolor{verylightgreen}317K & \cellcolor{verylightgreen}1M & \cellcolor{verylightgreen}6.6M & \cellcolor{verylightgreen}$10^{-5}$ & \cellcolor{verylightgreen}6 & \cellcolor{verylightgreen}0.27 & \cellcolor{verylightgreen}0.48 & \cellcolor{verylightgreen}0.31 & \cellcolor{verylightgreen}21 & \cellcolor{verylightgreen}8.3K & \cellcolor{verylightgreen}343 & \cellcolor{verylightgreen}114 & \cellcolor{verylightgreen}114 & \cellcolor{verylightgreen}114 & \cellcolor{verylightgreen}114 & \cellcolor{verylightgreen}114  \\ 
	 &   \cellcolor{verylightgreen}\textsf{ca-hollywood09}& \cellcolor{verylightgreen}1M & \cellcolor{verylightgreen}56.3M & \cellcolor{verylightgreen}14.7T & \cellcolor{verylightgreen}$10^{-5}$ & \cellcolor{verylightgreen}105 & \cellcolor{verylightgreen}0.35 & \cellcolor{verylightgreen}0.59 & \cellcolor{verylightgreen}0.31 & \cellcolor{verylightgreen}13.7K & \cellcolor{verylightgreen}3.9M & \cellcolor{verylightgreen}11.4K & \cellcolor{verylightgreen}2209 & \cellcolor{verylightgreen}2209 & \cellcolor{verylightgreen}2209 & \cellcolor{verylightgreen}2209 & \cellcolor{verylightgreen}2209  \\ 
	 &   \cellcolor{verylightgreen}\textsf{ca-netscience}& \cellcolor{verylightgreen}379 & \cellcolor{verylightgreen}914 & \cellcolor{verylightgreen}2.7K & \cellcolor{verylightgreen}0.01 & \cellcolor{verylightgreen}4 & \cellcolor{verylightgreen}-0.08 & \cellcolor{verylightgreen}0.58 & \cellcolor{verylightgreen}0.43 & \cellcolor{verylightgreen}7 & \cellcolor{verylightgreen}75 & \cellcolor{verylightgreen}34 & \cellcolor{verylightgreen}9 & \cellcolor{verylightgreen}9 & \cellcolor{verylightgreen}9 & \cellcolor{verylightgreen}9 & \cellcolor{verylightgreen}9  \\ 
	 &    \textsf{ca-sandi-auths}&  86 &  124 &  126 &  0.03 &  2 &  -0.26 &  0.33 &  0.27 &  1 &  7 &  12 &  5 &  5 &  4 &  5 &  6  \\ 
\midrule 
\multirow{9}{*}{\rotatebox{90}{\textbf { \sc interaction}}}&    \textsf{ia-email-EU}&  32.4K &  54.3K &  146.9K &  $10^{-4}$ &  3 &  -0.38 &  0.09 &  0.03 &  5 &  1.6K &  623 &  23 &  13 &  11 &  16 &  27  \\ 
	 &   \cellcolor{verylightblue}\textsf{ia-email-univ}& \cellcolor{verylightblue}1.1K & \cellcolor{verylightblue}5.4K & \cellcolor{verylightblue}16K & \cellcolor{verylightblue}$10^{-3}$ & \cellcolor{verylightblue}9 & \cellcolor{verylightblue}0.08 & \cellcolor{verylightblue}0.04 & \cellcolor{verylightblue}0.17 & \cellcolor{verylightblue}14 & \cellcolor{verylightblue}261 & \cellcolor{verylightblue}71 & \cellcolor{verylightblue}12 & \cellcolor{verylightblue}12 & \cellcolor{verylightblue}12 & \cellcolor{verylightblue}12 & \cellcolor{verylightblue}15  \\ 
	 &    \textsf{ia-enron-large}&  33.6K &  180.8K &  2.1M &  $10^{-4}$ &  10 &  -0.12 &  0.34 &  0.09 &  65 &  17.7K &  1.3K &  44 &  22 &  15 &  28 &  57  \\ 
	 &   \cellcolor{verylightred}\textsf{ia-enron-only}& \cellcolor{verylightred}143 & \cellcolor{verylightred}623 & \cellcolor{verylightred}2.6K & \cellcolor{verylightred}0.06 & \cellcolor{verylightred}8 & \cellcolor{verylightred}-0.02 & \cellcolor{verylightred}0.08 & \cellcolor{verylightred}0.36 & \cellcolor{verylightred}19 & \cellcolor{verylightred}125 & \cellcolor{verylightred}42 & \cellcolor{verylightred}10 & \cellcolor{verylightred}8 & \cellcolor{verylightred}8 & \cellcolor{verylightred}8 & \cellcolor{verylightred}11  \\ 
	 &    \textsf{ia-fb-messages}&  1.2K &  6.4K &  7.4K &  $10^{-3}$ &  10 &  -0.08 &  0.02 &  0.04 &  6 &  242 &  112 &  12 &  5 &  5 &  8 &  15  \\ 
	 &   \cellcolor{verylightred}\textsf{ia-infect-dublin}& \cellcolor{verylightred}410 & \cellcolor{verylightred}2.7K & \cellcolor{verylightred}21.3K & \cellcolor{verylightred}0.03 & \cellcolor{verylightred}13 & \cellcolor{verylightred}0.23 & \cellcolor{verylightred}0.05 & \cellcolor{verylightred}0.44 & \cellcolor{verylightred}52 & \cellcolor{verylightred}280 & \cellcolor{verylightred}50 & \cellcolor{verylightred}18 & \cellcolor{verylightred}16 & \cellcolor{verylightred}16 & \cellcolor{verylightred}16 & \cellcolor{verylightred}18  \\ 
	 &    \textsf{ia-infect-hyper}&  113 &  2.1K &  50.6K &  0.35 &  38 &  -0.12 &  0.00 &  0.50 &  448 &  1.7K &  98 &  29 &  18 &  15 &  19 &  28  \\ 
	 &    \textsf{ia-reality}&  6.8K &  7.6K &  1.2K &  $10^{-4}$ &  2 &  -0.68 &  0.01 &  0.00 &  0 &  52 &  261 &  6 &  5 &  4 &  5 &  7  \\ 
	 &    \textsf{ia-wiki-Talk}&  92.1K &  360.7K &  2.5M &  $10^{-5}$ &  7 &  -0.03 &  0.03 &  0.05 &  27 &  17.6K &  1.2K &  59 &  20 &  9 &  30 &  64  \\ 
\midrule 
\multirow{4}{*}{\rotatebox{90}{\textbf { \sc infra}}}&   \cellcolor{verylightred}\textsf{inf-USAir97}& \cellcolor{verylightred}332 & \cellcolor{verylightred}2.1K & \cellcolor{verylightred}36.5K & \cellcolor{verylightred}0.04 & \cellcolor{verylightred}12 & \cellcolor{verylightred}-0.21 & \cellcolor{verylightred}0.24 & \cellcolor{verylightred}0.40 & \cellcolor{verylightred}110 & \cellcolor{verylightred}1.4K & \cellcolor{verylightred}139 & \cellcolor{verylightred}27 & \cellcolor{verylightred}22 & \cellcolor{verylightred}22 & \cellcolor{verylightred}22 & \cellcolor{verylightred}31  \\ 
	 &   \cellcolor{verylightblue}\textsf{inf-power}& \cellcolor{verylightblue}4.9K & \cellcolor{verylightblue}6.5K & \cellcolor{verylightblue}1.9K & \cellcolor{verylightblue}$10^{-4}$ & \cellcolor{verylightblue}2 & \cellcolor{verylightblue}0.00 & \cellcolor{verylightblue}0.04 & \cellcolor{verylightblue}0.10 & \cellcolor{verylightblue}0 & \cellcolor{verylightblue}21 & \cellcolor{verylightblue}19 & \cellcolor{verylightblue}6 & \cellcolor{verylightblue}6 & \cellcolor{verylightblue}6 & \cellcolor{verylightblue}6 & \cellcolor{verylightblue}7  \\ 
	 &    \textsf{inf-roadNet-CA}&  1.9M &  2.7M &  361.4K &  $10^{-6}$ &  2 &  0.12 &  0.00 &  0.06 &  0 &  7 &  12 &  4 &  4 &  4 &  5 &  6  \\ 
	 &    \textsf{inf-roadNet-PA}&  1M &  1.5M &  201.3K &  $10^{-6}$ &  2 &  0.12 &  0.00 &  0.06 &  0 &  8 &  9 &  4 &  4 &  3 &  4 &  6  \\ 
\midrule 
\multirow{5}{*}{\rotatebox{90}{\textbf { \sc misc}}}&    \textsf{ASIC-320ks}&  321.6K &  1.5M &  5.9M &  $10^{-5}$ &  9 &  -0.05 &  0.00 &  0.11 &  19 &  2.2K &  822 &  9 &  18 &  5 &  6 &  8  \\ 
	 &    \textsf{IMDB-bi}&  896.3K &  3.7M &  13K &  $10^{-6}$ &  8 &  -0.05 &  0.00 &  0.00 &  0 &  78 &  1.5K &  24 &  3 &  3 &  11 &  24  \\ 
	 &    \textsf{Reuters911}&  13.3K &  148K &  3.5M &  $10^{-3}$ &  22 &  -0.11 &  0.16 &  0.11 &  263 &  69.8K &  2.2K &  74 &  40 &  26 &  38 &  77  \\ 
	 &   \cellcolor{verylightblue}\textsf{football}& \cellcolor{verylightblue}115 & \cellcolor{verylightblue}613 & \cellcolor{verylightblue}2.4K & \cellcolor{verylightblue}0.09 & \cellcolor{verylightblue}10 & \cellcolor{verylightblue}0.16 & \cellcolor{verylightblue}0.00 & \cellcolor{verylightblue}0.41 & \cellcolor{verylightblue}21 & \cellcolor{verylightblue}32 & \cellcolor{verylightblue}12 & \cellcolor{verylightblue}9 & \cellcolor{verylightblue}9 & \cellcolor{verylightblue}9 & \cellcolor{verylightblue}9 & \cellcolor{verylightblue}10  \\ 
	 &   \cellcolor{verylightblue}\textsf{lesmis}& \cellcolor{verylightblue}77 & \cellcolor{verylightblue}254 & \cellcolor{verylightblue}1.4K & \cellcolor{verylightblue}0.09 & \cellcolor{verylightblue}6 & \cellcolor{verylightblue}-0.17 & \cellcolor{verylightblue}0.34 & \cellcolor{verylightblue}0.50 & \cellcolor{verylightblue}18 & \cellcolor{verylightblue}82 & \cellcolor{verylightblue}36 & \cellcolor{verylightblue}10 & \cellcolor{verylightblue}10 & \cellcolor{verylightblue}10 & \cellcolor{verylightblue}10 & \cellcolor{verylightblue}12  \\ 
	 &   \cellcolor{verylightgreen}\textsf{rec-amazon}& \cellcolor{verylightgreen}91.8K & \cellcolor{verylightgreen}125.7K & \cellcolor{verylightgreen}103K & \cellcolor{verylightgreen}$10^{-5}$ & \cellcolor{verylightgreen}2 & \cellcolor{verylightgreen}0.19 & \cellcolor{verylightgreen}0.11 & \cellcolor{verylightgreen}0.35 & \cellcolor{verylightgreen}1 & \cellcolor{verylightgreen}9 & \cellcolor{verylightgreen}5 & \cellcolor{verylightgreen}5 & \cellcolor{verylightgreen}5 & \cellcolor{verylightgreen}5 & \cellcolor{verylightgreen}5 & \cellcolor{verylightgreen}5  \\ 
\midrule 
\multirow{3}{*}{\rotatebox{90}{\textbf { \sc rt}}}&   \cellcolor{verylightred}\textsf{rt-retweet-crawl}& \cellcolor{verylightred}1.1M & \cellcolor{verylightred}2.2M & \cellcolor{verylightred}525.9K & \cellcolor{verylightred}$10^{-6}$ & \cellcolor{verylightred}4 & \cellcolor{verylightred}-0.02 & \cellcolor{verylightred}0.01 & \cellcolor{verylightred}0.00 & \cellcolor{verylightred}0 & \cellcolor{verylightred}1.5K & \cellcolor{verylightred}5K & \cellcolor{verylightred}19 & \cellcolor{verylightred}13 & \cellcolor{verylightred}13 & \cellcolor{verylightred}13 & \cellcolor{verylightred}23  \\ 
	 &   \cellcolor{verylightblue}\textsf{rt-retweet}& \cellcolor{verylightblue}96 & \cellcolor{verylightblue}117 & \cellcolor{verylightblue}36 & \cellcolor{verylightblue}0.03 & \cellcolor{verylightblue}2 & \cellcolor{verylightblue}-0.18 & \cellcolor{verylightblue}0.03 & \cellcolor{verylightblue}0.07 & \cellcolor{verylightblue}0 & \cellcolor{verylightblue}6 & \cellcolor{verylightblue}17 & \cellcolor{verylightblue}4 & \cellcolor{verylightblue}4 & \cellcolor{verylightblue}4 & \cellcolor{verylightblue}4 & \cellcolor{verylightblue}5  \\ 
	 &    \textsf{rt-twitter-copen}&  761 &  1K &  447 &  $10^{-3}$ &  2 &  -0.10 &  0.05 &  0.06 &  1 &  27 &  37 &  5 &  4 &  4 &  5 &  8  \\ 
\midrule 
\end{tabularx}
\end{table}

\begin{table}[htb]
\caption{ 
\textbf{Network Statistics and Coloring Bounds (continued from Table~\ref{table:stats-sparse-ca}).}
}
\vspace{1mm}
\label{table:stats-sparse-social}
\centering\small\scriptsize
\fontsize{6.0}{6.5}\selectfont
\noindent
\begin{tabularx}{\linewidth}{ rr XXXHc rHcHX cccc cc}
\toprule
& & 
\multicolumn{10}{c}{\textbf{Graph measures}}& 
\multicolumn{4}{c}{\textbf{Bounds}}& 
\multicolumn{2}{c}{\textbf{Colors}}\\ 
 &
\textbf{graph}& 
$|V|$ & 
$|E|$ & 
$|T|$ & 
$\rho$ & 
$\bar{d}$ & 
${\rm r}$ & 
$\bar{\kappa}$ & 
$\kappa$ & 
$\tr_{\rm avg}$ & 
$\tr_{\rm max}$ & 
$\dmax$ & 
$K$+1 & 
$T$ & 
$\tilde{\omega}$ & 
$\coloring_{\min}$ &
$\coloring_{\max}$
\\ \midrule
\multirow{23}{*}{\rotatebox{90}{\textbf { \sc social networks}}}&    \textsf{soc-BlogCatalog}&  88.7K &  2M &  153M &  $10^{-4}$ &  47 &  -0.23 &  0.09 &  0.06 &  1.7K &  804.4K &  9.4K &  222 &  101 &  24 &  87 &  170  \\ 
	 &    \textsf{soc-FourSquare}&  639K &  3.2M &  64.9M &  $10^{-5}$ &  10 &  -0.71 &  0.03 &  0.00 &  102 &  1.9M &  106.2K &  64 &  38 &  25 &  34 &  47  \\ 
	 &    \textsf{soc-LiveMocha}&  104.1K &  2.1M &  10M &  $10^{-4}$ &  42 &  -0.15 &  0.01 &  0.01 &  97 &  36.9K &  2.9K &  93 &  27 &  10 &  34 &  76  \\ 
	 &    \textsf{soc-brightkite}&  56.7K &  212.9K &  1.4M &  $10^{-4}$ &  7 &  0.01 &  0.07 &  0.11 &  26 &  11.5K &  1.1K &  53 &  43 &  31 &  39 &  56  \\ 
	 &    \textsf{soc-buzznet}&  101.1K &  2.7M &  92.7M &  $10^{-4}$ &  54 &  2.85 &  0.01 &  0.03 &  917 &  1M &  64.2K &  154 &  59 &  21 &  62 &  125  \\ 
	 &    \textsf{soc-delicious}&  536.1K &  1.3M &  1.4M &  $10^{-6}$ &  5 &  -0.07 &  0.02 &  0.01 &  3 &  8K &  3.2K &  34 &  23 &  17 &  21 &  35  \\ 
	 &    \textsf{soc-digg}&  770.7K &  5.9M &  188M &  $10^{-5}$ &  15 &  -0.09 &  0.04 &  0.05 &  244 &  396K &  17.6K &  237 &  73 &  41 &  64 &  127  \\ 
	 &   \cellcolor{verylightblue}\textsf{soc-dolphins}& \cellcolor{verylightblue}62 & \cellcolor{verylightblue}159 & \cellcolor{verylightblue}285 & \cellcolor{verylightblue}0.08 & \cellcolor{verylightblue}5 & \cellcolor{verylightblue}-0.04 & \cellcolor{verylightblue}0.00 & \cellcolor{verylightblue}0.31 & \cellcolor{verylightblue}5 & \cellcolor{verylightblue}17 & \cellcolor{verylightblue}12 & \cellcolor{verylightblue}5 & \cellcolor{verylightblue}5 & \cellcolor{verylightblue}5 & \cellcolor{verylightblue}5 & \cellcolor{verylightblue}7  \\ 
	 &    \textsf{soc-douban}&  154.9K &  327.1K &  121K &  $10^{-5}$ &  4 &  -0.18 &  0.01 &  0.01 &  1 &  394 &  287 &  16 &  11 &  8 &  13 &  19  \\ 
	 &    \textsf{soc-epinions}&  26.5K &  100.1K &  479K &  $10^{-4}$ &  7 &  0.06 &  0.06 &  0.09 &  18 &  5.1K &  443 &  33 &  18 &  14 &  20 &  39  \\ 
	 &    \textsf{soc-flickr}&  513.9K &  3.1M &  176M &  $10^{-5}$ &  12 &  0.16 &  0.08 &  0.15 &  343 &  524K &  4.3K &  310 &  153 &  21 &  104 &  208  \\ 
	 &    \textsf{soc-flixster}&  2.5M &  7.9M &  23.6M &  $10^{-6}$ &  6 &  -0.32 &  0.05 &  0.01 &  9 &  15.1K &  1.4K &  69 &  47 &  29 &  40 &  75  \\ 
	 &   \cellcolor{verylightred}\textsf{soc-gowalla}& \cellcolor{verylightred}196.5K & \cellcolor{verylightred}950K & \cellcolor{verylightred}6.8M & \cellcolor{verylightred}$10^{-5}$ & \cellcolor{verylightred}9 & \cellcolor{verylightred}-0.03 & \cellcolor{verylightred}0.09 & \cellcolor{verylightred}0.02 & \cellcolor{verylightred}35 & \cellcolor{verylightred}93.8K & \cellcolor{verylightred}14.7K & \cellcolor{verylightred}52 & \cellcolor{verylightred}29 & \cellcolor{verylightred}29 & \cellcolor{verylightred}29 & \cellcolor{verylightred}64  \\ 
	 &   \cellcolor{verylightblue}\textsf{soc-karate}& \cellcolor{verylightblue}34 & \cellcolor{verylightblue}78 & \cellcolor{verylightblue}135 & \cellcolor{verylightblue}0.14 & \cellcolor{verylightblue}4 & \cellcolor{verylightblue}-0.48 & \cellcolor{verylightblue}0.32 & \cellcolor{verylightblue}0.26 & \cellcolor{verylightblue}4 & \cellcolor{verylightblue}18 & \cellcolor{verylightblue}17 & \cellcolor{verylightblue}5 & \cellcolor{verylightblue}5 & \cellcolor{verylightblue}5 & \cellcolor{verylightblue}5 & \cellcolor{verylightblue}6  \\ 
	 &    \textsf{soc-lastfm}&  1.1M &  4.5M &  11.8M &  $10^{-6}$ &  7 &  -0.14 &  0.03 &  0.01 &  10 &  38K &  5.1K &  71 &  23 &  14 &  24 &  57  \\ 
	 &   \cellcolor{verylightblue}\textsf{soc-livejournal}& \cellcolor{verylightblue}4M & \cellcolor{verylightblue}27.9M & \cellcolor{verylightblue}250.6M & \cellcolor{verylightblue}$10^{-6}$ & \cellcolor{verylightblue}13 & \cellcolor{verylightblue}0.27 & \cellcolor{verylightblue}0.09 & \cellcolor{verylightblue}0.14 & \cellcolor{verylightblue}62 & \cellcolor{verylightblue}79.7K & \cellcolor{verylightblue}2.6K & \cellcolor{verylightblue}214 & \cellcolor{verylightblue}214 & \cellcolor{verylightblue}214 & \cellcolor{verylightblue}214 & \cellcolor{verylightblue}218  \\ 
	 &    \textsf{soc-orkut}&  2.9M &  106.3M &  1.5T &  $10^{-5}$ &  70 &  0.02 &  0.01 &  0.04 &  525 &  1.3M &  27.4K &  231 &  75 &  37 &  83 &  190  \\ 
	 &    \textsf{soc-pokec}&  1.6M &  22.3M &  97.6M &  $10^{-5}$ &  27 &  0.00 &  0.02 &  0.05 &  60 &  29.2K &  14.8K &  48 &  29 &  29 &  30 &  62  \\ 
	 &    \textsf{soc-slashdot}&  70K &  358.6K &  1.2M &  $10^{-4}$ &  10 &  -0.07 &  0.03 &  0.03 &  17 &  13.3K &  2.5K &  54 &  35 &  17 &  34 &  60  \\ 
	 &    \textsf{soc-twitter-follows}&  404.7K &  713K &  88.6K &  $10^{-6}$ &  3 &  -0.88 &  0.01 &  0.00 &  0 &  1.6K &  626 &  29 &  6 &  6 &  7 &  14  \\ 
	 &   \cellcolor{verylightred}\textsf{soc-wiki-Vote}& \cellcolor{verylightred}889 & \cellcolor{verylightred}2.9K & \cellcolor{verylightred}6.3K & \cellcolor{verylightred}$10^{-3}$ & \cellcolor{verylightred}6 & \cellcolor{verylightred}-0.03 & \cellcolor{verylightred}0.04 & \cellcolor{verylightred}0.13 & \cellcolor{verylightred}7 & \cellcolor{verylightred}251 & \cellcolor{verylightred}102 & \cellcolor{verylightred}10 & \cellcolor{verylightred}7 & \cellcolor{verylightred}7 & \cellcolor{verylightred}7 & \cellcolor{verylightred}15  \\ 
	 &    \textsf{soc-youtube-snap}&  1.1M &  2.9M &  9.1M &  $10^{-6}$ &  5 &  -0.04 &  0.04 &  0.01 &  8 &  180K &  28.7K &  52 &  19 &  13 &  30 &  64  \\ 
	 &    \textsf{soc-youtube}&  495K &  1.9M &  7.3M &  $10^{-5}$ &  7 &  -0.03 &  0.05 &  0.01 &  15 &  151K &  25.4K &  50 &  19 &  11 &  28 &  61  \\ 
\midrule 
\multirow{17}{*}{\rotatebox{90}{\textbf { \sc facebook networks}}}&    \textsf{fb-A-anon}&  3M &  23.6M &  166M &  $10^{-6}$ &  15 &  -0.06 &  0.04 &  0.05 &  54 &  50.2K &  4.9K &  75 &  30 &  23 &  33 &  69  \\ 
	 &    \textsf{fb-B-anon}&  2.9M &  20.9M &  155.9M &  $10^{-6}$ &  14 &  -0.11 &  0.04 &  0.05 &  53 &  36.8K &  4.3K &  64 &  31 &  23 &  29 &  60  \\ 
	 &    \textsf{fb-Berkeley13}&  22.9K &  852.4K &  16.1M &  $10^{-3}$ &  74 &  0.01 &  0.01 &  0.11 &  703 &  69.5K &  3.4K &  65 &  47 &  39 &  48 &  84  \\ 
	 &    \textsf{fb-CMU}&  6.6K &  249.9K &  6.9M &  0.01 &  75 &  0.12 &  0.02 &  0.19 &  1K &  24K &  840 &  70 &  45 &  42 &  49 &  83  \\ 
	 &    \textsf{fb-Duke14}&  9.8K &  506.4K &  15.4M &  0.01 &  102 &  0.07 &  0.02 &  0.17 &  1.5K &  41.9K &  1.8K &  86 &  47 &  29 &  47 &  85  \\ 
	 &    \textsf{fb-Indiana}&  29.7K &  1.3M &  28.1M &  $10^{-3}$ &  87 &  0.13 &  0.01 &  0.14 &  948 &  37.2K &  1.3K &  77 &  53 &  43 &  52 &  91  \\ 
	 &    \textsf{fb-MIT}&  6.4K &  251.2K &  7.1M &  0.01 &  78 &  0.12 &  0.02 &  0.18 &  1.1K &  27.7K &  708 &  73 &  41 &  30 &  44 &  78  \\ 
	 &    \textsf{fb-OR}&  63.3K &  816.8K &  10.5M &  $10^{-4}$ &  25 &  0.18 &  0.04 &  0.15 &  166 &  19.4K &  1K &  53 &  36 &  28 &  36 &  63  \\ 
	 &    \textsf{fb-Penn94}&  41.5K &  1.3M &  21.6M &  $10^{-3}$ &  65 &  -0.00 &  0.01 &  0.10 &  521 &  68K &  4.4K &  63 &  48 &  43 &  47 &  78  \\ 
	 &    \textsf{fb-Stanford3}&  11.5K &  568.3K &  17.5M &  $10^{-3}$ &  98 &  0.10 &  0.02 &  0.16 &  1.5K &  33.1K &  1.1K &  92 &  60 &  47 &  58 &  90  \\ 
	 &    \textsf{fb-Texas84}&  36.3K &  1.5M &  33.5M &  $10^{-3}$ &  87 &  -0.00 &  0.01 &  0.10 &  922 &  141K &  6.3K &  82 &  62 &  44 &  57 &  100  \\ 
	 &    \textsf{fb-UCLA}&  20.4K &  747.6K &  15.3M &  $10^{-3}$ &  73 &  0.14 &  0.02 &  0.14 &  750 &  17.5K &  1.1K &  66 &  54 &  49 &  53 &  78  \\ 
	 &    \textsf{fb-UCSB37}&  14.9K &  482.2K &  9.2M &  $10^{-3}$ &  64 &  0.18 &  0.01 &  0.16 &  619 &  16.1K &  810 &  66 &  60 &  51 &  56 &  78  \\ 
	 &    \textsf{fb-UConn}&  17.2K &  604.8K &  10.2M &  $10^{-3}$ &  70 &  0.09 &  0.01 &  0.13 &  596 &  21.5K &  1.7K &  66 &  53 &  47 &  51 &  75  \\ 
	 &    \textsf{fb-UF}&  35.1K &  1.4M &  36.4M &  $10^{-3}$ &  83 &  -0.01 &  0.01 &  0.12 &  1K &  159K &  8.2K &  84 &  67 &  51 &  61 &  100  \\ 
	 &    \textsf{fb-UIllinois}&  30.7K &  1.2M &  28M &  $10^{-3}$ &  82 &  0.03 &  0.01 &  0.14 &  911 &  66.1K &  4.6K &  86 &  65 &  54 &  59 &  88  \\ 
	 &    \textsf{fb-Wisconsin87}&  23.8K &  835.9K &  14.5M &  $10^{-3}$ &  70 &  -0.00 &  0.01 &  0.12 &  612 &  46.7K &  3.4K &  61 &  42 &  34 &  42 &  71  \\ 
\midrule 
\multirow{7}{*}{\rotatebox{90}{\textbf { \sc technological}}}&    \textsf{tech-RL-caida}&  190K &  607K &  1.3M &  $10^{-5}$ &  6 &  0.02 &  0.06 &  0.06 &  7 &  6K &  1K &  33 &  19 &  15 &  18 &  34  \\ 
	 &    \textsf{tech-WHOIS}&  7.4K &  56.9K &  2.3M &  $10^{-3}$ &  15 &  -0.04 &  0.26 &  0.31 &  314 &  22.2K &  1K &  89 &  71 &  49 &  66 &  88  \\ 
	 &    \textsf{tech-as-caida07}&  26.4K &  53.3K &  109K &  $10^{-4}$ &  4 &  -0.19 &  0.16 &  0.01 &  4 &  3.8K &  2.6K &  23 &  16 &  9 &  18 &  30  \\ 
	 &    \textsf{tech-as-skitter}&  1.6M &  11M &  86.3M &  $10^{-6}$ &  13 &  -0.08 &  0.08 &  0.01 &  51 &  564.6K &  35.4K &  112 &  68 &  41 &  70 &  115  \\ 
	 &    \textsf{tech-internet-as}&  40.1K &  85.1K &  189K &  $10^{-4}$ &  4 &  -0.18 &  0.15 &  0.01 &  5 &  8.5K &  3.3K &  24 &  17 &  14 &  18 &  28  \\ 
	 &    \textsf{tech-p2p-gnutella}&  62.5K &  147K &  6K &  $10^{-5}$ &  4 &  -0.09 &  0.00 &  0.00 &  0 &  17 &  95 &  7 &  4 &  4 &  7 &  11  \\ 
	 &   \cellcolor{verylightblue}\textsf{tech-routers-rf}& \cellcolor{verylightblue}2.1K & \cellcolor{verylightblue}6.6K & \cellcolor{verylightblue}31.2K & \cellcolor{verylightblue}$10^{-3}$ & \cellcolor{verylightblue}6 & \cellcolor{verylightblue}0.02 & \cellcolor{verylightblue}0.11 & \cellcolor{verylightblue}0.23 & \cellcolor{verylightblue}15 & \cellcolor{verylightblue}588 & \cellcolor{verylightblue}109 & \cellcolor{verylightblue}16 & \cellcolor{verylightblue}16 & \cellcolor{verylightblue}16 & \cellcolor{verylightblue}16 & \cellcolor{verylightblue}20  \\ 
\midrule 
\multirow{13}{*}{\rotatebox{90}{\textbf { \sc web networks}}}&   \cellcolor{verylightgreen}\textsf{web-BerkStan}& \cellcolor{verylightgreen}12.3K & \cellcolor{verylightgreen}19.5K & \cellcolor{verylightgreen}30.9K & \cellcolor{verylightgreen}$10^{-4}$ & \cellcolor{verylightgreen}3 & \cellcolor{verylightgreen}0.12 & \cellcolor{verylightgreen}0.18 & \cellcolor{verylightgreen}0.28 & \cellcolor{verylightgreen}3 & \cellcolor{verylightgreen}384 & \cellcolor{verylightgreen}59 & \cellcolor{verylightgreen}29 & \cellcolor{verylightgreen}29 & \cellcolor{verylightgreen}29 & \cellcolor{verylightgreen}29 & \cellcolor{verylightgreen}29  \\ 
	 &   \cellcolor{verylightgreen}\textsf{web-arabic05}& \cellcolor{verylightgreen}163.5K & \cellcolor{verylightgreen}1.7M & \cellcolor{verylightgreen}65M & \cellcolor{verylightgreen}$10^{-4}$ & \cellcolor{verylightgreen}21 & \cellcolor{verylightgreen}0.15 & \cellcolor{verylightgreen}0.51 & \cellcolor{verylightgreen}0.95 & \cellcolor{verylightgreen}397 & \cellcolor{verylightgreen}5.8K & \cellcolor{verylightgreen}1.1K & \cellcolor{verylightgreen}102 & \cellcolor{verylightgreen}102 & \cellcolor{verylightgreen}102 & \cellcolor{verylightgreen}102 & \cellcolor{verylightgreen}102  \\ 
	 &   \cellcolor{verylightblue}\textsf{web-edu}& \cellcolor{verylightblue}3K & \cellcolor{verylightblue}6.4K & \cellcolor{verylightblue}30.1K & \cellcolor{verylightblue}$10^{-3}$ & \cellcolor{verylightblue}4 & \cellcolor{verylightblue}-0.17 & \cellcolor{verylightblue}0.07 & \cellcolor{verylightblue}0.27 & \cellcolor{verylightblue}10 & \cellcolor{verylightblue}523 & \cellcolor{verylightblue}104 & \cellcolor{verylightblue}30 & \cellcolor{verylightblue}30 & \cellcolor{verylightblue}30 & \cellcolor{verylightblue}30 & \cellcolor{verylightblue}31  \\ 
	 &   \cellcolor{verylightblue}\textsf{web-google}& \cellcolor{verylightblue}1.2K & \cellcolor{verylightblue}2.7K & \cellcolor{verylightblue}15.2K & \cellcolor{verylightblue}$10^{-3}$ & \cellcolor{verylightblue}4 & \cellcolor{verylightblue}-0.05 & \cellcolor{verylightblue}0.25 & \cellcolor{verylightblue}0.53 & \cellcolor{verylightblue}12 & \cellcolor{verylightblue}189 & \cellcolor{verylightblue}59 & \cellcolor{verylightblue}18 & \cellcolor{verylightblue}18 & \cellcolor{verylightblue}18 & \cellcolor{verylightblue}18 & \cellcolor{verylightblue}19  \\ 
	 &   \cellcolor{verylightgreen}\textsf{web-indochina04}& \cellcolor{verylightgreen}11.3K & \cellcolor{verylightgreen}47.6K & \cellcolor{verylightgreen}630.2K & \cellcolor{verylightgreen}$10^{-4}$ & \cellcolor{verylightgreen}8 & \cellcolor{verylightgreen}0.12 & \cellcolor{verylightgreen}0.48 & \cellcolor{verylightgreen}0.57 & \cellcolor{verylightgreen}55 & \cellcolor{verylightgreen}1.4K & \cellcolor{verylightgreen}199 & \cellcolor{verylightgreen}50 & \cellcolor{verylightgreen}50 & \cellcolor{verylightgreen}50 & \cellcolor{verylightgreen}50 & \cellcolor{verylightgreen}50  \\ 
	 &    \textsf{web-it04}&  509K &  7.1M &  1T &  $10^{-5}$ &  28 &  0.99 &  0.76 &  0.95 &  1.9K &  93.3K &  469 &  432 &  432 &  431 &  432 &  432  \\ 
	 &    \textsf{web-polblogs}&  643 &  2.2K &  9K &  0.01 &  7 &  -0.22 &  0.06 &  0.16 &  14 &  392 &  165 &  13 &  10 &  9 &  10 &  15  \\ 
	 &   \cellcolor{verylightgreen}\textsf{web-sk-2005}& \cellcolor{verylightgreen}121.4K & \cellcolor{verylightgreen}334.4K & \cellcolor{verylightgreen}2.9M & \cellcolor{verylightgreen}$10^{-5}$ & \cellcolor{verylightgreen}5 & \cellcolor{verylightgreen}0.08 & \cellcolor{verylightgreen}0.09 & \cellcolor{verylightgreen}0.47 & \cellcolor{verylightgreen}25 & \cellcolor{verylightgreen}3.4K & \cellcolor{verylightgreen}590 & \cellcolor{verylightgreen}82 & \cellcolor{verylightgreen}82 & \cellcolor{verylightgreen}82 & \cellcolor{verylightgreen}82 & \cellcolor{verylightgreen}82  \\ 
	 &    \textsf{web-spam}&  4.7K &  37.3K &  387K &  $10^{-3}$ &  15 &  0.00 &  0.08 &  0.15 &  81 &  6.2K &  477 &  36 &  23 &  20 &  22 &  42  \\ 
	 &   \cellcolor{verylightgreen}\textsf{web-uk-2005}& \cellcolor{verylightgreen}129K & \cellcolor{verylightgreen}11.7M & \cellcolor{verylightgreen}2.5T & \cellcolor{verylightgreen}$10^{-3}$ & \cellcolor{verylightgreen}181 & \cellcolor{verylightgreen}1.00 & \cellcolor{verylightgreen}0.98 & \cellcolor{verylightgreen}1.00 & \cellcolor{verylightgreen}19.3K & \cellcolor{verylightgreen}124.2K & \cellcolor{verylightgreen}850 & \cellcolor{verylightgreen}500 & \cellcolor{verylightgreen}500 & \cellcolor{verylightgreen}500 & \cellcolor{verylightgreen}500 & \cellcolor{verylightgreen}500  \\ 
	 &   \cellcolor{verylightgreen}\textsf{web-webbase01}& \cellcolor{verylightgreen}16K & \cellcolor{verylightgreen}25.5K & \cellcolor{verylightgreen}63.3K & \cellcolor{verylightgreen}$10^{-4}$ & \cellcolor{verylightgreen}3 & \cellcolor{verylightgreen}-0.10 & \cellcolor{verylightgreen}0.17 & \cellcolor{verylightgreen}0.02 & \cellcolor{verylightgreen}4 & \cellcolor{verylightgreen}1.3K & \cellcolor{verylightgreen}1.6K & \cellcolor{verylightgreen}33 & \cellcolor{verylightgreen}33 & \cellcolor{verylightgreen}33 & \cellcolor{verylightgreen}33 & \cellcolor{verylightgreen}33  \\ 
	 &   \cellcolor{verylightred}\textsf{web-wikipedia09}& \cellcolor{verylightred}1.8M & \cellcolor{verylightred}4.5M & \cellcolor{verylightred}6.6M & \cellcolor{verylightred}$10^{-6}$ & \cellcolor{verylightred}4 & \cellcolor{verylightred}0.05 & \cellcolor{verylightred}0.07 & \cellcolor{verylightred}0.05 & \cellcolor{verylightred}4 & \cellcolor{verylightred}12.4K & \cellcolor{verylightred}2.6K & \cellcolor{verylightred}67 & \cellcolor{verylightred}31 & \cellcolor{verylightred}31 & \cellcolor{verylightred}31 & \cellcolor{verylightred}32  \\ 
\midrule 
\end{tabularx}
\end{table}

\section{Results and Analysis}
\label{sec:results}

This section evaluates the proposed methods using a large collection of graphs.
In particular, we designed experiments to answer the following questions:

\smallskip
\begin{compactenum}[
\leftmargin=-10em]
  \setlength{\itemsep}{2pt}
  \setlength{\parskip}{0pt}
  \setlength{\parsep}{4pt}
\item \noindent  \textbf{\textcolor{theblue}{Section}~\ref{sec:accuracy}) Accuracy}. Are the proposed greedy coloring methods effective and accurate for social and information networks?
\item \textbf{\textcolor{theblue}{Section}~\ref{sec:scalability}) Scalability}. Do the methods scale for coloring large graphs?
\item \textbf{\textcolor{theblue}{Section}~\ref{sec:results-recolor}) Impact of \textsc{Recolor}ing}. Is the \textsc{recolor} method effective in reducing the number of colors used?
\item \textbf{\textcolor{theblue}{Section}~\ref{sec:results-bounds}) Utility of Bounds}. Are the lower and upper bounds useful and informative? 
\eenum
\smallskip

\noindent
For these experiments we used over 100+ networks of different types (i.e., social vs. biological), sizes, structural properties, and sparsity.
Our main focus was on a variety of \emph{large sparse networks} including social, biological, information, and technological networks\footnote{\tiny \url{http://www.networkrepository.com/}}.
Self-loops and any weights were discarded.
For comparison, we also used a variety of dense graphs including the DIMACs\footnote{\tiny \url{http://iridia.ulb.ac.be/~fmascia/maximum_clique/}} graph collection and the BHOSLIB\footnote{\tiny \url{http://www.nlsde.buaa.edu.cn/~kexu/benchmarks/graph-benchmarks.htm}} graph collection (benchmarks with hidden optimum solutions) which were generated from joining cliques together.

In this work, ties are broken as follows:
Given two vertices $v_i$ and $v_j$ where $\f(v_i) = \f(v_j)$, then $v_i$ is ordered before $v_j$ if $i > j$.
While the importance of tie-breaking was discussed in Section~\ref{sec:framework}, 
many results in the literature are difficult to reproduce as key details such as the tie-breaking strategy are left undefined.

\begin{table}[t!]
\caption{
\textbf{\foliofont Accuracy of coloring methods.} 
\xfigtextfont
We report the frequency (number of graphs) for which each algorithm performed the best overall.
Graphs for which all algorithms performed equally were discarded.
 }
 \label{table:accuracy-results}
\centering\small 
\xfigtextfont
\begin{tabularx}{0.70\linewidth}
{ H r !{\vrule width 0.30mm} X X !{\vrule width 0.30mm} X X X X X X X !{\vrule width 0.50mm} }
\TTT \BBB & 
& 
\multicolumn{2}{c !{\vrule width 0.30mm}}{} &
\multicolumn{7}{c !{\vrule width 0.50mm}}{{\bf \normalsize \xfigtextfont \TTT \BBB \tabtextfont Types of Sparse Graphs}} 
\\ 
& 
 {\normalsize \textbf{\tabtextfont Algorithm}} & 
 \rotatebox{90}{\textsc{\xfigtextfont \normalsize\tabtextfont Sparse}} & \rotatebox{90}{\textsc{\normalsize \xfigtextfont \normalsize\tabtextfont Dense}} &
 \rotatebox{90}{\textbf{\xfigtextfont Biological}} & \rotatebox{90}{\textbf{\xfigtextfont Collaboration}} & \rotatebox{90}{\textbf{\xfigtextfont Interaction}} & \rotatebox{90}{\textbf{\xfigtextfont Social networks}} & \rotatebox{90}{\textbf{\xfigtextfont Facebook networks}} & \rotatebox{90}{\textbf{\xfigtextfont Technological}} & \rotatebox{90}{\textbf{\xfigtextfont Web networks}}  \\ 
\hboldline
     &    
 
\cellcolor{color1and29} \textsc{rand}  &    
\cellcolor{color1and29} \text{0}  &    \cellcolor{color1and30} \text{7}    & 
\cellcolor{bio-color1and2} \text{0}  &    \cellcolor{ca-color1and4} \text{0}  &    \cellcolor{ia-color1and7} \text{0}  &    \cellcolor{soc-color1and11} \text{0}  &    \cellcolor{socfbcolor1and8} \text{0}  &    \cellcolor{tech-color1and2} \text{0}  &    \cellcolor{web-color1and4} \text{0}   \\ 

    &    \cellcolor{color2and29} \textsc{deg-vol}  &    
\cellcolor{color2and29} \text{13}  &    \cellcolor{color8and30} \text{29}  &
\cellcolor{bio-color2and2} \text{1}  &    \cellcolor{ca-color2and4} \text{3}  &    \cellcolor{ia-color3and7} \text{4}  &    \cellcolor{soc-color4and11} \text{3}  &    \cellcolor{socfbcolor1and8} \text{0}  &    \cellcolor{tech-color2and2} \text{1}  &    \cellcolor{web-color2and4} \text{1}   \\ 

    &    \cellcolor{color3and29} \textsc{dlf}  &    
     \cellcolor{color3and29} \text{14}  &    \cellcolor{color6and30} \text{23}  &
    \cellcolor{bio-color2and2} \text{1}  &    \cellcolor{ca-color3and4} \text{4}  &    \cellcolor{ia-color3and7} \text{4}  &    \cellcolor{soc-color3and11} \text{2}  &    \cellcolor{socfbcolor1and8} \text{0}  &    \cellcolor{tech-color3and2} \text{2}  &    \cellcolor{web-color2and4} \text{1}   \\ 

    &    \cellcolor{color3and29} \textsc{dist-two-dlf}  &    
 \cellcolor{color3and29} \text{14}  &    \cellcolor{color6and30} \text{23}  &    
    \cellcolor{bio-color2and2} \text{1}  &    \cellcolor{ca-color3and4} \text{4}  &    \cellcolor{ia-color3and7} \text{4}  &    \cellcolor{soc-color3and11} \text{2}  &    \cellcolor{socfbcolor1and8} \text{0}  &    \cellcolor{tech-color3and2} \text{2}  &    \cellcolor{web-color2and4} \text{1}   \\ 

    &    \cellcolor{color3and29} \textsc{dist-two-kcore}  &    
\cellcolor{color3and29} \text{14}  &    \cellcolor{color6and30} \text{23}  &    
    \cellcolor{bio-color2and2} \text{1}  &    \cellcolor{ca-color3and4} \text{4}  &    \cellcolor{ia-color3and7} \text{4}  &    \cellcolor{soc-color3and11} \text{2}  &    \cellcolor{socfbcolor1and8} \text{0}  &    \cellcolor{tech-color3and2} \text{2}  &    \cellcolor{web-color2and4} \text{1}   \\ 

    &    \cellcolor{color3and29} \textsc{ido}  &    
\cellcolor{color3and29} \text{14}  &    \cellcolor{color6and30} \text{23}  &    
    \cellcolor{bio-color2and2} \text{1}  &    \cellcolor{ca-color3and4} \text{4}  &    \cellcolor{ia-color3and7} \text{4}  &    \cellcolor{soc-color3and11} \text{2}  &    \cellcolor{socfbcolor1and8} \text{0}  &    \cellcolor{tech-color3and2} \text{2}  &    \cellcolor{web-color2and4} \text{1}   \\ 

    &    \cellcolor{color3and29} \textsc{dist-two-ido}  &    
  \cellcolor{color3and29} \text{14}  &    \cellcolor{color6and30} \text{23}  &    
    \cellcolor{bio-color2and2} \text{1}  &    \cellcolor{ca-color3and4} \text{4}  &    \cellcolor{ia-color3and7} \text{4}  &    \cellcolor{soc-color3and11} \text{2}  &    \cellcolor{socfbcolor1and8} \text{0}  &    \cellcolor{tech-color3and2} \text{2}  &    \cellcolor{web-color2and4} \text{1}   \\ 

    &    \cellcolor{color3and29} \textsc{natural}  &    
 \cellcolor{color3and29} \text{14}  &    \cellcolor{color10and30} \text{31}  &    
    \cellcolor{bio-color1and2} \text{0}  &    \cellcolor{ca-color2and4} \text{3}  &    \cellcolor{ia-color3and7} \text{4}  &    \cellcolor{soc-color5and11} \text{4}  &    \cellcolor{socfbcolor1and8} \text{0}  &    \cellcolor{tech-color3and2} \text{2}  &    \cellcolor{web-color2and4} \text{1}   \\ 

    &    \cellcolor{color3and29} \textsc{kcore}  &    
\cellcolor{color3and29} \text{14}  &    \cellcolor{color7and30} \text{26}  &    
    \cellcolor{bio-color2and2} \text{1}  &    \cellcolor{ca-color2and4} \text{3}  &    \cellcolor{ia-color4and7} \text{5}  &    \cellcolor{soc-color4and11} \text{3}  &    \cellcolor{socfbcolor1and8} \text{0}  &    \cellcolor{tech-color2and2} \text{1}  &    \cellcolor{web-color2and4} \text{1}   \\ 

    &    \cellcolor{color3and29} \textsc{deg}  &    
\cellcolor{color3and29} \text{14}  &    \cellcolor{color7and30} \text{26}  &    
    \cellcolor{bio-color2and2} \text{1}  &    \cellcolor{ca-color2and4} \text{3}  &    \cellcolor{ia-color4and7} \text{5}  &    \cellcolor{soc-color4and11} \text{3}  &    \cellcolor{socfbcolor1and8} \text{0}  &    \cellcolor{tech-color2and2} \text{1}  &    \cellcolor{web-color2and4} \text{1}   \\ 

    &    \cellcolor{color4and29} \textsc{tri}  &    
\cellcolor{color4and29} \text{15}  &    \cellcolor{color7and30} \text{26}  &    
    \cellcolor{bio-color2and2} \text{1}  &    \cellcolor{ca-color2and4} \text{3}  &    \cellcolor{ia-color4and7} \text{5}  &    \cellcolor{soc-color5and11} \text{4}  &    \cellcolor{socfbcolor1and8} \text{0}  &    \cellcolor{tech-color2and2} \text{1}  &    \cellcolor{web-color2and4} \text{1}   \\ 

    &    \cellcolor{color5and29} \textsc{kcore-deg}  &    
  \cellcolor{color5and29} \text{16}  &    \cellcolor{color7and30} \text{26}  &    
    \cellcolor{bio-color2and2} \text{1}  &    \cellcolor{ca-color2and4} \text{3}  &    \cellcolor{ia-color4and7} \text{5}  &    \cellcolor{soc-color5and11} \text{4}  &    \cellcolor{socfbcolor1and8} \text{0}  &    \cellcolor{tech-color3and2} \text{2}  &    \cellcolor{web-color2and4} \text{1}   \\ 

    &    \cellcolor{color5and29} \textsc{kcore-vol}  &    
\cellcolor{color5and29} \text{16}  &    \cellcolor{color7and30} \text{26}  &
\cellcolor{bio-color2and2} \text{1}  &    \cellcolor{ca-color2and4} \text{3}  &    \cellcolor{ia-color4and7} \text{5}  &    \cellcolor{soc-color5and11} \text{4}  &    \cellcolor{socfbcolor1and8} \text{0}  &    \cellcolor{tech-color3and2} \text{2}  &    \cellcolor{web-color2and4} \text{1}   \\ 
 
    &    \cellcolor{color5and29} \textsc{deg-kco-tri-vol}  &    
\cellcolor{color5and29} \text{16}  &    \cellcolor{color2and30} \text{12}  &    
    \cellcolor{bio-color2and2} \text{1}  &    \cellcolor{ca-color3and4} \text{4}  &    \cellcolor{ia-color4and7} \text{5}  &    \cellcolor{soc-color2and11} \text{1}  &    \cellcolor{socfbcolor1and8} \text{0}  &    \cellcolor{tech-color2and2} \text{1}  &    \cellcolor{web-color4and4} \text{4}  \\ 
  
    &    \cellcolor{color6and29} \textsc{kcore-tri-vol}  &    
     \cellcolor{color6and29} \text{25}  &    \cellcolor{color3and30} \text{15}  &
    \cellcolor{bio-color3and2} \text{2}  &    \cellcolor{ca-color3and4} \text{4}  &    \cellcolor{ia-color2and7} \text{3}  &    \cellcolor{soc-color6and11} \text{7}  &    \cellcolor{socfbcolor3and8} \text{6}  &    \cellcolor{tech-color2and2} \text{1}  &    \cellcolor{web-color3and4} \text{2}  \\ 

    &    \cellcolor{color7and29} \textsc{deg-tri}  &    
     \cellcolor{color7and29} \text{26}  &    \cellcolor{color9and30} \text{30}  &
    \cellcolor{bio-color2and2} \text{1}  &    \cellcolor{ca-color2and4} \text{3}  &    \cellcolor{ia-color2and7} \text{3}  &    \cellcolor{soc-color6and11} \text{7}  &    \cellcolor{socfbcolor4and8} \text{8}  &    \cellcolor{tech-color3and2} \text{2}  &    \cellcolor{web-color3and4} \text{2}   \\ 

    &    \cellcolor{color7and29} \textsc{kcore-tri}  &    
    \cellcolor{color7and29} \text{26}  &    \cellcolor{color8and30} \text{29}  &
    \cellcolor{bio-color2and2} \text{1}  &    \cellcolor{ca-color2and4} \text{3}  &    \cellcolor{ia-color2and7} \text{3}  &    \cellcolor{soc-color6and11} \text{7}  &    \cellcolor{socfbcolor4and8} \text{8}  &    \cellcolor{tech-color3and2} \text{2}  &    \cellcolor{web-color3and4} \text{2}   \\ 
 
    &    \cellcolor{color7and29} \textsc{deg-kcore-vol}  &    
    \cellcolor{color7and29} \text{26}  &    \cellcolor{color4and30} \text{16}  &
    \cellcolor{bio-color3and2} \text{2}  &    \cellcolor{ca-color3and4} \text{4}  &    \cellcolor{ia-color2and7} \text{3}  &    \cellcolor{soc-color8and11} \text{11}  &    \cellcolor{socfbcolor2and8} \text{2}  &    \cellcolor{tech-color3and2} \text{2}  &    \cellcolor{web-color3and4} \text{2}   \\ 

    &    \cellcolor{color8and29} \textsc{kcore-deg-tri}  &    
     \cellcolor{color8and29} \text{27}  &    \cellcolor{color5and30} \text{17}  &
\cellcolor{bio-color2and2} \text{1}  &    \cellcolor{ca-color3and4} \text{4}  &    \cellcolor{ia-color4and7} \text{5}  &    \cellcolor{soc-color8and11} \text{11}  &    \cellcolor{socfbcolor2and8} \text{2}  &    \cellcolor{tech-color3and2} \text{2}  &    \cellcolor{web-color3and4} \text{2}   \\ 

    &    \cellcolor{color9and29} \textsc{tri-vol}  &    
    \cellcolor{color9and29} \text{29}  &    \cellcolor{color11and30} \text{37}  &
\cellcolor{bio-color2and2} \text{1}  &    \cellcolor{ca-color2and4} \text{3}  &    \cellcolor{ia-color5and7} \text{7}  &    \cellcolor{soc-color7and11} \text{8}  &    \cellcolor{socfbcolor3and8} \text{6}  &    \cellcolor{tech-color1and2} \text{0}  &    \cellcolor{web-color4and4} \text{4}   \\ 

    &    \cellcolor{color9and29} \textsc{tcore-vol}  &    
\cellcolor{color9and29} \text{29}  &    \cellcolor{color11and30} \text{37}  &
\cellcolor{bio-color2and2} \text{1}  &    \cellcolor{ca-color2and4} \text{3}  &    \cellcolor{ia-color5and7} \text{7}  &    \cellcolor{soc-color7and11} \text{8}  &    \cellcolor{socfbcolor3and8} \text{6}  &    \cellcolor{tech-color1and2} \text{0}  &    \cellcolor{web-color4and4} \text{4}   \\ 

    &    \cellcolor{color9and29} \textsc{tcore-max}  &    
 \cellcolor{color9and29} \text{29}  &    \cellcolor{color11and30} \text{37}  &    
\cellcolor{bio-color2and2} \text{1}  &    \cellcolor{ca-color2and4} \text{3}  &    \cellcolor{ia-color5and7} \text{7}  &    \cellcolor{soc-color7and11} \text{8}  &    \cellcolor{socfbcolor3and8} \text{6}  &    \cellcolor{tech-color1and2} \text{0}  &    \cellcolor{web-color4and4} \text{4}   \\ 
\hline \toprule 
\end{tabularx} 
\end{table} 

\subsection{Accuracy}
\label{sec:accuracy}

As an error measure, 
we compute the frequency (i.e., number of graphs) for which each coloring method performed best overall, i.e., used the minimum number of colors.
If two methods used the minimum colors relative to the other methods, then the score of both are increased by one.
The graphs for which all methods achieved the best are ignored.
The proposed methods are evaluated below for use on 
(i) sparse/dense graphs and also 
(ii) for each type of large sparse network (i.e., social or information networks).

\smallskip \noindent {\normalsize \theoremfont Best Methods for Sparse and Dense Graphs:}
The methods are compared in Table~\ref{table:accuracy-results} (columns $2$ and $3$) independently on the basis of sparsity.
Notice the methods in the first column of Table~\ref{table:accuracy-results} are ranked and shaded according to their accuracy on sparse graphs (following an ascending order).
A few of our general findings from Table~\ref{table:accuracy-results} are discussed below.

\smallskip
\begin{compactenum}[{\scriptsize $\bullet$}]
  \setlength{\itemsep}{2pt}
  \setlength{\parskip}{0pt}
  \setlength{\parsep}{4pt}

\item Selecting the nodes uniformly at random (\textsc{RAND}) generally performs the worst for both sparse and dense graphs. 
This highlights the importance of selecting vertices that are more constrained in the number of possible colors first, which can't be achieved by random selection.

\item Nearly all the proposed methods (with the exception of \textsc{deg-vol}) gave fewer colors and found to be significantly better than the traditional degree-based methods.

\item As expected, the traditional degree-based methods are more suitable for dense graphs than sparse graphs. 
Nevertheless, the triangle and triangle-core methods performed the best on the majority of dense graphs.

\item In both sparse and dense graphs, we find that \textsc{tcore-max}/\textsc{vol}, and \textsc{tri-vol} gave the fewest colors overall. 

\item Interestingly, the natural order performed best on 31 of the dense graphs.
Further examination revealed that the majority of these cases are the BHOSLIB graphs.
These graphs are synthetically generated by forming $n$ distinct cliques and randomly connect pairs of cliques together. 
We found that the vertices in these cliques are ordered consecutively and thus give rise to this unexpected behavior found when using the natural order.

\end{compactenum}
\smallskip
\noindent
For additional insights, we provide the coloring bounds and various statistics for the DIMACs and BHOSLIB graph collections are provided in Table~\ref{table:stats-dense-dimacs} and Table~\ref{table:stats-dense-bhoslib}.
The coloring numbers from the various algorithms for the DIMACs and BHOSLIB graph collections are also shown in Table~\ref{table:results-colors-dense-dimacs} and Table~\ref{table:results-colors-dense-bhoslib}, respectively.
We find that in all cases, the proposed methods improve over the previous methods.
In some graphs, the proposed methods offer drastically better solutions with much fewer number of colors, for instance, see MANN-a81 which is currently an unsolved instance. 

{
\renewcommand{\tabcolsep}{1pt}
\begin{table}[h!]
\vspace{-2mm}
\caption{Colors used by the proposed methods.
For comparison, we used \textsc{rand}, \textsc{deg}, \textsc{ido}, and \textsc{dist-two-ido}.
We also provide the strong upper bounds and lower bound for additional insights.
For each network, we bold the best solution among all methods.
Note that we removed the less interesting networks (i.e., $\coloring_{\min} = \coloring_{\max}$, since those are effectively summarized in Table~\ref{table:stats-sparse-ca} and~\ref{table:stats-sparse-social}.
}
\vspace{-0mm}
\label{table:results-colors}
\centering\small\tiny
\fontsize{6.0}{6.5}\selectfont
\noindent
\begin{tabularx}{\linewidth}{ Hr Hc HHHH cHHH cccc  
rrrrr rrrrr rrrrr r}
\multicolumn{17}{c}{\textbf{\sectionfont Stats \& Bounds}} & 
\multicolumn{15}{c}{\textbf{\sectionfont Coloring Methods}} \\
 &
\textbf{graph}& 
$|V|$ & 
$|E|$ & 
$|T|$ & 
$\rho$ & 
$d_{\text{avg}}$ & 
${\rm r}$ & 
$\bar{\kappa}$ & 
$\kappa$ & 
$\tr_{\rm avg}$ & 
$\tr_{\rm max}$ & 
$\dmax$ & 
$K$+1 & 
$T$ & 
$\tilde{\omega}$ & 
\rotatebox{90}{\textsc{rand}}& \rotatebox{90}{\textsc{deg}}& \rotatebox{90}{\textsc{ido}}& \rotatebox{90}{\textsc{dist-two-ido}}& \rotatebox{90}{\textsc{triangles}}& \rotatebox{90}{\textsc{kcore-deg}}& \rotatebox{90}{\textsc{triangle-vol}}& \rotatebox{90}{\textsc{tcore-vol}}& \rotatebox{90}{\textsc{tcore-max}}& \rotatebox{90}{\textsc{deg-triangles}}& \rotatebox{90}{\textsc{kcore-triangles}}& \rotatebox{90}{\textsc{kcore-deg-tri}}& \rotatebox{90}{\textsc{deg-kcore-vol}}& \rotatebox{90}{\textsc{kcore-tri-vol}}& \rotatebox{90}{\textsc{deg-kcore-tri-vol}}\\ \midrule
 
	 & \TTZZ \BBZZ   \textsf{bio-dmela}&  7.3K &  25.5K &  8.6K &  $10^{-4}$ &  6 &  -0.05 &  0.00 &  0.01 &  1 &  225 &  190 &  12 &  7 &  7 &  12 &  9 &  9 &  9 &  9 &  9 &  9 &  9 &  9 &  9 &  9 &  9 &  \textbf{8}&  \textbf{8}&  9 & \\ 
\midrule
\multirow{9}{*}{\rotatebox{90}{\textbf{ \normalsize interaction}}}& \TTZZ \BBZZ   \textsf{ia-email-EU}&  32.4K &  54.3K &  146.9K &  $10^{-4}$ &  3 &  -0.38 &  0.09 &  0.03 &  5 &  1.6K &  623 &  23 &  13 &  11 &  23 &  19 &  19 &  19 &  18 &  18 &  17 &  17 &  17 &  17 &  17 &  17 &  17 &  17 &  \textbf{16}& \\ 
	 & \TTZZ \BBZZ   \textsf{ia-enron-large}&  33.6K &  180K &  2.1M &  $10^{-4}$ &  10 &  -0.12 &  0.34 &  0.09 &  65 &  17.7K &  1.3K &  44 &  22 &  15 &  40 &  31 &  31 &  31 &  30 &  30 &  \textbf{28}&  \textbf{28}&  \textbf{28}&  \textbf{28}&  \textbf{28}&  \textbf{28}&  \textbf{28}&  \textbf{28}&  37 & \\ 
	 & \TTZZ \BBZZ   \textsf{ia-fb-messages}&  1.2K &  6.4K &  7.4K &  $10^{-3}$ &  10 &  -0.08 &  0.02 &  0.04 &  6 &  242 &  112 &  12 &  5 &  5 &  12 &  9 &  9 &  9 &  \textbf{8}&  \textbf{8}&  \textbf{8}&  \textbf{8}&  \textbf{8}&  9 &  9 &  \textbf{8}&  9 &  \textbf{8}&  9 & \\ 
	 & \TTZZ \BBZZ  \cellcolor{verylightred}\textsf{ia-infect-dublin}& \cellcolor{verylightred}410 & \cellcolor{verylightred}2.7K & \cellcolor{verylightred}21.3K & \cellcolor{verylightred}0.03 & \cellcolor{verylightred}13 & \cellcolor{verylightred}0.23 & \cellcolor{verylightred}0.05 & \cellcolor{verylightred}0.44 & \cellcolor{verylightred}52 & \cellcolor{verylightred}280 & \cellcolor{verylightred}50 & \cellcolor{verylightred}18 & \cellcolor{verylightred}16 & \cellcolor{verylightred}16 & \cellcolor{verylightred}16 & \cellcolor{verylightred}17 & \cellcolor{verylightred}17 & \cellcolor{verylightred}17 & \cellcolor{verylightred}17 & \cellcolor{verylightred}17 & \cellcolor{verylightred}\textbf{16}& \cellcolor{verylightred}\textbf{16}& \cellcolor{verylightred}\textbf{16}& \cellcolor{verylightred}17 & \cellcolor{verylightred}17 & \cellcolor{verylightred}17 & \cellcolor{verylightred}17 & \cellcolor{verylightred}17 & \cellcolor{verylightred}\textbf{16}& \\ 
	 & \TTZZ \BBZZ   \textsf{ia-wiki-Talk}&  92.1K &  360K &  2.5M &  $10^{-5}$ &  7 &  -0.03 &  0.03 &  0.05 &  27 &  17.6K &  1.2K &  59 &  20 &  9 &  45 &  35 &  35 &  35 &  34 &  34 &  \textbf{30}&  \textbf{30}&  \textbf{30}&  31 &  31 &  31 &  31 &  31 &  40 & \\ 
\midrule
\multirow{21}{*}{\rotatebox{90}{\textbf{ \normalsize social networks}}}& \TTZZ \BBZZ   \textsf{soc-BlogCatalog}&  88.7K &  2M &  153.5M &  $10^{-4}$ &  47 &  -0.23 &  0.09 &  0.06 &  1.7K &  804.4K &  9.4K &  222 &  101 &  24 &  124 &  89 &  89 &  89 &  88 &  88 &  \textbf{87}&  \textbf{87}&  \textbf{87}&  88 &  90 &  109 &  108 &  115 &  117 & \\ 
	 & \TTZZ \BBZZ   \textsf{soc-LiveMocha}&  104K &  2.1M &  10M &  $10^{-4}$ &  42 &  -0.15 &  0.01 &  0.01 &  97 &  36.9K &  2.9K &  93 &  27 &  10 &  53 &  38 &  38 &  38 &  39 &  39 &  \textbf{34}&  \textbf{34}&  \textbf{34}&  36 &  36 &  37 &  38 &  36 &  45 & \\ 
	 & \TTZZ \BBZZ   \textsf{soc-brightkite}&  56.7K &  212K &  1.4M &  $10^{-4}$ &  7 &  0.01 &  0.07 &  0.11 &  26 &  11.5K &  1.1K &  53 &  43 &  31 &  49 &  40 &  40 &  40 &  40 &  40 &  \textbf{39}&  \textbf{39}&  \textbf{39}&  42 &  42 &  41 &  41 &  41 &  46 & \\ 
	 & \TTZZ \BBZZ   \textsf{soc-buzznet}&  101K &  2.7M &  92.7M &  $10^{-4}$ &  54 &  2.85 &  0.01 &  0.03 &  917 &  1M &  64.2K &  154 &  59 &  21 &  89 &  63 &  63 &  63 &  \textbf{62}&  \textbf{62}&  63 &  63 &  63 &  64 &  65 &  79 &  80 &  87 &  86 & \\ 
	 & \TTZZ \BBZZ   \textsf{soc-delicious}&  536K &  1.3M &  1.4M &  $10^{-6}$ &  5 &  -0.07 &  0.02 &  0.01 &  3 &  8K &  3.2K &  34 &  23 &  17 &  26 &  22 &  22 &  22 &  22 &  22 &  22 &  22 &  22 &  \textbf{21}&  \textbf{21}&  \textbf{21}&  \textbf{21}&  \textbf{21}&  22 & \\ 
	 & \TTZZ \BBZZ   \textsf{soc-digg}&  770K &  5.9M &  188M &  $10^{-5}$ &  15 &  -0.09 &  0.04 &  0.05 &  244 &  396.5K &  17.6K &  237 &  73 &  41 &  93 &  66 &  66 &  66 &  67 &  67 &  71 &  71 &  71 &  \textbf{64}&  \textbf{64}&  81 &  82 &  89 &  90 & \\ 
	 & \TTZZ \BBZZ   \textsf{soc-douban}&  154K &  327K &  121K &  $10^{-5}$ &  4 &  -0.18 &  0.01 &  0.01 &  1 &  394 &  287 &  16 &  11 &  8 &  17 &  14 &  14 &  14 &  \textbf{13}&  \textbf{13}&  \textbf{13}&  \textbf{13}&  \textbf{13}&  \textbf{13}&  \textbf{13}&  \textbf{13}&  \textbf{13}&  \textbf{13}&  \textbf{13}& \\ 
	 & \TTZZ \BBZZ   \textsf{soc-epinions}&  26.5K &  100K &  479K &  $10^{-4}$ &  7 &  0.06 &  0.06 &  0.09 &  18 &  5.1K &  443 &  33 &  18 &  14 &  30 &  25 &  25 &  25 &  25 &  25 &  \textbf{20}&  \textbf{20}&  \textbf{20}&  \textbf{20}&  \textbf{20}&  \textbf{20}&  \textbf{20}&  \textbf{20}&  21 & \\ 
	 & \TTZZ \BBZZ   \textsf{soc-flickr}&  513K &  3.1M &  176M &  $10^{-5}$ &  12 &  0.16 &  0.08 &  0.15 &  343 &  524.5K &  4.3K &  310 &  153 &  21 &  146 &  109 &  109 &  109 &  108 &  108 &  \textbf{104}&  \textbf{104}&  \textbf{104}&  105 &  106 &  129 &  126 &  138 &  142 & \\ 
	 & \TTZZ \BBZZ   \textsf{soc-flixster}&  2.5M &  7.9M &  23.6M &  $10^{-6}$ &  6 &  -0.32 &  0.05 &  0.01 &  9 &  15.1K &  1.4K &  69 &  47 &  29 &  57 &  47 &  47 &  47 &  47 &  47 &  47 &  47 &  47 &  44 &  44 &  \textbf{40}&  \textbf{40}&  \textbf{40}&  49 & \\ 
	 & \TTZZ \BBZZ  \cellcolor{verylightred}\textsf{soc-gowalla}& \cellcolor{verylightred}196K & \cellcolor{verylightred}950K & \cellcolor{verylightred}6.8M & \cellcolor{verylightred}$10^{-5}$ & \cellcolor{verylightred}9 & \cellcolor{verylightred}-0.03 & \cellcolor{verylightred}0.09 & \cellcolor{verylightred}0.02 & \cellcolor{verylightred}35 & \cellcolor{verylightred}93.8K & \cellcolor{verylightred}14.7K & \cellcolor{verylightred}52 & \cellcolor{verylightred}29 & \cellcolor{verylightred}29 & \cellcolor{verylightred}44 & \cellcolor{verylightred}30 & \cellcolor{verylightred}30 & \cellcolor{verylightred}30 & \cellcolor{verylightred}30 & \cellcolor{verylightred}30 & \cellcolor{verylightred}30 & \cellcolor{verylightred}30 & \cellcolor{verylightred}30 & \cellcolor{verylightred}30 & \cellcolor{verylightred}30 & \cellcolor{verylightred}\textbf{29}& \cellcolor{verylightred}\textbf{29}& \cellcolor{verylightred}\textbf{29}& \cellcolor{verylightred}37 & \\ 

	 & \TTZZ \BBZZ   \textsf{soc-lastfm}&  1.1M &  4.5M &  11.8M &  $10^{-6}$ &  7 &  -0.14 &  0.03 &  0.01 &  10 &  38K &  5.1K &  71 &  23 &  14 &  43 &  26 &  26 &  26 &  26 &  26 &  27 &  27 &  27 &  \textbf{24}&  \textbf{24}&  28 &  28 &  27 &  40 & \\ 
	 & \TTZZ \BBZZ   \textsf{soc-pokec}&  1.6M &  22.3M &  97.6M &  $10^{-5}$ &  27 &  0.00 &  0.02 &  0.05 &  60 &  29.2K &  14.8K &  48 &  29 &  29 &  43 &  33 &  33 &  33 &  33 &  33 &  33 &  33 &  33 &  33 &  33 &  \textbf{30}&  \textbf{30}&  \textbf{30}&  38 & \\ 
	 & \TTZZ \BBZZ   \textsf{soc-slashdot}&  70K &  358K &  1.2M &  $10^{-4}$ &  10 &  -0.07 &  0.03 &  0.03 &  17 &  13.3K &  2.5K &  54 &  35 &  17 &  44 &  39 &  39 &  39 &  40 &  40 &  \textbf{34}&  \textbf{34}&  \textbf{34}&  35 &  35 &  35 &  35 &  35 &  43 & \\ 
	 & \TTZZ \BBZZ   \textsf{soc-twitter-fol}&  404K &  713K &  88.6K &  $10^{-6}$ &  3 &  -0.88 &  0.01 &  0.00 &  0 &  1.6K &  626 &  29 &  6 &  6 &  13 &  8 &  8 &  8 &  \textbf{7}&  \textbf{7}&  \textbf{7}&  \textbf{7}&  \textbf{7}&  8 &  8 &  9 &  12 &  12 &  8 & \\ 
	 & \TTZZ \BBZZ  \cellcolor{verylightred}\textsf{soc-wiki-Vote}& \cellcolor{verylightred}889 & \cellcolor{verylightred}2.9K & \cellcolor{verylightred}6.3K & \cellcolor{verylightred}$10^{-3}$ & \cellcolor{verylightred}6 & \cellcolor{verylightred}-0.03 & \cellcolor{verylightred}0.04 & \cellcolor{verylightred}0.13 & \cellcolor{verylightred}7 & \cellcolor{verylightred}251 & \cellcolor{verylightred}102 & \cellcolor{verylightred}10 & \cellcolor{verylightred}7 & \cellcolor{verylightred}7 & \cellcolor{verylightred}11 & \cellcolor{verylightred}9 & \cellcolor{verylightred}9 & \cellcolor{verylightred}9 & \cellcolor{verylightred}9 & \cellcolor{verylightred}9 & \cellcolor{verylightred}8 & \cellcolor{verylightred}8 & \cellcolor{verylightred}8 & \cellcolor{verylightred}8 & \cellcolor{verylightred}8 & \cellcolor{verylightred}\textbf{7}& \cellcolor{verylightred}\textbf{7}& \cellcolor{verylightred}8 & \cellcolor{verylightred}8 & \\  
	 & \TTZZ \BBZZ   \textsf{soc-youtube}&  495K &  1.9M &  7.3M &  $10^{-5}$ &  7 &  -0.03 &  0.05 &  0.01 &  15 &  151K &  25.4K &  50 &  19 &  11 &  42 &  32 &  32 &  32 &  32 &  32 &  30 &  30 &  30 &  30 &  30 &  \textbf{28}&  \textbf{28}&  29 &  37 & \\ 
\midrule
\multirow{17}{*}{\rotatebox{90}{\textbf{ \normalsize facebook networks}}}& \TTZZ \BBZZ   \textsf{fb-A-anon}&  3M &  23.6M &  166.8M &  $10^{-6}$ &  15 &  -0.06 &  0.04 &  0.05 &  54 &  50.2K &  4.9K &  75 &  30 &  23 &  52 &  35 &  35 &  35 &  35 &  35 &  34 &  34 &  34 &  \textbf{33}&  \textbf{33}&  34 &  34 &  35 &  45 & \\ 
	 & \TTZZ \BBZZ   \textsf{fb-B-anon}&  2.9M &  20.9M &  155M &  $10^{-6}$ &  14 &  -0.11 &  0.04 &  0.05 &  53 &  36.8K &  4.3K &  64 &  31 &  23 &  47 &  30 &  30 &  30 &  30 &  30 &  30 &  30 &  30 &  \textbf{29}&  \textbf{29}&  \textbf{29}&  \textbf{29}&  30 &  41 & \\ 
	 & \TTZZ \BBZZ   \textsf{fb-Berkeley13}&  22.9K &  852K &  16.1M &  $10^{-3}$ &  74 &  0.01 &  0.01 &  0.11 &  703 &  69.5K &  3.4K &  65 &  47 &  39 &  57 &  49 &  49 &  49 &  49 &  49 &  50 &  50 &  50 &  \textbf{48}&  \textbf{48}&  49 &  49 &  49 &  56 & \\ 
	 & \TTZZ \BBZZ   \textsf{fb-CMU}&  6.6K &  249K &  6.9M &  0.01 &  75 &  0.12 &  0.02 &  0.19 &  1K &  24K &  840 &  70 &  45 &  42 &  58 &  50 &  50 &  50 &  50 &  50 &  \textbf{49}&  \textbf{49}&  \textbf{49}&  51 &  51 &  50 &  50 &  51 &  55 & \\ 
	 & \TTZZ \BBZZ   \textsf{fb-Duke14}&  9.8K &  506K &  15.4M &  0.01 &  102 &  0.07 &  0.02 &  0.17 &  1.5K &  41.9K &  1.8K &  86 &  47 &  29 &  64 &  56 &  56 &  56 &  55 &  55 &  \textbf{47}&  \textbf{47}&  \textbf{47}&  52 &  52 &  49 &  49 &  49 &  61 & \\ 
	 & \TTZZ \BBZZ   \textsf{fb-Indiana}&  29.7K &  1.3M &  28.1M &  $10^{-3}$ &  87 &  0.13 &  0.01 &  0.14 &  948 &  37.2K &  1.3K &  77 &  53 &  43 &  66 &  58 &  58 &  58 &  58 &  58 &  56 &  56 &  56 &  54 &  54 &  54 &  54 &  \textbf{52}&  62 & \\ 
	 & \TTZZ \BBZZ   \textsf{fb-MIT}&  6.4K &  251K &  7.1M &  0.01 &  78 &  0.12 &  0.02 &  0.18 &  1.1K &  27.7K &  708 &  73 &  41 &  30 &  59 &  50 &  50 &  50 &  48 &  48 &  \textbf{44}&  \textbf{44}&  \textbf{44}&  46 &  46 &  46 &  46 &  47 &  55 & \\ 
	 & \TTZZ \BBZZ   \textsf{fb-OR}&  63.3K &  816K &  10.5M &  $10^{-4}$ &  25 &  0.18 &  0.04 &  0.15 &  166 &  19.4K &  1K &  53 &  36 &  28 &  46 &  41 &  41 &  41 &  41 &  41 &  37 &  37 &  37 &  37 &  37 &  37 &  37 &  \textbf{36}&  44 & \\ 
	 & \TTZZ \BBZZ   \textsf{fb-Penn94}&  41.5K &  1.3M &  21.6M &  $10^{-3}$ &  65 &  -0.00 &  0.01 &  0.10 &  521 &  68K &  4.4K &  63 &  48 &  43 &  56 &  52 &  52 &  52 &  53 &  53 &  48 &  48 &  48 &  50 &  50 &  48 &  48 &  \textbf{47}&  52 & \\ 
	 & \TTZZ \BBZZ   \textsf{fb-Stanford3}&  11.5K &  568K &  17.5M &  $10^{-3}$ &  98 &  0.10 &  0.02 &  0.16 &  1.5K &  33.1K &  1.1K &  92 &  60 &  47 &  68 &  63 &  63 &  63 &  63 &  63 &  59 &  59 &  59 &  \textbf{58}&  \textbf{58}&  59 &  59 &  60 &  67 & \\ 
	 & \TTZZ \BBZZ   \textsf{fb-Texas84}&  36.3K &  1.5M &  33.5M &  $10^{-3}$ &  87 &  -0.00 &  0.01 &  0.10 &  922 &  141K &  6.3K &  82 &  62 &  44 &  74 &  64 &  64 &  64 &  64 &  64 &  \textbf{57}&  \textbf{57}&  \textbf{57}&  60 &  60 &  60 &  60 &  61 &  71 & \\ 
	 & \TTZZ \BBZZ   \textsf{fb-UCLA}&  20.4K &  747K &  15.3M &  $10^{-3}$ &  73 &  0.14 &  0.02 &  0.14 &  750 &  17.5K &  1.1K &  66 &  54 &  49 &  61 &  54 &  54 &  54 &  56 &  56 &  \textbf{53}&  \textbf{53}&  \textbf{53}&  54 &  54 &  54 &  54 &  54 &  57 & \\ 
	 & \TTZZ \BBZZ   \textsf{fb-UCSB37}&  14.9K &  482K &  9.2M &  $10^{-3}$ &  64 &  0.18 &  0.01 &  0.16 &  619 &  16.1K &  810 &  66 &  60 &  51 &  63 &  59 &  59 &  59 &  60 &  60 &  \textbf{56}&  \textbf{56}&  \textbf{56}&  \textbf{56}&  \textbf{56}&  \textbf{56}&  \textbf{56}&  \textbf{56}&  62 & \\ 
	 & \TTZZ \BBZZ   \textsf{fb-UConn}&  17.2K &  604K &  10.2M &  $10^{-3}$ &  70 &  0.09 &  0.01 &  0.13 &  596 &  21.5K &  1.7K &  66 &  53 &  47 &  60 &  56 &  56 &  56 &  57 &  57 &  56 &  56 &  56 &  \textbf{51}&  \textbf{51}&  52 &  52 &  52 &  57 & \\ 
	 & \TTZZ \BBZZ   \textsf{fb-UF}&  35.1K &  1.4M &  36.4M &  $10^{-3}$ &  83 &  -0.01 &  0.01 &  0.12 &  1K &  159K &  8.2K &  84 &  67 &  51 &  75 &  66 &  66 &  66 &  65 &  65 &  64 &  64 &  64 &  \textbf{61}&  \textbf{61}&  62 &  62 &  \textbf{61}&  72 & \\ 
	 & \TTZZ \BBZZ   \textsf{fb-UIllinois}&  30.7K &  1.2M &  28M &  $10^{-3}$ &  82 &  0.03 &  0.01 &  0.14 &  911 &  66.1K &  4.6K &  86 &  65 &  54 &  72 &  64 &  64 &  64 &  63 &  63 &  62 &  62 &  62 &  \textbf{59}&  \textbf{59}&  60 &  60 &  61 &  68 & \\ 
	 & \TTZZ \BBZZ   \textsf{fb-Wisconsin87}&  23.8K &  835K &  14.5M &  $10^{-3}$ &  70 &  -0.00 &  0.01 &  0.12 &  612 &  46.7K &  3.4K &  61 &  42 &  34 &  54 &  48 &  48 &  48 &  47 &  47 &  46 &  46 &  46 &  44 &  44 &  43 &  43 &  \textbf{42}&  51 & \\ 
\midrule
\multirow{7}{*}{\rotatebox{90}{\textbf{ \normalsize technological}}}& \TTZZ \BBZZ   \textsf{tech-RL-caida}&  190K &  607K &  1.3M &  $10^{-5}$ &  6 &  0.02 &  0.06 &  0.06 &  7 &  6K &  1K &  33 &  19 &  15 &  25 &  20 &  20 &  20 &  20 &  20 &  20 &  20 &  20 &  19 &  19 &  19 &  19 &  19 &  \textbf{18}& \\ 
	 & \TTZZ \BBZZ   \textsf{tech-WHOIS}&  7.4K &  56.9K &  2.3M &  $10^{-3}$ &  15 &  -0.04 &  0.26 &  0.31 &  314 &  22.2K &  1K &  89 &  71 &  49 &  72 &  67 &  67 &  67 &  67 &  \textbf{66}&  67 &  67 &  67 &  \textbf{66}&  \textbf{66}&  67 &  67 &  \textbf{66}&  72 & \\ 
	 & \TTZZ \BBZZ   \textsf{tech-as-skitter}&  1.6M &  11M &  86.3M &  $10^{-6}$ &  13 &  -0.08 &  0.08 &  0.01 &  51 &  564.6K &  35.4K &  112 &  68 &  41 &  81 &  71 &  71 &  71 &  71 &  71 &  71 &  71 &  71 &  \textbf{70}&  \textbf{70}&  77 &  76 &  73 &  74 & \\ 
	 & \TTZZ \BBZZ   \textsf{tech-internet-as}&  40.1K &  85.1K &  189K &  $10^{-4}$ &  4 &  -0.18 &  0.15 &  0.01 &  5 &  8.5K &  3.3K &  24 &  17 &  14 &  22 &  19 &  19 &  19 &  19 &  19 &  19 &  19 &  19 &  19 &  19 &  \textbf{18}&  \textbf{18}&  19 &  20 & \\ 
	 & \TTZZ \BBZZ  \cellcolor{verylightblue}\textsf{tech-routers-rf}& \cellcolor{verylightblue}2.1K & \cellcolor{verylightblue}6.6K & \cellcolor{verylightblue}31.2K & \cellcolor{verylightblue}$10^{-3}$ & \cellcolor{verylightblue}6 & \cellcolor{verylightblue}0.02 & \cellcolor{verylightblue}0.11 & \cellcolor{verylightblue}0.23 & \cellcolor{verylightblue}15 & \cellcolor{verylightblue}588 & \cellcolor{verylightblue}109 & \cellcolor{verylightblue}16 & \cellcolor{verylightblue}16 & \cellcolor{verylightblue}16 & \cellcolor{verylightblue}18 & \cellcolor{verylightblue}17 & \cellcolor{verylightblue}17 & \cellcolor{verylightblue}17 & \cellcolor{verylightblue}\textbf{16}& \cellcolor{verylightblue}\textbf{16}& \cellcolor{verylightblue}17 & \cellcolor{verylightblue}17 & \cellcolor{verylightblue}17 & \cellcolor{verylightblue}17 & \cellcolor{verylightblue}17 & \cellcolor{verylightblue}\textbf{16}& \cellcolor{verylightblue}\textbf{16}& \cellcolor{verylightblue}17 & \cellcolor{verylightblue}17 & \\ 
\midrule
	 & \TTZZ \BBZZ   \textsf{web-polblogs}&  643 &  2.2K &  9K &  0.01 &  7 &  -0.22 &  0.06 &  0.16 &  14 &  392 &  165 &  13 &  10 &  9 &  12 &  11 &  11 &  11 &  11 &  11 &  \textbf{10}&  \textbf{10}&  \textbf{10}&  11 &  11 &  \textbf{10}&  \textbf{10}&  11 &  \textbf{10}& \\ 
	 & \TTZZ \BBZZ   \textsf{web-spam}&  4.7K &  37.3K &  387K &  $10^{-3}$ &  15 &  0.00 &  0.08 &  0.15 &  81 &  6.2K &  477 &  36 &  23 &  20 &  31 &  24 &  24 &  24 &  24 &  24 &  \textbf{22}&  \textbf{22}&  \textbf{22}&  23 &  23 &  24 &  24 &  24 &  \textbf{22}& \\ 
\midrule
\end{tabularx}
\vspace{-8mm}
\end{table}
}

\smallskip \noindent {\normalsize \theoremfont Best Methods: From Social to Information Networks:}
The sparse graphs are examined further by their respective types (i.e., social networks).
For each network of a specific type, we apply the coloring methods in Table~\ref{tab:selection-criterion} and measure their accuracy just as before.
This allows us to determine the coloring methods that are most accurate for each type of network.
The results are shown in Table~\ref{table:accuracy-results} (columns $4$ to $10$).
The greedy coloring methods are ranked and colored according to their overall rank shown previously in the first two columns of Table~\ref{table:accuracy-results}.

\begin{compactenum}[{\scriptsize $\bullet$}]
  \setlength{\itemsep}{2pt}
  \setlength{\parskip}{0pt}
  \setlength{\parsep}{4pt}

\item In nearly all types of networks, the proposed methods are more accurate than the traditional degree-based methods (i.e., use fewer colors).

\item For social and Facebook networks, the triangle and triangle-core methods performed the best (i.e., accuracy), using fewer number of colors.
\end{compactenum}

\smallskip
\noindent
In the majority of cases, we found that the proposed methods are significantly better than the traditional degree-based methods (i.e., \textsc{ido}, \textsc{deg}) at $p < 0.01$ level.
More specifically, greedy coloring methods that use triangle properties or triangle-core based methods significantly improve over the other methods, resulting in a better coloring with fewer number of colors.
In addition, the colors used by the proposed methods for each network are compared in Table~\ref{table:results-colors}.


{
\renewcommand{\tabcolsep}{1pt}
\begin{table}[h!]
\caption{
Comparing the coloring algorithms by runtime. 
The runtime in seconds is reported for each graph in the collection. 
Graphs with insignificant coloring runtimes were removed for brevity (sec. $<$ 0.1) 
}
\vspace{0mm}
\label{table:runtime-coloring-algs-soc}
\centering\small\tiny
\fontsize{6.0}{7.0}\selectfont
\noindent
\begin{tabularx}{\linewidth}{ l HX HHHH HHHHH HHXX XXXXXXXXXXXXXX}

\textbf{graph} & 
$|V|$ & 
$|E|$ & 
$|T|$ & 
$\rho$ & 
$d_{\text{avg}}$ & 
${\rm r}$ & 
$\bar{\kappa}$ & 
$\kappa$ & 
$\tr_{\rm avg}$ & 
$\tr_{\rm max}$ & 
$\dmax$ & 
$K$+1 & 
$T$ & 
$\tilde{\omega}$ & 
\rotatebox{90}{\textsc{deg}} & 
\rotatebox{90}{\textsc{ido}} & 
\rotatebox{90}{\textsc{dist-two-ido}} & 
\rotatebox{90}{\textsc{triangles}} & 
\rotatebox{90}{\textsc{kcore-deg}} & 
\rotatebox{90}{\textsc{triangle-vol}} & 
\rotatebox{90}{\textsc{triangle-core-vol}} & 
\rotatebox{90}{\textsc{triangle-core-max}} & 
\rotatebox{90}{\textsc{deg-triangles}} & 
\rotatebox{90}{\textsc{kcore-triangles}} & 
\rotatebox{90}{\textsc{kcore-deg-tri}} & 
\rotatebox{90}{\textsc{deg-kcore-vol}} & 
\rotatebox{90}{\textsc{kcore-triangle-vol}} & 
\rotatebox{90}{\textsc{deg-kcore-tri-vol}} 
\\ \bottomrule 
 
 \TTZZ \BBZZ   \textsf{soc-BlogCatalog} &  88K &  2M &  153M &  $10^{-4}$ &  47 &  -0.23 &  0.09 &  0.06 &  1.7K &  804K &  9.4K &  222 &  101 &  24 &  0.1 &  0.1 &  0.1 &  0.5 &  0.6 &  1.1 &  1.1 &  1.2 &  1.2 &  1.3 &  1.5 &  1.6 &  1.6 &  1.8 \\ 
 \TTZZ \BBZZ   \textsf{soc-FourSquare} &  639K &  3.2M &  64M &  $10^{-5}$ &  10 &  -0.71 &  0.03 &  0.00 &  102 &  1.9M &  106K &  64 &  38 &  25 &  0.5 &  0.7 &  0.8 &  2.5 &  3.5 &  8.9 &  10.3 &  6.4 &  7.4 &  7.6 &  8.3 &  8.5 &  10.5 &  9.6 \\ 
 \TTZZ \BBZZ   \textsf{soc-LiveMocha} &  104K &  2.1M &  10M &  $10^{-4}$ &  42 &  -0.15 &  0.01 &  0.01 &  97 &  36K &  2.9K &  93 &  27 &  10 &  0.1 &  0.2 &  0.2 &  0.7 &  0.9 &  1.6 &  1.6 &  1.2 &  1.7 &  1.9 &  1.4 &  1.5 &  2.4 &  1.7 \\ 
 \TTZZ \BBZZ   \textsf{soc-buzznet} &  101K &  2.7M &  92M &  $10^{-4}$ &  54 &  2.85 &  0.01 &  0.03 &  917 &  1M &  64K &  154 &  59 &  21 &  0.2 &  0.2 &  0.2 &  1.0 &  1.4 &  2.0 &  2.2 &  2.3 &  2.6 &  2.9 &  2.9 &  3.0 &  3.0 &  3.2 \\ 
 \TTZZ \BBZZ   \textsf{soc-delicious} &  536K &  1.3M &  1.4M &  $10^{-6}$ &  5 &  -0.07 &  0.02 &  0.01 &  3 &  8K &  3.2K &  34 &  23 &  17 &  0.5 &  0.6 &  0.6 &  2.3 &  3.0 &  4.5 &  5.5 &  5.9 &  6.1 &  7.6 &  7.5 &  8.1 &  8.2 &  8.6 \\ 
 \TTZZ \BBZZ   \textsf{soc-digg} &  770K &  5.9M &  188M &  $10^{-5}$ &  15 &  -0.09 &  0.04 &  0.05 &  244 &  396K &  17K &  237 &  73 &  41 &  0.8 &  1.0 &  1.3 &  3.3 &  4.4 &  8.3 &  7.5 &  9.6 &  10.8 &  13.8 &  12.8 &  12.3 &  14.6 &  15.8 \\ 
 \TTZZ \BBZZ   \textsf{soc-douban} &  154K &  327K &  121K &  $10^{-5}$ &  4 &  -0.18 &  0.01 &  0.01 &  1 &  394 &  287 &  16 &  11 &  8 &  0.1 &  0.1 &  0.1 &  0.6 &  0.9 &  1.1 &  1.4 &  1.5 &  1.7 &  1.8 &  2.0 &  2.2 &  2.2 &  2.4 \\ 
 \TTZZ \BBZZ   \textsf{soc-flickr} &  513K &  3.1M &  176M &  $10^{-5}$ &  12 &  0.16 &  0.08 &  0.15 &  343 &  524K &  4.3K &  310 &  153 &  21 &  0.5 &  0.6 &  0.7 &  2.4 &  3.2 &  4.5 &  4.5 &  6.0 &  6.0 &  7.5 &  8.4 &  7.3 &  8.3 &  9.0 \\ 
 \TTZZ \BBZZ   \textsf{soc-flixster} &  2.5M &  7.9M &  23M &  $10^{-6}$ &  6 &  -0.32 &  0.05 &  0.01 &  9 &  15K &  1.4K &  69 &  47 &  29 &  2.9 &  3.6 &  3.6 &  10.0 &  13.4 &  23.2 &  24.3 &  29.2 &  28.1 &  31.8 &  33.7 &  39.2 &  40.1 &  42.7 \\ 
 \TTZZ \BBZZ   \textsf{soc-lastfm} &  1.1M &  4.5M &  11M &  $10^{-6}$ &  7 &  -0.14 &  0.03 &  0.01 &  10 &  38K &  5.1K &  71 &  23 &  14 &  1.1 &  1.4 &  1.4 &  6.1 &  5.6 &  12.7 &  11.4 &  13.6 &  13.7 &  13.1 &  14.4 &  15.0 &  17.1 &  19.8 \\ 
 \TTZZ \BBZZ   \textsf{soc-pokec} &  1.6M &  22M &  97M &  $10^{-5}$ &  27 &  0.00 &  0.02 &  0.05 &  60 &  29K &  14K &  48 &  29 &  29 &  3.7 &  4.8 &  4.0 &  12.1 &  15.1 &  33.9 &  35.2 &  29.5 &  41.8 &  44.1 &  36.9 &  51.9 &  43.7 &  46.7 \\ 
 \TTZZ \BBZZ   \textsf{soc-slashdot} &  70K &  358K &  1.2M &  $10^{-4}$ &  10 &  -0.07 &  0.03 &  0.03 &  17 &  13K &  2.5K &  54 &  35 &  17 &  - &  - &  - &  0.2 &  0.2 &  0.4 &  0.5 &  0.5 &  0.5 &  0.6 &  0.6 &  0.7 &  0.7 &  0.8 \\ 
 \TTZZ \BBZZ   \textsf{soc-twitter-foll} &  404K &  713K &  88K &  $10^{-6}$ &  3 &  -0.88 &  0.01 &  0.00 &  0 &  1.6K &  626 &  29 &  6 &  6 &  0.3 &  0.3 &  0.3 &  1.8 &  1.5 &  2.3 &  2.9 &  3.0 &  3.3 &  3.6 &  3.8 &  5.2 &  4.6 &  4.9 \\ 
 \TTZZ \BBZZ   \textsf{soc-youtube} &  495K &  1.9M &  7.3M &  $10^{-5}$ &  7 &  -0.03 &  0.05 &  0.01 &  15 &  151K &  25K &  50 &  19 &  11 &  0.4 &  0.4 &  0.4 &  1.7 &  2.0 &  3.3 &  3.6 &  4.0 &  4.6 &  5.0 &  5.4 &  6.0 &  6.6 &  6.5 \\ 
\midrule
 \TTZZ \BBZZ   \textsf{fb-Berkeley13} &  22K &  852K &  16M &  $10^{-3}$ &  74 &  0.01 &  0.01 &  0.11 &  703 &  69K &  3.4K &  65 &  47 &  39 &  0.1 &  0.1 &  0.1 &  0.2 &  0.2 &  0.5 &  0.5 &  0.6 &  0.6 &  0.6 &  0.7 &  0.8 &  0.8 &  0.8 \\  
 \TTZZ \BBZZ   \textsf{fb-Indiana} &  29K &  1.3M &  28M &  $10^{-3}$ &  87 &  0.13 &  0.01 &  0.14 &  948 &  37K &  1.3K &  77 &  53 &  43 &  - &  0.1 &  0.1 &  0.2 &  0.3 &  0.4 &  0.4 &  0.5 &  0.5 &  0.5 &  0.6 &  0.6 &  0.6 &  0.7 \\ 
 \TTZZ \BBZZ   \textsf{fb-OR} &  63K &  816K &  10M &  $10^{-4}$ &  25 &  0.18 &  0.04 &  0.15 &  166 &  19K &  1K &  53 &  36 &  28 &  0.1 &  0.1 &  0.1 &  0.4 &  0.3 &  0.8 &  0.9 &  1.0 &  0.7 &  1.1 &  0.8 &  1.2 &  1.3 &  1.4 \\ 
 \TTZZ \BBZZ   \textsf{fb-Penn94} &  41K &  1.3M &  21M &  $10^{-3}$ &  65 &  -0.00 &  0.01 &  0.10 &  521 &  68K &  4.4K &  63 &  48 &  43 &  0.1 &  0.1 &  0.1 &  0.4 &  0.4 &  0.9 &  0.5 &  0.6 &  1.1 &  0.7 &  1.3 &  1.4 &  0.9 &  0.9 \\ 
 \TTZZ \BBZZ   \textsf{fb-UF} &  35K &  1.4M &  36M &  $10^{-3}$ &  83 &  -0.01 &  0.01 &  0.12 &  1K &  159K &  8.2K &  84 &  67 &  51 &  0.1 &  0.1 &  0.1 &  0.5 &  0.4 &  0.6 &  0.7 &  0.8 &  0.6 &  0.7 &  0.7 &  0.8 &  1.4 &  0.9 \\  
\midrule
 \TTZZ \BBZZ   \textsf{tech-RL-caida} &  190K &  607K &  1.3M &  $10^{-5}$ &  6 &  0.02 &  0.06 &  0.06 &  7 &  6K &  1K &  33 &  19 &  15 &  0.2 &  0.2 &  0.2 &  0.8 &  1.0 &  1.9 &  1.9 &  2.0 &  2.2 &  2.1 &  2.6 &  3.0 &  3.1 &  3.3 \\  
 \TTZZ \BBZZ   \textsf{tech-as-caida} &  26K &  53K &  109K &  $10^{-4}$ &  4 &  -0.19 &  0.16 &  0.01 &  4 &  3.8K &  2.6K &  23 &  16 &  9 &  0.6 &  0.5 &  0.4 &  1.1 &  1.0 &  2.5 &  2.8 &  2.5 &  2.6 &  2.7 &  2.9 &  3.5 &  3.5 &  2.3 \\ 
 \TTZZ \BBZZ   \textsf{tech-as-skitter} &  1.6M &  11M &  86M &  $10^{-6}$ &  13 &  -0.08 &  0.08 &  0.01 &  51 &  564K &  35K &  112 &  68 &  41 &  5.2 &  5.3 &  5.3 &  17.8 &  13.9 &  47.5 &  25.9 &  31.9 &  31.2 &  33.8 &  40.4 &  37.4 &  41.1 &  44.7 \\  
 \TTZZ \BBZZ   \textsf{tech-p2p-gnutell} &  62K &  147K &  6K &  $10^{-5}$ &  4 &  -0.09 &  0.00 &  0.00 &  0 &  17 &  95 &  7 &  4 &  4 &  - &  0.1 &  0.1 &  0.2 &  0.3 &  0.5 &  0.5 &  0.6 &  0.7 &  0.8 &  0.8 &  0.9 &  0.9 &  0.9 \\ 
\midrule 
 \TTZZ \BBZZ   \textsf{web-arabic} &  163K &  1.7M &  65M &  $10^{-4}$ &  21 &  0.15 &  0.51 &  0.95 &  397 &  5.8K &  1.1K &  102 &  102 &  102 &  0.1 &  0.2 &  0.2 &  0.7 &  1.0 &  1.7 &  1.9 &  1.7 &  2.2 &  2.3 &  2.4 &  2.7 &  3.0 &  3.1 \\  
 \TTZZ \BBZZ   \textsf{web-it} &  509K &  7.1M &  1T &  $10^{-5}$ &  28 &  0.99 &  0.76 &  0.95 &  1.9K &  93K &  469 &  432 &  432 &  431 &  0.6 &  0.8 &  0.7 &  2.7 &  4.1 &  6.9 &  5.6 &  6.9 &  7.7 &  8.6 &  8.6 &  9.4 &  9.7 &  8.8 \\ 
 \TTZZ \BBZZ   \textsf{web-italycnr} &  325K &  3.1M &  96M &  $10^{-5}$ &  19 &  -0.19 &  0.21 &  0.01 &  295 &  720K &  18K &  162 &  18278 &  84 &  0.3 &  0.4 &  0.3 &  1.5 &  2.1 &  3.5 &  3.6 &  3.1 &  4.4 &  4.6 &  5.0 &  5.1 &  5.3 &  5.5 \\ 
 \TTZZ \BBZZ   \textsf{web-sk} &  121K &  334K &  2.9M &  $10^{-5}$ &  5 &  0.08 &  0.09 &  0.47 &  25 &  3.4K &  590 &  82 &  82 &  82 &  0.1 &  0.1 &  0.1 &  0.4 &  0.5 &  0.8 &  0.8 &  0.9 &  1.4 &  1.2 &  1.3 &  1.4 &  1.5 &  1.9 \\ 
 \TTZZ \BBZZ   \textsf{web-uk} &  129K &  11M &  2.5T &  $10^{-3}$ &  181 &  1.00 &  0.98 &  1.00 &  19K &  124K &  850 &  500 &  500 &  500 &  0.4 &  0.4 &  0.4 &  1.5 &  2.1 &  3.7 &  3.8 &  4.4 &  3.9 &  4.2 &  4.5 &  5.5 &  6.0 &  6.2 \\  
 \TTZZ \BBZZ   \textsf{web-wikipedia} &  1.8M &  4.5M &  6.6M &  $10^{-6}$ &  4 &  0.05 &  0.07 &  0.05 &  4 &  12K &  2.6K &  67 &  31 &  31 &  2.3 &  2.6 &  2.7 &  9.3 &  11.0 &  16.9 &  18.0 &  20.2 &  21.4 &  26.3 &  25.6 &  30.2 &  29.0 &  32.6 \\ 
\bottomrule
\end{tabularx} 
\end{table}
} 

\subsection{Scalability}
\label{sec:scalability}
Now, we evaluate the scalability of the proposed methods.
In particular, do the methods scale as the size of the graph increases (i.e., number of vertices and edges)?
To answer this question, we use the proposed greedy coloring methods to color a variety of networks including both large sparse social and information networks as well as a variety of dense graphs.
Figure~\ref{fig:scalability} plots the size of the graph versus the runtime in seconds (both are logged).
Overall, we find the proposed greedy coloring methods scale linearly with the size of the graph.
Moreover this holds for both large sparse and dense networks.
Nevertheless, coloring dense graphs is found to be slightly faster with less variance in the runtime, as compared to social networks which exhibit slightly more variance in the runtime of graphs that are approximately equal size.

We also compare the wall clock time (i.e., runtime in seconds) between a representative set of methods on a variety of networks.
Results are provided in Table~\ref{table:runtime-coloring-algs-soc}.
For brevity, we removed the graphs for which all methods took less than 0.1 seconds to color.
Not surprisingly, the simple degree-based methods (distance-1 and 2) are the fastest to compute.

\begin{figure}[h!]
\centering
\includegraphics[width=3.0in]{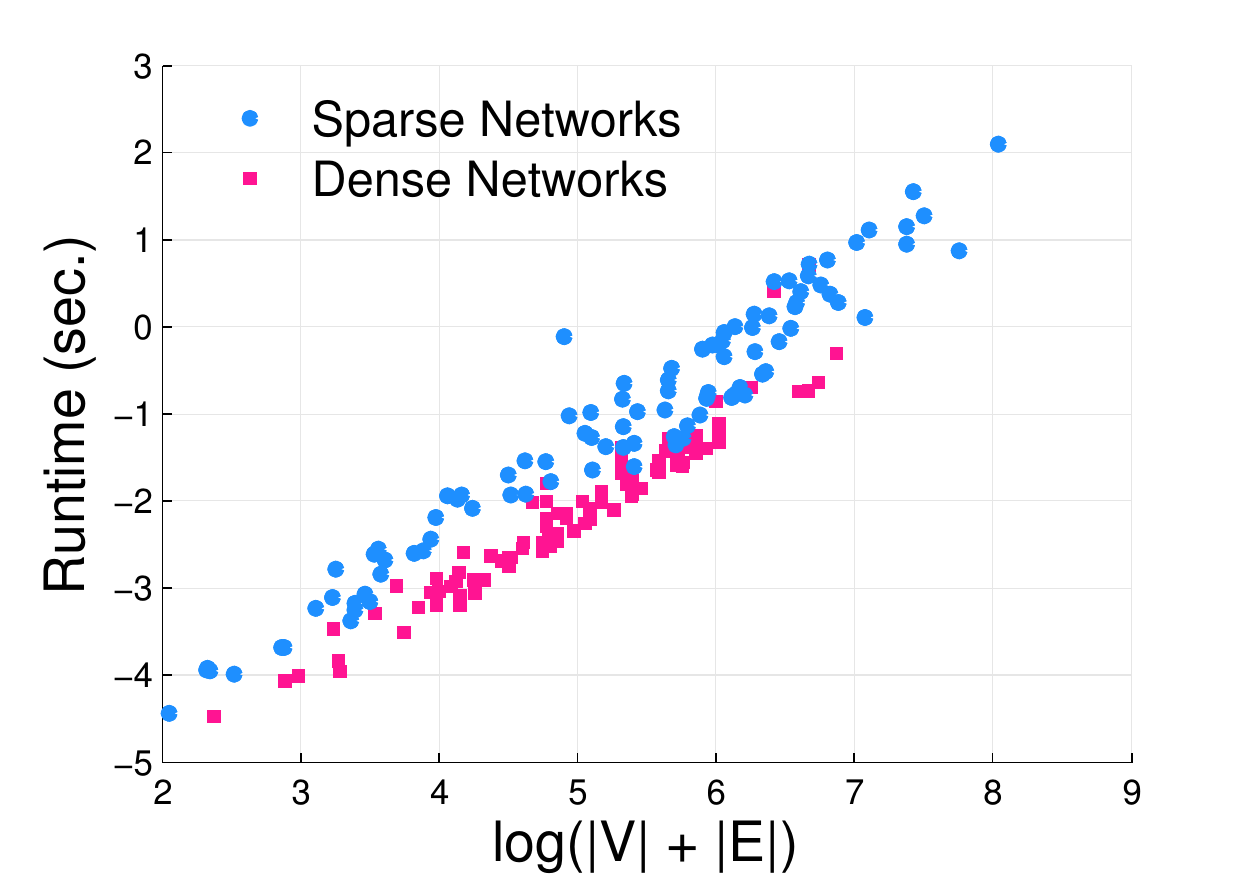}
\caption{
{\foliofont Scalability of the Proposed Coloring Methods.}
\xfigtextfont
The x-axis represents the log of the size of the graph whereas the y-axis is the log runtime (in seconds).
For both large sparse and dense networks, find that the proposed methods scale linearly as the size of the graph increases and thus practical for a variety of applications.
}
\label{fig:scalability}
\end{figure}

These results indicate that in practice, the proposed methods are fast, scaling linearly as the size of the graph increases.
Hence, these methods are well-suited for use in a variety of applications including network analysis, relational machine learning, sampling, among many others.
See Section~\ref{sec:neigh-experiments} for details on the scalability of the neighborhood coloring methods.

\begin{table}[b!]
\caption{
\footlinefont
\xfigtextfont
\textbf{\foliofont Recolor Statistics}. 
\xfigtextfont
We compare the variants that use \textsc{recolor} to those that do not.
The statistics in the table are computed over all graphs and greedy coloring methods.
The max and mean improvement are measured as the maximum/average difference between the number of colors used before and after recoloring.
}
\label{table:recolor}
\xfigtextfont
\centering
\xfigtextfont
\begin{tabular}
{ r | c c   c c }
\scriptsize
\TTT \BBB & 
\multicolumn{2}{c }{\textbf{\subsectionfont Percentage}} &
\multicolumn{2}{c }{\textbf{\subsectionfont Difference}} 
\\
\footnotesize
\TTT \BBB & 
\rotatebox{0}{\textbf{\theoremfont Improved}} &
\rotatebox{0}{\textbf{\theoremfont Same}} &
\rotatebox{0}{{\textbf{\theoremfont Max Diff.}}} & 
\rotatebox{0}{{\textbf{\theoremfont Mean Diff.}}}
\\ 
\hline
\footnotesize

\TTT \BBB
 {\textbf{\theoremfont Sparse}}  & 40.9\% & 59.1\% &  11 & 1.01 \\
 \TTT \BBB
 {\textbf{\theoremfont Dense}} & 84.4\% & 14.6\% &  313 & 14.65 \\
\end{tabular} 
\end{table}

\subsection{Effectiveness of Recolor }
\label{sec:results-recolor}
This section investigates the effectiveness of the \textsc{recolor} method.
In particular, how often does it reduce the number of colors?
For this, we investigate and compare greedy coloring variants that utilize \textsc{recolor} to the methods that do not.
Given a graph $G$ and a vertex ordering $\pi$ from one of the proposed selection strategies in Section~\ref{sec:ordering}, 
we color the graph using the basic coloring framework (Algorithm~\ref{alg:greedy-coloring}) and then we color the graph again using the \textsc{recolor} method.
From these two colorings, we measure the difference in the number of colors (after recoloring and before recoloring) and number of times the \textsc{recolor} method improved over the basic method.
The results are shown in Table~\ref{table:recolor}.
Note that the statistics are computed over all graphs and greedy coloring methods, including the methods that do not perform well (i.e., degree-based methods).
Note that the maximum improvement (i.e, Max Diff. in Table~\ref{table:recolor}) and average improvement (i.e, Mean Diff. in Table~\ref{table:recolor}) are measured as the maximum/average difference between the number of colors used before and after recoloring.

In sparse graphs, the recolor method results in fewer colors $40.9\%$ of the time whereas the improvement for dense graphs is $84.4\%$.
We find that the improvement for dense graphs is much larger since the number of colors initially (before recoloring) used on average is usually far from the optimal number. 
Note that for sparse graphs, this includes the graphs where the greedy coloring methods was able to find the optimal number of colors (and thus, it is impossible for \textsc{recolor} to improve over the basic coloring).
Additionally, the sparse graphs use fewer colors than the dense graphs and also the number of colors used from the greedy coloring methods tends to be closer to the optimal.
These results indicate that \textsc{recolor} is both \textit{fast} and \textit{effective} for reducing the number of colors used by any of the proposed methods.

In addition, we also provide results for both recolor and basic variants on a variety of large sparse real-world networks, see Table~\ref{table:recolor-results-sparse-social} and Table~\ref{table:recolor-results-sparse-fb}.
These can be used to infer additional insights.
In Table~\ref{table:recolor-results-dense-dimacs} and Table~\ref{table:recolor-results-dense-bhoslib}, we also compare the recolor variant to the faster but less accurate basic coloring variant of each coloring method for the DIMACs and BHOSLIB graph collections.

{
\renewcommand{\tabcolsep}{1pt}
\begin{table}[h!]
\caption{
Recolor variant is compared to the faster but less accurate basic variant for each of the proposed methods.
We also include four previous methods for comparison.
For each graph, the top row is the results from the basic variant whereas the bottom row is from the recolor variant.
}
\vspace{1mm}
\label{table:recolor-results-sparse-social}
\centering\small\tiny
\fontsize{6.0}{7.0}\selectfont
\noindent

\begin{tabularx}{\linewidth}{ Hl Hc HHHH HHHH cXXX 
 rrrrr rrrrr rrrrr r HH}
 
\multicolumn{17}{c}{\textbf{\sectionfont Stats \& Bounds}} & 
\multicolumn{15}{c}{\textbf{\sectionfont Coloring Methods}} \\
 &
\textbf{graph}& 
$|V|$ & 
$|E|$ & 
$|T|$ & 
$\rho$ & 
$d_{\text{avg}}$ & 
${\rm r}$ & 
$\bar{\kappa}$ & 
$\kappa$ & 
$\tr_{\rm avg}$ & 
$\tr_{\rm max}$ & 
$\dmax$ & 
$K$+1 & 
$T$ & 
$\tilde{\omega}$ & 
\rotatebox{90}{\textsc{rand}}& \rotatebox{90}{\textsc{deg}}& \rotatebox{90}{\textsc{ido}}& \rotatebox{90}{\textsc{dist-two-ido}}& \rotatebox{90}{\textsc{triangles}}& \rotatebox{90}{\textsc{kcore-deg}}& \rotatebox{90}{\textsc{triangle-vol}}& \rotatebox{90}{\textsc{triangle-core-vol}}& \rotatebox{90}{\textsc{triangle-core-max}}& \rotatebox{90}{\textsc{deg-triangles}}& \rotatebox{90}{\textsc{kcore-triangles}}& \rotatebox{90}{\textsc{kcore-deg-tri}}& \rotatebox{90}{\textsc{deg-kcore-vol}}& \rotatebox{90}{\textsc{kcore-triangle-vol}}& \rotatebox{90}{\textsc{deg-kcore-triangle-vol}}\\ 
\bottomrule

\multirow{21}{*}{\rotatebox{90}{\textbf{ \normalsize social networks}}}& \TTZZ \BBZZ   \multirow{2}{*}{\textsf{soc-BlogCat.}}&\multirow{2}{*}{88.7K}&\multirow{2}{*}{2M}&\multirow{2}{*}{153.5M}&$10^{-4}$ &\multirow{2}{*}{47}&\multirow{2}{*}{-0.23}&\multirow{2}{*}{0.09}&\multirow{2}{*}{0.06}&\multirow{2}{*}{1.7K}&\multirow{2}{*}{804.4K}&\multirow{2}{*}{9.4K}&\multirow{2}{*}{222}&\multirow{2}{*}{101}&\multirow{2}{*}{24}&124 &89 &89 &89 &88 &88 &\textbf{87}&\textbf{87}&\textbf{87}&88 &90 &109 &108 &115 &117 & \\ 
	 & \TTZZ \BBZZ  && & & & & & & & & & & & & &118 &85 &85 &85 &\textbf{83}&84 &84 &84 &84 &85 &85 &102 &102 &110 &112 & \\ 
  	 \cline{2-31}	 & \TTZZ \BBZZ   \multirow{2}{*}{\textsf{soc-LiveMo.}}&\multirow{2}{*}{104K}&\multirow{2}{*}{2.1M}&\multirow{2}{*}{10M}&$10^{-4}$ &\multirow{2}{*}{42}&\multirow{2}{*}{-0.15}&\multirow{2}{*}{0.01}&\multirow{2}{*}{0.01}&\multirow{2}{*}{97}&\multirow{2}{*}{36.9K}&\multirow{2}{*}{2.9K}&\multirow{2}{*}{93}&\multirow{2}{*}{27}&\multirow{2}{*}{10}&53 &38 &38 &38 &39 &39 &\textbf{34}&\textbf{34}&\textbf{34}&36 &36 &37 &38 &36 &45 & \\ 
	 & \TTZZ \BBZZ  && & & & & & & & & & & & & &50 &36 &36 &36 &36 &36 &\textbf{30}&\textbf{30}&\textbf{30}&33 &33 &36 &37 &33 &42 & \\ 

  	 \cline{2-31}	 & \TTZZ \BBZZ   \multirow{2}{*}{\textsf{soc-buzznet}}&\multirow{2}{*}{101K}&\multirow{2}{*}{2.7M}&\multirow{2}{*}{92.7M}&$10^{-4}$ &\multirow{2}{*}{54}&\multirow{2}{*}{2.85}&\multirow{2}{*}{0.01}&\multirow{2}{*}{0.03}&\multirow{2}{*}{917}&\multirow{2}{*}{1M}&\multirow{2}{*}{64.2K}&\multirow{2}{*}{154}&\multirow{2}{*}{59}&\multirow{2}{*}{21}&89 &63 &63 &63 &\textbf{62}&\textbf{62}&63 &63 &63 &64 &65 &79 &80 &87 &86 & \\ 
	 & \TTZZ \BBZZ  && & & & & & & & & & & & & &85 &59 &59 &59 &59 &\textbf{58}&59 &59 &59 &59 &\textbf{58}&74 &75 &82 &83 & \\ 
  	 \cline{2-31}	 & \TTZZ \BBZZ   \multirow{2}{*}{\textsf{soc-delicious}}&\multirow{2}{*}{536K}&\multirow{2}{*}{1.3M}&\multirow{2}{*}{1.4M}&$10^{-6}$ &\multirow{2}{*}{5}&\multirow{2}{*}{-0.07}&\multirow{2}{*}{0.02}&\multirow{2}{*}{0.01}&\multirow{2}{*}{3}&\multirow{2}{*}{8K}&\multirow{2}{*}{3.2K}&\multirow{2}{*}{34}&\multirow{2}{*}{23}&\multirow{2}{*}{17}&26 &22 &22 &22 &22 &22 &22 &22 &22 &\textbf{21}&\textbf{21}&\textbf{21}&\textbf{21}&\textbf{21}&22 & \\ 
	 & \TTZZ \BBZZ  && & & & & & & & & & & & & &25 &22 &22 &22 &22 &22 &22 &22 &22 &\textbf{21}&\textbf{21}&\textbf{21}&\textbf{21}&\textbf{21}&\textbf{21}& \\ 
  	 \cline{2-31}	 & \TTZZ \BBZZ   \multirow{2}{*}{\textsf{soc-digg}}&\multirow{2}{*}{770K}&\multirow{2}{*}{5.9M}&\multirow{2}{*}{188.1M}&$10^{-5}$ &\multirow{2}{*}{15}&\multirow{2}{*}{-0.09}&\multirow{2}{*}{0.04}&\multirow{2}{*}{0.05}&\multirow{2}{*}{244}&\multirow{2}{*}{396.5K}&\multirow{2}{*}{17.6K}&\multirow{2}{*}{237}&\multirow{2}{*}{73}&\multirow{2}{*}{41}&93 &66 &66 &66 &67 &67 &71 &71 &71 &\textbf{64}&\textbf{64}&81 &82 &89 &90 & \\ 
	 & \TTZZ \BBZZ  && & & & & & & & & & & & & &88 &63 &63 &63 &63 &63 &63 &63 &63 &\textbf{61}&\textbf{61}&75 &77 &80 &83 & \\ 
  	 \cline{2-31}	 & \TTZZ \BBZZ   \multirow{2}{*}{\textsf{soc-douban}}&\multirow{2}{*}{154K}&\multirow{2}{*}{327K}&\multirow{2}{*}{121.8K}&$10^{-5}$ &\multirow{2}{*}{4}&\multirow{2}{*}{-0.18}&\multirow{2}{*}{0.01}&\multirow{2}{*}{0.01}&\multirow{2}{*}{1}&\multirow{2}{*}{394}&\multirow{2}{*}{287}&\multirow{2}{*}{16}&\multirow{2}{*}{11}&\multirow{2}{*}{8}&17 &14 &14 &14 &\textbf{13}&\textbf{13}&\textbf{13}&\textbf{13}&\textbf{13}&\textbf{13}&\textbf{13}&\textbf{13}&\textbf{13}&\textbf{13}&\textbf{13}& \\ 
	 & \TTZZ \BBZZ  && & & & & & & & & & & & & &17 &14 &14 &14 &13 &13 &13 &13 &13 &13 &13 &13 &13 &13 &\textbf{12}& \\ 
  	 \cline{2-31}	 & \TTZZ \BBZZ   \multirow{2}{*}{\textsf{soc-epinions}}&\multirow{2}{*}{26.5K}&\multirow{2}{*}{100K}&\multirow{2}{*}{479.1K}&$10^{-4}$ &\multirow{2}{*}{7}&\multirow{2}{*}{0.06}&\multirow{2}{*}{0.06}&\multirow{2}{*}{0.09}&\multirow{2}{*}{18}&\multirow{2}{*}{5.1K}&\multirow{2}{*}{443}&\multirow{2}{*}{33}&\multirow{2}{*}{18}&\multirow{2}{*}{14}&30 &25 &25 &25 &25 &25 &\textbf{20}&\textbf{20}&\textbf{20}&\textbf{20}&\textbf{20}&\textbf{20}&\textbf{20}&\textbf{20}&21 & \\ 
	 & \TTZZ \BBZZ  && & & & & & & & & & & & & &28 &22 &22 &22 &22 &22 &\textbf{20}&\textbf{20}&\textbf{20}&\textbf{20}&\textbf{20}&\textbf{20}&\textbf{20}&\textbf{20}&\textbf{20}& \\ 
  	 \cline{2-31}	 & \TTZZ \BBZZ   \multirow{2}{*}{\textsf{soc-flickr}}&\multirow{2}{*}{513K}&\multirow{2}{*}{3.1M}&\multirow{2}{*}{176.3M}&$10^{-5}$ &\multirow{2}{*}{12}&\multirow{2}{*}{0.16}&\multirow{2}{*}{0.08}&\multirow{2}{*}{0.15}&\multirow{2}{*}{343}&\multirow{2}{*}{524.5K}&\multirow{2}{*}{4.3K}&\multirow{2}{*}{310}&\multirow{2}{*}{153}&\multirow{2}{*}{21}&146 &109 &109 &109 &108 &108 &\textbf{104}&\textbf{104}&\textbf{104}&105 &106 &129 &126 &138 &142 & \\ 
	 & \TTZZ \BBZZ  && & & & & & & & & & & & & &138 &104 &104 &104 &102 &102 &100 &100 &100 &100 &\textbf{99}&118 &119 &127 &131 & \\ 
  	 \cline{2-31}	 & \TTZZ \BBZZ   \multirow{2}{*}{\textsf{soc-flixster}}&\multirow{2}{*}{2.5M}&\multirow{2}{*}{7.9M}&\multirow{2}{*}{23.6M}&$10^{-6}$ &\multirow{2}{*}{6}&\multirow{2}{*}{-0.32}&\multirow{2}{*}{0.05}&\multirow{2}{*}{0.01}&\multirow{2}{*}{9}&\multirow{2}{*}{15.1K}&\multirow{2}{*}{1.4K}&\multirow{2}{*}{69}&\multirow{2}{*}{47}&\multirow{2}{*}{29}&57 &47 &47 &47 &47 &47 &47 &47 &47 &44 &44 &\textbf{40}&\textbf{40}&\textbf{40}&49 & \\ 
	 & \TTZZ \BBZZ  && & & & & & & & & & & & & &53 &46 &46 &46 &46 &46 &46 &46 &46 &42 &42 &\textbf{38}&\textbf{38}&\textbf{38}&46 & \\ 
  	 \cline{2-31}	 & \TTZZ \BBZZ   \multirow{2}{*}{\textsf{soc-gowalla}}&\multirow{2}{*}{196K}&\multirow{2}{*}{950K}&\multirow{2}{*}{6.8M}&$10^{-5}$ &\multirow{2}{*}{9}&\multirow{2}{*}{-0.03}&\multirow{2}{*}{0.09}&\multirow{2}{*}{0.02}&\multirow{2}{*}{35}&\multirow{2}{*}{93.8K}&\multirow{2}{*}{14.7K}&\multirow{2}{*}{52}&\multirow{2}{*}{29}&\multirow{2}{*}{29}&44 &30 &30 &30 &30 &30 &30 &30 &30 &30 &30 &\textbf{29}&\textbf{29}&\textbf{29}&37 & \\ 
	 & \TTZZ \BBZZ  && & & & & & & & & & & & & &42 &30 &30 &30 &\textbf{29}&\textbf{29}&\textbf{29}&\textbf{29}&\textbf{29}&\textbf{29}&\textbf{29}&\textbf{29}&\textbf{29}&\textbf{29}&37 & \\ 
  	 \cline{2-31}	 & \TTZZ \BBZZ   \multirow{2}{*}{\textsf{soc-lastfm}}&\multirow{2}{*}{1.1M}&\multirow{2}{*}{4.5M}&\multirow{2}{*}{11.8M}&$10^{-6}$ &\multirow{2}{*}{7}&\multirow{2}{*}{-0.14}&\multirow{2}{*}{0.03}&\multirow{2}{*}{0.01}&\multirow{2}{*}{10}&\multirow{2}{*}{38K}&\multirow{2}{*}{5.1K}&\multirow{2}{*}{71}&\multirow{2}{*}{23}&\multirow{2}{*}{14}&43 &26 &26 &26 &26 &26 &27 &27 &27 &\textbf{24}&\textbf{24}&28 &28 &27 &40 & \\ 
	 & \TTZZ \BBZZ  && & & & & & & & & & & & & &39 &26 &26 &26 &26 &26 &25 &25 &25 &\textbf{23}&\textbf{23}&28 &28 &27 &36 & \\ 
  	 \cline{2-31}	 & \TTZZ \BBZZ   \multirow{2}{*}{\textsf{soc-pokec}}&\multirow{2}{*}{1.6M}&\multirow{2}{*}{22.3M}&\multirow{2}{*}{97.6M}&$10^{-5}$ &\multirow{2}{*}{27}&\multirow{2}{*}{0.00}&\multirow{2}{*}{0.02}&\multirow{2}{*}{0.05}&\multirow{2}{*}{60}&\multirow{2}{*}{29.2K}&\multirow{2}{*}{14.8K}&\multirow{2}{*}{48}&\multirow{2}{*}{29}&\multirow{2}{*}{29}&43 &33 &33 &33 &33 &33 &33 &33 &33 &33 &33 &\textbf{30}&\textbf{30}&\textbf{30}&38 & \\ 
	 & \TTZZ \BBZZ  && & & & & & & & & & & & & &41 &31 &31 &31 &31 &31 &31 &31 &31 &32 &32 &\textbf{29}&\textbf{29}&\textbf{29}&37 & \\ 
  	 \cline{2-31}	 & \TTZZ \BBZZ   \multirow{2}{*}{\textsf{soc-slashdot}}&\multirow{2}{*}{70K}&\multirow{2}{*}{358K}&\multirow{2}{*}{1.2M}&$10^{-4}$ &\multirow{2}{*}{10}&\multirow{2}{*}{-0.07}&\multirow{2}{*}{0.03}&\multirow{2}{*}{0.03}&\multirow{2}{*}{17}&\multirow{2}{*}{13.3K}&\multirow{2}{*}{2.5K}&\multirow{2}{*}{54}&\multirow{2}{*}{35}&\multirow{2}{*}{17}&44 &39 &39 &39 &40 &40 &\textbf{34}&\textbf{34}&\textbf{34}&35 &35 &35 &35 &35 &43 & \\ 
	 & \TTZZ \BBZZ  && & & & & & & & & & & & & &42 &36 &36 &36 &36 &36 &\textbf{31}&\textbf{31}&\textbf{31}&32 &32 &32 &32 &32 &41 & \\ 
  	 \cline{2-31}	 & \TTZZ \BBZZ   \multirow{2}{*}{\textsf{soc-wiki-Vote}}&\multirow{2}{*}{889}&\multirow{2}{*}{2.9K}&\multirow{2}{*}{6.3K}&$10^{-3}$ &\multirow{2}{*}{6}&\multirow{2}{*}{-0.03}&\multirow{2}{*}{0.04}&\multirow{2}{*}{0.13}&\multirow{2}{*}{7}&\multirow{2}{*}{251}&\multirow{2}{*}{102}&\multirow{2}{*}{10}&\multirow{2}{*}{7}&\multirow{2}{*}{7}&11 &9 &9 &9 &9 &9 &8 &8 &8 &8 &8 &\textbf{7}&\textbf{7}&8 &8 & \\ 
	 & \TTZZ \BBZZ  && & & & & & & & & & & & & &10 &8 &8 &8 &8 &8 &8 &8 &8 &8 &8 &\textbf{7}&\textbf{7}&\textbf{7}&8 & \\ 
  	 \cline{2-31}	 & \TTZZ \BBZZ   \multirow{2}{*}{\textsf{soc-youtu-sn}}&\multirow{2}{*}{1.1M}&\multirow{2}{*}{2.9M}&\multirow{2}{*}{9.1M}&$10^{-6}$ &\multirow{2}{*}{5}&\multirow{2}{*}{-0.04}&\multirow{2}{*}{0.04}&\multirow{2}{*}{0.01}&\multirow{2}{*}{8}&\multirow{2}{*}{180.8K}&\multirow{2}{*}{28.7K}&\multirow{2}{*}{52}&\multirow{2}{*}{19}&\multirow{2}{*}{13}&44 &33 &33 &33 &33 &33 &33 &33 &33 &31 &31 &\textbf{30}&\textbf{30}&31 &40 & \\ 
	 & \TTZZ \BBZZ  && & & & & & & & & & & & & &42 &31 &31 &31 &31 &31 &31 &31 &31 &30 &30 &\textbf{28}&\textbf{28}&29 &37 & \\ 
  	 \cline{2-31}	 & \TTZZ \BBZZ   \multirow{2}{*}{\textsf{soc-youtube}}&\multirow{2}{*}{495K}&\multirow{2}{*}{1.9M}&\multirow{2}{*}{7.3M}&$10^{-5}$ &\multirow{2}{*}{7}&\multirow{2}{*}{-0.03}&\multirow{2}{*}{0.05}&\multirow{2}{*}{0.01}&\multirow{2}{*}{15}&\multirow{2}{*}{151K}&\multirow{2}{*}{25.4K}&\multirow{2}{*}{50}&\multirow{2}{*}{19}&\multirow{2}{*}{11}&42 &32 &32 &32 &32 &32 &30 &30 &30 &30 &30 &\textbf{28}&\textbf{28}&29 &37 & \\ 
	 & \TTZZ \BBZZ  && & & & & & & & & & & & & &40 &29 &29 &29 &30 &30 &29 &29 &29 &28 &28 &\textbf{27}&\textbf{27}&28 &36 & \\ 
  	 \cline{2-31}
  	 \end{tabularx}
\end{table}
}

{
\renewcommand{\tabcolsep}{2.2pt}
\begin{table}[h!]
\caption{
Recolor variant is compared to the faster but less accurate basic variant for each of the proposed methods (continued from Table~\ref{table:recolor-results-sparse-social}).
We also include four previous methods for comparison.
For each graph, the top row is the results from the basic variant whereas the bottom row is from the recolor variant.
}
\vspace{1mm}
\label{table:recolor-results-sparse-fb}
\centering\small\tiny
\fontsize{6.0}{7.0}\selectfont
\noindent

\begin{tabularx}{\linewidth}{ Hl Hc HHHH HHHH XXXX 
 rrrrr rrrrr rrrrr r HH}

\multicolumn{17}{c}{\textbf{\sectionfont Stats \& Bounds}} & 
\multicolumn{15}{c}{\textbf{\sectionfont Coloring Methods}} \\
 &
\textbf{graph}& 
$|V|$ & 
$|E|$ & 
$|T|$ & 
$\rho$ & 
$d_{\text{avg}}$ & 
${\rm r}$ & 
$\bar{\kappa}$ & 
$\kappa$ & 
$\tr_{\rm avg}$ & 
$\tr_{\rm max}$ & 
$\dmax$ & 
$K$+1 & 
$T$ & 
$\tilde{\omega}$ & 
\rotatebox{90}{\textsc{rand}}& \rotatebox{90}{\textsc{deg}}& \rotatebox{90}{\textsc{ido}}& \rotatebox{90}{\textsc{dist-two-ido}}& \rotatebox{90}{\textsc{triangles}}& \rotatebox{90}{\textsc{kcore-deg}}& \rotatebox{90}{\textsc{triangle-vol}}& \rotatebox{90}{\textsc{triangle-core-vol}}& \rotatebox{90}{\textsc{triangle-core-max}}& \rotatebox{90}{\textsc{deg-triangles}}& \rotatebox{90}{\textsc{kcore-triangles}}& \rotatebox{90}{\textsc{kcore-deg-tri}}& \rotatebox{90}{\textsc{deg-kcore-vol}}& \rotatebox{90}{\textsc{kcore-triangle-vol}}& \rotatebox{90}{\textsc{deg-kcore-triangle-vol}}\\ 
\bottomrule

\multirow{9}{*}{\rotatebox{90}{\textbf{ \normalsize interaction}}}& \TTZZ \BBZZ   \multirow{2}{*}{\textsf{ia-email-EU}}&\multirow{2}{*}{32.4K}&\multirow{2}{*}{54.3K}&\multirow{2}{*}{146.9K}&$10^{-4}$ &\multirow{2}{*}{3}&\multirow{2}{*}{-0.38}&\multirow{2}{*}{0.09}&\multirow{2}{*}{0.03}&\multirow{2}{*}{5}&\multirow{2}{*}{1.6K}&\multirow{2}{*}{623}&\multirow{2}{*}{23}&\multirow{2}{*}{13}&\multirow{2}{*}{11}&23 &19 &19 &19 &18 &18 &17 &17 &17 &17 &17 &17 &17 &17 &\textbf{16}& \\ 
	 & \TTZZ \BBZZ  && & & & & & & & & & & & & &22 &17 &17 &17 &18 &18 &17 &17 &17 &16 &16 &16 &16 &16 &\textbf{15}& \\ 
  	 \cline{2-31}	 & \TTZZ \BBZZ   \multirow{2}{*}{\textsf{ia-enron-large}}&\multirow{2}{*}{33.6K}&\multirow{2}{*}{180K}&\multirow{2}{*}{2.1M}&$10^{-4}$ &\multirow{2}{*}{10}&\multirow{2}{*}{-0.12}&\multirow{2}{*}{0.34}&\multirow{2}{*}{0.09}&\multirow{2}{*}{65}&\multirow{2}{*}{17.7K}&\multirow{2}{*}{1.3K}&\multirow{2}{*}{44}&\multirow{2}{*}{22}&\multirow{2}{*}{15}&40 &31 &31 &31 &30 &30 &\textbf{28}&\textbf{28}&\textbf{28}&\textbf{28}&\textbf{28}&\textbf{28}&\textbf{28}&\textbf{28}&37 & \\ 
	 & \TTZZ \BBZZ  && & & & & & & & & & & & & &38 &29 &29 &29 &29 &29 &\textbf{26}&\textbf{26}&\textbf{26}&27 &27 &27 &27 &28 &34 & \\ 
  	 \cline{2-31}	 & \TTZZ \BBZZ   \multirow{2}{*}{\textsf{ia-fb-messages}}&\multirow{2}{*}{1.2K}&\multirow{2}{*}{6.4K}&\multirow{2}{*}{7.4K}&$10^{-3}$ &\multirow{2}{*}{10}&\multirow{2}{*}{-0.08}&\multirow{2}{*}{0.02}&\multirow{2}{*}{0.04}&\multirow{2}{*}{6}&\multirow{2}{*}{242}&\multirow{2}{*}{112}&\multirow{2}{*}{12}&\multirow{2}{*}{5}&\multirow{2}{*}{5}&12 &9 &9 &9 &\textbf{8}&\textbf{8}&\textbf{8}&\textbf{8}&\textbf{8}&9 &9 &\textbf{8}&9 &\textbf{8}&9 & \\ 
	 & \TTZZ \BBZZ  && & & & & & & & & & & & & &12 &9 &9 &9 &\textbf{8}&\textbf{8}&\textbf{8}&\textbf{8}&\textbf{8}&\textbf{8}&\textbf{8}&\textbf{8}&9 &9 &9 & \\ 

  	 \cline{2-31}	 & \TTZZ \BBZZ   \multirow{2}{*}{\textsf{ia-wiki-Talk}}&\multirow{2}{*}{92.1K}&\multirow{2}{*}{360K}&\multirow{2}{*}{2.5M}&$10^{-5}$ &\multirow{2}{*}{7}&\multirow{2}{*}{-0.03}&\multirow{2}{*}{0.03}&\multirow{2}{*}{0.05}&\multirow{2}{*}{27}&\multirow{2}{*}{17.6K}&\multirow{2}{*}{1.2K}&\multirow{2}{*}{59}&\multirow{2}{*}{20}&\multirow{2}{*}{9}&45 &35 &35 &35 &34 &34 &\textbf{30}&\textbf{30}&\textbf{30}&31 &31 &31 &31 &31 &40 & \\ 
	 & \TTZZ \BBZZ  && & & & & & & & & & & & & &43 &33 &33 &33 &32 &32 &\textbf{28}&\textbf{28}&\textbf{28}&\textbf{28}&\textbf{28}&29 &29 &\textbf{28}&37 & \\ 
  	 \cline{2-31}

\multirow{17}{*}{\rotatebox{90}{\textbf{ \normalsize facebook networks}}}&  \TTZZ \BBZZ   \multirow{2}{*}{\textsf{fb-B-anon}}&\multirow{2}{*}{2.9M}&\multirow{2}{*}{20.9M}&\multirow{2}{*}{155.9M}&$10^{-6}$ &\multirow{2}{*}{14}&\multirow{2}{*}{-0.11}&\multirow{2}{*}{0.04}&\multirow{2}{*}{0.05}&\multirow{2}{*}{53}&\multirow{2}{*}{36.8K}&\multirow{2}{*}{4.3K}&\multirow{2}{*}{64}&\multirow{2}{*}{31}&\multirow{2}{*}{23}&47 &30 &30 &30 &30 &30 &30 &30 &30 &\textbf{29}&\textbf{29}&\textbf{29}&\textbf{29}&30 &41 & \\ 
	 & \TTZZ \BBZZ  && & & & & & & & & & & & & &45 &28 &28 &28 &28 &28 &28 &28 &28 &28 &28 &\textbf{27}&\textbf{27}&28 &37 & \\ 
  	 \cline{2-31}	 & \TTZZ \BBZZ   \multirow{2}{*}{\textsf{fb-Berkeley13}}&\multirow{2}{*}{22.9K}&\multirow{2}{*}{852K}&\multirow{2}{*}{16.1M}&$10^{-3}$ &\multirow{2}{*}{74}&\multirow{2}{*}{0.01}&\multirow{2}{*}{0.01}&\multirow{2}{*}{0.11}&\multirow{2}{*}{703}&\multirow{2}{*}{69.5K}&\multirow{2}{*}{3.4K}&\multirow{2}{*}{65}&\multirow{2}{*}{47}&\multirow{2}{*}{39}&57 &49 &49 &49 &49 &49 &50 &50 &50 &\textbf{48}&\textbf{48}&49 &49 &49 &56 & \\ 
	 & \TTZZ \BBZZ  && & & & & & & & & & & & & &55 &47 &47 &47 &49 &49 &48 &48 &48 &47 &47 &\textbf{46}&\textbf{46}&\textbf{46}&52 & \\ 
  	 \cline{2-31}	 & \TTZZ \BBZZ   \multirow{2}{*}{\textsf{fb-CMU}}&\multirow{2}{*}{6.6K}&\multirow{2}{*}{249K}&\multirow{2}{*}{6.9M}&0.01 &\multirow{2}{*}{75}&\multirow{2}{*}{0.12}&\multirow{2}{*}{0.02}&\multirow{2}{*}{0.19}&\multirow{2}{*}{1K}&\multirow{2}{*}{24K}&\multirow{2}{*}{840}&\multirow{2}{*}{70}&\multirow{2}{*}{45}&\multirow{2}{*}{42}&58 &50 &50 &50 &50 &50 &\textbf{49}&\textbf{49}&\textbf{49}&51 &51 &50 &50 &51 &55 & \\ 
	 & \TTZZ \BBZZ  && & & & & & & & & & & & & &56 &49 &49 &49 &48 &48 &\textbf{47}&\textbf{47}&\textbf{47}&\textbf{47}&\textbf{47}&49 &49 &\textbf{47}&53 & \\ 
  	 \cline{2-31}	 & \TTZZ \BBZZ   \multirow{2}{*}{\textsf{fb-Duke14}}&\multirow{2}{*}{9.8K}&\multirow{2}{*}{506K}&\multirow{2}{*}{15.4M}&0.01 &\multirow{2}{*}{102}&\multirow{2}{*}{0.07}&\multirow{2}{*}{0.02}&\multirow{2}{*}{0.17}&\multirow{2}{*}{1.5K}&\multirow{2}{*}{41.9K}&\multirow{2}{*}{1.8K}&\multirow{2}{*}{86}&\multirow{2}{*}{47}&\multirow{2}{*}{29}&64 &56 &56 &56 &55 &55 &\textbf{47}&\textbf{47}&\textbf{47}&52 &52 &49 &49 &49 &61 & \\ 
	 & \TTZZ \BBZZ  && & & & & & & & & & & & & &61 &50 &50 &50 &50 &50 &\textbf{45}&\textbf{45}&\textbf{45}&46 &46 &47 &47 &47 &54 & \\ 
  	 \cline{2-31}	 & \TTZZ \BBZZ   \multirow{2}{*}{\textsf{fb-Indiana}}&\multirow{2}{*}{29.7K}&\multirow{2}{*}{1.3M}&\multirow{2}{*}{28.1M}&$10^{-3}$ &\multirow{2}{*}{87}&\multirow{2}{*}{0.13}&\multirow{2}{*}{0.01}&\multirow{2}{*}{0.14}&\multirow{2}{*}{948}&\multirow{2}{*}{37.2K}&\multirow{2}{*}{1.3K}&\multirow{2}{*}{77}&\multirow{2}{*}{53}&\multirow{2}{*}{43}&66 &58 &58 &58 &58 &58 &56 &56 &56 &54 &54 &54 &54 &\textbf{52}&62 & \\ 
	 & \TTZZ \BBZZ  && & & & & & & & & & & & & &62 &55 &55 &55 &54 &54 &52 &52 &52 &\textbf{51}&\textbf{51}&\textbf{51}&\textbf{51}&\textbf{51}&57 & \\ 
  	 \cline{2-31}	 & \TTZZ \BBZZ   \multirow{2}{*}{\textsf{fb-MIT}}&\multirow{2}{*}{6.4K}&\multirow{2}{*}{251K}&\multirow{2}{*}{7.1M}&0.01 &\multirow{2}{*}{78}&\multirow{2}{*}{0.12}&\multirow{2}{*}{0.02}&\multirow{2}{*}{0.18}&\multirow{2}{*}{1.1K}&\multirow{2}{*}{27.7K}&\multirow{2}{*}{708}&\multirow{2}{*}{73}&\multirow{2}{*}{41}&\multirow{2}{*}{30}&59 &50 &50 &50 &48 &48 &\textbf{44}&\textbf{44}&\textbf{44}&46 &46 &46 &46 &47 &55 & \\ 
	 & \TTZZ \BBZZ  && & & & & & & & & & & & & &57 &47 &47 &47 &44 &44 &\textbf{43}&\textbf{43}&\textbf{43}&\textbf{43}&\textbf{43}&44 &44 &\textbf{43}&50 & \\ 
  	 \cline{2-31}	 & \TTZZ \BBZZ   \multirow{2}{*}{\textsf{fb-OR}}&\multirow{2}{*}{63.3K}&\multirow{2}{*}{816K}&\multirow{2}{*}{10.5M}&$10^{-4}$ &\multirow{2}{*}{25}&\multirow{2}{*}{0.18}&\multirow{2}{*}{0.04}&\multirow{2}{*}{0.15}&\multirow{2}{*}{166}&\multirow{2}{*}{19.4K}&\multirow{2}{*}{1K}&\multirow{2}{*}{53}&\multirow{2}{*}{36}&\multirow{2}{*}{28}&46 &41 &41 &41 &41 &41 &37 &37 &37 &37 &37 &37 &37 &\textbf{36}&44 & \\ 
	 & \TTZZ \BBZZ  && & & & & & & & & & & & & &44 &37 &37 &37 &38 &38 &\textbf{34}&\textbf{34}&\textbf{34}&\textbf{34}&\textbf{34}&\textbf{34}&\textbf{34}&\textbf{34}&42 & \\ 
  	 \cline{2-31}	 & \TTZZ \BBZZ   \multirow{2}{*}{\textsf{fb-Penn94}}&\multirow{2}{*}{41.5K}&\multirow{2}{*}{1.3M}&\multirow{2}{*}{21.6M}&$10^{-3}$ &\multirow{2}{*}{65}&\multirow{2}{*}{-0.00}&\multirow{2}{*}{0.01}&\multirow{2}{*}{0.10}&\multirow{2}{*}{521}&\multirow{2}{*}{68K}&\multirow{2}{*}{4.4K}&\multirow{2}{*}{63}&\multirow{2}{*}{48}&\multirow{2}{*}{43}&56 &52 &52 &52 &53 &53 &48 &48 &48 &50 &50 &48 &48 &\textbf{47}&52 & \\ 
	 & \TTZZ \BBZZ  && & & & & & & & & & & & & &53 &49 &49 &49 &49 &49 &47 &47 &47 &\textbf{46}&\textbf{46}&\textbf{46}&\textbf{46}&\textbf{46}&51 & \\ 
  	 \cline{2-31}	 & \TTZZ \BBZZ   \multirow{2}{*}{\textsf{fb-Stanford3}}&\multirow{2}{*}{11.5K}&\multirow{2}{*}{568K}&\multirow{2}{*}{17.5M}&$10^{-3}$ &\multirow{2}{*}{98}&\multirow{2}{*}{0.10}&\multirow{2}{*}{0.02}&\multirow{2}{*}{0.16}&\multirow{2}{*}{1.5K}&\multirow{2}{*}{33.1K}&\multirow{2}{*}{1.1K}&\multirow{2}{*}{92}&\multirow{2}{*}{60}&\multirow{2}{*}{47}&68 &63 &63 &63 &63 &63 &59 &59 &59 &\textbf{58}&\textbf{58}&59 &59 &60 &67 & \\ 
	 & \TTZZ \BBZZ  && & & & & & & & & & & & & &64 &58 &58 &58 &58 &58 &\textbf{55}&\textbf{55}&\textbf{55}&\textbf{55}&\textbf{55}&\textbf{55}&\textbf{55}&56 &61 & \\ 
  	 \cline{2-31}	 & \TTZZ \BBZZ   \multirow{2}{*}{\textsf{fb-Texas84}}&\multirow{2}{*}{36.3K}&\multirow{2}{*}{1.5M}&\multirow{2}{*}{33.5M}&$10^{-3}$ &\multirow{2}{*}{87}&\multirow{2}{*}{-0.00}&\multirow{2}{*}{0.01}&\multirow{2}{*}{0.10}&\multirow{2}{*}{922}&\multirow{2}{*}{141K}&\multirow{2}{*}{6.3K}&\multirow{2}{*}{82}&\multirow{2}{*}{62}&\multirow{2}{*}{44}&74 &64 &64 &64 &64 &64 &\textbf{57}&\textbf{57}&\textbf{57}&60 &60 &60 &60 &61 &71 & \\ 
	 & \TTZZ \BBZZ  && & & & & & & & & & & & & &71 &62 &62 &62 &61 &61 &\textbf{56}&\textbf{56}&\textbf{56}&57 &57 &57 &57 &\textbf{56}&65 & \\ 

  	 \cline{2-31}	 & \TTZZ \BBZZ   \multirow{2}{*}{\textsf{fb-UCSB37}}&\multirow{2}{*}{14.9K}&\multirow{2}{*}{482K}&\multirow{2}{*}{9.2M}&$10^{-3}$ &\multirow{2}{*}{64}&\multirow{2}{*}{0.18}&\multirow{2}{*}{0.01}&\multirow{2}{*}{0.16}&\multirow{2}{*}{619}&\multirow{2}{*}{16.1K}&\multirow{2}{*}{810}&\multirow{2}{*}{66}&\multirow{2}{*}{60}&\multirow{2}{*}{51}&63 &59 &59 &59 &60 &60 &\textbf{56}&\textbf{56}&\textbf{56}&\textbf{56}&\textbf{56}&\textbf{56}&\textbf{56}&\textbf{56}&62 & \\ 
	 & \TTZZ \BBZZ  && & & & & & & & & & & & & &62 &56 &56 &56 &57 &58 &55 &55 &55 &55 &55 &55 &55 &\textbf{54}&59 & \\ 
  	 \cline{2-31}	 & \TTZZ \BBZZ   \multirow{2}{*}{\textsf{fb-UConn}}&\multirow{2}{*}{17.2K}&\multirow{2}{*}{604K}&\multirow{2}{*}{10.2M}&$10^{-3}$ &\multirow{2}{*}{70}&\multirow{2}{*}{0.09}&\multirow{2}{*}{0.01}&\multirow{2}{*}{0.13}&\multirow{2}{*}{596}&\multirow{2}{*}{21.5K}&\multirow{2}{*}{1.7K}&\multirow{2}{*}{66}&\multirow{2}{*}{53}&\multirow{2}{*}{47}&60 &56 &56 &56 &57 &57 &56 &56 &56 &\textbf{51}&\textbf{51}&52 &52 &52 &57 & \\ 
	 & \TTZZ \BBZZ  && & & & & & & & & & & & & &58 &51 &51 &51 &53 &53 &51 &51 &51 &\textbf{50}&\textbf{50}&\textbf{50}&\textbf{50}&\textbf{50}&54 & \\ 
  	 \cline{2-31}	 & \TTZZ \BBZZ   \multirow{2}{*}{\textsf{fb-UF}}&\multirow{2}{*}{35.1K}&\multirow{2}{*}{1.4M}&\multirow{2}{*}{36.4M}&$10^{-3}$ &\multirow{2}{*}{83}&\multirow{2}{*}{-0.01}&\multirow{2}{*}{0.01}&\multirow{2}{*}{0.12}&\multirow{2}{*}{1K}&\multirow{2}{*}{159K}&\multirow{2}{*}{8.2K}&\multirow{2}{*}{84}&\multirow{2}{*}{67}&\multirow{2}{*}{51}&75 &66 &66 &66 &65 &65 &64 &64 &64 &\textbf{61}&\textbf{61}&62 &62 &\textbf{61}&72 & \\ 
	 & \TTZZ \BBZZ  && & & & & & & & & & & & & &71 &62 &62 &62 &62 &62 &60 &60 &60 &\textbf{59}&\textbf{59}&61 &61 &\textbf{59}&66 & \\ 
  	 \cline{2-31}	 & \TTZZ \BBZZ   \multirow{2}{*}{\textsf{fb-UIllinois}}&\multirow{2}{*}{30.7K}&\multirow{2}{*}{1.2M}&\multirow{2}{*}{28M}&$10^{-3}$ &\multirow{2}{*}{82}&\multirow{2}{*}{0.03}&\multirow{2}{*}{0.01}&\multirow{2}{*}{0.14}&\multirow{2}{*}{911}&\multirow{2}{*}{66.1K}&\multirow{2}{*}{4.6K}&\multirow{2}{*}{86}&\multirow{2}{*}{65}&\multirow{2}{*}{54}&72 &64 &64 &64 &63 &63 &62 &62 &62 &\textbf{59}&\textbf{59}&60 &60 &61 &68 & \\ 
	 & \TTZZ \BBZZ  && & & & & & & & & & & & & &68 &60 &60 &60 &61 &61 &\textbf{58}&\textbf{58}&\textbf{58}&\textbf{58}&\textbf{58}&\textbf{58}&\textbf{58}&\textbf{58}&64 & \\ 
  	 \cline{2-31}	 & \TTZZ \BBZZ   \multirow{2}{*}{\textsf{fb-Wisconsin87}}&\multirow{2}{*}{23.8K}&\multirow{2}{*}{835K}&\multirow{2}{*}{14.5M}&$10^{-3}$ &\multirow{2}{*}{70}&\multirow{2}{*}{-0.00}&\multirow{2}{*}{0.01}&\multirow{2}{*}{0.12}&\multirow{2}{*}{612}&\multirow{2}{*}{46.7K}&\multirow{2}{*}{3.4K}&\multirow{2}{*}{61}&\multirow{2}{*}{42}&\multirow{2}{*}{34}&54 &48 &48 &48 &47 &47 &46 &46 &46 &44 &44 &43 &43 &\textbf{42}&51 & \\ 
	 & \TTZZ \BBZZ  && & & & & & & & & & & & & &52 &45 &45 &45 &45 &45 &43 &43 &43 &\textbf{42}&\textbf{42}&\textbf{42}&\textbf{42}&\textbf{42}&48 & \\ 
  	 \cline{2-31}

  	 \multirow{7}{*}{\rotatebox{90}{\textbf{ \normalsize technological}}}& \TTZZ \BBZZ   \multirow{2}{*}{\textsf{tech-RL-caida}}&\multirow{2}{*}{190K}&\multirow{2}{*}{607K}&\multirow{2}{*}{1.3M}&$10^{-5}$ &\multirow{2}{*}{6}&\multirow{2}{*}{0.02}&\multirow{2}{*}{0.06}&\multirow{2}{*}{0.06}&\multirow{2}{*}{7}&\multirow{2}{*}{6K}&\multirow{2}{*}{1K}&\multirow{2}{*}{33}&\multirow{2}{*}{19}&\multirow{2}{*}{15}&25 &20 &20 &20 &20 &20 &20 &20 &20 &19 &19 &19 &19 &19 &\textbf{18}& \\ 
	 & \TTZZ \BBZZ  && & & & & & & & & & & & & &24 &19 &19 &19 &19 &19 &19 &19 &19 &19 &19 &18 &18 &\textbf{17}&18 & \\ 
  	 \cline{2-31}	 & \TTZZ \BBZZ   \multirow{2}{*}{\textsf{tech-WHOIS}}&\multirow{2}{*}{7.4K}&\multirow{2}{*}{56.9K}&\multirow{2}{*}{2.3M}&$10^{-3}$ &\multirow{2}{*}{15}&\multirow{2}{*}{-0.04}&\multirow{2}{*}{0.26}&\multirow{2}{*}{0.31}&\multirow{2}{*}{314}&\multirow{2}{*}{22.2K}&\multirow{2}{*}{1K}&\multirow{2}{*}{89}&\multirow{2}{*}{71}&\multirow{2}{*}{49}&72 &67 &67 &67 &67 &\textbf{66}&67 &67 &67 &\textbf{66}&\textbf{66}&67 &67 &\textbf{66}&72 & \\ 
	 & \TTZZ \BBZZ  && & & & & & & & & & & & & &66 &62 &62 &62 &61 &61 &\textbf{60}&\textbf{60}&\textbf{60}&63 &63 &62 &62 &63 &64 & \\ 
  	 \cline{2-31}	 & \TTZZ \BBZZ   \multirow{2}{*}{\textsf{tech-internet-as}}&\multirow{2}{*}{40.1K}&\multirow{2}{*}{85.1K}&\multirow{2}{*}{189.8K}&$10^{-4}$ &\multirow{2}{*}{4}&\multirow{2}{*}{-0.18}&\multirow{2}{*}{0.15}&\multirow{2}{*}{0.01}&\multirow{2}{*}{5}&\multirow{2}{*}{8.5K}&\multirow{2}{*}{3.3K}&\multirow{2}{*}{24}&\multirow{2}{*}{17}&\multirow{2}{*}{14}&22 &19 &19 &19 &19 &19 &19 &19 &19 &19 &19 &\textbf{18}&\textbf{18}&19 &20 & \\ 
	 & \TTZZ \BBZZ  && & & & & & & & & & & & & &21 &18 &18 &18 &18 &18 &18 &18 &18 &18 &18 &\textbf{17}&\textbf{17}&18 &18 & \\ 
  	 \cline{2-31}	 & \TTZZ \BBZZ   \multirow{2}{*}{\textsf{tech-routers-rf}}&\multirow{2}{*}{2.1K}&\multirow{2}{*}{6.6K}&\multirow{2}{*}{31.2K}&$10^{-3}$ &\multirow{2}{*}{6}&\multirow{2}{*}{0.02}&\multirow{2}{*}{0.11}&\multirow{2}{*}{0.23}&\multirow{2}{*}{15}&\multirow{2}{*}{588}&\multirow{2}{*}{109}&\multirow{2}{*}{16}&\multirow{2}{*}{16}&\multirow{2}{*}{16}&18 &17 &17 &17 &\textbf{16}&\textbf{16}&17 &17 &17 &17 &17 &\textbf{16}&\textbf{16}&17 &17 & \\ 
	 & \TTZZ \BBZZ  && & & & & & & & & & & & & &16 &17 &17 &17 &\textbf{16}&\textbf{16}&17 &17 &17 &17 &17 &\textbf{16}&\textbf{16}&17 &\textbf{16}& \\ 
  	 \cline{2-31}
  	 
  	 \end{tabularx}
\end{table}
}

\subsection{Bounds \& Provably Optimal Coloring}
\label{sec:results-bounds}
This section describes two ways to leverage the bounds.
Results are then provided in Table~\ref{table:stats-sparse-ca} and Table~\ref{table:stats-sparse-social} for a representative set of graphs from the collection.

First, the lower bound can be used to verify that the coloring from a greedy method is optimal.
Let $\lb(G)$ be a lower bound of $\Chromatic(G)$ (i.e., optimal number of colors), 
then we have the following simple relationship:
\[
\lb(G) \leq \omega(G) \leq \Chromatic(G) \leq \coloring(G,\pi) \leq \dmax(G)+1
\]
\noindent where $\coloring(G,\pi)$ is the number of colors from a greedy coloring that uses $\pi$ and $\dmax(G)+1$ is the maximum degree of $G$.
Consequently, if $\lb(G) = \coloring(G,\pi)$, then 
as a result of the above, we know $\coloring(G,\pi)$ must be optimal.

Second, we may also use the bounds to characterize the accuracy of a greedy coloring method or prove that a solution is \textit{not optimal}.
For instance, suppose ${\rm \lb(G)} \leq K(G) \leq \coloring(G,\pi)$, then we know $\coloring(G,\pi)$ is not optimal.

We find the optimal number of colors is directly obtained and verified via both lower and upper bounds for nearly all collaboration networks and web graphs as shown in Table~\ref{table:stats-sparse-ca}.
In 6 of the 13 collaboration networks, we found that 
$\lb(G) = \coloring_{\min}(G,\pi) = \coloring_{\max}(G,\pi) = K(G)+1 = T(G)$
where $\coloring_{\min}(G,\pi)$ and $\coloring_{\max}(G,\pi)$ are the min and max number of colors used by any of the coloring methods.
This implies that the ordering is insignificant for these networks as all the methods resulted in a coloring that is provably optimal.
Notably, from the 7 other networks, 5 of them differ only in $\coloring_{\max}(G,\pi)$.
In addition, many other interesting observations and insights may be drawn from Table~\ref{table:stats-sparse-social} and Table~\ref{table:stats-sparse-ca}.

To summarize we find that:
(i) For some types of information networks, the proposed greedy coloring methods produce an optimal coloring.
(ii) The upper and lower bounds are effective for proving the optimality or suboptimality of a solution from a greedy coloring heuristic.
(iii) For the majority of graphs that are \textit{significantly} skewed and power-lawed, 
the optimal number of colors is directly obtained and verified via both lower and upper bounds.

\section{Finding Colorful Neighborhoods}
\label{sec:neigh-coloring}
Given a large graph or a collection of neighborhood subgraphs, how can we define a domain-independent basis that succinctly characterizes the common structural properties of the neighborhood subgraphs?
For this task, we define the problem of coloring local neighborhood subgraphs and propose a fast parallel flexible approach for solving it. Formally, a neighborhood subgraph can be defined as the induced subgraph centered around a vertex $v$ and induced by all neighbors of $v$.
Our parallel neighborhood coloring framework makes heavy use of the proposed coloring methods from Table~\ref{tab:selection-criterion} as well as the basic coloring variant in Alg~\ref{alg:greedy-coloring} and the more accurate recolor variant shown in Alg~\ref{alg:greedy-coloring-repair}.
In particular, we propose parallel methods for coloring neighborhoods that are (i) fast and scalable for large networks, (ii) space-efficient, (iii) flexible for a variety of applications, (iv) and accurate, finding in many cases nearly optimal or provably optimal solutions.

One of the main observations we make is that neighborhoods that are colored using a relatively few number of colors are not well connected, with low clustering and a small number of triangles.
To understand this fundamental finding and the key intuition, we provide a series of simple neighborhood colorings shown in Figure~\ref{fig:neigh-coloring-examples}.
We also observe that neighborhoods that are colored using a relatively large number of colors have large clustering coefficients and usually contain large cliques relative in size to the other neighborhood colorings.
Therefore, the set of neighborhood colorings is an important fundamental graph property, giving a number of key insights into the structural properties of the network at large and its local neighborhoods.
In a similar manner as we have demonstrated above, one can also use neighborhood coloring to draw a number of other interesting insights and ultimately use it for characterizing the structure and behavior of many types of large networks.
Besides these key benefits, we demonstrate that neighborhood coloring is fast and scalable to compute for large networks, and more specifically, it is linear in the number of edges.
This is clearly much faster than computing the frequency of vertex/edge triangles~\cite{rossi2014pakdd} or counting the frequency of other subgraph patterns and motifs~\cite{prvzulj2007biological,shervashidze2009efficient,rahman2012graft}.
We also show that it is straightforward to parallelize for both shared-memory (CPU and GPU) and distributed architectures.

\begin{figure*}[t!]
\centering
\hspace{-4mm} 
\subfigure[Star]{\label{fig:neigh-coloring-star} \includegraphics[width=0.24\linewidth,bb=0 0 121 104]{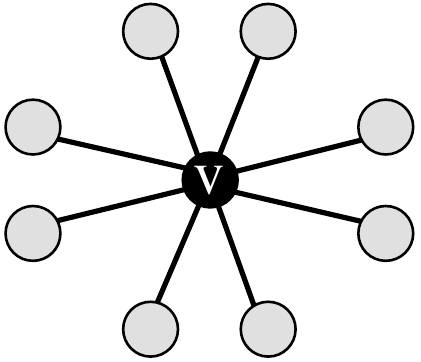}}
\hspace{-4mm} \hfill
\subfigure[Star w/ Triangles]{\label{fig:neigh-coloring-star-to-clique1} \includegraphics[width=0.24\linewidth,bb=0 0 121 104]{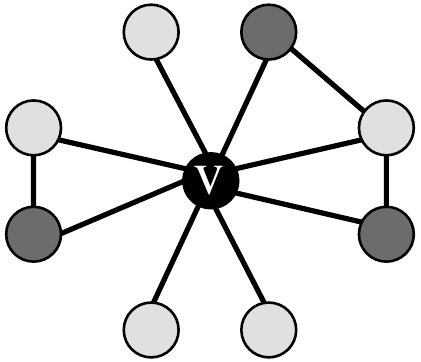}}
\hspace{-4mm} \hfill
\subfigure[Small Cliques]{\label{fig:neigh-coloring-star-to-clique1} \includegraphics[width=0.24\linewidth,bb=0 0 121 104]{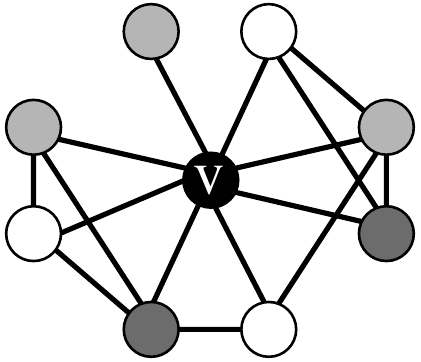}}
\hspace{-4mm} \hfill
\subfigure[Large Clique]{\label{fig:neigh-coloring-star-to-clique1} \includegraphics[width=0.24\linewidth,bb=0 0 121 104]{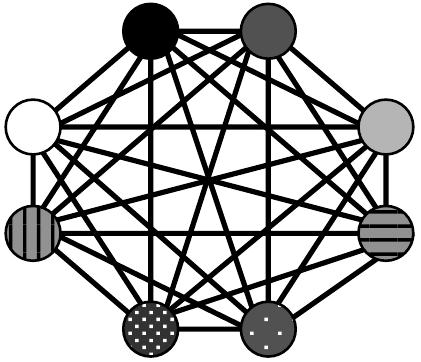}}
\hspace{-4mm} \hfill
\caption{
\textbf{Neighborhood Coloring Extremes: From Stars to Cliques.}
For (a)-(c), the vertex $v$ in the center is the vertex in which the neighborhood was induced, thus the other vertices are in the set $\N(v)$.
In (a) the vertex neighborhood is a simple star--no connections between the neighbors of $v$, and thus can be colored using only 2 colors.
The neighborhood subgraph in (b) is essentially a star with a few neighbors of $v$ with edges among each other, thus, forming triangles.
Similarly, in (c) we find more neighbors forming connections among each other giving rise numerous triangles and two cliques of size 4.
Finally, the neighborhood subgraph in (d) represents a single large clique. 
Node $v$ was removed for clarity.
These neighborhood subgraphs go from the least constrained neighborhood representing a star (a) to the most constrained neighborhood representing a clique (d).
The neighborhood subgraphs shown in (b) and (c) are better representatives of neighborhood subgraphs found in large real-world networks (e.g., Facebook or other social networks).
Note that in reality, the vertex $v$ in which the neighborhood subgraph corresponds may be removed from the coloring, since $v$ must be connected to every other vertex.
Thus, the neighborhood subgraphs above are (k-1)-colorable when $v$ is removed.
}
\label{fig:neigh-coloring-examples}
\end{figure*}

Local neighborhood coloring consists of assigning a color to every vertex in a vertex neighborhood such that no two vertices linked by an edge share the same color while minimizing the number of colors used.
The most colorful neighborhood is the one that requires the maximum number of colors.
In this section, we propose parallel methods for coloring neighborhoods that are (i) fast and scalable for large networks, (ii) space-efficient, (iii) flexible for a variety of applications, (iv) and accurate, finding in many cases nearly optimal or provably optimal solutions.

The neighborhood colorings may be useful for finding better communities, especially in local community detection methods~\cite{malliaros2012fast}.
Besides community methods, neighborhood coloring may also be used in prediction tasks such as detecting anomalous patterns in graphs, see~\cite{akoglu2010oddball} for one such egonet-based method.
Other prediction tasks such as relational classification may also benefit from neighborhood coloring.
For instance, one may construct a set of node features such as from these neighborhood colorings to improve the accuracy of classification (e.g., number of colors, largest independent set).

The results of our neighborhood coloring have direct and immediate implications on exact algorithms for the maximum clique problem~\cite{bomze1999maximum,prosser2012}.
In fact, the most successful approaches have used coloring as a bound, but vary in the ordering and method used~\cite{konc2007improved,SanSegundo,mcr-tomita:2007,tomitaefficient, mcs-tomita2010simple,rossi2013parallel-cliques}.
For instance, suppose vertex neighborhoods are searched in parallel, similar to our heuristic in Alg~\ref{alg:heuristic}, then the neighborhood coloring results may be used for bounding the search space in branch-n-bound algorithms, for pruning entire neighborhoods directly, and for ordering vertices via the number of colors from the vertex neighborhood coloring, among many other vertex level features that could be derived from such a set of neighborhood colorings.
These may enhance recent parallel algorithms such as \textsc{pmc}~\cite{rossi2014pmc-www} that utilizes degeneracy ordering giving a worst-case runtime of $O(2^{d/4})$ on sparse graphs with bounded degeneracy.
In addition, super-linear speedups may become more frequent using the ordering from neighborhood coloring/pruning, i.e., these were observed using \textsc{pmc}~\cite{rossi2014pmc-www} and later confirmed again using a parallel version of \textsc{mcs}~\cite{mcs-tomita2010simple,mccreesh2013multi}.
Nevertheless, the set of vertex neighborhood colorings may also be used for pruning in other ego-centric search methods.
They also provide a basis for a variety of ordering methods which may have applications, e.g., graph compression~\cite{boldi2004webgraph}.

\subsection{Problem Formulation}
Our focus is on coloring vertex neighborhoods.
Let $\N(v) = \{v\} \cup \{u : (u,v) \in E\}$ be the closed neighborhood of a vertex $v$ and we define $H_v$ as the neighborhood subgraph induced from $\N(v)$, consisting of $v$, the neighbors of $v$, and any edges between them.
Suppose $H_v$ is a neighborhood subgraph of $G$ and $G$ is $k$-colorable, then $H_v$ must also be $k$-colorable.
Consequently, if $H_v$ is a subgraph of $G$, then $\Chromatic(H) \leq \Chromatic(G)$.

The \textit{local chromatic number} of $G$ is the maximum number of colors appearing in the closed neighborhood (subgraph) of a vertex minimized over all proper colorings. 
More formally,
\[ 		\localcoloring(G) = \min_{c} \; \; \max_{v \in V} \; |\{ c(u) : u \in \N(v) \}| 		\]
\noindent where the minimum is taken over all proper colorings $c$ and $\localcoloring(G)$ is the number of colors appearing in the most colorful closed neighborhood of a vertex.
Clearly, $\localcoloring(G) \leq \Chromatic(G)$ and we find for large real-world graphs (i.e., social and information networks~\cite{mislove-2007-socialnetworks,nkahmed:tkdd13}) these two numbers are usually close.
Despite this result, we note that for general graphs $\localcoloring(G)$ may be small while $\Chromatic(G)$ can be arbitrarily large~\cite{erdos1986coloring,godsil2001algebraic}.

We relax the strict requirement above from considering all proper colorings to considering only a single proper coloring for each neighborhood. 
In particular, this article proposes a {\rm \textit{framework}} of \textit{local greedy coloring methods} designed for dense and large sparse graphs found in real-world (e.g., social networks).
Given a neighborhood subgraph of $v$ denoted $H_v$ and a graph property $\f(\cdot)$, let $\f(H_v) = \vx$ where $\vx \in \mathbb{R}^{n}$ is a vector of vertex weights and $x_i$ is the value of vertex $u_i \in \N(v)$.
Using the weight vector $\vx$ as a basis for ordering the vertices in the closed neighborhood, we denote this ordering as $\pi_v = \{u_1,u_2,...\}$.
Further, let $\coloring(\Hg,\pi_v)$ be the number of colors used by a local greedy coloring algorithm that uses the ordering $\pi_v$ to color $\Hg$.
Consequently, an \textit{approximation} of the local chromatic number of $G$ is defined as:
\[ 		\localcoloring(G,\Pi) =  \max_{v \in V} \;  \coloring(\N[v],\pi_v)	\]
\noindent where  $\localcoloring(G, \Pi)$ is the \textit{maximum number of colors} used by a \textit{local greedy coloring method} that uses the set of neighborhood vertex orderings $\Pi = \{\pi_{v_1}, \pi_{v_2},...,\pi_{v_n}\}$.
Intuitively, the above gives rise to the following relationship:
\[ 		\mc(G) \leq \Chromatic(G) \leq \localcoloring(G,\Pi) 	\]
Also, if we consider a vertex neighborhood subgraph $\Hg_v$, then:
\[ \mc(\Hg_v) \leq \Chromatic(\Hg_v) \leq \coloring(\Hg_v,\pi_v) \leq \dmax(\Hg_v)+1 \]
\noindent where $\mc(\Hg_v)$ is the size of the maximum clique,
$\Chromatic(\Hg_v)$ is the optimal number of colors required to color $\Hg_v$ (minimized over all proper colorings of $\N(v)$), 
and $\coloring(\Hg_v,\pi_v)$ is the number of colors from a greedy coloring of $\N(v)$ using $\pi_v \in \Pi$.

\begin{figure}[t!]
\hspace{-4mm} 
\subfigure[Representative Graphs]{\label{fig:speedup-graphs} \includegraphics[width=0.56\linewidth]{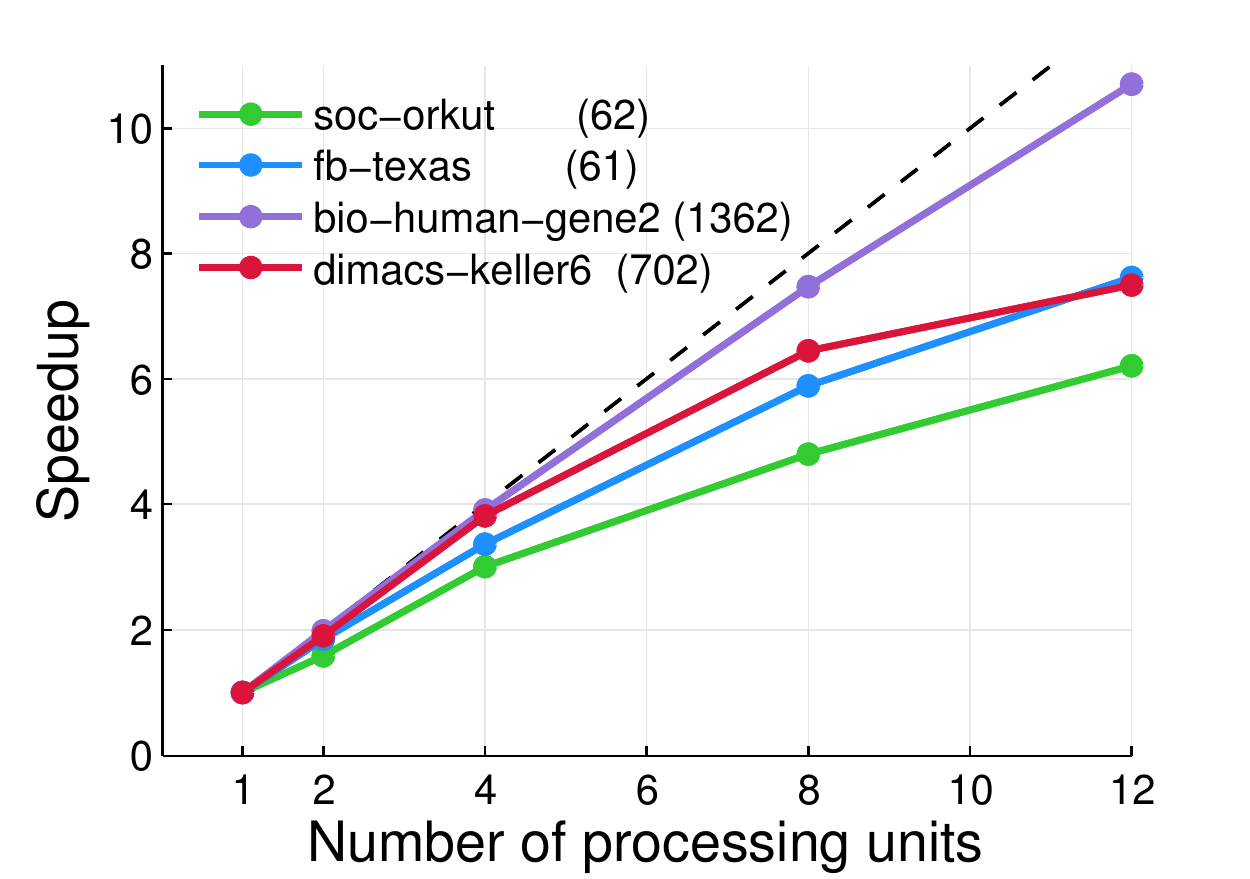}}
\hspace{-24mm} \hfill
\subfigure[Coloring Variants]{\label{fig:speedup-variants} \includegraphics[width=0.56\linewidth]{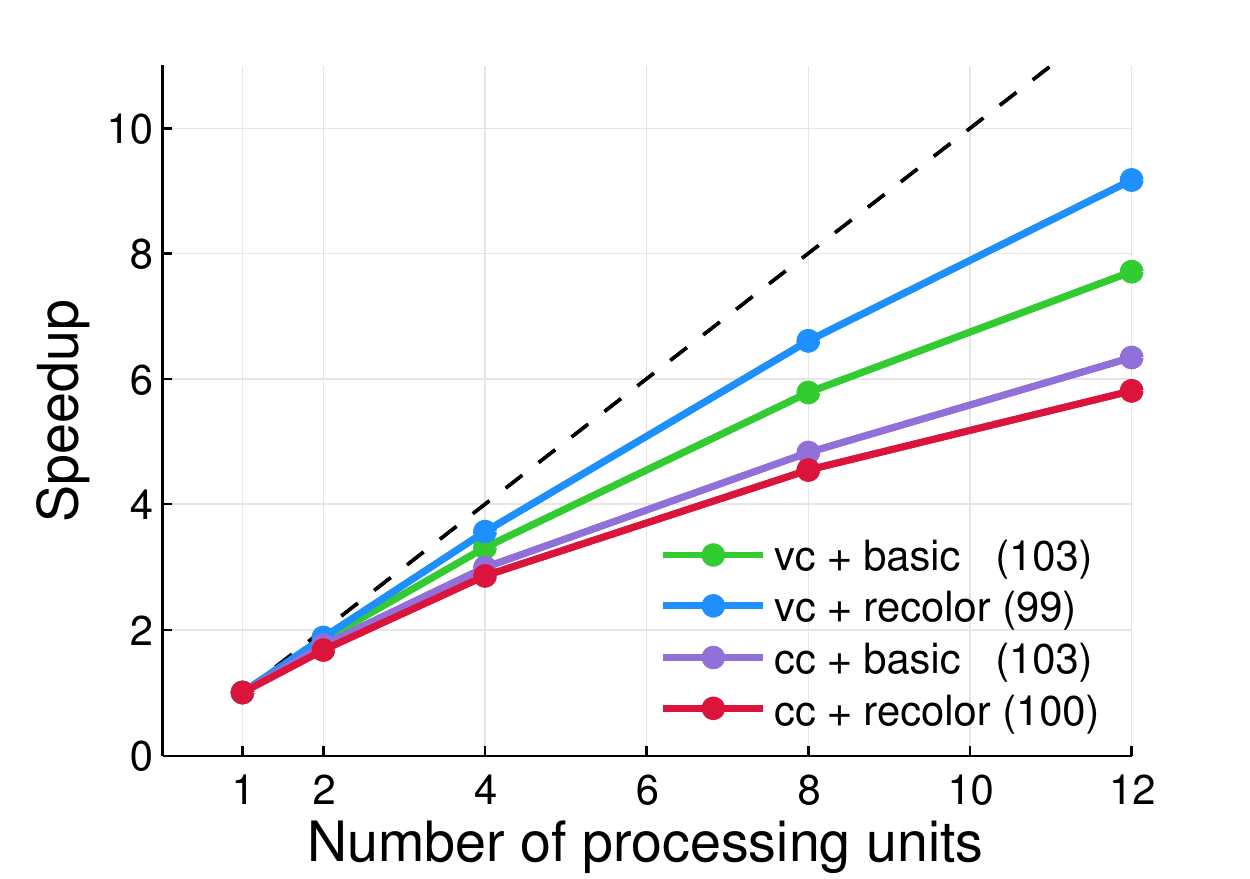}}
\hspace{-8mm}
\caption{
\textbf{Scalability.}
The speedup of our methods on different types of graphs are shown in Fig.~\ref{fig:speedup-graphs}, 
whereas the speedup of different coloring variants for soc-flickr are shown in Fig.~\ref{fig:speedup-variants}.
It is clear that all proposed variants are scalable for large graphs, while vertex-centric coloring (vc) using \textsc{recolor} scales slightly better than the others for the large sparse flickr social network.
Processing units are cores (one thread per core).
}
\label{fig:speedup}
\end{figure}

\subsection{Neighborhood Coloring}\label{sec:neigh-coloring-alg}
The parallel framework is shown in Alg~\ref{alg:greedy-coloring-neighborhoods}.
Here, $\bound(\cdot)$ is assumed to be normalized with respect to cliques, hence, $\bound(v) = K(v)+1$.
This allows us to generalize the algorithm over any arbitrary upper bound.

\begin{figure}[t!]
\begin{center}
\begin{minipage}{0.9\linewidth}
\begin{algorithm}[H]
\small
\caption{Parallel Neighborhood Coloring Framework
}
\label{alg:greedy-coloring-neighborhoods}
{\algfontsize
\bspacing
\begin{algorithmic}[1]
\State Initialize data structures 																					\label{algline:initialize}		
\State Compute upper bounds $\bound(G)$ 																\label{algline:lc-ub}
\State Obtain a lower bound $\omega(G) \leftarrow \textsc{HeuClique}(G)$			\label{algline:lc-lb}
\State Prune vertices and edges from $G$ (explicitly)													\label{algline:lc-pruning1}
\State Obtain a vertex ordering $\pi=\{v_1,...,v_n\}$ 													\label{algline:lc-global-ordering}
\For{{\bf each} $v_i$ in an ordering $\pi$ {\bf in parallel} }											\label{algline:lc-parallel-for}

		\If{$\bound(v_i) > \maxcolor$}																				\label{algline:lc-pruning2}
				\State $P \leftarrow \{v_i\}$																			\label{algline:lc-init-P}
				\For{$w \in N(v_i)$}																						\label{algline:lc-for-neigh}
						\If{$\bound(w) > \maxcolor$}						
																																	\label{algline:lc-pruning3}
								$P \leftarrow P \cup \{w\}$															\label{algline:lc-add-neigh-to-P}
						\EndIf
				\EndFor
				\State Set $\vx$ to be the computed graph property $\f(P)$							\label{algline:lc-local-graph-property}
				\State Order vertices in $P$ using $\vx$ 														\label{algline:lc-local-ordering}
				\State $k \leftarrow$ \textsc{ColoringVariant}($G$, $P$)							\label{algline:lc-coloring-variant}
				\If{$k > \maxcolor$} $\maxcolor \leftarrow k$		\EndIf								\label{algline:lc-lock}
		\EndIf
\EndFor
\State {\bf return} $\localcoloring(G,\pi) \leftarrow \maxcolor$
\end{algorithmic}
\espacing
}
\end{algorithm}
\end{minipage}
\end{center}
\end{figure}

Upper and lower bounds are computed in Line~\ref{algline:lc-ub} and \ref{algline:lc-lb}, respectively, and used for pruning in Line~\ref{algline:lc-pruning1}.
The vertices \textit{remaining} in $G$ are ordered in Line~\ref{algline:lc-global-ordering}, and then each vertex neighborhood in that order are colored (Line~\ref{algline:lc-parallel-for}).
For each vertex in order, we first try to avoid coloring $v_i$ by checking if the local vertex upper bound $\bound(v_i)$ is smaller than $\maxcolor$.
If not, then lines~\ref{algline:lc-init-P}--\ref{algline:lc-add-neigh-to-P} form the "reduced" set $P$ of neighboring vertices.
In line~\ref{algline:lc-local-ordering}, we obtain the local vertex ordering $\pi_{v_i}$ by ordering the vertices in $P$ using an arbitrary property $\f(P)$ computed in line~\ref{algline:lc-local-graph-property}.
Next, the subgraph $H_v$ induced by the ordered vertex set $P$ are colored using a coloring variant and color assignment/search strategy (Line~\ref{algline:lc-coloring-variant}).
Finally, line~\ref{algline:lc-lock} updates the maximum number of neighborhood colors required, if necessary.

Note that the three pruning steps are shown in Line~\ref{algline:lc-pruning1}, \ref{algline:lc-pruning2}, and Line~\ref{algline:lc-pruning3}, respectively.
If the goal is to compute $\localcoloring(G,\pi)$, then the pruning steps can significantly reduce the search space leading to faster and more accurate colorings.
For the problem of computing the complete set of neighborhood colorings, then we can simply avoid using the pruning steps.
In other words, the pruning steps and their utility are application dependent, and thus may be turned on/off accordingly.
We also note that these pruning steps are also useful for finding the max clique, computing a graph property for which the upper and lower bounds apply, and for finding dense subgraphs, among many other tasks.

In addition to the coloring variants from Section~\ref{sec:framework} and~\ref{sec:recolor}, we also investigate two types of search procedures for coloring (i.e., color-centric and vertex-centric) that differ in their implementation, but may result in significantly different runtimes depending on the structural properties of the input graph.
In particular, the search procedure in the basic and recolor variants may be performed by searching color-classes (i.e., the independent sets) or by searching the vertex neighborhoods (i.e., adjacent vertices) and thus, we term these search procedures as color-centric and vertex-centric, respectively.

\begin{figure}[h!]
\centering
\hspace{-7mm}
\subfigure
{\label{fig:ccdf-number-local-coloring} \includegraphics[width=0.56\textwidth]{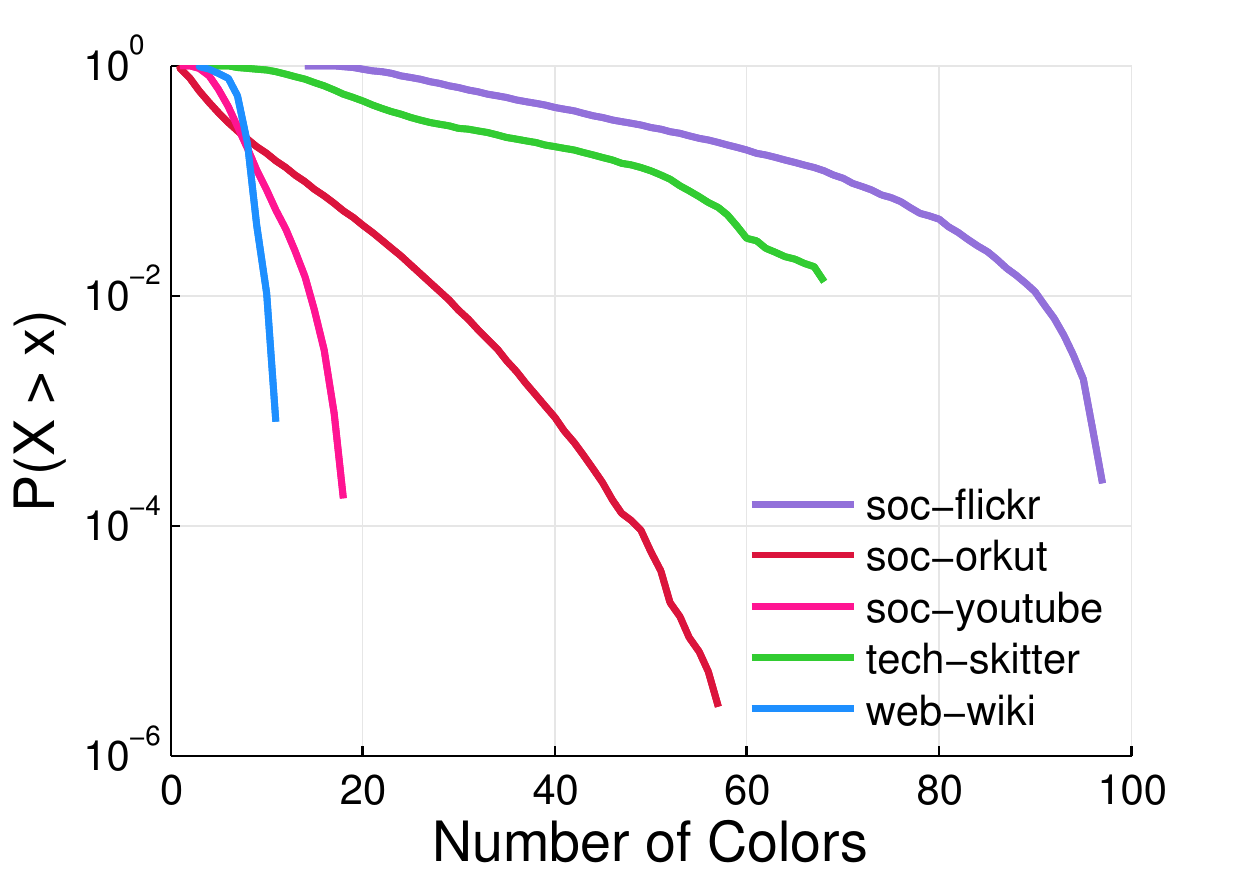}}
\hspace{-24mm} \hfill
\subfigure
{\label{fig:ccdf-indep-local-coloring} \includegraphics[width=0.56\textwidth]{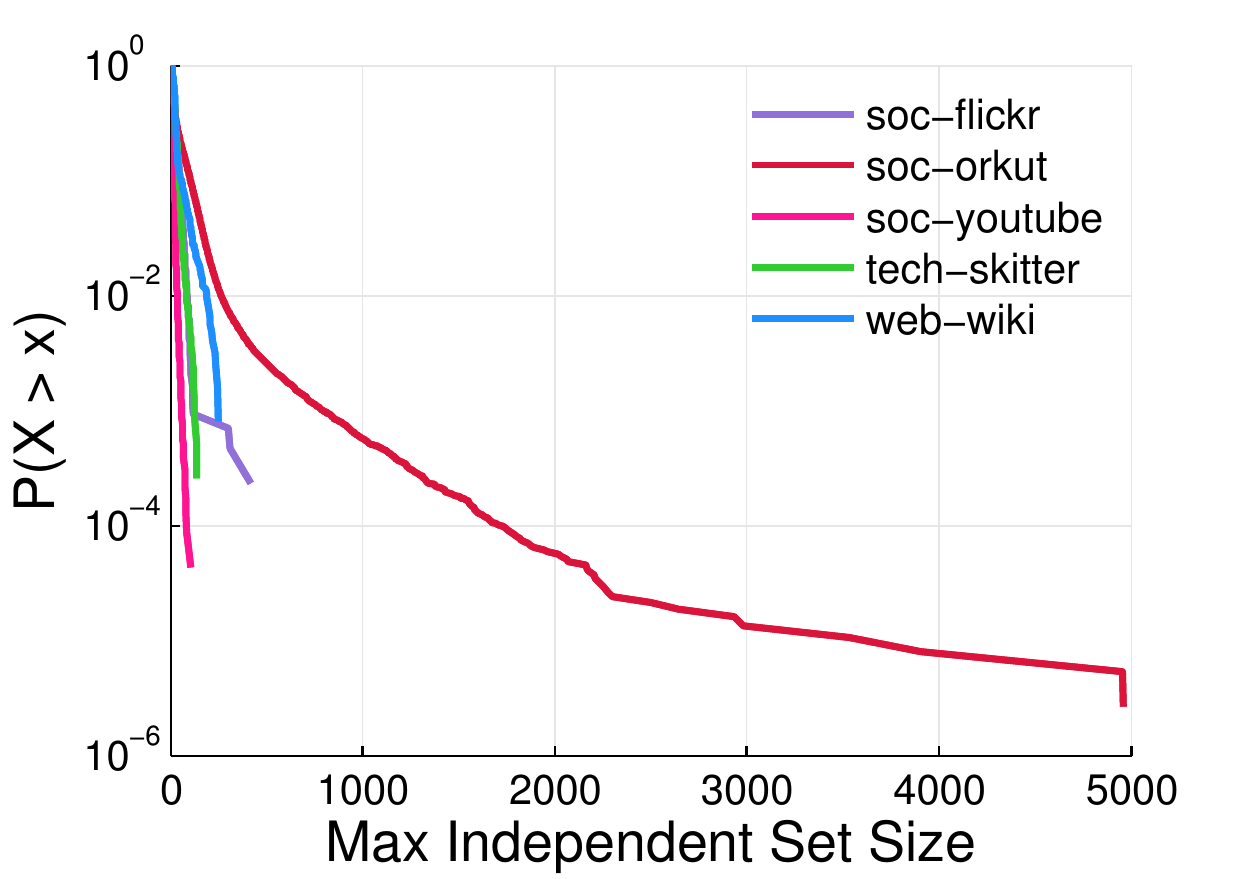}}
\hspace{-8mm}
\caption{
\textbf{Properties of the Neighborhood Colorings.}
Using the parallel neighborhood coloring algorithm, we color each vertex-induced neighborhood and record the number of colors used for that neighborhood as well as the maximum independent set size (i.e., largest such coloring class given by the neighborhood coloring of that vertex).
We use the complementary cumulative distribution function (CCDF) to study the coloring properties of a few large sparse real-world networks.
The max independent set size is with respect to the coloring (largest such coloring class).
}
\label{fig:ccdf-local-coloring}
\end{figure}

\subsubsection{Parallelization}\label{sec:parallel-alg}
The neighborhood coloring problem is parallelized by considering each neighborhood subgraph as independent and coloring each of these subgraphs in parallel.
We use dynamic scheduling and assign each processing unit a single neighborhood at a time.
This helps ensure the vertex neighborhoods are colored in approximately the correct order.
Our approach requires a single lock to ensure that the largest number of colors used thus far is consistent and avoid potential race conditions when updating it (see line~\ref{algline:lc-lock}).
Importantly, as soon as a processing unit updates $\maxcolor$, we immediately broadcast it to all other processing units.
We observed that this can significantly improve performance as the tighter lower bound may be used for additional pruning or result in terminating a search early.
As an aside, if the pruning rules are used, then two subsequent runs may result in slightly different $\localcoloring(G,\pi)$.
This is due to possible variations in the global vertex ordering which determines the underlying order in which the neighborhoods are colored. 

The parallel framework has many other advantages.
For instance, each processing unit only requires a neighborhood subgraph and therefore the framework is space efficient for streaming or graphs too large to reside in memory and thus a good candidate for GPU parallelization as well.

\begin{table}[t!]
\caption{\textbf{Upper and Lower Bounds of the Chromatic Number for the Graphs.}
We denote $\coloring(G,\pi)$ as the maximum number of colors used from the set of neighborhood colorings.
Note that $\coloring(G,\pi)$ is computed using none of the pruning steps and thus is larger than if pruning is used.
}
\vspace{1mm}
\label{table:neigh-coloring-datasets}
\centering\small\scriptsize
\begin{tabularx}{1.0\linewidth}{ Hr XXXHX cH cHX XXXX H XH} 
\toprule
& & 
\multicolumn{10}{c}{\textbf{Graph stats}} & 
\multicolumn{4}{c}{\textbf{Bounds}} & 
&
\\ 
 &
\textbf{graph} & 
$|V|$ & 
$|E|$ & 
$|T|$ & 
$\rho$ & 
$\bar{d}$ & 
${\rm r}$ & 
$\bar{\kappa}$ & 
$\kappa$ & 
$\tr_{\rm avg}$ & 
$\tr_{\rm max}$ & 
$\dmax$ & 
$K$+1 & 
$T$ & 
$\tilde{\omega}$ & 
$\omega$ &
$\coloring(G,\pi)$ &
$L_{\max}$ 
\\ \midrule
	 &    \textsf{soc-flickr} &  513K &  3.1M &  176M &  $10^{-5}$ &  12 &  0.16 &  0.08 &  0.15 &  343 &  524K &  4.3K &  310 &  153 &  21 & 58 &  104 &  208  \\ 
	 &    \textsf{soc-orkut} &  2.9M &  106M &  1.5B &  $10^{-5}$ &  70 &  0.02 &  0.01 &  0.04 &  525 &  1.3M &  27.4K &  231 &  75 &  37 & 47 &  83 &  190  \\ 
	 &    \textsf{soc-youtube} &  495K &  1.9M &  7.3M &  $10^{-5}$ &  7 &  -0.03 &  0.05 &  0.01 &  15 &  151K &  25.4K &  50 &  19 &  11 & 17 &  28 &  61  \\ 
	 &    \textsf{tech-as-skitter} &  1.6M &  11M &  86.3M &  $10^{-6}$ &  13 &  -0.08 &  0.08 &  0.01 &  51 &  564K &  35.4K &  112 &  68 &  41  & 68 &  70 &  115  \\ 
	 &    \textsf{bio-human-gene2} &  14K &  9M &  14.7B &  $0.09$ &  1.2K &  0.8 &  0.65 &  0.59 &  1M &  6.9M &  7.2K &  1903 &  1681 &  1267 & $\geq1276$ &  1329 &  1329  \\ 
	 &    \textsf{keller6} &  3.3K &  4.6M &  10.3B &  0.82 &  2.7K &  -0.02 &  0.00 &  0.82 &  3M &  3.5M &  2.9K &  2691 &  2084 &  45 & $-$ &  148 &  783  \\ 
\midrule
\end{tabularx} 
\vspace{-4mm}
\end{table} 

\subsection{Experiments}\label{sec:neigh-experiments}
We now analyze the effectiveness of our approach on a variety of real-world networks.
The network statistics including lower and upper bounds are provided in Figure~\ref{table:neigh-coloring-datasets}.

A number of observations are made from the experiments.
First and foremost, the scalability of our parallel framework is demonstrated in Figure~\ref{fig:speedup} where we observe that significant speedups are possible across a range of different types of graphs and coloring variants.
Importantly, Fig.~\ref{fig:speedup-graphs} demonstrates the scalability of our methods on a diverse set of graphs, from large sparse graphs (e.g., social and biological networks) to dense networks found in scientific computing.
Besides density, these graphs are known to contain very different structural properties.
We used the large Orkut social network that is sparse and power-lawed, the sparse Facebook Texas network, a slightly more dense biological network of a human gene,
and a very dense unsolved instance from the clique/coloring DIMAC's challenge.
The last two graphs were found to be more difficult to obtain nearly optimal local colorings.
Nevertheless, these two graphs, but especially the human gene, scale slightly stronger than the more sparse networks.
These graphs were colored using the basic coloring method with color-centric search and no pruning.
Further, vertices were ordered globally by kcore-vol ($\f(v) = \sum_{w \in \N(v)} K(w)$) and ordered locally using kcore-deg-vol and both orderings are from largest to smallest for simplicity.

Finally, we also investigated the scalability of a few different coloring variants using the large sparse flickr social network, see Figure~\ref{fig:speedup-variants} for details.
In particular, all the proposed variants are shown to scale well for large graphs, while vertex-centric coloring (vc) using \textsc{recolor} scales slightly better than the others. 
Similar results were also observed using other types of graphs and methods.

\begin{table}[b!]
\centering
\footnotesize
\caption{
\textbf{Comparing the Space of Neighborhood Coloring Methods}.
We evaluate a representative set of methods from the framework.
The local coloring number denoted $\localcoloring(G,\pi)$ is given for each of the variations.
We present results for a representative sample of methods from the framework.
In all methods, the local ordering is from largest to smallest, whereas the global ordering is from smallest to largest.
For the global ordering we used \textsc{kcore-vol} for simplicity, while varying the local ordering method.
}
\label{table:varying-framework}
\scriptsize
\begin{tabularx}{0.6\linewidth}
{
c c !{\vrule width 0.50mm} 
X | X | X | X | X | X | X | X
}

& \TTT \BBB   & \multicolumn{8}{c}{\textbf{\large Ordering Techniques}} \\

\TTT \BBB  \textbf{\normalsize Variant} 
& \rotatebox{65}{\textbf{\normalsize Pruning}}
& \multicolumn{1}{c}{\Rot{75}{\textsc{Rand}}}
& \multicolumn{1}{c}{\Rot{75}{\textsc{Kcore}}}
& \multicolumn{1}{c}{\Rot{75}{\textsc{Triangle}}}
& \multicolumn{1}{c}{\Rot{75}{\textsc{Tcore}}}
& \multicolumn{1}{c}{\Rot{75}{\textsc{Kcore-vol}}}
& \multicolumn{1}{c}{\Rot{75}{\textsc{Tri-vol}}}
& \multicolumn{1}{c}{\Rot{75}{\textsc{Tcore-vol}}}
& \multicolumn{1}{c}{\Rot{75}{\textsc{Deg-Kcore-vol}}}
\\
\hboldline
\TTT \BBB \multirow{2}{*}{\textbf{\normalsize Basic}} 
& {\large \cm} 
& 60 
& 58 & 54 & 58 
& 54 & 54 & 58 & 54 
\\ \cline{2-10} 

\TTT \BBB & {\large \xm} 
& 68 
& 62 & 58 & 62 
& 57 & 57 & 62 & 56 
\\ \cline{1-10} 

\multirow{2}{*}{\textbf{\normalsize Recolor}} 
\TTT \BBB & {\large \cm} 
& 57 
& 57 & 53 & 57 
& 53 & 52 & 57 & 52 
\\ \cline{2-10} 

\TTT \BBB & {\large \xm} 
& 67 
& 61 & 56 & 61 
& 57 & 56 & 61 & 56 
\\ \hboldlinesmall 
\end{tabularx}
\end{table}

\renewcommand{\figszhist}{\includegraphics[width=0.35\linewidth]}
\renewcommand{\hspacehist}{{\hspace{-4mm}}}
\begin{figure}[t!]
\centering
\hspacehist \subfigure
{\label{fig:hist-colors-bio} \figszhist {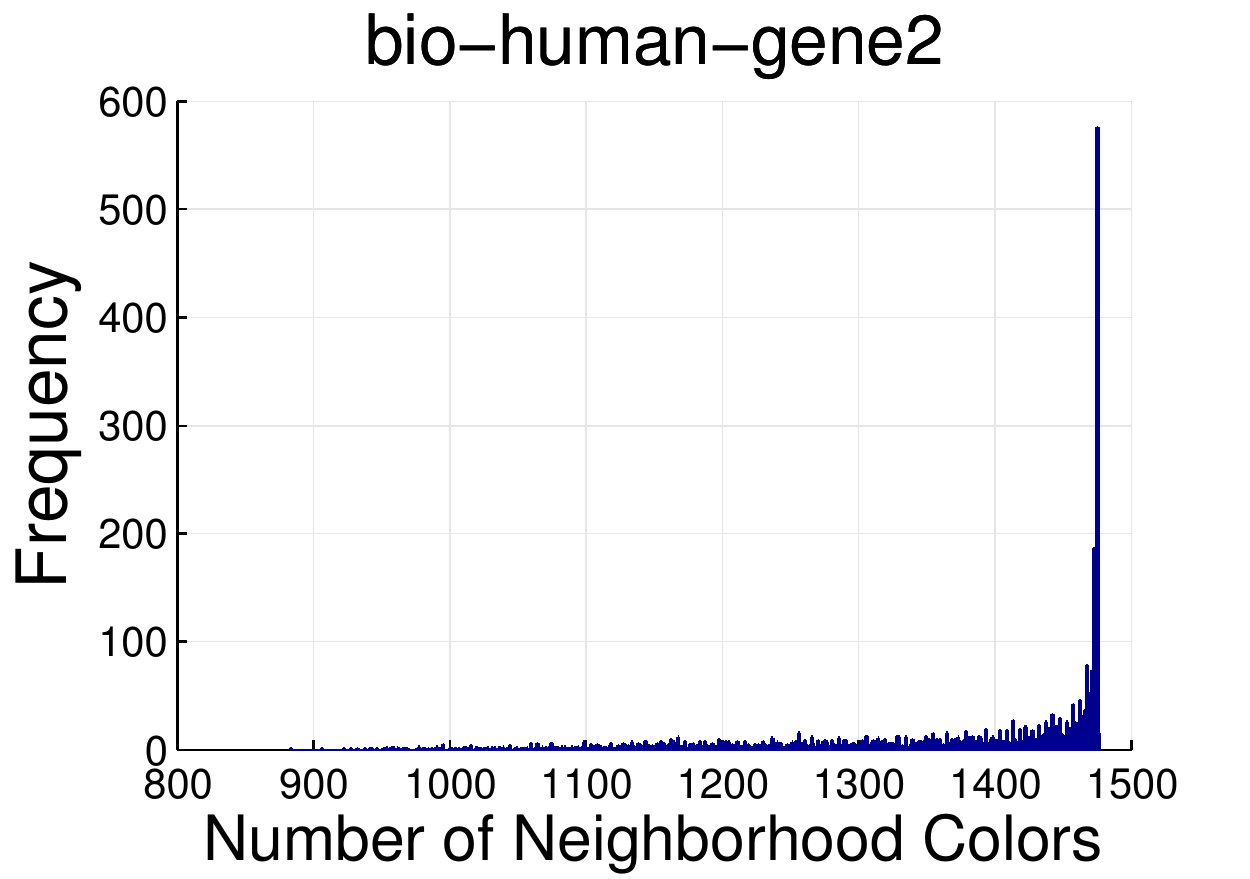}}
\hfill \hspacehist \subfigure
{\label{fig:hist-indep-bio} \figszhist{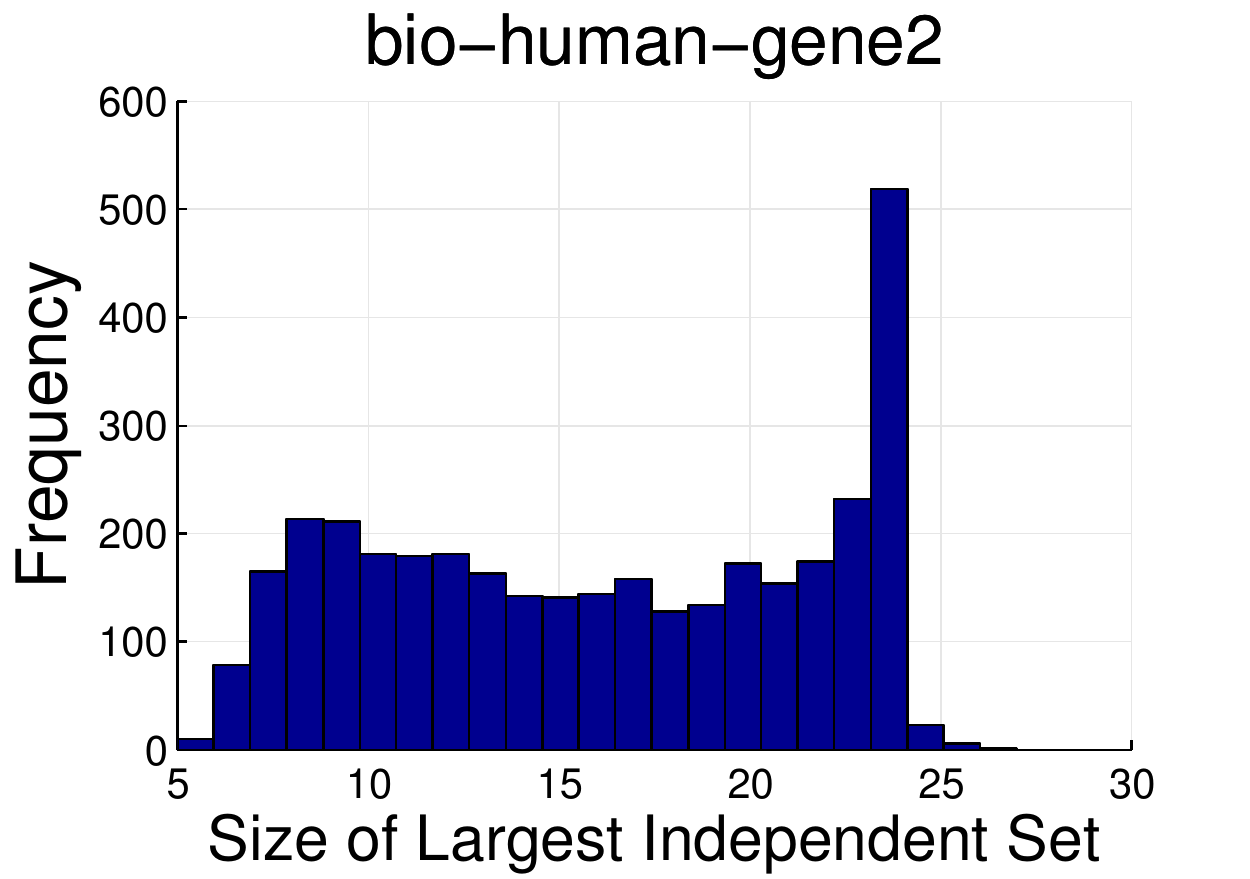}}
\hfill \hspacehist \subfigure
{\label{fig:hist-colors-vs-indep-bio} \figszhist{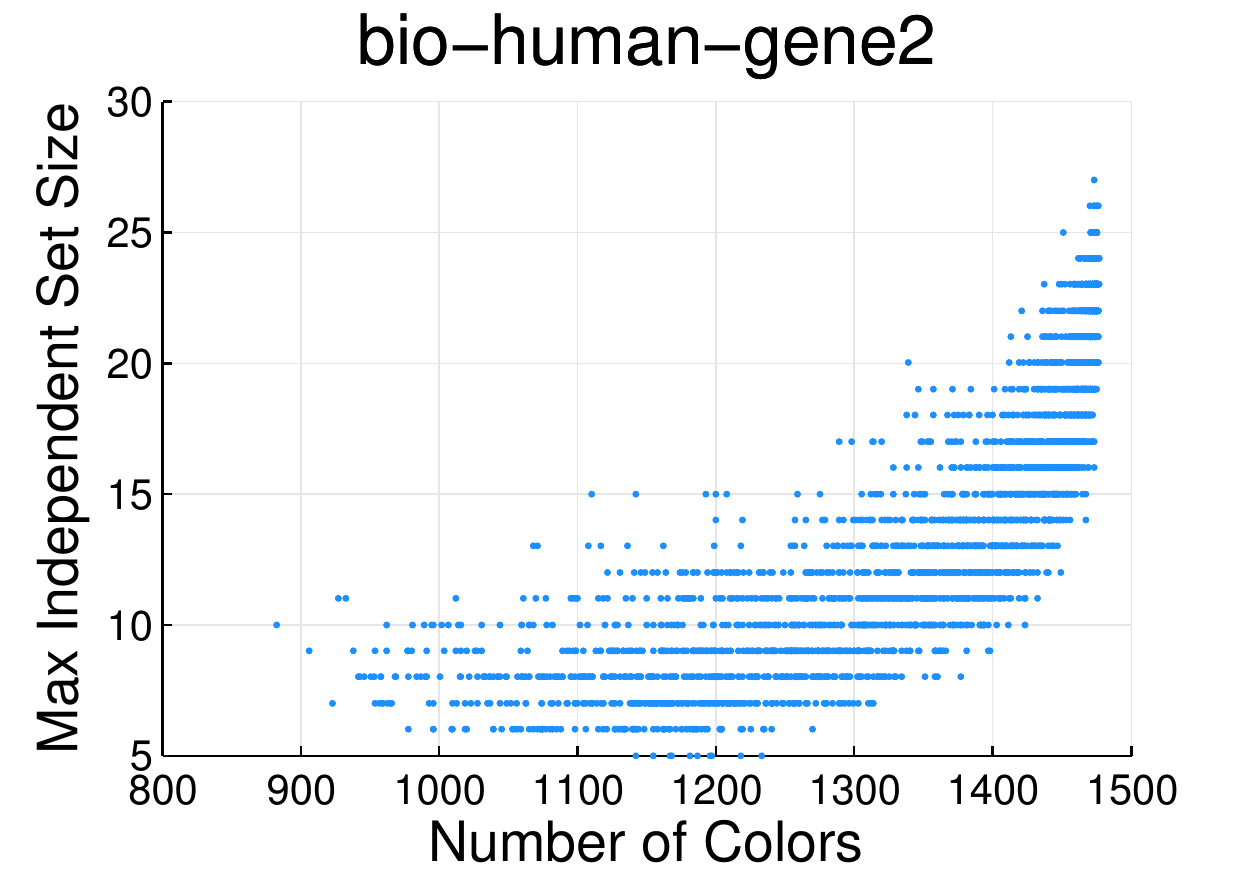}}
\hspacehist

\hspacehist\subfigure
{\label{fig:hist-colors-keller6} \figszhist{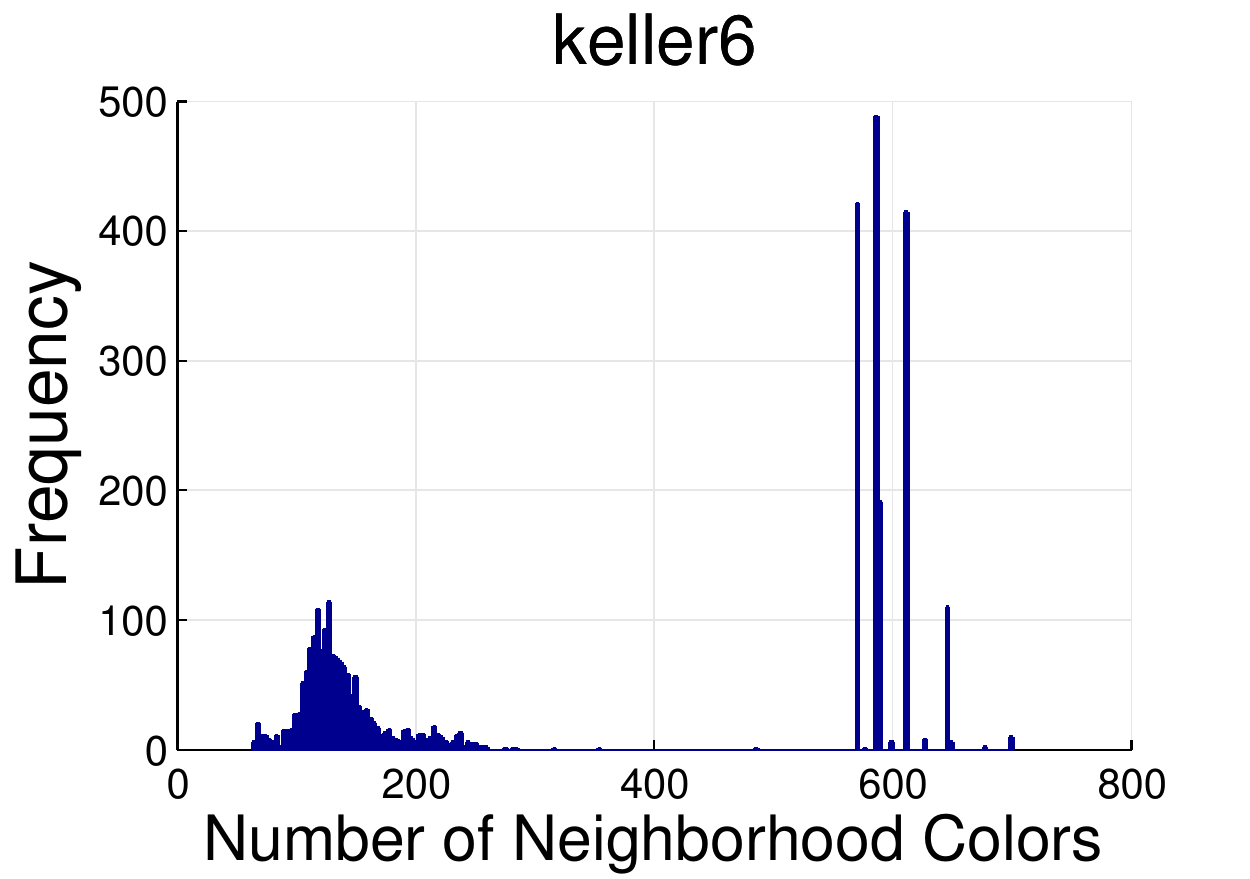}}
\hfill \hspacehist \subfigure
{\label{fig:hist-indep-keller6} \figszhist{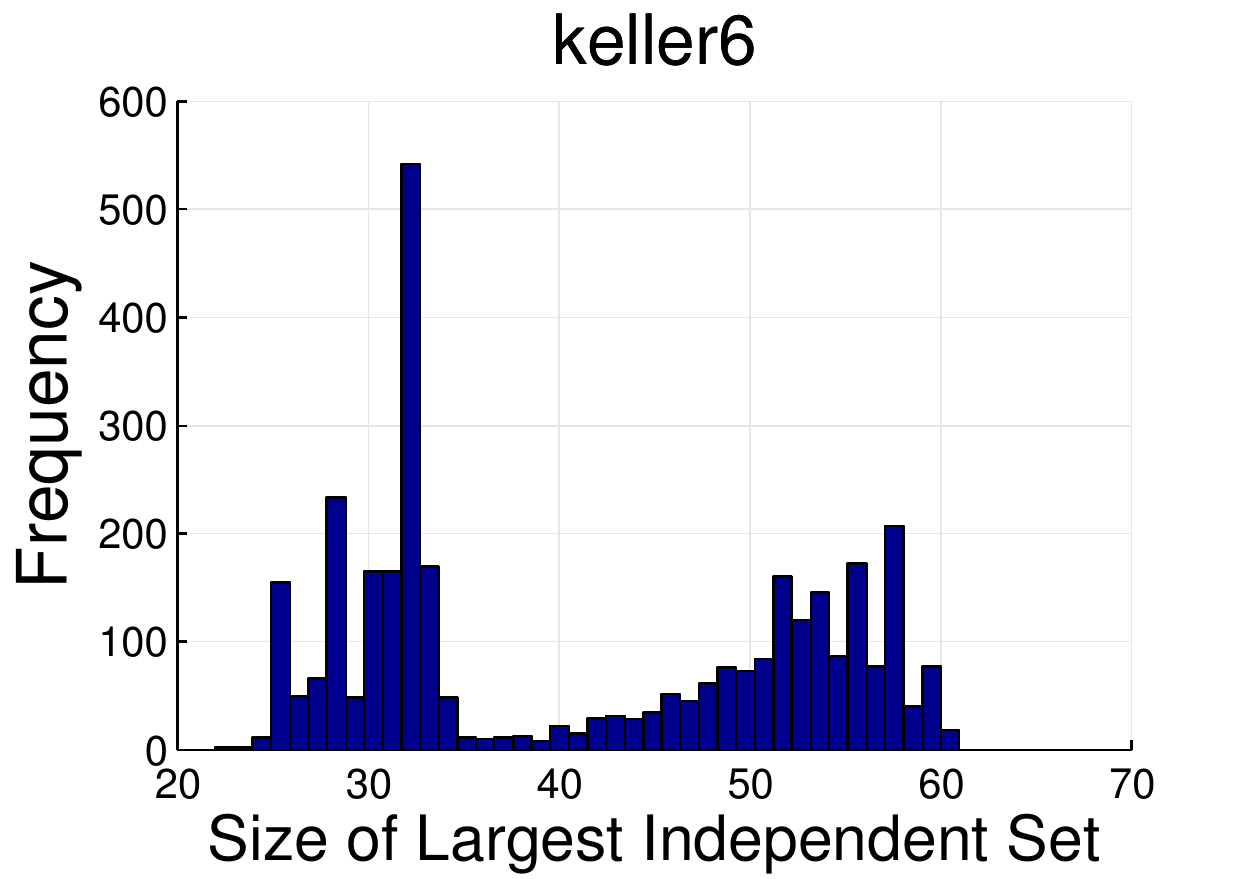}}
\hfill \hspacehist \subfigure
{\label{fig:hist-colors-vs-indep-keller6} \figszhist{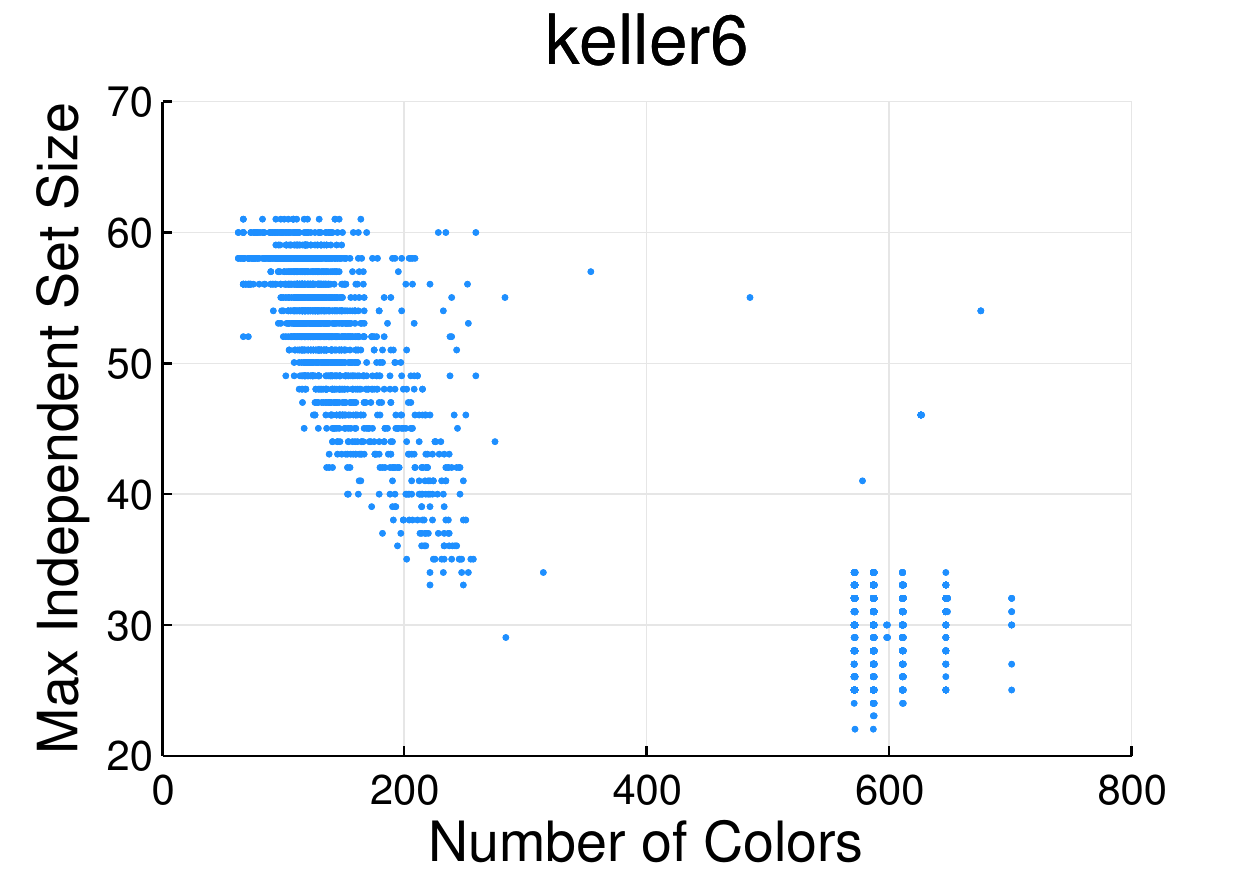}}
\hspacehist
\caption{
\textbf{Characterizing and comparing the various types of networks using statistics from neighborhood coloring.}
The number of colors used to color each of the neighborhoods are shown along with the size of the largest independent set in the coloring of the neighborhoods.
}
\label{fig:hist-colors}
\end{figure}

Now, we investigate a representative sample of coloring methods from the large space defined by the framework.
For this experiment, we use the three pruning steps and order the vertices globally using \textsc{kcore-vol} and are searched from smallest to largest.
The vertices in each neighborhood are ordered from maximum to minimum and thus the vertices more constrained in their choice of color are assigned colors early allowing more flexibility in the color assignment whereas vertices that are not as constrained take lower precedence in their color assignment since these vertices are usually easily assigned to a color.
In both global and local ordering, ties are broken using vertex ids such that if $\f(v) = \f(u)$ and $v > u$, then $v$ is ordered before $u$.

The results from a single graph  (soc-flickr) are shown in Table~\ref{table:varying-framework}, others were removed for brevity.
The first row represents the family of methods that use the basic variant with pruning, whereas the second row uses no pruning.
Likewise, the third and fourth rows use recolor with pruning and without it, respectively.
There are several interesting observations. First, the coloring number from the recolor variant is at least as accurate and usually better than the basic variant. This result is independent of whether pruning is used or not and it shows how much improvement can be achieved by using the recolor variant. Second, pruning is generally effective in obtaining a better coloring number. Note that using both pruning and the recolor variant clearly improves on the basic coloring method (without pruning or recolor). For example, using the \textsc{Tri-vol} method, we get a $\approx 9\%$ improvement in the number of colors, when we apply both pruning and recolor variant. Finally, we observe that the \textsc{Tri-vol} and \textsc{Deg-Kcore-vol} methods perform the best among all other methods (minimum number of colors). These results are consistent with the pervious discussion in Section~\ref{sec:results}.

\begin{figure*}[t!]
\hspacehist \subfigure
{\label{fig:hist-colors-soc-youtube} \figszhist{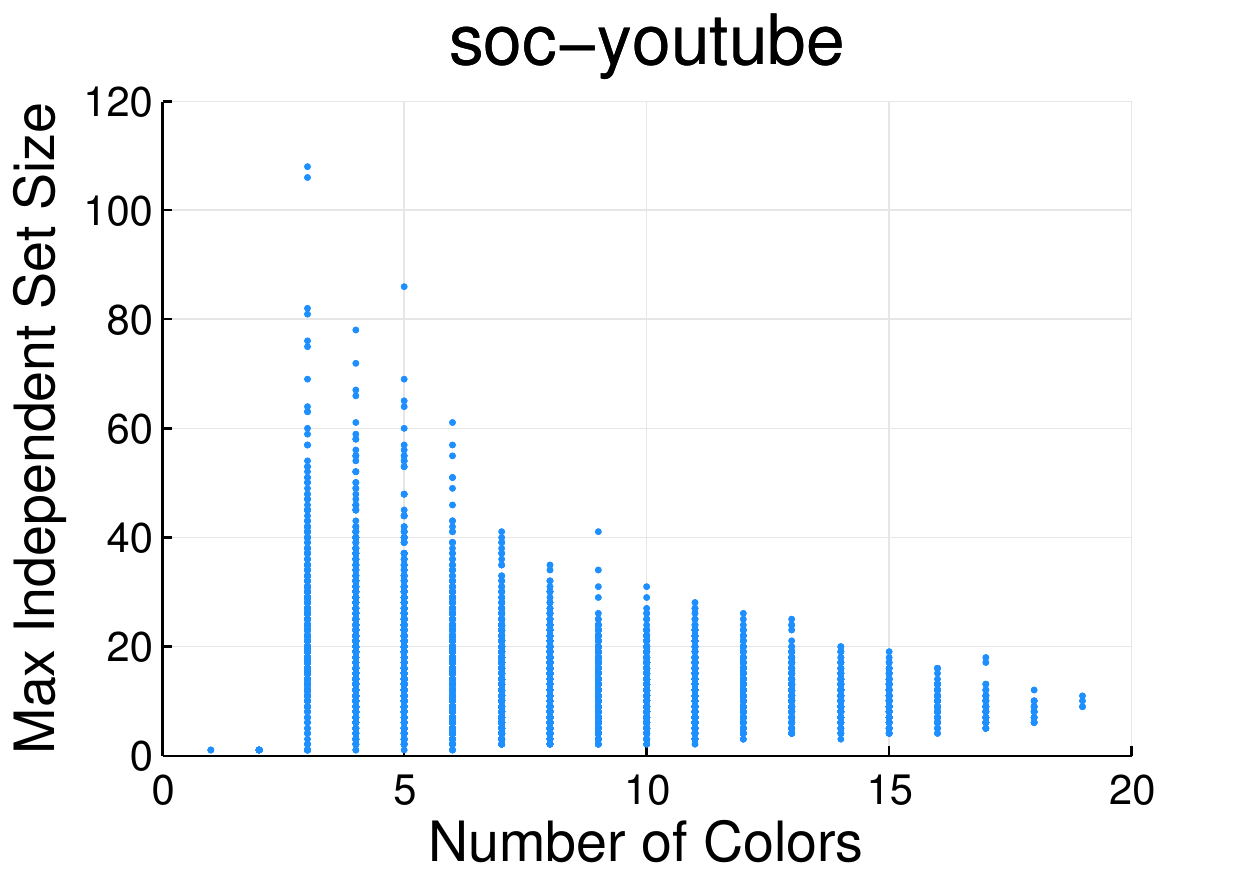}}
\hfill
\hspacehist \subfigure
{\label{fig:hist-colors-tech-skitter} \figszhist{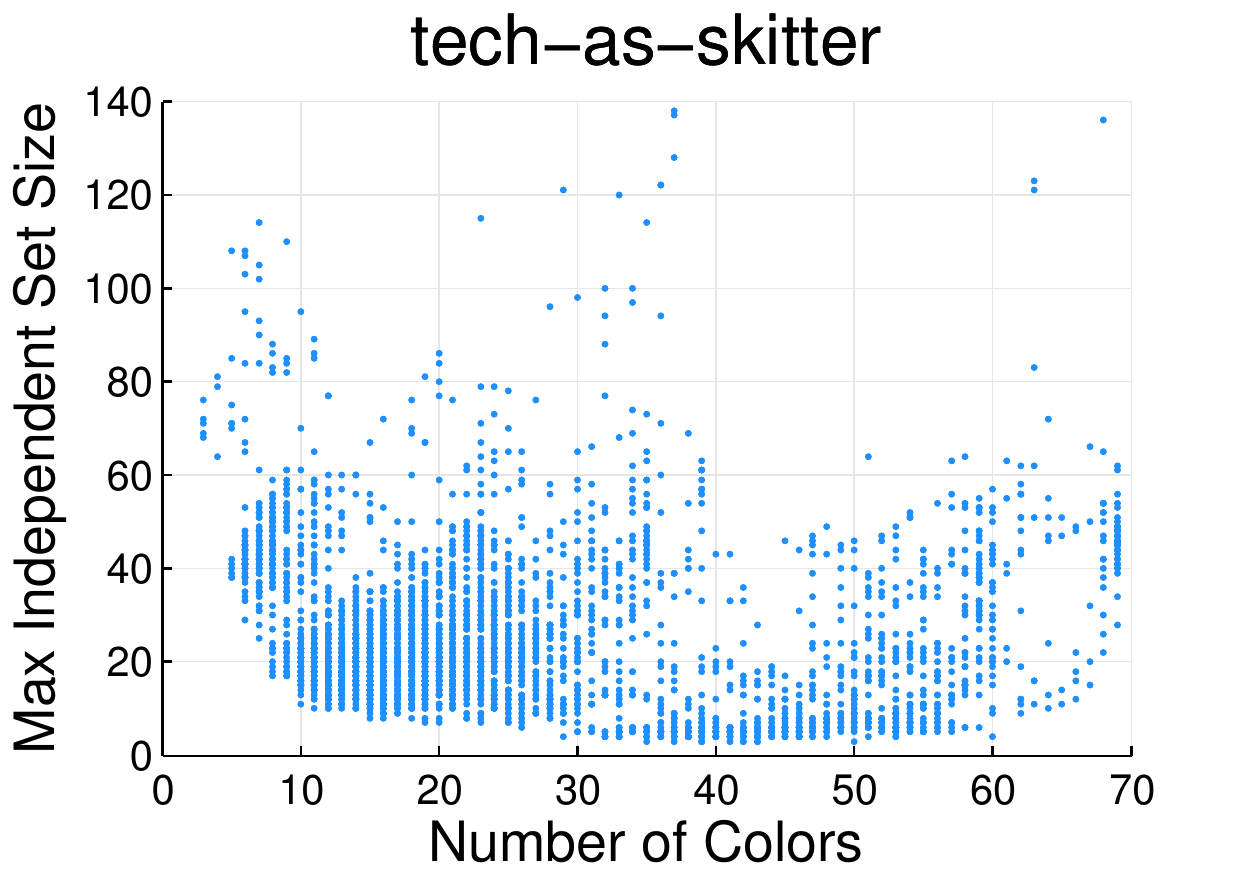}}
\hfill
\hspacehist \subfigure
{\label{fig:hist-colors-soc-orkut} \includegraphics[width=1.52in,bb=0 0 335 252]{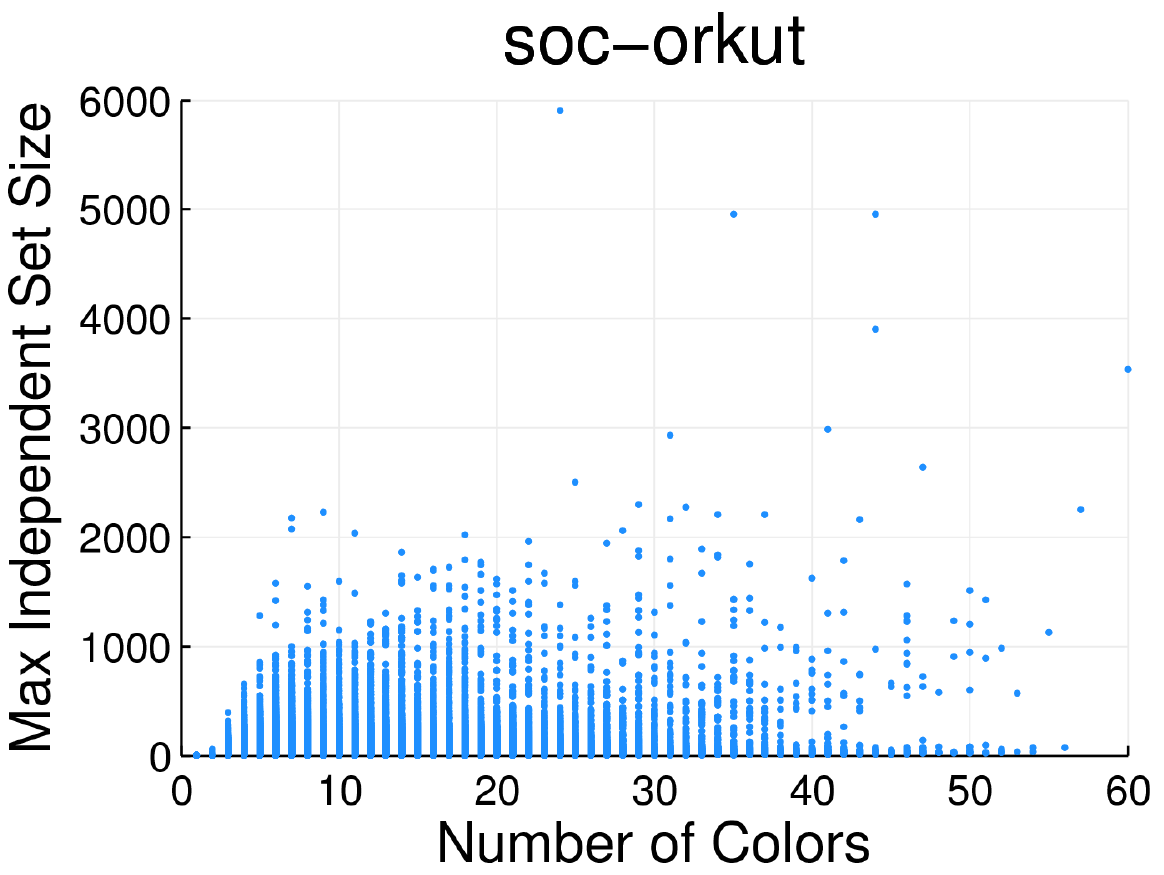}}
\caption{
\textbf{Exploring the relationship between two statistics from the neighborhood coloring.}
The number of colors used in each local coloring is compared with the size of the largest independent set from that coloring.
}
\label{fig:local-coloring-color-vs-indep}
\end{figure*}

In this section, we use neighborhood coloring to characterize the various types of networks as well as gain insight into the structural properties of the networks.
We view the neighborhood coloring as a process for discovering meaningful features that capture some underlying properties of the graph that arise from the notion of coloring.
From this, we first derive two vertex features. Specifically, for each vertex $v$, the first feature represents the number of colors used in the neighborhood coloring of the vertex $v$, and the second feature represents the size of the maximum independent set resulting from the neighborhood coloring of the vertex $v$.
Figure~\ref{fig:ccdf-local-coloring} shows the complementary cumulative distribution (CCDF) of these features across all the nodes in the graph. We observe that those graphs that are denser and more clustered (such as soc-flickr) typically use many colors for neighborhood coloring of the vertices. For example, the soc-flickr dataset uses $100$ colors to color the largest vertex neighborhood in the graph. On the other hand, graphs that are more sparse and less clustered (such as soc-youtube) typically use fewer colors for neighborhood coloring of the vertices. Further, we observe that the maximum independent set size is inversely proportional to the maximum number of neighborhood colors. Clearly, this observation is due to the rate of dependence among the graph vertices. For example, the soc-flickr dataset has a small independent set size $\approx 400$ vertices. On the other hand, a graph that is as large and as sparse as soc-orkut typically has a large independent set size $\approx 5000$ vertices. These observations show how significant the two features (number of colors and maximum independent set size) for capturing the underlying structural properties of various types of graphs. Note that in Figure~\ref{fig:ccdf-local-coloring}, we show only some of the datasets as examples, and we omit the others for brevity.

As an aside, egonet-based clique methods were proposed for sparse graphs~\cite{rossi2013parallel-cliques} and sampling methods and estimators based on egonets were developed in the same spirit~\cite{gjoka2013estimating,nkahmed:tkdd13}.
One may also straightforwardly use egonets to obtain an accurate estimate of the distribution of local coloring numbers.

In Figure~\ref{fig:hist-colors}, we focus the attention on the other denser graphs, bio-human-gene2 and keller6. Figure~\ref{fig:hist-colors} shows the histograms of the number of colors, maximum independent set size, and the correlation between them, for both bio-human-gene2 and keller6 graphs. We observe that the histograms of the number of colors and maximum independent set size are highly skewed. For example,  bio-human-gene2 graph shows that $600$ vertices uses more than $1400$ colors for their neighborhood. Moreover, the size of the maximum independent set size has a small range ($5$-$25$). The keller6 graph, however, is one of the clique DIMAC's challenge graphs. We observe that the histogram of the keller6 graph consists of two groups, one group with small number of colors ($ < 200$), and the other group with higher number ($ \approx 600$) of colors. This observation is clearly shown in the histogram of the maximum independent set size. 
Similarly, we show the correlation plots between the number of colors and the maximum independent set size for several datasets in Figure~\ref{fig:local-coloring-color-vs-indep}. The observations are similar to what we discussed before.

\section{Conclusion}
\label{sec:conclusion}
Despite the obvious practical importance of graph coloring,
existing works have not systematically investigated or designed methods for large complex networks.
In this work, we defined a unified framework that can serve as a fundamental basis for studying coloring on large networks.
Using this framework, we proposed three classes of fast and accurate methods including social-based, multi-property based, and egonet-based methods.
We demonstrated the effectiveness of the proposed methods on over 100+ networks and among 7 different types of networks (e.g., social, technological networks).
In the majority of cases, we found these methods to be more accurate than other widely used heuristics that have been used for coloring in other domains.
Importantly, we find that the solutions obtained from our methods are nearly optimal and sometimes provably optimal for certain types of networks.
Furthermore, the coloring methods were shown to be effective for the task of finding graph outliers as well as predicting the type of graph (e.g., social vs. biological network).
We also investigated the problem of coloring neighborhood subgraphs and proposed a parallel algorithm that leverages the proposed unified framework and methods.
One key finding is that neighborhoods that are colored using a relatively few number of colors are not well connected, with low clustering and a small number of triangles.
While neighborhood colorings that use a relatively large number of colors have large clustering coefficients and usually contain large cliques.
In future work, we plan to explore the neighborhood coloring further as it has proven to provide a number of key insights into the structural properties of the network and neighborhoods at large, while also fast to compute for large networks.
Overall, this work demonstrated the practical significance, accuracy, and scalability of our methods for coloring and analyzing large complex networks.

\section*{Acknowledgment}
We thank the anonymous reviewers for their constructive and helpful comments.
This material is based upon work supported by the National Science Foundation GRFP Fellowship under Grant No. DGE-1333468.

\vspace*{-1mm}
\setlength{\bibsep}{2.0pt}
\fontsize{8.0}{10.0} \selectfont
\bibliographystyle{spbasic}      
\bibliography{coloring-revised}

\pagebreak\newpage

\section*{Appendix}

\begin{table}[htb]
\caption{
\textbf{Network Statistics and Coloring Bounds for DIMACs.}
Recall $\rho$ is the density, $\bar{d}$ is the average degree, and ${\rm r}$ is the assortativity coefficient.
The global clustering coefficient is denoted by $\kappa$, $|T|$ is the total number of triangles, and $tr_{\rm avg}$ and $tr_{\rm \max}$ are the maximum and average number of triangles incident on a vertex, respectively.
The lower bound from the heuristic clique finder is denoted $\tilde{\omega}$.
For the upper bounds, we denote $K$ as the maximum k-core and similarly, we denote the  maximum triangle-core by $T$.
The maximum and minimum number of colors among all coloring methods are denoted $\coloring_{\max}$ and $\coloring_{\min}$, respectively.
}
\label{table:stats-dense-dimacs}
\centering\small\scriptsize
\fontsize{6.0}{7.0}\selectfont
\noindent
\begin{tabularx}{\linewidth}{ rr lllHl HHlHl XXXX XX}
\toprule
& & 
\multicolumn{10}{c}{\textbf{Graph measures}}& 
\multicolumn{4}{c}{\textbf{Bounds}}& 
\multicolumn{2}{c}{\textbf{Colors}}\\ 
 &
\textbf{graph}& 
$|V|$ & 
$|E|$ & 
$|T|$ & 
$\rho$ & 
$\bar{d}$ & 
${\rm r}$ & 
$\bar{\kappa}$ & 
$\kappa$ & 
$\tr_{\rm avg}$ & 
$\tr_{\rm max}$ & 
$\dmax$ & 
$K$+1 & 
$T$ & 
$\tilde{\omega}$ & 
$\coloring_{\min}$ &
$\coloring_{\max}$
\\ \midrule
	 &    \textsf{C1000-9}&  1K &  450K &  364.6M &  0.90 &  900 &  -0.00 &  0.00 &  0.90 &  364.6K &  385K &  925 &  875 &  764 &  51 &  311 &  327  \\ 
	 &    \textsf{C125-9}&  125 &  6.9K &  691.8K &  0.90 &  111 &  -0.02 &  0.00 &  0.90 &  5.5K &  6.3K &  119 &  103 &  86 &  27 &  54 &  59  \\ 
	 &    \textsf{C2000-5}&  2K &  999.8K &  499.7M &  0.50 &  999 &  -0.00 &  0.00 &  0.50 &  249.8K &  287.8K &  1K &  941 &  435 &  14 &  217 &  231  \\ 
	 &    \textsf{C2000-9}&  2K &  1.7M &  2.9T &  0.90 &  1.7K &  -0.00 &  0.00 &  0.90 &  1.4M &  1.5M &  1.8K &  1759 &  1549 &  59 &  570 &  603  \\ 
	 &    \textsf{C250-9}&  250 &  27.9K &  5.6M &  0.90 &  223 &  -0.01 &  0.00 &  0.90 &  22.4K &  24.8K &  236 &  211 &  181 &  36 &  92 &  103  \\ 
	 &    \textsf{C4000-5}&  4K &  4M &  4T &  0.50 &  2K &  -0.00 &  0.00 &  0.50 &  1M &  1.1M &  2.1K &  1910 &  899 &  15 &  391 &  408  \\ 
	 &    \textsf{C500-9}&  500 &  112.3K &  45.3M &  0.90 &  449 &  -0.00 &  0.00 &  0.90 &  90.7K &  98.4K &  468 &  433 &  373 &  44 &  168 &  183  \\ 
\midrule 
	 &    \textsf{DSJC1000-5}&  1K &  249.8K &  62.3M &  0.50 &  499 &  -0.00 &  0.00 &  0.50 &  62.3K &  75.9K &  551 &  460 &  207 &  13 &  120 &  130  \\ 
	 &    \textsf{DSJC500-5}&  500 &  62.6K &  7.8M &  0.50 &  250 &  -0.01 &  0.00 &  0.50 &  15.7K &  20.5K &  286 &  226 &  99 &  11 &  67 &  75  \\ 
\midrule 
	 &    \textsf{MANN-a27}&  378 &  70.5K &  26M &  0.99 &  373 &  0.00 &  0.00 &  0.99 &  68.8K &  69K &  374 &  365 &  352 &  125 &  138 &  144  \\ 
	 &    \textsf{MANN-a45}&  1K &  533.1K &  546.6M &  1.00 &  1K &  0.00 &  0.00 &  1.00 &  528.1K &  529K &  1K &  1013 &  991 &  341 &  367 &  375  \\ 
	 &    \textsf{MANN-a81}&  3.3K &  5.5M &  18.2T &  1.00 &  3.3K &  0.00 &  0.00 &  1.00 &  5.4M &  5.4M &  3.3K &  3281 &  3241 &  1096 &  1134 &  1161  \\ 
	 &    \textsf{MANN-a9}&  45 &  918 &  33.7K &  0.93 &  40 &  0.00 &  0.00 &  0.92 &  750 &  757 &  41 &  41 &  37 &  16 &  19 &  21  \\ 
\midrule 
	 &    \textsf{brock200-1}&  200 &  14.8K &  1.6M &  0.75 &  148 &  -0.00 &  0.00 &  0.75 &  8.1K &  10.1K &  165 &  135 &  91 &  17 &  54 &  59  \\ 
	 &    \textsf{brock200-2}&  200 &  9.8K &  479.6K &  0.50 &  98 &  -0.01 &  0.00 &  0.49 &  2.3K &  3.1K &  114 &  85 &  35 &  9 &  33 &  37  \\ 
	 &    \textsf{brock200-3}&  200 &  12K &  873.3K &  0.61 &  120 &  -0.01 &  0.00 &  0.61 &  4.3K &  5.4K &  134 &  106 &  56 &  12 &  41 &  46  \\ 
	 &    \textsf{brock200-4}&  200 &  13K &  1.1M &  0.66 &  130 &  -0.02 &  0.00 &  0.66 &  5.6K &  7K &  147 &  118 &  68 &  14 &  44 &  52  \\ 
	 &    \textsf{brock400-1}&  400 &  59.7K &  13.3M &  0.75 &  298 &  -0.00 &  0.00 &  0.75 &  33.2K &  38.1K &  320 &  278 &  192 &  20 &  96 &  107  \\ 
	 &    \textsf{brock400-2}&  400 &  59.7K &  13.3M &  0.75 &  298 &  -0.00 &  0.00 &  0.75 &  33.3K &  40.1K &  328 &  279 &  193 &  20 &  97 &  104  \\ 
	 &    \textsf{brock400-3}&  400 &  59.6K &  13.2M &  0.75 &  298 &  -0.01 &  0.00 &  0.75 &  33.2K &  38.5K &  322 &  279 &  192 &  20 &  95 &  105  \\ 
	 &    \textsf{brock400-4}&  400 &  59.7K &  13.3M &  0.75 &  298 &  -0.01 &  0.00 &  0.75 &  33.3K &  39.6K &  326 &  278 &  193 &  22 &  95 &  107  \\ 
	 &    \textsf{brock800-1}&  800 &  207.5K &  69.7M &  0.65 &  518 &  -0.00 &  0.00 &  0.65 &  87.2K &  101.4K &  560 &  488 &  292 &  17 &  139 &  149  \\ 
	 &    \textsf{brock800-2}&  800 &  208.1K &  70.4M &  0.65 &  520 &  -0.00 &  0.00 &  0.65 &  88K &  104K &  566 &  487 &  292 &  18 &  139 &  150  \\ 
	 &    \textsf{brock800-3}&  800 &  207.3K &  69.6M &  0.65 &  518 &  -0.00 &  0.00 &  0.65 &  87K &  100.8K &  558 &  484 &  289 &  17 &  138 &  148  \\ 
	 &    \textsf{brock800-4}&  800 &  207.6K &  69.9M &  0.65 &  519 &  -0.00 &  0.00 &  0.65 &  87.4K &  103.2K &  565 &  486 &  291 &  17 &  141 &  146  \\ 
\midrule 
	 &    \textsf{c-fat200-1}&  200 &  1.5K &  16.2K &  0.08 &  15 &  0.97 &  0.00 &  0.73 &  81 &  100 &  17 &  15 &  12 &  12 &  13 &  16  \\ 
	 &   \cellcolor{verylightred}\textsf{c-fat200-2}& \cellcolor{verylightred}200 & \cellcolor{verylightred}3.2K & \cellcolor{verylightred}76.7K & \cellcolor{verylightred}0.16 & \cellcolor{verylightred}32 & \cellcolor{verylightred}0.84 & \cellcolor{verylightred}0.00 & \cellcolor{verylightred}0.76 & \cellcolor{verylightred}384 & \cellcolor{verylightred}429 & \cellcolor{verylightred}34 & \cellcolor{verylightred}33 & \cellcolor{verylightred}24 & \cellcolor{verylightred}24 & \cellcolor{verylightred}24 & \cellcolor{verylightred}28  \\ 
	 &    \textsf{c-fat200-5}&  200 &  8.4K &  546.2K &  0.43 &  84 &  0.73 &  0.00 &  0.77 &  2.7K &  2.8K &  86 &  84 &  58 &  58 &  69 &  85  \\ 
	 &   \cellcolor{verylightred}\textsf{c-fat500-1}& \cellcolor{verylightred}500 & \cellcolor{verylightred}4.4K & \cellcolor{verylightred}55.7K & \cellcolor{verylightred}0.04 & \cellcolor{verylightred}17 & \cellcolor{verylightred}0.98 & \cellcolor{verylightred}0.00 & \cellcolor{verylightred}0.74 & \cellcolor{verylightred}111 & \cellcolor{verylightred}141 & \cellcolor{verylightred}20 & \cellcolor{verylightred}18 & \cellcolor{verylightred}14 & \cellcolor{verylightred}14 & \cellcolor{verylightred}14 & \cellcolor{verylightred}18  \\ 
	 &   \cellcolor{verylightred}\textsf{c-fat500-10}& \cellcolor{verylightred}500 & \cellcolor{verylightred}46.6K & \cellcolor{verylightred}6.6M & \cellcolor{verylightred}0.37 & \cellcolor{verylightred}186 & \cellcolor{verylightred}0.80 & \cellcolor{verylightred}0.00 & \cellcolor{verylightred}0.77 & \cellcolor{verylightred}13.3K & \cellcolor{verylightred}13.6K & \cellcolor{verylightred}188 & \cellcolor{verylightred}186 & \cellcolor{verylightred}126 & \cellcolor{verylightred}126 & \cellcolor{verylightred}126 & \cellcolor{verylightred}188  \\ 
	 &   \cellcolor{verylightred}\textsf{c-fat500-2}& \cellcolor{verylightred}500 & \cellcolor{verylightred}9.1K & \cellcolor{verylightred}247K & \cellcolor{verylightred}0.07 & \cellcolor{verylightred}36 & \cellcolor{verylightred}0.97 & \cellcolor{verylightred}0.00 & \cellcolor{verylightred}0.76 & \cellcolor{verylightred}494 & \cellcolor{verylightred}534 & \cellcolor{verylightred}38 & \cellcolor{verylightred}36 & \cellcolor{verylightred}26 & \cellcolor{verylightred}26 & \cellcolor{verylightred}26 & \cellcolor{verylightred}33  \\ 
	 &   \cellcolor{verylightred}\textsf{c-fat500-5}& \cellcolor{verylightred}500 & \cellcolor{verylightred}23.1K & \cellcolor{verylightred}1.6M & \cellcolor{verylightred}0.19 & \cellcolor{verylightred}92 & \cellcolor{verylightred}0.89 & \cellcolor{verylightred}0.00 & \cellcolor{verylightred}0.77 & \cellcolor{verylightred}3.2K & \cellcolor{verylightred}3.4K & \cellcolor{verylightred}95 & \cellcolor{verylightred}93 & \cellcolor{verylightred}64 & \cellcolor{verylightred}64 & \cellcolor{verylightred}64 & \cellcolor{verylightred}79  \\ 
\midrule 
	 &    \textsf{gen200-p0-9-44}&  200 &  17.9K &  2.8M &  0.90 &  179 &  -0.01 &  0.00 &  0.90 &  14.3K &  16.1K &  190 &  168 &  141 &  34 &  65 &  81  \\ 
	 &    \textsf{gen200-p0-9-55}&  200 &  17.9K &  2.8M &  0.90 &  179 &  -0.01 &  0.00 &  0.90 &  14.3K &  16.1K &  190 &  167 &  142 &  34 &  72 &  83  \\ 
	 &    \textsf{gen400-p0-9-55}&  400 &  71.8K &  23.1M &  0.90 &  359 &  -0.01 &  0.00 &  0.90 &  57.9K &  63.1K &  375 &  337 &  287 &  41 &  120 &  147  \\ 
	 &    \textsf{gen400-p0-9-65}&  400 &  71.8K &  23.1M &  0.90 &  359 &  -0.02 &  0.00 &  0.90 &  57.9K &  64.1K &  378 &  337 &  286 &  41 &  129 &  150  \\ 
	 &    \textsf{gen400-p0-9-75}&  400 &  71.8K &  23.1M &  0.90 &  359 &  -0.01 &  0.00 &  0.90 &  57.9K &  64.8K &  380 &  337 &  287 &  45 &  125 &  154  \\ 
\midrule 
	 &   \cellcolor{verylightyellow}\textsf{hamming10-2}& \cellcolor{verylightyellow}1K & \cellcolor{verylightyellow}518.6K & \cellcolor{verylightyellow}519.7M & \cellcolor{verylightyellow}0.99 & \cellcolor{verylightyellow}1K & \cellcolor{verylightyellow}1.00 & \cellcolor{verylightyellow}0.00 & \cellcolor{verylightyellow}0.99 & \cellcolor{verylightyellow}507.5K & \cellcolor{verylightyellow}507.5K & \cellcolor{verylightyellow}1K & \cellcolor{verylightyellow}1014 & \cellcolor{verylightyellow}1004 & \cellcolor{verylightyellow}512 & \cellcolor{verylightyellow}512 & \cellcolor{verylightyellow}540  \\ 
	 &    \textsf{hamming10-4}&  1K &  434.1K &  301.8M &  0.83 &  848 &  0.00 &  0.00 &  0.82 &  294.7K &  294.7K &  848 &  849 &  674 &  32 &  105 &  128  \\ 
	 &   \cellcolor{verylightyellow}\textsf{hamming6-2}& \cellcolor{verylightyellow}64 & \cellcolor{verylightyellow}1.8K & \cellcolor{verylightyellow}92.1K & \cellcolor{verylightyellow}0.90 & \cellcolor{verylightyellow}57 & \cellcolor{verylightyellow}0.00 & \cellcolor{verylightyellow}0.00 & \cellcolor{verylightyellow}0.90 & \cellcolor{verylightyellow}1.4K & \cellcolor{verylightyellow}1.4K & \cellcolor{verylightyellow}57 & \cellcolor{verylightyellow}58 & \cellcolor{verylightyellow}52 & \cellcolor{verylightyellow}32 & \cellcolor{verylightyellow}32 & \cellcolor{verylightyellow}35  \\ 
	 &    \textsf{hamming6-4}&  64 &  704 &  2.8K &  0.35 &  22 &  0.00 &  0.00 &  0.19 &  45 &  45 &  22 &  23 &  8 &  4 &  8 &  9  \\ 
	 &   \cellcolor{verylightyellow}\textsf{hamming8-2}& \cellcolor{verylightyellow}256 & \cellcolor{verylightyellow}31.6K & \cellcolor{verylightyellow}7.5M & \cellcolor{verylightyellow}0.97 & \cellcolor{verylightyellow}247 & \cellcolor{verylightyellow}0.00 & \cellcolor{verylightyellow}0.00 & \cellcolor{verylightyellow}0.97 & \cellcolor{verylightyellow}29.4K & \cellcolor{verylightyellow}29.4K & \cellcolor{verylightyellow}247 & \cellcolor{verylightyellow}248 & \cellcolor{verylightyellow}240 & \cellcolor{verylightyellow}128 & \cellcolor{verylightyellow}128 & \cellcolor{verylightyellow}138  \\ 
	 &    \textsf{hamming8-4}&  256 &  20.8K &  2M &  0.64 &  163 &  0.00 &  0.00 &  0.60 &  7.8K &  7.8K &  163 &  164 &  82 &  16 &  29 &  33  \\ 
\midrule 
\end{tabularx}
\end{table}

\begin{table}[htb]
\caption{
\textbf{Statistics and Bounds for DIMACs (cont. from Table~\ref{table:stats-dense-dimacs}).}
Recall $\rho$ is the density, $\bar{d}$ is the average degree, and ${\rm r}$ is the assortativity coefficient.
The global clustering coefficient is denoted by $\kappa$, $|T|$ is the total number of triangles, and $tr_{\rm avg}$ and $tr_{\rm \max}$ are the maximum and average number of triangles incident on a vertex, respectively.
The lower bound from the heuristic clique finder is denoted $\tilde{\omega}$.
For the upper bounds, we denote $K$ as the maximum k-core and similarly, we denote the  maximum triangle-core by $T$.
The maximum and minimum number of colors among all coloring methods are denoted $\coloring_{\max}$ and $\coloring_{\min}$, respectively.
}
\vspace{1mm}
\label{table:stats-dense-bhoslib}
\centering\small\scriptsize
\fontsize{6.0}{7.0}\selectfont
\noindent
\begin{tabularx}{\linewidth}{ rr lllHl HHlHl XXXX XX}
\toprule
& & 
\multicolumn{10}{c}{\textbf{Graph measures}}& 
\multicolumn{4}{c}{\textbf{Bounds}}& 
\multicolumn{2}{c}{\textbf{Colors}}\\ 
 &
\textbf{graph}& 
$|V|$ & 
$|E|$ & 
$|T|$ & 
$\rho$ & 
$\bar{d}$ & 
${\rm r}$ & 
$\bar{\kappa}$ & 
$\kappa$ & 
$\tr_{\rm avg}$ & 
$\tr_{\rm max}$ & 
$\dmax$ & 
$K$+1 & 
$T$ & 
$\tilde{\omega}$ & 
$\coloring_{\min}$ &
$\coloring_{\max}$
\\ \midrule
	 &    \textsf{johnson16-2-4}&  120 &  5.4K &  360.3K &  0.76 &  91 &  0.00 &  0.00 &  0.73 &  3K &  3K &  91 &  92 &  68 &  8 &  14 &  15  \\ 
	 &    \textsf{johnson32-2-4}&  496 &  107.8K &  40.7M &  0.88 &  435 &  0.00 &  0.00 &  0.87 &  82.2K &  82.2K &  435 &  436 &  380 &  16 &  30 &  34  \\ 
	 &    \textsf{johnson8-2-4}&  28 &  210 &  1.2K &  0.56 &  15 &  1.00 &  0.00 &  0.43 &  45 &  45 &  15 &  16 &  8 &  4 &  6 &  7  \\ 
	 &    \textsf{johnson8-4-4}&  70 &  1.8K &  71.8K &  0.77 &  53 &  1.00 &  0.00 &  0.74 &  1K &  1K &  53 &  54 &  38 &  14 &  19 &  22  \\ 
\midrule 
	 &    \textsf{keller4}&  171 &  9.4K &  649.7K &  0.65 &  110 &  -0.07 &  0.00 &  0.63 &  3.8K &  4.7K &  124 &  103 &  54 &  9 &  24 &  37  \\ 
	 &    \textsf{keller5}&  776 &  225.9K &  98M &  0.75 &  582 &  -0.03 &  0.00 &  0.75 &  126.3K &  151.2K &  638 &  561 &  379 &  22 &  61 &  176  \\ 
	 &    \textsf{keller6}&  3.3K &  4.6M &  10.3T &  0.82 &  2.7K &  -0.02 &  0.00 &  0.82 &  3M &  3.5M &  2.9K &  2691 &  2084 &  45 &  148 &  783  \\ 
\midrule 
	 &    \textsf{ph1000-1}&  1K &  122.2K &  9.2M &  0.24 &  244 &  -0.09 &  0.00 &  0.28 &  9.2K &  22.6K &  408 &  164 &  47 &  9 &  55 &  78  \\ 
	 &    \textsf{ph1000-2}&  1K &  244.7K &  73.9M &  0.49 &  489 &  -0.10 &  0.00 &  0.57 &  73.9K &  157.6K &  766 &  328 &  196 &  33 &  116 &  181  \\ 
	 &    \textsf{ph1000-3}&  1K &  371.7K &  211.3M &  0.74 &  743 &  -0.01 &  0.00 &  0.76 &  211.3K &  301.7K &  895 &  610 &  388 &  49 &  194 &  269  \\ 
	 &    \textsf{ph1500-1}&  1.5K &  284.9K &  34.3M &  0.25 &  379 &  -0.09 &  0.00 &  0.29 &  22.8K &  53.2K &  614 &  253 &  75 &  10 &  78 &  110  \\ 
	 &    \textsf{ph1500-2}&  1.5K &  568.9K &  273.5M &  0.51 &  758 &  -0.09 &  0.00 &  0.58 &  182.3K &  372.3K &  1.1K &  505 &  314 &  44 &  167 &  266  \\ 
	 &    \textsf{ph1500-3}&  1.5K &  847.2K &  741.1M &  0.75 &  1.1K &  -0.01 &  0.00 &  0.77 &  494.1K &  675.7K &  1.3K &  930 &  597 &  60 &  281 &  388  \\ 
	 &    \textsf{ph300-3}&  300 &  33.3K &  5.6M &  0.74 &  222 &  -0.02 &  0.00 &  0.76 &  18.8K &  26.7K &  267 &  181 &  118 &  26 &  73 &  94  \\ 
	 &    \textsf{ph500-1}&  500 &  31.5K &  1.2M &  0.25 &  126 &  -0.09 &  0.00 &  0.29 &  2.5K &  5.7K &  204 &  87 &  25 &  9 &  34 &  48  \\ 
	 &    \textsf{ph500-2}&  500 &  62.9K &  9.9M &  0.50 &  251 &  -0.10 &  0.00 &  0.58 &  19.9K &  40.8K &  389 &  171 &  102 &  32 &  68 &  104  \\ 
	 &    \textsf{ph500-3}&  500 &  93.8K &  27.1M &  0.75 &  375 &  -0.01 &  0.00 &  0.77 &  54.3K &  77.2K &  452 &  304 &  197 &  39 &  111 &  150  \\ 
	 &    \textsf{ph700-1}&  700 &  60.9K &  3.3M &  0.25 &  174 &  -0.10 &  0.00 &  0.29 &  4.7K &  11.5K &  286 &  118 &  34 &  8 &  42 &  60  \\ 
	 &    \textsf{ph700-2}&  700 &  121.7K &  26.6M &  0.50 &  347 &  -0.10 &  0.00 &  0.58 &  38K &  79.2K &  539 &  236 &  143 &  26 &  91 &  141  \\ 
	 &    \textsf{ph700-3}&  700 &  183K &  73.6M &  0.75 &  522 &  -0.01 &  0.00 &  0.76 &  105.1K &  147.9K &  627 &  427 &  273 &  40 &  145 &  201  \\ 
\midrule 
	 &    \textsf{s1000}&  1K &  250.5K &  86.3M &  0.50 &  501 &  0.67 &  0.00 &  0.69 &  86.3K &  100.7K &  550 &  465 &  399 &  10 &  15 &  46  \\ 
	 &    \textsf{s200-0-7-1}&  200 &  13.9K &  1.4M &  0.70 &  139 &  0.07 &  0.00 &  0.73 &  7K &  8.5K &  155 &  126 &  93 &  16 &  35 &  52  \\ 
	 &    \textsf{s200-0-7-2}&  200 &  13.9K &  1.4M &  0.70 &  139 &  0.26 &  0.00 &  0.74 &  7.2K &  9.7K &  164 &  123 &  112 &  14 &  18 &  40  \\ 
	 &    \textsf{s200-0-9-1}&  200 &  17.9K &  2.8M &  0.90 &  179 &  0.04 &  0.00 &  0.90 &  14.4K &  16.3K &  191 &  163 &  134 &  49 &  71 &  97  \\ 
	 &    \textsf{s200-0-9-2}&  200 &  17.9K &  2.8M &  0.90 &  179 &  -0.00 &  0.00 &  0.90 &  14.3K &  15.8K &  188 &  170 &  143 &  34 &  76 &  89  \\ 
	 &    \textsf{s200-0-9-3}&  200 &  17.9K &  2.8M &  0.90 &  179 &  -0.01 &  0.00 &  0.90 &  14.3K &  15.6K &  187 &  170 &  145 &  31 &  69 &  80  \\ 
	 &    \textsf{s400-0-5-1}&  400 &  39.9K &  5.2M &  0.50 &  199 &  0.60 &  0.00 &  0.66 &  13K &  15.8K &  225 &  184 &  154 &  8 &  13 &  29  \\ 
	 &    \textsf{s400-0-7-1}&  400 &  55.8K &  11.3M &  0.70 &  279 &  -0.00 &  0.00 &  0.73 &  28.4K &  32.3K &  301 &  262 &  182 &  22 &  71 &  82  \\ 
	 &    \textsf{s400-0-7-2}&  400 &  55.8K &  11.2M &  0.70 &  279 &  0.03 &  0.00 &  0.73 &  28.2K &  32.8K &  304 &  260 &  179 &  18 &  51 &  71  \\ 
	 &    \textsf{s400-0-7-3}&  400 &  55.8K &  11.1M &  0.70 &  279 &  0.18 &  0.00 &  0.72 &  27.9K &  33.6K &  307 &  254 &  182 &  16 &  22 &  63  \\ 
	 &    \textsf{s400-0-9-1}&  400 &  71.8K &  23.1M &  0.90 &  359 &  -0.00 &  0.00 &  0.90 &  57.9K &  62.7K &  374 &  345 &  294 &  57 &  151 &  168  \\ 
	 &    \textsf{sr200-0-7}&  200 &  13.8K &  1.3M &  0.70 &  138 &  -0.02 &  0.00 &  0.70 &  6.6K &  8.8K &  161 &  125 &  78 &  16 &  48 &  55  \\ 
	 &    \textsf{sr200-0-9}&  200 &  17.8K &  2.8M &  0.90 &  178 &  -0.01 &  0.00 &  0.90 &  14.2K &  15.9K &  189 &  167 &  141 &  34 &  76 &  85  \\ 
	 &    \textsf{sr400-0-5}&  400 &  39.9K &  3.9M &  0.50 &  199 &  -0.01 &  0.00 &  0.50 &  9.9K &  13.5K &  233 &  178 &  77 &  10 &  56 &  64  \\ 
	 &    \textsf{sr400-0-7}&  400 &  55.8K &  10.9M &  0.70 &  279 &  -0.00 &  0.00 &  0.70 &  27.2K &  33.5K &  310 &  259 &  164 &  17 &  86 &  94  \\ 
\midrule 
\end{tabularx}
\end{table}

\begin{table}[htb]
\caption{
\textbf{Statistics and Bounds for BHOSLIB.}
Recall $\rho$ is the density, $\bar{d}$ is the average degree, and ${\rm r}$ is the assortativity coefficient.
The global clustering coefficient is denoted by $\kappa$, $|T|$ is the total number of triangles, and $tr_{\rm avg}$ and $tr_{\rm \max}$ are the maximum and average number of triangles incident on a vertex, respectively.
The lower bound from the heuristic clique finder is denoted $\tilde{\omega}$.
For the upper bounds, we denote $K$ as the maximum k-core and similarly, we denote the  maximum triangle-core by $T$.
The maximum and minimum number of colors among all coloring methods are denoted $\coloring_{\max}$ and $\coloring_{\min}$, respectively.
}
\vspace{1mm}
\label{table:stats-dense-bhoslib}
\centering\small\scriptsize
\fontsize{6.0}{7.0}\selectfont
\noindent
\begin{tabularx}{\linewidth}{ rr lllHl HHlHl XXXX XX}
\toprule
& & 
\multicolumn{10}{c}{\textbf{Graph measures}}& 
\multicolumn{4}{c}{\textbf{Bounds}}& 
\multicolumn{2}{c}{\textbf{Colors}}\\ 
 &
\textbf{graph}& 
$|V|$ & 
$|E|$ & 
$|T|$ & 
$\rho$ & 
$\bar{d}$ & 
${\rm r}$ & 
$\bar{\kappa}$ & 
$\kappa$ & 
$\tr_{\rm avg}$ & 
$\tr_{\rm max}$ & 
$\dmax$ & 
$K$+1 & 
$T$ & 
$\tilde{\omega}$ & 
$\coloring_{\min}$ &
$\coloring_{\max}$
\\ \midrule
	 &    \textsf{frb100-40}&  4K &  7.4M &  25.5T &  0.93 &  3.7K &  -0.01 &  0.00 &  0.93 &  6.3M &  6.9M &  3.8K &  3572 &  3468 &  78 &  106 &  558  \\ 
	 &    \textsf{frb30-15-1}&  450 &  83.1K &  25.2M &  0.82 &  369 &  -0.03 &  0.00 &  0.82 &  56K &  67.4K &  407 &  340 &  257 &  25 &  41 &  90  \\ 
	 &    \textsf{frb30-15-2}&  450 &  83.1K &  25.1M &  0.82 &  369 &  -0.02 &  0.00 &  0.82 &  55.9K &  66.7K &  404 &  338 &  257 &  24 &  36 &  93  \\ 
	 &    \textsf{frb30-15-3}&  450 &  83.2K &  25.2M &  0.82 &  369 &  -0.03 &  0.00 &  0.82 &  56.1K &  65.2K &  400 &  337 &  254 &  25 &  38 &  95  \\ 
	 &    \textsf{frb30-15-4}&  450 &  83.1K &  25.2M &  0.82 &  369 &  -0.03 &  0.00 &  0.82 &  56K &  65.7K &  401 &  340 &  255 &  25 &  36 &  94  \\ 
	 &    \textsf{frb30-15-5}&  450 &  83.2K &  25.2M &  0.82 &  369 &  -0.05 &  0.00 &  0.82 &  56.1K &  66.4K &  403 &  333 &  254 &  24 &  34 &  87  \\ 
	 &    \textsf{frb35-17-1}&  595 &  148.8K &  62.5M &  0.84 &  500 &  -0.04 &  0.00 &  0.84 &  105.1K &  123.9K &  544 &  463 &  361 &  29 &  41 &  118  \\ 
	 &    \textsf{frb35-17-2}&  595 &  148.8K &  62.5M &  0.84 &  500 &  -0.03 &  0.00 &  0.84 &  105.1K &  122.2K &  541 &  465 &  362 &  28 &  41 &  113  \\ 
	 &    \textsf{frb35-17-3}&  595 &  148.7K &  62.5M &  0.84 &  500 &  -0.04 &  0.00 &  0.84 &  105.1K &  126.2K &  549 &  451 &  352 &  28 &  38 &  113  \\ 
	 &    \textsf{frb35-17-4}&  595 &  148.8K &  62.6M &  0.84 &  500 &  -0.04 &  0.00 &  0.84 &  105.3K &  131.1K &  560 &  456 &  354 &  30 &  41 &  118  \\ 
	 &    \textsf{frb35-17-5}&  595 &  148.5K &  62.2M &  0.84 &  499 &  -0.02 &  0.00 &  0.84 &  104.5K &  126.3K &  550 &  461 &  355 &  29 &  45 &  115  \\ 
	 &    \textsf{frb40-19-1}&  760 &  247.1K &  137.5M &  0.86 &  650 &  -0.03 &  0.00 &  0.86 &  180.9K &  210.6K &  703 &  595 &  465 &  32 &  45 &  136  \\ 
	 &    \textsf{frb40-19-2}&  760 &  247.1K &  137.5M &  0.86 &  650 &  -0.03 &  0.00 &  0.86 &  180.9K &  209.9K &  702 &  598 &  477 &  33 &  45 &  145  \\ 
	 &    \textsf{frb40-19-3}&  760 &  247.3K &  137.6M &  0.86 &  650 &  -0.02 &  0.00 &  0.86 &  181.1K &  210.2K &  702 &  613 &  491 &  31 &  41 &  141  \\ 
	 &    \textsf{frb40-19-4}&  760 &  246.8K &  136.8M &  0.86 &  649 &  -0.03 &  0.00 &  0.85 &  180K &  203.9K &  692 &  600 &  481 &  32 &  50 &  144  \\ 
	 &    \textsf{frb40-19-5}&  760 &  246.8K &  136.8M &  0.86 &  649 &  -0.03 &  0.00 &  0.85 &  180K &  203.6K &  691 &  596 &  476 &  32 &  43 &  143  \\ 
	 &    \textsf{frb45-21-1}&  945 &  386.8K &  274.2M &  0.87 &  818 &  -0.02 &  0.00 &  0.87 &  290.1K &  331.5K &  876 &  769 &  626 &  36 &  51 &  187  \\ 
	 &    \textsf{frb45-21-2}&  945 &  387.4K &  275.3M &  0.87 &  819 &  -0.03 &  0.00 &  0.87 &  291.3K &  326.9K &  870 &  769 &  625 &  34 &  48 &  174  \\ 
	 &    \textsf{frb45-21-3}&  945 &  387.7K &  276.2M &  0.87 &  820 &  -0.02 &  0.00 &  0.87 &  292.2K &  329.6K &  872 &  764 &  624 &  35 &  48 &  182  \\ 
	 &    \textsf{frb45-21-4}&  945 &  387.4K &  275.7M &  0.87 &  820 &  -0.03 &  0.00 &  0.87 &  291.7K &  331.2K &  875 &  757 &  618 &  35 &  50 &  170  \\ 
	 &    \textsf{frb45-21-5}&  945 &  387.4K &  275.4M &  0.87 &  820 &  -0.03 &  0.00 &  0.87 &  291.4K &  330.5K &  874 &  771 &  629 &  37 &  49 &  175  \\ 
	 &    \textsf{frb50-23-1}&  1.1K &  580.6K &  514.4M &  0.88 &  1K &  -0.02 &  0.00 &  0.88 &  447.3K &  496.1K &  1K &  950 &  786 &  39 &  52 &  202  \\ 
	 &    \textsf{frb50-23-2}&  1.1K &  579.8K &  512.4M &  0.88 &  1K &  -0.02 &  0.00 &  0.88 &  445.6K &  504.6K &  1K &  936 &  771 &  39 &  53 &  204  \\ 
	 &    \textsf{frb50-23-3}&  1.1K &  579.6K &  511.5M &  0.88 &  1K &  -0.02 &  0.00 &  0.88 &  444.8K &  508K &  1K &  953 &  792 &  40 &  55 &  210  \\ 
	 &    \textsf{frb50-23-4}&  1.1K &  580.4K &  513.9M &  0.88 &  1K &  -0.03 &  0.00 &  0.88 &  446.9K &  502.6K &  1K &  950 &  786 &  40 &  58 &  196  \\ 
	 &    \textsf{frb50-23-5}&  1.1K &  580.6K &  514.6M &  0.88 &  1K &  -0.03 &  0.00 &  0.88 &  447.5K &  510.8K &  1K &  949 &  790 &  41 &  56 &  200  \\ 
	 &    \textsf{frb53-24-1}&  1.2K &  714.1K &  707.5M &  0.88 &  1.1K &  -0.01 &  0.00 &  0.88 &  556.2K &  619.3K &  1.1K &  1054 &  881 &  43 &  58 &  220  \\ 
	 &    \textsf{frb53-24-2}&  1.2K &  714K &  707.1M &  0.88 &  1.1K &  -0.02 &  0.00 &  0.88 &  555.9K &  615.3K &  1.1K &  1057 &  884 &  43 &  57 &  228  \\ 
	 &    \textsf{frb53-24-3}&  1.2K &  714.2K &  707.7M &  0.88 &  1.1K &  -0.02 &  0.00 &  0.88 &  556.3K &  616.5K &  1.1K &  1050 &  878 &  42 &  61 &  216  \\ 
	 &    \textsf{frb53-24-4}&  1.2K &  714K &  707M &  0.88 &  1.1K &  -0.02 &  0.00 &  0.88 &  555.8K &  622.1K &  1.1K &  1063 &  890 &  43 &  59 &  217  \\ 
	 &    \textsf{frb53-24-5}&  1.2K &  714.1K &  707.2M &  0.88 &  1.1K &  -0.02 &  0.00 &  0.88 &  555.9K &  633K &  1.1K &  1071 &  900 &  42 &  58 &  217  \\ 
	 &    \textsf{frb59-26-1}&  1.5K &  1M &  1.2T &  0.89 &  1.3K &  -0.01 &  0.00 &  0.89 &  834.1K &  921K &  1.4K &  1282 &  1084 &  48 &  63 &  255  \\ 
	 &    \textsf{frb59-26-2}&  1.5K &  1M &  1.2T &  0.89 &  1.3K &  -0.02 &  0.00 &  0.89 &  835K &  914.6K &  1.4K &  1285 &  1086 &  46 &  61 &  247  \\ 
	 &    \textsf{frb59-26-3}&  1.5K &  1M &  1.2T &  0.89 &  1.3K &  -0.02 &  0.00 &  0.89 &  835.2K &  942.7K &  1.4K &  1296 &  1098 &  45 &  64 &  256  \\ 
	 &    \textsf{frb59-26-4}&  1.5K &  1M &  1.2T &  0.89 &  1.3K &  -0.01 &  0.00 &  0.89 &  833.1K &  916.6K &  1.4K &  1284 &  1085 &  48 &  61 &  259  \\ 
	 &    \textsf{frb59-26-5}&  1.5K &  1M &  1.2T &  0.89 &  1.3K &  -0.02 &  0.00 &  0.89 &  835.5K &  937.8K &  1.4K &  1302 &  1105 &  46 &  67 &  256  \\ 
\midrule 
\end{tabularx}
\end{table}

{
\renewcommand{\tabcolsep}{1pt}
\begin{table}[htb]
\caption{Colors used by the proposed methods for the DIMACs graph collection.
For comparison, we used \textsc{rand}, \textsc{deg}, \textsc{ido}, and \textsc{dist-two-ido}.
We also included a few of the stronger upper bounds along with our lower bound to get a better understanding and insight of the networks and the colorings possible from them.
For each network, we bold the best solution among all methods.
Note that we removed the less interesting networks (i.e., $\coloring_{\min} = \coloring_{\max}$, since those are effectively summarized in previous tables).
Also, in \textsf{MANN-a81} ``-'' denotes that the solution was no better than random, i.e., $1161$.
}
\vspace{1mm}
\label{table:results-colors-dense-dimacs}
\centering\small\tiny
\fontsize{6.0}{6.5}\selectfont
\noindent

\begin{tabularx}{\linewidth}{ Hl HH HHHH HHHH XXXX  
ccccc ccccc ccccc c}
\multicolumn{17}{c}{\textbf{\sectionfont Stats \& Bounds}} & 
\multicolumn{15}{c}{\textbf{\sectionfont Coloring Methods}} \\
 &
\textbf{graph}& 
$|V|$ & 
$|E|$ & 
$|T|$ & 
$\rho$ & 
$d_{\text{avg}}$ & 
${\rm r}$ & 
$\bar{\kappa}$ & 
$\kappa$ & 
$\tr_{\rm avg}$ & 
$\tr_{\rm max}$ & 
$\dmax$ & 
$K$+1 & 
$T$ & 
$\tilde{\omega}$ & 
\rotatebox{90}{\textsc{rand}}& \rotatebox{90}{\textsc{deg}}& \rotatebox{90}{\textsc{ido}}& \rotatebox{90}{\textsc{dist-two-ido}}& \rotatebox{90}{\textsc{triangles}}& \rotatebox{90}{\textsc{kcore-deg}}& \rotatebox{90}{\textsc{triangle-vol}}& \rotatebox{90}{\textsc{triangle-core-vol}}& \rotatebox{90}{\textsc{triangle-core-max}}& \rotatebox{90}{\textsc{deg-triangles}}& \rotatebox{90}{\textsc{kcore-triangles}}& \rotatebox{90}{\textsc{kcore-deg-tri}}& \rotatebox{90}{\textsc{deg-kcore-vol}}& \rotatebox{90}{\textsc{kcore-tri-vol}}& \rotatebox{90}{\textsc{deg-kcore-tri-vol}}\\ \bottomrule
 
	 & \TTZZ \BBZZ   \textsf{C1000-9}&  1K &  450K &  364M &  0.90 &  900 &  -0.00 &  0.00 &  0.90 &  364.6K &  385K &  925 &  875 &  764 &  51 &  319 &  315 &  315 &  315 &  317 &  317 &  319 &  319 &  319 &  319 &  319 &  317 &  \textbf{311}&  318 &  319 & \\ 
	 & \TTZZ \BBZZ   \textsf{C2000-9}&  2K &  1.7M &  2.9T &  0.90 &  1.7K &  -0.00 &  0.00 &  0.90 &  1.4M &  1.5M &  1.8K &  1759 &  1549 &  59 &  585 &  576 &  576 &  576 &  573 &  573 &  \textbf{570}&  \textbf{570}&  \textbf{570}&  \textbf{570}&  \textbf{570}&  582 &  583 &  574 &  578 & \\ 
	 & \TTZZ \BBZZ   \textsf{C4000-5}&  4K &  4M &  4T &  0.50 &  2K &  -0.00 &  0.00 &  0.50 &  1M &  1.1M &  2.1K &  1910 &  899 &  15 &  402 &  395 &  395 &  395 &  396 &  396 &  397 &  397 &  397 &  \textbf{391}&  395 &  398 &  398 &  395 &  393 & \\ 
	 & \TTZZ \BBZZ   \textsf{C500-9}&  500 &  112.3K &  45.3M &  0.90 &  449 &  -0.00 &  0.00 &  0.90 &  90.7K &  98.4K &  468 &  433 &  373 &  44 &  177 &  170 &  170 &  170 &  171 &  171 &  \textbf{168}&  \textbf{168}&  \textbf{168}&  172 &  172 &  170 &  170 &  172 &  174 & \\ 
\midrule
	 & \TTZZ \BBZZ   \textsf{DSJC1000-5}&  1K &  249.8K &  62.3M &  0.50 &  499 &  -0.00 &  0.00 &  0.50 &  62.3K &  75.9K &  551 &  460 &  207 &  13 &  127 &  122 &  122 &  122 &  122 &  122 &  123 &  123 &  123 &  \textbf{120}&  \textbf{120}&  122 &  125 &  125 &  124 & \\ 
\midrule
	 & \TTZZ \BBZZ   \textsf{MANN-a81}&  3.3K &  5.5M &  18.2T &  1.00 &  3.3K &  0.00 &  0.00 &  1.00 &  5.4M &  5.4M &  3.3K &  3281 &  3241 &  1096 &  1161 &  - &  - &  - &  - &  - &  - &  - &  - &  - &  - &  \textbf{1134}&  - &  \textbf{1134}&  \textbf{1134}& \\ 
\midrule
	 & \TTZZ \BBZZ   \textsf{brock200-1}&  200 &  14.8K &  1.6M &  0.75 &  148 &  -0.00 &  0.00 &  0.75 &  8.1K &  10.1K &  165 &  135 &  91 &  17 &  58 &  56 &  56 &  56 &  57 &  57 &  57 &  57 &  57 &  56 &  56 &  \textbf{54}&  \textbf{54}&  56 &  58 & \\ 
	 & \TTZZ \BBZZ   \textsf{brock200-3}&  200 &  12K &  873K &  0.61 &  120 &  -0.01 &  0.00 &  0.61 &  4.3K &  5.4K &  134 &  106 &  56 &  12 &  44 &  43 &  43 &  43 &  \textbf{41}&  \textbf{41}&  42 &  42 &  42 &  43 &  43 &  \textbf{41}&  \textbf{41}&  43 &  43 & \\ 
	 & \TTZZ \BBZZ   \textsf{brock400-2}&  400 &  59.7K &  13.3M &  0.75 &  298 &  -0.00 &  0.00 &  0.75 &  33.3K &  40.1K &  328 &  279 &  193 &  20 &  100 &  100 &  100 &  100 &  99 &  99 &  \textbf{97}&  \textbf{97}&  \textbf{97}&  99 &  99 &  \textbf{97}&  100 &  100 &  99 & \\ 
	 & \TTZZ \BBZZ   \textsf{brock400-3}&  400 &  59.6K &  13.2M &  0.75 &  298 &  -0.01 &  0.00 &  0.75 &  33.2K &  38.5K &  322 &  279 &  192 &  20 &  101 &  96 &  96 &  96 &  98 &  98 &  \textbf{95}&  \textbf{95}&  \textbf{95}&  97 &  97 &  97 &  96 &  96 &  98 & \\ 
	 & \TTZZ \BBZZ   \textsf{brock800-1}&  800 &  207.5K &  69.7M &  0.65 &  518 &  -0.00 &  0.00 &  0.65 &  87.2K &  101.4K &  560 &  488 &  292 &  17 &  144 &  142 &  142 &  142 &  142 &  142 &  \textbf{139}&  \textbf{139}&  \textbf{139}&  144 &  144 &  144 &  142 &  144 &  142 & \\ 
	 & \TTZZ \BBZZ   \textsf{brock800-3}&  800 &  207.3K &  69.6M &  0.65 &  518 &  -0.00 &  0.00 &  0.65 &  87K &  100.8K &  558 &  484 &  289 &  17 &  145 &  142 &  142 &  142 &  \textbf{138}&  \textbf{138}&  141 &  141 &  141 &  140 &  140 &  142 &  141 &  140 &  142 & \\ 
	 & \TTZZ \BBZZ   \textsf{brock800-4}&  800 &  207.6K &  69.9M &  0.65 &  519 &  -0.00 &  0.00 &  0.65 &  87.4K &  103.2K &  565 &  486 &  291 &  17 &  144 &  143 &  143 &  143 &  142 &  142 &  \textbf{141}&  \textbf{141}&  \textbf{141}&  142 &  142 &  144 &  144 &  144 &  \textbf{141}& \\ 
\midrule
	 & \TTZZ \BBZZ   \textsf{gen-9-44}&  200 &  17.9K &  2.8M &  0.90 &  179 &  -0.01 &  0.00 &  0.90 &  14.3K &  16.1K &  190 &  168 &  141 &  34 &  74 &  66 &  66 &  66 &  66 &  66 &  \textbf{65}&  \textbf{65}&  \textbf{65}&  \textbf{65}&  \textbf{65}&  \textbf{65}&  \textbf{65}&  \textbf{65}&  73 & \\ 
	 & \TTZZ \BBZZ   \textsf{gen-9-55}&  400 &  71.8K &  23.1M &  0.90 &  359 &  -0.01 &  0.00 &  0.90 &  57.9K &  63.1K &  375 &  337 &  287 &  41 &  136 &  123 &  123 &  123 &  126 &  126 &  \textbf{120}&  \textbf{120}&  \textbf{120}&  121 &  121 &  131 &  130 &  130 &  135 & \\ 
	 & \TTZZ \BBZZ   \textsf{gen-9-65}&  400 &  71.8K &  23.1M &  0.90 &  359 &  -0.02 &  0.00 &  0.90 &  57.9K &  64.1K &  378 &  337 &  286 &  41 &  139 &  131 &  131 &  131 &  \textbf{129}&  \textbf{129}&  138 &  138 &  138 &  136 &  136 &  140 &  138 &  138 &  144 & \\ 
	 & \TTZZ \BBZZ   \textsf{gen-9-75}&  400 &  71.8K &  23.1M &  0.90 &  359 &  -0.01 &  0.00 &  0.90 &  57.9K &  64.8K &  380 &  337 &  287 &  45 &  144 &  135 &  135 &  135 &  134 &  134 &  \textbf{126}&  \textbf{126}&  \textbf{126}&  129 &  129 &  138 &  136 &  136 &  144 & \\ 
\midrule
	 & \TTZZ \BBZZ   \textsf{joh32-2-4}&  496 &  107K &  40.7M &  0.88 &  435 &  0.00 &  0.00 &  0.87 &  82.2K &  82.2K &  435 &  436 &  380 &  16 &  34 &  31 &  31 &  31 &  \textbf{30}&  \textbf{30}&  \textbf{30}&  \textbf{30}&  \textbf{30}&  \textbf{30}&  \textbf{30}&  \textbf{30}&  \textbf{30}&  \textbf{30}&  \textbf{30}& \\ 
	 & \TTZZ \BBZZ   \textsf{joh8-2-4}&  28 &  210 &  1.2K &  0.56 &  15 &  1.00 &  0.00 &  0.43 &  45 &  45 &  15 &  16 &  8 &  4 &  7 &  7 &  7 &  7 &  \textbf{6}&  \textbf{6}&  \textbf{6}&  \textbf{6}&  \textbf{6}&  \textbf{6}&  \textbf{6}&  \textbf{6}&  \textbf{6}&  \textbf{6}&  \textbf{6}& \\ 
\midrule
	 & \TTZZ \BBZZ   \textsf{ph1000-2}&  1K &  244K &  73.9M &  0.49 &  489 &  -0.10 &  0.00 &  0.57 &  73.9K &  157.6K &  766 &  328 &  196 &  33 &  147 &  117 &  117 &  117 &  118 &  118 &  \textbf{116}&  \textbf{116}&  \textbf{116}&  \textbf{116}&  \textbf{116}&  134 &  132 &  142 &  140 & \\ 
	 & \TTZZ \BBZZ   \textsf{ph1000-3}&  1K &  371K &  211.3M &  0.74 &  743 &  -0.01 &  0.00 &  0.76 &  211.3K &  301.7K &  895 &  610 &  388 &  49 &  228 &  199 &  199 &  199 &  \textbf{194}&  \textbf{194}&  199 &  199 &  199 &  \textbf{194}&  \textbf{194}&  209 &  207 &  214 &  220 & \\ 
	 & \TTZZ \BBZZ   \textsf{ph1500-1}&  1.5K &  284.9K &  34.3M &  0.25 &  379 &  -0.09 &  0.00 &  0.29 &  22.8K &  53.2K &  614 &  253 &  75 &  10 &  96 &  79 &  79 &  79 &  \textbf{78}&  \textbf{78}&  79 &  79 &  79 &  79 &  79 &  95 &  92 &  98 &  98 & \\ 
	 & \TTZZ \BBZZ   \textsf{ph1500-3}&  1.5K &  847.2K &  741.1M &  0.75 &  1.1K &  -0.01 &  0.00 &  0.77 &  494.1K &  675.7K &  1.3K &  930 &  597 &  60 &  328 &  285 &  285 &  285 &  284 &  284 &  \textbf{281}&  \textbf{281}&  \textbf{281}&  283 &  283 &  295 &  290 &  298 &  304 & \\ 
	 & \TTZZ \BBZZ   \textsf{ph500-1}&  500 &  31.5K &  1.2M &  0.25 &  126 &  -0.09 &  0.00 &  0.29 &  2.5K &  5.7K &  204 &  87 &  25 &  9 &  42 &  35 &  35 &  35 &  \textbf{34}&  \textbf{34}&  \textbf{34}&  \textbf{34}&  \textbf{34}&  35 &  35 &  36 &  36 &  35 &  37 & \\ 
	 & \TTZZ \BBZZ   \textsf{ph700-1}&  700 &  60.9K &  3.3M &  0.25 &  174 &  -0.10 &  0.00 &  0.29 &  4.7K &  11.5K &  286 &  118 &  34 &  8 &  53 &  44 &  44 &  44 &  44 &  44 &  \textbf{42}&  \textbf{42}&  \textbf{42}&  43 &  43 &  43 &  43 &  45 &  49 & \\ 
\midrule
	 & \TTZZ \BBZZ   \textsf{s1000}&  1K &  250.5K &  86.3M &  0.50 &  501 &  0.67 &  0.00 &  0.69 &  86.3K &  100.7K &  550 &  465 &  399 &  10 &  45 &  27 &  27 &  27 &  \textbf{15}&  \textbf{15}&  \textbf{15}&  \textbf{15}&  \textbf{15}&  \textbf{15}&  \textbf{15}&  29 &  29 &  33 &  37 & \\ 
	 & \TTZZ \BBZZ   \textsf{s200-0-7-1}&  200 &  13.9K &  1.4M &  0.70 &  139 &  0.07 &  0.00 &  0.73 &  7K &  8.5K &  155 &  126 &  93 &  16 &  48 &  46 &  46 &  46 &  42 &  42 &  \textbf{35}&  \textbf{35}&  \textbf{35}&  37 &  37 &  36 &  36 &  37 &  36 & \\ 
	 & \TTZZ \BBZZ   \textsf{s200-0-9-1}&  200 &  17.9K &  2.8M &  0.90 &  179 &  0.04 &  0.00 &  0.90 &  14.4K &  16.3K &  191 &  163 &  134 &  49 &  88 &  77 &  77 &  77 &  77 &  77 &  \textbf{71}&  \textbf{71}&  \textbf{71}&  74 &  74 &  74 &  74 &  74 &  84 & \\ 
	 & \TTZZ \BBZZ   \textsf{s200-0-9-2}&  200 &  17.9K &  2.8M &  0.90 &  179 &  -0.00 &  0.00 &  0.90 &  14.3K &  15.8K &  188 &  170 &  143 &  34 &  84 &  78 &  78 &  78 &  \textbf{76}&  \textbf{76}&  79 &  79 &  79 &  82 &  82 &  82 &  82 &  82 &  82 & \\ 
	 & \TTZZ \BBZZ   \textsf{s200-0-9-3}&  200 &  17.9K &  2.8M &  0.90 &  179 &  -0.01 &  0.00 &  0.90 &  14.3K &  15.6K &  187 &  170 &  145 &  31 &  77 &  70 &  70 &  70 &  70 &  70 &  73 &  73 &  73 &  73 &  73 &  73 &  73 &  73 &  \textbf{69}& \\ 
	 & \TTZZ \BBZZ   \textsf{s400-0-5-1}&  400 &  39.9K &  5.2M &  0.50 &  199 &  0.60 &  0.00 &  0.66 &  13K &  15.8K &  225 &  184 &  154 &  8 &  28 &  20 &  20 &  20 &  \textbf{13}&  \textbf{13}&  \textbf{13}&  \textbf{13}&  \textbf{13}&  \textbf{13}&  \textbf{13}&  \textbf{13}&  \textbf{13}&  \textbf{13}&  29 & \\ 
	 & \TTZZ \BBZZ   \textsf{s400-0-7-1}&  400 &  55.8K &  11.3M &  0.70 &  279 &  -0.00 &  0.00 &  0.73 &  28.4K &  32.3K &  301 &  262 &  182 &  22 &  78 &  82 &  82 &  82 &  78 &  78 &  81 &  81 &  81 &  79 &  79 &  80 &  79 &  79 &  \textbf{77}& \\ 
	 & \TTZZ \BBZZ   \textsf{s400-0-7-2}&  400 &  55.8K &  11.2M &  0.70 &  279 &  0.03 &  0.00 &  0.73 &  28.2K &  32.8K &  304 &  260 &  179 &  18 &  70 &  69 &  69 &  69 &  64 &  64 &  \textbf{51}&  \textbf{51}&  \textbf{51}&  62 &  62 &  70 &  70 &  63 &  65 & \\ 
	 & \TTZZ \BBZZ   \textsf{s400-0-7-3}&  400 &  55.8K &  11.1M &  0.70 &  279 &  0.18 &  0.00 &  0.72 &  27.9K &  33.6K &  307 &  254 &  182 &  16 &  63 &  35 &  35 &  35 &  24 &  24 &  \textbf{22}&  \textbf{22}&  \textbf{22}&  \textbf{22}&  \textbf{22}&  35 &  35 &  \textbf{22}&  25 & \\ 
	 & \TTZZ \BBZZ   \textsf{s400-0-9-1}&  400 &  71.8K &  23.1M &  0.90 &  359 &  -0.00 &  0.00 &  0.90 &  57.9K &  62.7K &  374 &  345 &  294 &  57 &  157 &  160 &  160 &  160 &  158 &  158 &  \textbf{151}&  \textbf{151}&  \textbf{151}&  153 &  153 &  154 &  154 &  154 &  164 & \\ 
	 & \TTZZ \BBZZ   \textsf{sr400-0-7}&  400 &  55.8K &  10.9M &  0.70 &  279 &  -0.00 &  0.00 &  0.70 &  27.2K &  33.5K &  310 &  259 &  164 &  17 &  92 &  90 &  90 &  90 &  88 &  88 &  \textbf{86}&  \textbf{86}&  \textbf{86}&  88 &  88 &  94 &  94 &  88 &  89 & \\ 
\bottomrule
\end{tabularx}
\end{table}
\vspace{-8mm}
}

{
\renewcommand{\tabcolsep}{1pt}
\begin{table}[htb]
\caption{Colors used by the proposed methods for the BHOSLIB graph collection.
For comparison, we used \textsc{rand}, \textsc{deg}, \textsc{ido}, and \textsc{dist-two-ido}.
We also included a few of the stronger upper bounds along with our lower bound to get a better understanding and insight of the networks and the colorings possible from them.
For each network, we bold the best solution among all methods.
Note that we removed the less interesting networks (i.e., $\coloring_{\min} = \coloring_{\max}$, since those are effectively summarized in Table~\ref{table:stats-sparse-ca} and~\ref{table:stats-sparse-social}.
}
\vspace{1mm}
\label{table:results-colors-dense-bhoslib}
\centering\small\tiny
\fontsize{6.0}{7.0}\selectfont
\noindent

\begin{tabularx}{\linewidth}{ HX HH HHHH HHHH XXXX  
rrrrr rrrrr rrrrr r}
\multicolumn{17}{c}{\textbf{\sectionfont Stats \& Bounds}} & 
\multicolumn{15}{c}{\textbf{\sectionfont Coloring Methods}} \\
 &
\textbf{graph}& 
$|V|$ & 
$|E|$ & 
$|T|$ & 
$\rho$ & 
$d_{\text{avg}}$ & 
${\rm r}$ & 
$\bar{\kappa}$ & 
$\kappa$ & 
$\tr_{\rm avg}$ & 
$\tr_{\rm max}$ & 
$\dmax$ & 
$K$+1 & 
$T$ & 
$\tilde{\omega}$ & 
\rotatebox{90}{\textsc{rand}}& \rotatebox{90}{\textsc{deg}}& \rotatebox{90}{\textsc{ido}}& \rotatebox{90}{\textsc{dist-two-ido}}& \rotatebox{90}{\textsc{triangles}}& \rotatebox{90}{\textsc{kcore-deg}}& \rotatebox{90}{\textsc{triangle-vol}}& \rotatebox{90}{\textsc{triangle-core-vol}}& \rotatebox{90}{\textsc{triangle-core-max}}& \rotatebox{90}{\textsc{deg-triangles}}& \rotatebox{90}{\textsc{kcore-triangles}}& \rotatebox{90}{\textsc{kcore-deg-tri}}& \rotatebox{90}{\textsc{deg-kcore-vol}}& \rotatebox{90}{\textsc{kcore-triangle-vol}}& \rotatebox{90}{\textsc{deg-kcore-triangle-vol}}\\ 
\midrule
	 & \TTZZ \BBZZ   \textsf{f30-15-2}&  450 &  83.1K &  25.1M &  0.82 &  369 &  -0.02 &  0.00 &  0.82 &  55.9K &  66.7K &  404 &  338 &  257 &  24 &  93 &  47 &  47 &  47 &  47 &  47 &  \textbf{43}&  \textbf{43}&  \textbf{43}&  44 &  44 &  60 &  58 &  61 &  83 & \\ 
	 & \TTZZ \BBZZ   \textsf{f30-15-3}&  450 &  83.2K &  25.2M &  0.82 &  369 &  -0.03 &  0.00 &  0.82 &  56.1K &  65.2K &  400 &  337 &  254 &  25 &  92 &  50 &  50 &  50 &  49 &  49 &  52 &  52 &  52 &  \textbf{46}&  \textbf{46}&  49 &  54 &  62 &  95 & \\ 
	 & \TTZZ \BBZZ   \textsf{f35-17-1}&  595 &  148.8K &  62.5M &  0.84 &  500 &  -0.04 &  0.00 &  0.84 &  105.1K &  123.9K &  544 &  463 &  361 &  29 &  118 &  60 &  60 &  60 &  60 &  60 &  70 &  70 &  70 &  \textbf{59}&  \textbf{59}&  115 &  109 &  106 &  118 & \\ 
	 & \TTZZ \BBZZ   \textsf{f35-17-2}&  595 &  148.8K &  62.5M &  0.84 &  500 &  -0.03 &  0.00 &  0.84 &  105.1K &  122.2K &  541 &  465 &  362 &  28 &  107 &  65 &  65 &  65 &  \textbf{64}&  \textbf{64}&  69 &  69 &  69 &  70 &  70 &  97 &  101 &  108 &  108 & \\ 
	 & \TTZZ \BBZZ   \textsf{f35-17-3}&  595 &  148.7K &  62.5M &  0.84 &  500 &  -0.04 &  0.00 &  0.84 &  105.1K &  126.2K &  549 &  451 &  352 &  28 &  112 &  56 &  56 &  56 &  56 &  56 &  \textbf{49}&  \textbf{49}&  \textbf{49}&  58 &  58 &  94 &  92 &  102 &  103 & \\ 
	 & \TTZZ \BBZZ   \textsf{f35-17-4}&  595 &  148.8K &  62.6M &  0.84 &  500 &  -0.04 &  0.00 &  0.84 &  105.3K &  131.1K &  560 &  456 &  354 &  30 &  112 &  67 &  67 &  67 &  66 &  66 &  \textbf{47}&  \textbf{47}&  \textbf{47}&  \textbf{47}&  \textbf{47}&  91 &  90 &  95 &  117 & \\ 
	 & \TTZZ \BBZZ   \textsf{f35-17-5}&  595 &  148.5K &  62.2M &  0.84 &  499 &  -0.02 &  0.00 &  0.84 &  104.5K &  126.3K &  550 &  461 &  355 &  29 &  114 &  62 &  62 &  62 &  61 &  61 &  \textbf{58}&  \textbf{58}&  \textbf{58}&  62 &  62 &  98 &  100 &  110 &  112 & \\ 
	 & \TTZZ \BBZZ   \textsf{f40-19-1}&  760 &  247.1K &  137.5M &  0.86 &  650 &  -0.03 &  0.00 &  0.86 &  180.9K &  210.6K &  703 &  595 &  465 &  32 &  136 &  55 &  55 &  55 &  55 &  55 &  53 &  53 &  53 &  \textbf{50}&  \textbf{50}&  119 &  120 &  124 &  129 & \\ 
	 & \TTZZ \BBZZ   \textsf{f40-19-3}&  760 &  247.3K &  137.6M &  0.86 &  650 &  -0.02 &  0.00 &  0.86 &  181.1K &  210.2K &  702 &  613 &  491 &  31 &  141 &  71 &  71 &  71 &  71 &  71 &  76 &  76 &  76 &  \textbf{68}&  \textbf{68}&  136 &  131 &  137 &  128 & \\ 
	 & \TTZZ \BBZZ   \textsf{f40-19-4}&  760 &  246.8K &  136.8M &  0.86 &  649 &  -0.03 &  0.00 &  0.85 &  180K &  203.9K &  692 &  600 &  481 &  32 &  144 &  85 &  85 &  85 &  \textbf{84}&  \textbf{84}&  94 &  94 &  94 &  91 &  91 &  130 &  125 &  139 &  139 & \\ 
	 & \TTZZ \BBZZ   \textsf{f40-19-5}&  760 &  246.8K &  136.8M &  0.86 &  649 &  -0.03 &  0.00 &  0.85 &  180K &  203.6K &  691 &  596 &  476 &  32 &  135 &  71 &  71 &  71 &  \textbf{70}&  \textbf{70}&  73 &  73 &  73 &  83 &  83 &  139 &  134 &  128 &  143 & \\ 
	 & \TTZZ \BBZZ   \textsf{f45-21-1}&  945 &  386.8K &  274.2M &  0.87 &  818 &  -0.02 &  0.00 &  0.87 &  290.1K &  331.5K &  876 &  769 &  626 &  36 &  174 &  100 &  100 &  100 &  100 &  100 &  \textbf{97}&  \textbf{97}&  \textbf{97}&  103 &  103 &  167 &  162 &  166 &  162 & \\ 
	 & \TTZZ \BBZZ   \textsf{f45-21-3}&  945 &  387.7K &  276.2M &  0.87 &  820 &  -0.02 &  0.00 &  0.87 &  292.2K &  329.6K &  872 &  764 &  624 &  35 &  169 &  80 &  80 &  80 &  82 &  82 &  \textbf{79}&  \textbf{79}&  \textbf{79}&  82 &  82 &  148 &  141 &  154 &  149 & \\ 
	 & \TTZZ \BBZZ   \textsf{f45-21-4}&  945 &  387.4K &  275.7M &  0.87 &  820 &  -0.03 &  0.00 &  0.87 &  291.7K &  331.2K &  875 &  757 &  618 &  35 &  169 &  79 &  79 &  79 &  80 &  80 &  \textbf{70}&  \textbf{70}&  \textbf{70}&  77 &  77 &  140 &  145 &  149 &  157 & \\ 
	 & \TTZZ \BBZZ   \textsf{f50-23-1}&  1.1K &  580.6K &  514.4M &  0.88 &  1K &  -0.02 &  0.00 &  0.88 &  447.3K &  496.1K &  1K &  950 &  786 &  39 &  202 &  91 &  91 &  91 &  90 &  90 &  77 &  77 &  77 &  \textbf{73}&  \textbf{73}&  181 &  167 &  182 &  185 & \\ 
	 & \TTZZ \BBZZ   \textsf{f50-23-2}&  1.1K &  579.8K &  512.4M &  0.88 &  1K &  -0.02 &  0.00 &  0.88 &  445.6K &  504.6K &  1K &  936 &  771 &  39 &  194 &  98 &  98 &  98 &  98 &  98 &  \textbf{69}&  \textbf{69}&  \textbf{69}&  99 &  99 &  189 &  166 &  193 &  194 & \\ 
	 & \TTZZ \BBZZ   \textsf{f50-23-3}&  1.1K &  579.6K &  511.5M &  0.88 &  1K &  -0.02 &  0.00 &  0.88 &  444.8K &  508K &  1K &  953 &  792 &  40 &  201 &  93 &  93 &  93 &  119 &  119 &  \textbf{89}&  \textbf{89}&  \textbf{89}&  \textbf{89}&  \textbf{89}&  187 &  150 &  192 &  187 & \\ 
	 & \TTZZ \BBZZ   \textsf{f50-23-4}&  1.1K &  580.4K &  513.9M &  0.88 &  1K &  -0.03 &  0.00 &  0.88 &  446.9K &  502.6K &  1K &  950 &  786 &  40 &  195 &  107 &  107 &  107 &  \textbf{92}&  \textbf{92}&  94 &  94 &  94 &  110 &  110 &  178 &  170 &  195 &  183 & \\ 
	 & \TTZZ \BBZZ   \textsf{f50-23-5}&  1.1K &  580.6K &  514.6M &  0.88 &  1K &  -0.03 &  0.00 &  0.88 &  447.5K &  510.8K &  1K &  949 &  790 &  41 &  199 &  79 &  79 &  79 &  78 &  78 &  81 &  81 &  81 &  \textbf{69}&  \textbf{69}&  181 &  164 &  180 &  190 & \\ 
	 & \TTZZ \BBZZ   \textsf{f53-24-1}&  1.2K &  714.1K &  707.5M &  0.88 &  1.1K &  -0.01 &  0.00 &  0.88 &  556.2K &  619.3K &  1.1K &  1054 &  881 &  43 &  220 &  69 &  69 &  69 &  \textbf{68}&  \textbf{68}&  98 &  98 &  98 &  92 &  92 &  194 &  175 &  197 &  195 & \\ 
	 & \TTZZ \BBZZ   \textsf{f53-24-2}&  1.2K &  714K &  707.1M &  0.88 &  1.1K &  -0.02 &  0.00 &  0.88 &  555.9K &  615.3K &  1.1K &  1057 &  884 &  43 &  216 &  103 &  103 &  103 &  104 &  104 &  \textbf{95}&  \textbf{95}&  \textbf{95}&  105 &  105 &  187 &  163 &  175 &  209 & \\ 
	 & \TTZZ \BBZZ   \textsf{f53-24-3}&  1.2K &  714.2K &  707.7M &  0.88 &  1.1K &  -0.02 &  0.00 &  0.88 &  556.3K &  616.5K &  1.1K &  1050 &  878 &  42 &  211 &  97 &  97 &  97 &  97 &  97 &  104 &  104 &  104 &  \textbf{92}&  \textbf{92}&  185 &  168 &  185 &  208 & \\ 
	 & \TTZZ \BBZZ   \textsf{f53-24-4}&  1.2K &  714K &  707M &  0.88 &  1.1K &  -0.02 &  0.00 &  0.88 &  555.8K &  622.1K &  1.1K &  1063 &  890 &  43 &  215 &  104 &  104 &  104 &  102 &  102 &  \textbf{84}&  \textbf{84}&  \textbf{84}&  105 &  105 &  188 &  168 &  195 &  209 & \\ 
	 & \TTZZ \BBZZ   \textsf{f53-24-5}&  1.2K &  714.1K &  707.2M &  0.88 &  1.1K &  -0.02 &  0.00 &  0.88 &  555.9K &  633K &  1.1K &  1071 &  900 &  42 &  217 &  122 &  122 &  122 &  124 &  124 &  \textbf{98}&  \textbf{98}&  \textbf{98}&  101 &  101 &  186 &  181 &  195 &  200 & \\ 
	 & \TTZZ \BBZZ   \textsf{f59-26-1}&  1.5K &  1M &  1.2T &  0.89 &  1.3K &  -0.01 &  0.00 &  0.89 &  834.1K &  921K &  1.4K &  1282 &  1084 &  48 &  254 &  108 &  108 &  108 &  109 &  109 &  105 &  105 &  105 &  \textbf{102}&  \textbf{102}&  208 &  208 &  226 &  226 & \\ 
\bottomrule
\end{tabularx}
\end{table}
}

{
\renewcommand{\tabcolsep}{1.3pt}
\begin{table}[h!]
\caption{
Comparing the recolor variant to the basic coloring variant that is faster but less accurate for the DIMACs graph collection. 
In particular, we provide the basic and recolor results for each of the proposed coloring methods along with the previous methods for comparison.
}
\vspace{1mm}
\label{table:recolor-results-dense-dimacs}
\centering\small\tiny
\fontsize{6.0}{7.0}\selectfont
\noindent

\begin{tabularx}{\linewidth}{ Hl HH HHHH HHHH XXXX 
 rrrrr rrrrr rrrrr r HH}
 
\multicolumn{17}{c}{\textbf{\sectionfont Stats \& Bounds}} & 
\multicolumn{15}{c}{\textbf{\sectionfont Coloring Methods}} \\
 &
\textbf{graph} \quad \quad & 
$|V|$ & 
$|E|$ & 
$|T|$ & 
$\rho$ & 
$d_{\text{avg}}$ & 
${\rm r}$ & 
$\bar{\kappa}$ & 
$\kappa$ & 
$\tr_{\rm avg}$ & 
$\tr_{\rm max}$ & 
$\dmax$ & 
$K$+1 & 
$T$ & 
$\tilde{\omega}$ & 
\rotatebox{90}{\textsc{rand}}& \rotatebox{90}{\textsc{deg}}& \rotatebox{90}{\textsc{ido}}& \rotatebox{90}{\textsc{dist-two-ido}}& \rotatebox{90}{\textsc{triangles}}& \rotatebox{90}{\textsc{kcore-deg}}& \rotatebox{90}{\textsc{triangle-vol}}& \rotatebox{90}{\textsc{triangle-core-vol}}& \rotatebox{90}{\textsc{triangle-core-max}}& \rotatebox{90}{\textsc{deg-triangles}}& \rotatebox{90}{\textsc{kcore-triangles}}& \rotatebox{90}{\textsc{kcore-deg-tri}}& \rotatebox{90}{\textsc{deg-kcore-vol}}& \rotatebox{90}{\textsc{kcore-triangle-vol}}& \rotatebox{90}{\textsc{deg-kcore-triangle-vol}}\\ 
\bottomrule
 
	 & \TTZZ \BBZZ   \multirow{2}{*}{\textsf{C1000-9}}&\multirow{2}{*}{1K}&\multirow{2}{*}{450K}&\multirow{2}{*}{364.6M}&0.90 &\multirow{2}{*}{900}&\multirow{2}{*}{-0.00}&\multirow{2}{*}{0.00}&\multirow{2}{*}{0.90}&\multirow{2}{*}{364.6K}&\multirow{2}{*}{385K}&\multirow{2}{*}{925}&\multirow{2}{*}{875}&\multirow{2}{*}{764}&\multirow{2}{*}{51}&319 &315 &315 &315 &317 &317 &319 &319 &319 &319 &319 &317 &\textbf{311}&318 &319 & \\ 
	 & \TTZZ \BBZZ  && & & & & & & & & & & & & &294 &291 &291 &291 &289 &289 &291 &291 &291 &\textbf{288}&\textbf{288}&289 &291 &289 &\textbf{288}& \\ 
  	 \cline{2-31}	 & \TTZZ \BBZZ   \multirow{2}{*}{\textsf{C2000-9}}&\multirow{2}{*}{2K}&\multirow{2}{*}{1.7M}&\multirow{2}{*}{2.9T}&0.90 &\multirow{2}{*}{1.7K}&\multirow{2}{*}{-0.00}&\multirow{2}{*}{0.00}&\multirow{2}{*}{0.90}&\multirow{2}{*}{1.4M}&\multirow{2}{*}{1.5M}&\multirow{2}{*}{1.8K}&\multirow{2}{*}{1759}&\multirow{2}{*}{1549}&\multirow{2}{*}{59}&585 &576 &576 &576 &573 &573 &\textbf{570}&\textbf{570}&\textbf{570}&\textbf{570}&\textbf{570}&582 &583 &574 &578 & \\ 
	 & \TTZZ \BBZZ  && & & & & & & & & & & & & &534 &529 &529 &529 &531 &531 &533 &533 &533 &533 &533 &526 &527 &528 &\textbf{525}& \\ 
  	 \cline{2-31}	 & \TTZZ \BBZZ   \multirow{2}{*}{\textsf{C4000-5}}&\multirow{2}{*}{4K}&\multirow{2}{*}{4M}&\multirow{2}{*}{4T}&0.50 &\multirow{2}{*}{2K}&\multirow{2}{*}{-0.00}&\multirow{2}{*}{0.00}&\multirow{2}{*}{0.50}&\multirow{2}{*}{1M}&\multirow{2}{*}{1.1M}&\multirow{2}{*}{2.1K}&\multirow{2}{*}{1910}&\multirow{2}{*}{899}&\multirow{2}{*}{15}&402 &395 &395 &395 &396 &396 &397 &397 &397 &\textbf{391}&395 &398 &398 &395 &393 & \\ 
	 & \TTZZ \BBZZ  && & & & & & & & & & & & & &379 &374 &374 &374 &375 &375 &\textbf{372}&\textbf{372}&\textbf{372}&375 &\textbf{372}&375 &375 &\textbf{372}&373 & \\ 
  	 \cline{2-31}	 & \TTZZ \BBZZ   \multirow{2}{*}{\textsf{C500-9}}&\multirow{2}{*}{500}&\multirow{2}{*}{112.3K}&\multirow{2}{*}{45.3M}&0.90 &\multirow{2}{*}{449}&\multirow{2}{*}{-0.00}&\multirow{2}{*}{0.00}&\multirow{2}{*}{0.90}&\multirow{2}{*}{90.7K}&\multirow{2}{*}{98.4K}&\multirow{2}{*}{468}&\multirow{2}{*}{433}&\multirow{2}{*}{373}&\multirow{2}{*}{44}&177 &170 &170 &170 &171 &171 &\textbf{168}&\textbf{168}&\textbf{168}&172 &172 &170 &170 &172 &174 & \\ 
	 & \TTZZ \BBZZ  && & & & & & & & & & & & & &161 &160 &160 &160 &158 &159 &158 &158 &158 &159 &158 &\textbf{156}&157 &160 &\textbf{156}& \\ 
  	 \cline{2-31}	 & \TTZZ \BBZZ   \multirow{2}{*}{\textsf{brock200-3}}&\multirow{2}{*}{200}&\multirow{2}{*}{12K}&\multirow{2}{*}{873.3K}&0.61 &\multirow{2}{*}{120}&\multirow{2}{*}{-0.01}&\multirow{2}{*}{0.00}&\multirow{2}{*}{0.61}&\multirow{2}{*}{4.3K}&\multirow{2}{*}{5.4K}&\multirow{2}{*}{134}&\multirow{2}{*}{106}&\multirow{2}{*}{56}&\multirow{2}{*}{12}&44 &43 &43 &43 &\textbf{41}&\textbf{41}&42 &42 &42 &43 &43 &\textbf{41}&\textbf{41}&43 &43 & \\ 
	 & \TTZZ \BBZZ  && & & & & & & & & & & & & &40 &39 &39 &39 &39 &39 &38 &38 &38 &\textbf{37}&\textbf{37}&39 &39 &39 &40 & \\ 
  	 \cline{2-31}	 & \TTZZ \BBZZ   \multirow{2}{*}{\textsf{brock400-2}}&\multirow{2}{*}{400}&\multirow{2}{*}{59.7K}&\multirow{2}{*}{13.3M}&0.75 &\multirow{2}{*}{298}&\multirow{2}{*}{-0.00}&\multirow{2}{*}{0.00}&\multirow{2}{*}{0.75}&\multirow{2}{*}{33.3K}&\multirow{2}{*}{40.1K}&\multirow{2}{*}{328}&\multirow{2}{*}{279}&\multirow{2}{*}{193}&\multirow{2}{*}{20}&100 &100 &100 &100 &99 &99 &\textbf{97}&\textbf{97}&\textbf{97}&99 &99 &\textbf{97}&100 &100 &99 & \\ 
	 & \TTZZ \BBZZ  && & & & & & & & & & & & & &94 &92 &92 &92 &92 &92 &\textbf{90}&\textbf{90}&\textbf{90}&\textbf{90}&91 &\textbf{90}&\textbf{90}&\textbf{90}&\textbf{90}& \\ 
  	 \cline{2-31}	 & \TTZZ \BBZZ   \multirow{2}{*}{\textsf{brock400-3}}&\multirow{2}{*}{400}&\multirow{2}{*}{59.6K}&\multirow{2}{*}{13.2M}&0.75 &\multirow{2}{*}{298}&\multirow{2}{*}{-0.01}&\multirow{2}{*}{0.00}&\multirow{2}{*}{0.75}&\multirow{2}{*}{33.2K}&\multirow{2}{*}{38.5K}&\multirow{2}{*}{322}&\multirow{2}{*}{279}&\multirow{2}{*}{192}&\multirow{2}{*}{20}&101 &96 &96 &96 &98 &98 &\textbf{95}&\textbf{95}&\textbf{95}&97 &97 &97 &96 &96 &98 & \\ 
	 & \TTZZ \BBZZ  && & & & & & & & & & & & & &91 &91 &91 &91 &90 &90 &\textbf{88}&\textbf{88}&\textbf{88}&90 &\textbf{88}&\textbf{88}&90 &90 &90 & \\ 
  	 \cline{2-31}	 & \TTZZ \BBZZ   \multirow{2}{*}{\textsf{brock800-1}}&\multirow{2}{*}{800}&\multirow{2}{*}{207.5K}&\multirow{2}{*}{69.7M}&0.65 &\multirow{2}{*}{518}&\multirow{2}{*}{-0.00}&\multirow{2}{*}{0.00}&\multirow{2}{*}{0.65}&\multirow{2}{*}{87.2K}&\multirow{2}{*}{101.4K}&\multirow{2}{*}{560}&\multirow{2}{*}{488}&\multirow{2}{*}{292}&\multirow{2}{*}{17}&144 &142 &142 &142 &142 &142 &\textbf{139}&\textbf{139}&\textbf{139}&144 &144 &144 &142 &144 &142 & \\ 
	 & \TTZZ \BBZZ  && & & & & & & & & & & & & &133 &132 &132 &132 &131 &131 &132 &132 &132 &\textbf{130}&\textbf{130}&\textbf{130}&\textbf{130}&131 &131 & \\ 
  	 \cline{2-31}	 & \TTZZ \BBZZ   \multirow{2}{*}{\textsf{brock800-3}}&\multirow{2}{*}{800}&\multirow{2}{*}{207.3K}&\multirow{2}{*}{69.6M}&0.65 &\multirow{2}{*}{518}&\multirow{2}{*}{-0.00}&\multirow{2}{*}{0.00}&\multirow{2}{*}{0.65}&\multirow{2}{*}{87K}&\multirow{2}{*}{100.8K}&\multirow{2}{*}{558}&\multirow{2}{*}{484}&\multirow{2}{*}{289}&\multirow{2}{*}{17}&145 &142 &142 &142 &\textbf{138}&\textbf{138}&141 &141 &141 &140 &140 &142 &141 &140 &142 & \\ 
	 & \TTZZ \BBZZ  && & & & & & & & & & & & & &132 &132 &132 &132 &132 &132 &\textbf{130}&\textbf{130}&\textbf{130}&132 &133 &132 &\textbf{130}&131 &132 & \\ 
  	 \cline{2-31}	 & \TTZZ \BBZZ   \multirow{2}{*}{\textsf{brock800-4}}&\multirow{2}{*}{800}&\multirow{2}{*}{207.6K}&\multirow{2}{*}{69.9M}&0.65 &\multirow{2}{*}{519}&\multirow{2}{*}{-0.00}&\multirow{2}{*}{0.00}&\multirow{2}{*}{0.65}&\multirow{2}{*}{87.4K}&\multirow{2}{*}{103.2K}&\multirow{2}{*}{565}&\multirow{2}{*}{486}&\multirow{2}{*}{291}&\multirow{2}{*}{17}&144 &143 &143 &143 &142 &142 &\textbf{141}&\textbf{141}&\textbf{141}&142 &142 &144 &144 &144 &\textbf{141}& \\ 
	 & \TTZZ \BBZZ  && & & & & & & & & & & & & &135 &131 &131 &131 &\textbf{130}&\textbf{130}&132 &132 &132 &\textbf{130}&132 &132 &134 &133 &132 & \\ 
  	 \cline{2-31}
	 & \TTZZ \BBZZ   \multirow{2}{*}{\textsf{gen-9-44}}&\multirow{2}{*}{200}&\multirow{2}{*}{17.9K}&\multirow{2}{*}{2.8M}&0.90 &\multirow{2}{*}{179}&\multirow{2}{*}{-0.01}&\multirow{2}{*}{0.00}&\multirow{2}{*}{0.90}&\multirow{2}{*}{14.3K}&\multirow{2}{*}{16.1K}&\multirow{2}{*}{190}&\multirow{2}{*}{168}&\multirow{2}{*}{141}&\multirow{2}{*}{34}&74 &66 &66 &66 &66 &66 &\textbf{65}&\textbf{65}&\textbf{65}&\textbf{65}&\textbf{65}&\textbf{65}&\textbf{65}&\textbf{65}&73 & \\ 
	 & \TTZZ \BBZZ  && & & & & & & & & & & & & &65 &55 &55 &55 &55 &55 &\textbf{53}&\textbf{53}&\textbf{53}&59 &59 &58 &58 &59 &63 & \\ 
  	 \cline{2-31}	 & \TTZZ \BBZZ   \multirow{2}{*}{\textsf{gen-9-55}}&\multirow{2}{*}{400}&\multirow{2}{*}{71.8K}&\multirow{2}{*}{23.1M}&0.90 &\multirow{2}{*}{359}&\multirow{2}{*}{-0.01}&\multirow{2}{*}{0.00}&\multirow{2}{*}{0.90}&\multirow{2}{*}{57.9K}&\multirow{2}{*}{63.1K}&\multirow{2}{*}{375}&\multirow{2}{*}{337}&\multirow{2}{*}{287}&\multirow{2}{*}{41}&136 &123 &123 &123 &126 &126 &\textbf{120}&\textbf{120}&\textbf{120}&121 &121 &131 &130 &130 &135 & \\ 
	 & \TTZZ \BBZZ  && & & & & & & & & & & & & &101 &101 &101 &101 &99 &103 &104 &104 &104 &\textbf{92}&\textbf{92}&107 &106 &106 &107 & \\ 
  	 \cline{2-31}	 & \TTZZ \BBZZ   \multirow{2}{*}{\textsf{gen-9-65}}&\multirow{2}{*}{400}&\multirow{2}{*}{71.8K}&\multirow{2}{*}{23.1M}&0.90 &\multirow{2}{*}{359}&\multirow{2}{*}{-0.02}&\multirow{2}{*}{0.00}&\multirow{2}{*}{0.90}&\multirow{2}{*}{57.9K}&\multirow{2}{*}{64.1K}&\multirow{2}{*}{378}&\multirow{2}{*}{337}&\multirow{2}{*}{286}&\multirow{2}{*}{41}&139 &131 &131 &131 &\textbf{129}&\textbf{129}&138 &138 &138 &136 &136 &140 &138 &138 &144 & \\ 
	 & \TTZZ \BBZZ  && & & & & & & & & & & & & &119 &105 &105 &105 &104 &104 &110 &110 &110 &103 &\textbf{101}&112 &122 &122 &112 & \\ 
  	 \cline{2-31}	 & \TTZZ \BBZZ   \multirow{2}{*}{\textsf{gen-9-75}}&\multirow{2}{*}{400}&\multirow{2}{*}{71.8K}&\multirow{2}{*}{23.1M}&0.90 &\multirow{2}{*}{359}&\multirow{2}{*}{-0.01}&\multirow{2}{*}{0.00}&\multirow{2}{*}{0.90}&\multirow{2}{*}{57.9K}&\multirow{2}{*}{64.8K}&\multirow{2}{*}{380}&\multirow{2}{*}{337}&\multirow{2}{*}{287}&\multirow{2}{*}{45}&144 &135 &135 &135 &134 &134 &\textbf{126}&\textbf{126}&\textbf{126}&129 &129 &138 &136 &136 &144 & \\ 
	 & \TTZZ \BBZZ  && & & & & & & & & & & & & &124 &118 &118 &118 &114 &115 &114 &114 &114 &\textbf{113}&\textbf{113}&118 &120 &120 &126 & \\ 
  	 \cline{2-31}
	 & \TTZZ \BBZZ   \multirow{2}{*}{\textsf{ph1000-2}}&\multirow{2}{*}{1K}&\multirow{2}{*}{244.7K}&\multirow{2}{*}{73.9M}&0.49 &\multirow{2}{*}{489}&\multirow{2}{*}{-0.10}&\multirow{2}{*}{0.00}&\multirow{2}{*}{0.57}&\multirow{2}{*}{73.9K}&\multirow{2}{*}{157.6K}&\multirow{2}{*}{766}&\multirow{2}{*}{328}&\multirow{2}{*}{196}&\multirow{2}{*}{33}&147 &117 &117 &117 &118 &118 &\textbf{116}&\textbf{116}&\textbf{116}&\textbf{116}&\textbf{116}&134 &132 &142 &140 & \\ 
	 & \TTZZ \BBZZ  && & & & & & & & & & & & & &136 &113 &113 &113 &\textbf{112}&113 &\textbf{112}&\textbf{112}&\textbf{112}&\textbf{112}&\textbf{112}&124 &126 &128 &135 & \\ 
  	 \cline{2-31}	 & \TTZZ \BBZZ   \multirow{2}{*}{\textsf{ph1000-3}}&\multirow{2}{*}{1K}&\multirow{2}{*}{371.7K}&\multirow{2}{*}{211.3M}&0.74 &\multirow{2}{*}{743}&\multirow{2}{*}{-0.01}&\multirow{2}{*}{0.00}&\multirow{2}{*}{0.76}&\multirow{2}{*}{211.3K}&\multirow{2}{*}{301.7K}&\multirow{2}{*}{895}&\multirow{2}{*}{610}&\multirow{2}{*}{388}&\multirow{2}{*}{49}&228 &199 &199 &199 &\textbf{194}&\textbf{194}&199 &199 &199 &\textbf{194}&\textbf{194}&209 &207 &214 &220 & \\ 
	 & \TTZZ \BBZZ  && & & & & & & & & & & & & &208 &188 &188 &188 &188 &189 &188 &188 &188 &\textbf{185}&186 &195 &193 &196 &202 & \\ 
  	 \cline{2-31}	 & \TTZZ \BBZZ   \multirow{2}{*}{\textsf{ph1500-1}}&\multirow{2}{*}{1.5K}&\multirow{2}{*}{284.9K}&\multirow{2}{*}{34.3M}&0.25 &\multirow{2}{*}{379}&\multirow{2}{*}{-0.09}&\multirow{2}{*}{0.00}&\multirow{2}{*}{0.29}&\multirow{2}{*}{22.8K}&\multirow{2}{*}{53.2K}&\multirow{2}{*}{614}&\multirow{2}{*}{253}&\multirow{2}{*}{75}&\multirow{2}{*}{10}&96 &79 &79 &79 &\textbf{78}&\textbf{78}&79 &79 &79 &79 &79 &95 &92 &98 &98 & \\ 
	 & \TTZZ \BBZZ  && & & & & & & & & & & & & &90 &76 &76 &76 &76 &76 &76 &76 &76 &\textbf{75}&76 &89 &88 &91 &92 & \\ 

  	 \cline{2-31}
	 & \TTZZ \BBZZ   \multirow{2}{*}{\textsf{s1000}}&\multirow{2}{*}{1K}&\multirow{2}{*}{250.5K}&\multirow{2}{*}{86.3M}&0.50 &\multirow{2}{*}{501}&\multirow{2}{*}{0.67}&\multirow{2}{*}{0.00}&\multirow{2}{*}{0.69}&\multirow{2}{*}{86.3K}&\multirow{2}{*}{100.7K}&\multirow{2}{*}{550}&\multirow{2}{*}{465}&\multirow{2}{*}{399}&\multirow{2}{*}{10}&45 &27 &27 &27 &\textbf{15}&\textbf{15}&\textbf{15}&\textbf{15}&\textbf{15}&\textbf{15}&\textbf{15}&29 &29 &33 &37 & \\ 
	 & \TTZZ \BBZZ  && & & & & & & & & & & & & &38 &24 &24 &24 &\textbf{15}&\textbf{15}&\textbf{15}&\textbf{15}&\textbf{15}&\textbf{15}&\textbf{15}&26 &29 &28 &32 & \\ 
  	 \cline{2-31}	 & \TTZZ \BBZZ   \multirow{2}{*}{\textsf{s200-0-7-1}}&\multirow{2}{*}{200}&\multirow{2}{*}{13.9K}&\multirow{2}{*}{1.4M}&0.70 &\multirow{2}{*}{139}&\multirow{2}{*}{0.07}&\multirow{2}{*}{0.00}&\multirow{2}{*}{0.73}&\multirow{2}{*}{7K}&\multirow{2}{*}{8.5K}&\multirow{2}{*}{155}&\multirow{2}{*}{126}&\multirow{2}{*}{93}&\multirow{2}{*}{16}&48 &46 &46 &46 &42 &42 &\textbf{35}&\textbf{35}&\textbf{35}&37 &37 &36 &36 &37 &36 & \\ 
	 & \TTZZ \BBZZ  && & & & & & & & & & & & & &43 &38 &38 &38 &40 &40 &\textbf{35}&\textbf{35}&\textbf{35}&36 &36 &39 &39 &36 &37 & \\ 

  	 \cline{2-31}	 & \TTZZ \BBZZ   \multirow{2}{*}{\textsf{s200-0-9-2}}&\multirow{2}{*}{200}&\multirow{2}{*}{17.9K}&\multirow{2}{*}{2.8M}&0.90 &\multirow{2}{*}{179}&\multirow{2}{*}{-0.00}&\multirow{2}{*}{0.00}&\multirow{2}{*}{0.90}&\multirow{2}{*}{14.3K}&\multirow{2}{*}{15.8K}&\multirow{2}{*}{188}&\multirow{2}{*}{170}&\multirow{2}{*}{143}&\multirow{2}{*}{34}&84 &78 &78 &78 &\textbf{76}&\textbf{76}&79 &79 &79 &82 &82 &82 &82 &82 &82 & \\ 
	 & \TTZZ \BBZZ  && & & & & & & & & & & & & &74 &74 &74 &74 &\textbf{71}&\textbf{71}&74 &74 &74 &73 &\textbf{71}&\textbf{71}&\textbf{71}&73 &72 & \\ 
	 
  	 \cline{2-31}	 & \TTZZ \BBZZ   \multirow{2}{*}{\textsf{s400-0-5-1}}&\multirow{2}{*}{400}&\multirow{2}{*}{39.9K}&\multirow{2}{*}{5.2M}&0.50 &\multirow{2}{*}{199}&\multirow{2}{*}{0.60}&\multirow{2}{*}{0.00}&\multirow{2}{*}{0.66}&\multirow{2}{*}{13K}&\multirow{2}{*}{15.8K}&\multirow{2}{*}{225}&\multirow{2}{*}{184}&\multirow{2}{*}{154}&\multirow{2}{*}{8}&28 &20 &20 &20 &\textbf{13}&\textbf{13}&\textbf{13}&\textbf{13}&\textbf{13}&\textbf{13}&\textbf{13}&\textbf{13}&\textbf{13}&\textbf{13}&29 & \\ 
	 & \TTZZ \BBZZ  && & & & & & & & & & & & & &23 &19 &19 &19 &\textbf{14}&\textbf{14}&\textbf{14}&\textbf{14}&\textbf{14}&\textbf{14}&\textbf{14}&\textbf{14}&\textbf{14}&\textbf{14}&23 & \\ 
  	 \cline{2-31}	 & \TTZZ \BBZZ   \multirow{2}{*}{\textsf{s400-0-7-1}}&\multirow{2}{*}{400}&\multirow{2}{*}{55.8K}&\multirow{2}{*}{11.3M}&0.70 &\multirow{2}{*}{279}&\multirow{2}{*}{-0.00}&\multirow{2}{*}{0.00}&\multirow{2}{*}{0.73}&\multirow{2}{*}{28.4K}&\multirow{2}{*}{32.3K}&\multirow{2}{*}{301}&\multirow{2}{*}{262}&\multirow{2}{*}{182}&\multirow{2}{*}{22}&78 &82 &82 &82 &78 &78 &81 &81 &81 &79 &79 &80 &79 &79 &\textbf{77}& \\ 
	 & \TTZZ \BBZZ  && & & & & & & & & & & & & &65 &69 &69 &69 &\textbf{64}&\textbf{64}&\textbf{64}&\textbf{64}&\textbf{64}&69 &68 &67 &67 &66 &68 & \\ 
  	 \cline{2-31}	 & \TTZZ \BBZZ   \multirow{2}{*}{\textsf{s400-0-7-2}}&\multirow{2}{*}{400}&\multirow{2}{*}{55.8K}&\multirow{2}{*}{11.2M}&0.70 &\multirow{2}{*}{279}&\multirow{2}{*}{0.03}&\multirow{2}{*}{0.00}&\multirow{2}{*}{0.73}&\multirow{2}{*}{28.2K}&\multirow{2}{*}{32.8K}&\multirow{2}{*}{304}&\multirow{2}{*}{260}&\multirow{2}{*}{179}&\multirow{2}{*}{18}&70 &69 &69 &69 &64 &64 &\textbf{51}&\textbf{51}&\textbf{51}&62 &62 &70 &70 &63 &65 & \\ 
	 & \TTZZ \BBZZ  && & & & & & & & & & & & & &58 &52 &52 &52 &51 &51 &51 &51 &51 &50 &50 &55 &55 &53 &\textbf{49}& \\ 
  	 \cline{2-31}	 & \TTZZ \BBZZ   \multirow{2}{*}{\textsf{s400-0-7-3}}&\multirow{2}{*}{400}&\multirow{2}{*}{55.8K}&\multirow{2}{*}{11.1M}&0.70 &\multirow{2}{*}{279}&\multirow{2}{*}{0.18}&\multirow{2}{*}{0.00}&\multirow{2}{*}{0.72}&\multirow{2}{*}{27.9K}&\multirow{2}{*}{33.6K}&\multirow{2}{*}{307}&\multirow{2}{*}{254}&\multirow{2}{*}{182}&\multirow{2}{*}{16}&63 &35 &35 &35 &24 &24 &\textbf{22}&\textbf{22}&\textbf{22}&\textbf{22}&\textbf{22}&35 &35 &\textbf{22}&25 & \\ 
	 & \TTZZ \BBZZ  && & & & & & & & & & & & & &46 &30 &30 &30 &25 &25 &\textbf{22}&\textbf{22}&\textbf{22}&\textbf{22}&\textbf{22}&37 &37 &\textbf{22}&24 & \\ 
  	 \cline{2-31}	 & \TTZZ \BBZZ   \multirow{2}{*}{\textsf{s400-0-9-1}}&\multirow{2}{*}{400}&\multirow{2}{*}{71.8K}&\multirow{2}{*}{23.1M}&0.90 &\multirow{2}{*}{359}&\multirow{2}{*}{-0.00}&\multirow{2}{*}{0.00}&\multirow{2}{*}{0.90}&\multirow{2}{*}{57.9K}&\multirow{2}{*}{62.7K}&\multirow{2}{*}{374}&\multirow{2}{*}{345}&\multirow{2}{*}{294}&\multirow{2}{*}{57}&157 &160 &160 &160 &158 &158 &\textbf{151}&\textbf{151}&\textbf{151}&153 &153 &154 &154 &154 &164 & \\ 
	 & \TTZZ \BBZZ  && & & & & & & & & & & & & &139 &141 &141 &141 &141 &141 &\textbf{136}&\textbf{136}&\textbf{136}&\textbf{136}&\textbf{136}&138 &138 &138 &137 & \\ 
  	 \cline{2-31}	 & \TTZZ \BBZZ   \multirow{2}{*}{\textsf{sr400-0-7}}&\multirow{2}{*}{400}&\multirow{2}{*}{55.8K}&\multirow{2}{*}{10.9M}&0.70 &\multirow{2}{*}{279}&\multirow{2}{*}{-0.00}&\multirow{2}{*}{0.00}&\multirow{2}{*}{0.70}&\multirow{2}{*}{27.2K}&\multirow{2}{*}{33.5K}&\multirow{2}{*}{310}&\multirow{2}{*}{259}&\multirow{2}{*}{164}&\multirow{2}{*}{17}&92 &90 &90 &90 &88 &88 &\textbf{86}&\textbf{86}&\textbf{86}&88 &88 &94 &94 &88 &89 & \\ 
	 & \TTZZ \BBZZ  && & & & & & & & & & & & & &84 &82 &82 &82 &81 &81 &81 &81 &81 &81 &\textbf{80}&86 &84 &83 &81 & \\ 
  	 \cline{2-31}
\end{tabularx}
\end{table}
}

{
\renewcommand{\tabcolsep}{1.4pt}
\begin{table}[h!]
\caption{
Comparing the recolor variant to the basic coloring variant that is faster but less accurate for the DIMACs graph collection. 
In particular, we provide the basic and recolor results for each of the proposed coloring methods along with the previous methods for comparison.
}
\vspace{1mm}
\label{table:recolor-results-dense-bhoslib}
\centering\small\tiny
\fontsize{6.0}{7.0}\selectfont
\noindent

\begin{tabularx}{\linewidth}{ Hl HH HHHH HHHH XXXX 
 rrrrr rrrrr rrrrr r HH}
 
\multicolumn{17}{c}{\textbf{\sectionfont Stats \& Bounds}} & 
\multicolumn{15}{c}{\textbf{\sectionfont Coloring Methods}} \\
 &
\textbf{graph} \quad \quad & 
$|V|$ & 
$|E|$ & 
$|T|$ & 
$\rho$ & 
$d_{\text{avg}}$ & 
${\rm r}$ & 
$\bar{\kappa}$ & 
$\kappa$ & 
$\tr_{\rm avg}$ & 
$\tr_{\rm max}$ & 
$\dmax$ & 
$K$+1 & 
$T$ & 
$\tilde{\omega}$ & 
\rotatebox{90}{\textsc{rand}}& \rotatebox{90}{\textsc{deg}}& \rotatebox{90}{\textsc{ido}}& \rotatebox{90}{\textsc{dist-two-ido}}& \rotatebox{90}{\textsc{triangles}}& \rotatebox{90}{\textsc{kcore-deg}}& \rotatebox{90}{\textsc{triangle-vol}}& \rotatebox{90}{\textsc{triangle-core-vol}}& \rotatebox{90}{\textsc{triangle-core-max}}& \rotatebox{90}{\textsc{deg-triangles}}& \rotatebox{90}{\textsc{kcore-triangles}}& \rotatebox{90}{\textsc{kcore-deg-tri}}& \rotatebox{90}{\textsc{deg-kcore-vol}}& \rotatebox{90}{\textsc{kcore-triangle-vol}}& \rotatebox{90}{\textsc{deg-kcore-triangle-vol}}\\ \bottomrule
  	 
	 & \TTZZ \BBZZ   \multirow{2}{*}{\textsf{f30-15-2}}&\multirow{2}{*}{450}&\multirow{2}{*}{83.1K}&\multirow{2}{*}{25.1M}&0.82 &\multirow{2}{*}{369}&\multirow{2}{*}{-0.02}&\multirow{2}{*}{0.00}&\multirow{2}{*}{0.82}&\multirow{2}{*}{55.9K}&\multirow{2}{*}{66.7K}&\multirow{2}{*}{404}&\multirow{2}{*}{338}&\multirow{2}{*}{257}&\multirow{2}{*}{24}&93 &47 &47 &47 &47 &47 &\textbf{43}&\textbf{43}&\textbf{43}&44 &44 &60 &58 &61 &83 & \\ 
	 & \TTZZ \BBZZ  && & & & & & & & & & & & & &64 &36 &36 &36 &36 &36 &\textbf{35}&\textbf{35}&\textbf{35}&42 &42 &49 &47 &49 &60 & \\ 
  	 \cline{2-31}	 & \TTZZ \BBZZ   \multirow{2}{*}{\textsf{f30-15-3}}&\multirow{2}{*}{450}&\multirow{2}{*}{83.2K}&\multirow{2}{*}{25.2M}&0.82 &\multirow{2}{*}{369}&\multirow{2}{*}{-0.03}&\multirow{2}{*}{0.00}&\multirow{2}{*}{0.82}&\multirow{2}{*}{56.1K}&\multirow{2}{*}{65.2K}&\multirow{2}{*}{400}&\multirow{2}{*}{337}&\multirow{2}{*}{254}&\multirow{2}{*}{25}&92 &50 &50 &50 &49 &49 &52 &52 &52 &\textbf{46}&\textbf{46}&49 &54 &62 &95 & \\ 
	 & \TTZZ \BBZZ  && & & & & & & & & & & & & &66 &39 &39 &39 &\textbf{38}&\textbf{38}&46 &46 &46 &\textbf{38}&\textbf{38}&40 &43 &50 &61 & \\ 
  	 \cline{2-31}	 & \TTZZ \BBZZ   \multirow{2}{*}{\textsf{f35-17-1}}&\multirow{2}{*}{595}&\multirow{2}{*}{148.8K}&\multirow{2}{*}{62.5M}&0.84 &\multirow{2}{*}{500}&\multirow{2}{*}{-0.04}&\multirow{2}{*}{0.00}&\multirow{2}{*}{0.84}&\multirow{2}{*}{105.1K}&\multirow{2}{*}{123.9K}&\multirow{2}{*}{544}&\multirow{2}{*}{463}&\multirow{2}{*}{361}&\multirow{2}{*}{29}&118 &60 &60 &60 &60 &60 &70 &70 &70 &\textbf{59}&\textbf{59}&115 &109 &106 &118 & \\ 
	 & \TTZZ \BBZZ  && & & & & & & & & & & & & &84 &54 &54 &54 &53 &53 &48 &48 &48 &\textbf{43}&\textbf{43}&78 &81 &81 &85 & \\ 
  	 \cline{2-31}	 & \TTZZ \BBZZ   \multirow{2}{*}{\textsf{f35-17-2}}&\multirow{2}{*}{595}&\multirow{2}{*}{148.8K}&\multirow{2}{*}{62.5M}&0.84 &\multirow{2}{*}{500}&\multirow{2}{*}{-0.03}&\multirow{2}{*}{0.00}&\multirow{2}{*}{0.84}&\multirow{2}{*}{105.1K}&\multirow{2}{*}{122.2K}&\multirow{2}{*}{541}&\multirow{2}{*}{465}&\multirow{2}{*}{362}&\multirow{2}{*}{28}&107 &65 &65 &65 &\textbf{64}&\textbf{64}&69 &69 &69 &70 &70 &97 &101 &108 &108 & \\ 
	 & \TTZZ \BBZZ  && & & & & & & & & & & & & &80 &51 &51 &51 &48 &48 &\textbf{46}&\textbf{46}&\textbf{46}&\textbf{46}&\textbf{46}&72 &78 &84 &76 & \\ 
  	 \cline{2-31}	 & \TTZZ \BBZZ   \multirow{2}{*}{\textsf{f35-17-3}}&\multirow{2}{*}{595}&\multirow{2}{*}{148.7K}&\multirow{2}{*}{62.5M}&0.84 &\multirow{2}{*}{500}&\multirow{2}{*}{-0.04}&\multirow{2}{*}{0.00}&\multirow{2}{*}{0.84}&\multirow{2}{*}{105.1K}&\multirow{2}{*}{126.2K}&\multirow{2}{*}{549}&\multirow{2}{*}{451}&\multirow{2}{*}{352}&\multirow{2}{*}{28}&112 &56 &56 &56 &56 &56 &\textbf{49}&\textbf{49}&\textbf{49}&58 &58 &94 &92 &102 &103 & \\ 
	 & \TTZZ \BBZZ  && & & & & & & & & & & & & &79 &50 &50 &50 &50 &50 &\textbf{45}&\textbf{45}&\textbf{45}&46 &46 &63 &65 &67 &70 & \\ 
  	 \cline{2-31}	 & \TTZZ \BBZZ   \multirow{2}{*}{\textsf{f35-17-4}}&\multirow{2}{*}{595}&\multirow{2}{*}{148.8K}&\multirow{2}{*}{62.6M}&0.84 &\multirow{2}{*}{500}&\multirow{2}{*}{-0.04}&\multirow{2}{*}{0.00}&\multirow{2}{*}{0.84}&\multirow{2}{*}{105.3K}&\multirow{2}{*}{131.1K}&\multirow{2}{*}{560}&\multirow{2}{*}{456}&\multirow{2}{*}{354}&\multirow{2}{*}{30}&112 &67 &67 &67 &66 &66 &\textbf{47}&\textbf{47}&\textbf{47}&\textbf{47}&\textbf{47}&91 &90 &95 &117 & \\ 
	 & \TTZZ \BBZZ  && & & & & & & & & & & & & &84 &52 &52 &52 &51 &51 &\textbf{41}&\textbf{41}&\textbf{41}&47 &47 &68 &71 &66 &75 & \\ 
  	 \cline{2-31}	 & \TTZZ \BBZZ   \multirow{2}{*}{\textsf{f35-17-5}}&\multirow{2}{*}{595}&\multirow{2}{*}{148.5K}&\multirow{2}{*}{62.2M}&0.84 &\multirow{2}{*}{499}&\multirow{2}{*}{-0.02}&\multirow{2}{*}{0.00}&\multirow{2}{*}{0.84}&\multirow{2}{*}{104.5K}&\multirow{2}{*}{126.3K}&\multirow{2}{*}{550}&\multirow{2}{*}{461}&\multirow{2}{*}{355}&\multirow{2}{*}{29}&114 &62 &62 &62 &61 &61 &\textbf{58}&\textbf{58}&\textbf{58}&62 &62 &98 &100 &110 &112 & \\ 
	 & \TTZZ \BBZZ  && & & & & & & & & & & & & &80 &52 &52 &52 &52 &52 &\textbf{44}&\textbf{44}&\textbf{44}&\textbf{44}&\textbf{44}&61 &65 &79 &77 & \\ 
  	 \cline{2-31}	 & \TTZZ \BBZZ   \multirow{2}{*}{\textsf{f40-19-1}}&\multirow{2}{*}{760}&\multirow{2}{*}{247.1K}&\multirow{2}{*}{137.5M}&0.86 &\multirow{2}{*}{650}&\multirow{2}{*}{-0.03}&\multirow{2}{*}{0.00}&\multirow{2}{*}{0.86}&\multirow{2}{*}{180.9K}&\multirow{2}{*}{210.6K}&\multirow{2}{*}{703}&\multirow{2}{*}{595}&\multirow{2}{*}{465}&\multirow{2}{*}{32}&136 &55 &55 &55 &55 &55 &53 &53 &53 &\textbf{50}&\textbf{50}&119 &120 &124 &129 & \\ 
	 & \TTZZ \BBZZ  && & & & & & & & & & & & & &96 &48 &48 &48 &47 &47 &46 &46 &46 &\textbf{44}&\textbf{44}&83 &92 &90 &88 & \\ 
  	 \cline{2-31}	 & \TTZZ \BBZZ   \multirow{2}{*}{\textsf{f40-19-4}}&\multirow{2}{*}{760}&\multirow{2}{*}{246.8K}&\multirow{2}{*}{136.8M}&0.86 &\multirow{2}{*}{649}&\multirow{2}{*}{-0.03}&\multirow{2}{*}{0.00}&\multirow{2}{*}{0.85}&\multirow{2}{*}{180K}&\multirow{2}{*}{203.9K}&\multirow{2}{*}{692}&\multirow{2}{*}{600}&\multirow{2}{*}{481}&\multirow{2}{*}{32}&144 &85 &85 &85 &\textbf{84}&\textbf{84}&94 &94 &94 &91 &91 &130 &125 &139 &139 & \\ 
	 & \TTZZ \BBZZ  && & & & & & & & & & & & & &92 &63 &63 &63 &60 &60 &\textbf{53}&\textbf{53}&\textbf{53}&61 &61 &83 &85 &87 &91 & \\ 
  	 \cline{2-31}	 & \TTZZ \BBZZ   \multirow{2}{*}{\textsf{f40-19-5}}&\multirow{2}{*}{760}&\multirow{2}{*}{246.8K}&\multirow{2}{*}{136.8M}&0.86 &\multirow{2}{*}{649}&\multirow{2}{*}{-0.03}&\multirow{2}{*}{0.00}&\multirow{2}{*}{0.85}&\multirow{2}{*}{180K}&\multirow{2}{*}{203.6K}&\multirow{2}{*}{691}&\multirow{2}{*}{596}&\multirow{2}{*}{476}&\multirow{2}{*}{32}&135 &71 &71 &71 &\textbf{70}&\textbf{70}&73 &73 &73 &83 &83 &139 &134 &128 &143 & \\ 
	 & \TTZZ \BBZZ  && & & & & & & & & & & & & &105 &50 &50 &50 &\textbf{49}&\textbf{49}&51 &51 &51 &51 &51 &81 &94 &85 &93 & \\ 
  	 \cline{2-31}	 & \TTZZ \BBZZ   \multirow{2}{*}{\textsf{f45-21-3}}&\multirow{2}{*}{945}&\multirow{2}{*}{387.7K}&\multirow{2}{*}{276.2M}&0.87 &\multirow{2}{*}{820}&\multirow{2}{*}{-0.02}&\multirow{2}{*}{0.00}&\multirow{2}{*}{0.87}&\multirow{2}{*}{292.2K}&\multirow{2}{*}{329.6K}&\multirow{2}{*}{872}&\multirow{2}{*}{764}&\multirow{2}{*}{624}&\multirow{2}{*}{35}&169 &80 &80 &80 &82 &82 &\textbf{79}&\textbf{79}&\textbf{79}&82 &82 &148 &141 &154 &149 & \\ 
	 & \TTZZ \BBZZ  && & & & & & & & & & & & & &98 &63 &63 &63 &67 &67 &59 &59 &59 &56 &\textbf{55}&90 &90 &91 &104 & \\ 
  	 \cline{2-31}	 & \TTZZ \BBZZ   \multirow{2}{*}{\textsf{f45-21-4}}&\multirow{2}{*}{945}&\multirow{2}{*}{387.4K}&\multirow{2}{*}{275.7M}&0.87 &\multirow{2}{*}{820}&\multirow{2}{*}{-0.03}&\multirow{2}{*}{0.00}&\multirow{2}{*}{0.87}&\multirow{2}{*}{291.7K}&\multirow{2}{*}{331.2K}&\multirow{2}{*}{875}&\multirow{2}{*}{757}&\multirow{2}{*}{618}&\multirow{2}{*}{35}&169 &79 &79 &79 &80 &80 &\textbf{70}&\textbf{70}&\textbf{70}&77 &77 &140 &145 &149 &157 & \\ 
	 & \TTZZ \BBZZ  && & & & & & & & & & & & & &112 &61 &61 &61 &61 &61 &\textbf{55}&\textbf{55}&\textbf{55}&57 &57 &98 &105 &109 &98 & \\ 
  	 \cline{2-31}	 & \TTZZ \BBZZ   \multirow{2}{*}{\textsf{f50-23-1}}&\multirow{2}{*}{1.1K}&\multirow{2}{*}{580.6K}&\multirow{2}{*}{514.4M}&0.88 &\multirow{2}{*}{1K}&\multirow{2}{*}{-0.02}&\multirow{2}{*}{0.00}&\multirow{2}{*}{0.88}&\multirow{2}{*}{447.3K}&\multirow{2}{*}{496.1K}&\multirow{2}{*}{1K}&\multirow{2}{*}{950}&\multirow{2}{*}{786}&\multirow{2}{*}{39}&202 &91 &91 &91 &90 &90 &77 &77 &77 &\textbf{73}&\textbf{73}&181 &167 &182 &185 & \\ 
	 & \TTZZ \BBZZ  && & & & & & & & & & & & & &125 &63 &63 &63 &63 &63 &60 &60 &60 &\textbf{59}&\textbf{59}&106 &110 &117 &112 & \\ 
  	 \cline{2-31}	 & \TTZZ \BBZZ   \multirow{2}{*}{\textsf{f50-23-4}}&\multirow{2}{*}{1.1K}&\multirow{2}{*}{580.4K}&\multirow{2}{*}{513.9M}&0.88 &\multirow{2}{*}{1K}&\multirow{2}{*}{-0.03}&\multirow{2}{*}{0.00}&\multirow{2}{*}{0.88}&\multirow{2}{*}{446.9K}&\multirow{2}{*}{502.6K}&\multirow{2}{*}{1K}&\multirow{2}{*}{950}&\multirow{2}{*}{786}&\multirow{2}{*}{40}&195 &107 &107 &107 &\textbf{92}&\textbf{92}&94 &94 &94 &110 &110 &178 &170 &195 &183 & \\ 
	 & \TTZZ \BBZZ  && & & & & & & & & & & & & &126 &79 &79 &79 &77 &78 &75 &75 &75 &\textbf{74}&\textbf{74}&107 &105 &112 &119 & \\ 
  	 \cline{2-31}	 & \TTZZ \BBZZ   \multirow{2}{*}{\textsf{f50-23-5}}&\multirow{2}{*}{1.1K}&\multirow{2}{*}{580.6K}&\multirow{2}{*}{514.6M}&0.88 &\multirow{2}{*}{1K}&\multirow{2}{*}{-0.03}&\multirow{2}{*}{0.00}&\multirow{2}{*}{0.88}&\multirow{2}{*}{447.5K}&\multirow{2}{*}{510.8K}&\multirow{2}{*}{1K}&\multirow{2}{*}{949}&\multirow{2}{*}{790}&\multirow{2}{*}{41}&199 &79 &79 &79 &78 &78 &81 &81 &81 &\textbf{69}&\textbf{69}&181 &164 &180 &190 & \\ 
	 & \TTZZ \BBZZ  && & & & & & & & & & & & & &128 &63 &63 &63 &\textbf{60}&\textbf{60}&67 &67 &67 &61 &61 &109 &105 &104 &114 & \\ 
  	 \cline{2-31}	 & \TTZZ \BBZZ   \multirow{2}{*}{\textsf{f53-24-1}}&\multirow{2}{*}{1.2K}&\multirow{2}{*}{714.1K}&\multirow{2}{*}{707.5M}&0.88 &\multirow{2}{*}{1.1K}&\multirow{2}{*}{-0.01}&\multirow{2}{*}{0.00}&\multirow{2}{*}{0.88}&\multirow{2}{*}{556.2K}&\multirow{2}{*}{619.3K}&\multirow{2}{*}{1.1K}&\multirow{2}{*}{1054}&\multirow{2}{*}{881}&\multirow{2}{*}{43}&220 &69 &69 &69 &\textbf{68}&\textbf{68}&98 &98 &98 &92 &92 &194 &175 &197 &195 & \\ 
	 & \TTZZ \BBZZ  && & & & & & & & & & & & & &130 &62 &62 &62 &61 &61 &62 &62 &62 &\textbf{59}&\textbf{59}&106 &110 &125 &128 & \\ 

  	 \cline{2-31}	 & \TTZZ \BBZZ   \multirow{2}{*}{\textsf{f53-24-3}}&\multirow{2}{*}{1.2K}&\multirow{2}{*}{714.2K}&\multirow{2}{*}{707.7M}&0.88 &\multirow{2}{*}{1.1K}&\multirow{2}{*}{-0.02}&\multirow{2}{*}{0.00}&\multirow{2}{*}{0.88}&\multirow{2}{*}{556.3K}&\multirow{2}{*}{616.5K}&\multirow{2}{*}{1.1K}&\multirow{2}{*}{1050}&\multirow{2}{*}{878}&\multirow{2}{*}{42}&211 &97 &97 &97 &97 &97 &104 &104 &104 &\textbf{92}&\textbf{92}&185 &168 &185 &208 & \\ 
	 & \TTZZ \BBZZ  && & & & & & & & & & & & & &144 &72 &72 &72 &72 &72 &73 &73 &73 &\textbf{70}&\textbf{70}&116 &113 &113 &116 & \\ 
  	 \cline{2-31}	 & \TTZZ \BBZZ   \multirow{2}{*}{\textsf{f53-24-4}}&\multirow{2}{*}{1.2K}&\multirow{2}{*}{714K}&\multirow{2}{*}{707M}&0.88 &\multirow{2}{*}{1.1K}&\multirow{2}{*}{-0.02}&\multirow{2}{*}{0.00}&\multirow{2}{*}{0.88}&\multirow{2}{*}{555.8K}&\multirow{2}{*}{622.1K}&\multirow{2}{*}{1.1K}&\multirow{2}{*}{1063}&\multirow{2}{*}{890}&\multirow{2}{*}{43}&215 &104 &104 &104 &102 &102 &\textbf{84}&\textbf{84}&\textbf{84}&105 &105 &188 &168 &195 &209 & \\ 
	 & \TTZZ \BBZZ  && & & & & & & & & & & & & &125 &67 &67 &67 &\textbf{66}&\textbf{66}&73 &73 &73 &68 &68 &112 &109 &105 &113 & \\ 
  	 \cline{2-31}	 & \TTZZ \BBZZ   \multirow{2}{*}{\textsf{f53-24-5}}&\multirow{2}{*}{1.2K}&\multirow{2}{*}{714.1K}&\multirow{2}{*}{707.2M}&0.88 &\multirow{2}{*}{1.1K}&\multirow{2}{*}{-0.02}&\multirow{2}{*}{0.00}&\multirow{2}{*}{0.88}&\multirow{2}{*}{555.9K}&\multirow{2}{*}{633K}&\multirow{2}{*}{1.1K}&\multirow{2}{*}{1071}&\multirow{2}{*}{900}&\multirow{2}{*}{42}&217 &122 &122 &122 &124 &124 &\textbf{98}&\textbf{98}&\textbf{98}&101 &101 &186 &181 &195 &200 & \\ 
	 & \TTZZ \BBZZ  && & & & & & & & & & & & & &129 &89 &89 &89 &89 &89 &78 &78 &78 &\textbf{68}&\textbf{68}&114 &122 &122 &108 & \\ 
  	 \cline{2-31}	 & \TTZZ \BBZZ   \multirow{2}{*}{\textsf{f59-26-1}}&\multirow{2}{*}{1.5K}&\multirow{2}{*}{1M}&\multirow{2}{*}{1.2T}&0.89 &\multirow{2}{*}{1.3K}&\multirow{2}{*}{-0.01}&\multirow{2}{*}{0.00}&\multirow{2}{*}{0.89}&\multirow{2}{*}{834.1K}&\multirow{2}{*}{921K}&\multirow{2}{*}{1.4K}&\multirow{2}{*}{1282}&\multirow{2}{*}{1084}&\multirow{2}{*}{48}&254 &108 &108 &108 &109 &109 &105 &105 &105 &\textbf{102}&\textbf{102}&208 &208 &226 &226 & \\ 
	 & \TTZZ \BBZZ  && & & & & & & & & & & & & &143 &92 &92 &92 &100 &100 &\textbf{74}&\textbf{74}&\textbf{74}&\textbf{74}&\textbf{74}&130 &130 &137 &133 & \\ 
  	 \cline{2-31}
\end{tabularx}
\end{table}
}

\end{document}